\documentclass[a4paper,fleqn,usenatbib]{mnras}

\usepackage{newtxtext,newtxmath}

\usepackage[T1]{fontenc}
\usepackage{ae,aecompl}
\usepackage[official]{eurosym}
\usepackage{euscript}

\usepackage{graphicx}	
\usepackage{amsmath}	
\usepackage{amssymb}	

\usepackage{tikz}
\usetikzlibrary{arrows,shapes,automata,positioning}
\tikzstyle{every picture}+=[remember picture]

\usepackage{theorem}
\usepackage{empheq}
\usepackage{algorithm}
\usepackage[noend]{algpseudocode}
\makeatletter
\def\BState{\State\hskip-\ALG@thistlm}
\makeatother

\newcommand{\dom}{\ensuremath{\text{\rm dom}}\,}
\newcommand{\prox}{\ensuremath{\text{\rm prox}}}
\newcommand{\argmin}[1]{\ensuremath{\underset{#1}{\text{\rm argmin \,}}}}

\newcommand{\real}{\ensuremath{\text{\rm Re}}}
\newcommand{\imag}{\ensuremath{\text{\rm Im}}}

\newcommand{\menge}[2]{\big\{{#1}~\big |~{#2}\big\}} 
\newcommand{\Menge}[2]{\left\{{#1}~\Big|~{#2}\right\}}

\renewcommand{\leq}{\ensuremath{\leqslant}}
\renewcommand{\geq}{\ensuremath{\geqslant}}
\renewcommand{\le}{\ensuremath{\leqslant}}
\renewcommand{\ge}{\ensuremath{\geqslant}}
\newcommand{\minimize}[2]{\ensuremath{\underset{\substack{{#1}}}%
{\text{\rm minimize}}\;\;#2 }}

\newcommand{\bb}{\ensuremath{\boldsymbol{b}}}

\newcommand{\db}{\ensuremath{\boldsymbol{d}}}

\newcommand{\lb}{\ensuremath{\boldsymbol{l}}}

\newcommand{\ub}{\ensuremath{\boldsymbol{u}}}
\newcommand{\vb}{\ensuremath{\boldsymbol{v}}}
\newcommand{\wb}{\ensuremath{\boldsymbol{w}}}
\newcommand{\xb}{\ensuremath{\boldsymbol{x}}}
\newcommand{\yb}{\ensuremath{\boldsymbol{y}}}
\newcommand{\zb}{\ensuremath{\boldsymbol{z}}}
\newcommand{\ds}{\ensuremath{{\mathsf{d}}}}
\newcommand{\es}{\ensuremath{{\mathsf{e}}}}
\newcommand{\is}{\ensuremath{{\mathsf{i}}}}

\newcommand{\Bbs}{\ensuremath{\boldsymbol{\mathsf{B}}}}
\newcommand{\Cbs}{\ensuremath{\boldsymbol{\mathsf{C}}}}
\newcommand{\Dbs}{\ensuremath{\boldsymbol{\mathsf{D}}}}
\newcommand{\Fbs}{\ensuremath{\boldsymbol{\mathsf{F}}}}
\newcommand{\Gbs}{\ensuremath{\boldsymbol{\mathsf{G}}}}
\newcommand{\Hbs}{\ensuremath{\boldsymbol{\mathsf{H}}}}
\newcommand{\Ibs}{\ensuremath{\boldsymbol{\mathsf{I}}}}

\newcommand{\Ubs}{\ensuremath{\boldsymbol{\mathsf{U}}}}

\newcommand{\Wbs}{\ensuremath{\boldsymbol{\mathsf{W}}}}
\newcommand{\Xbs}{\ensuremath{\boldsymbol{\mathsf{X}}}}
\newcommand{\Ybs}{\ensuremath{\boldsymbol{\mathsf{Y}}}}

\newcommand{\Es}{\ensuremath{{\mathsf{E}}}}

\newcommand{\Xs}{\ensuremath{{\mathsf{X}}}}
\newcommand{\Ys}{\ensuremath{{\mathsf{Y}}}}

\newcommand{\unb}{\ensuremath{\boldsymbol{1}}}

\newcommand{\epsilonb}{\ensuremath{\boldsymbol{\epsilon}}}

\newcommand{\Psib}{\ensuremath{\boldsymbol{\Psi}}}

\newcommand{\Dc}{\ensuremath{\mathcal{D}}}

\newcommand{\Gc}{\ensuremath{\mathcal{G}}}

\newcommand{\Lc}{\ensuremath{\mathcal{L}}}

\newcommand{\Sc}{\ensuremath{\mathcal{S}}}
\newcommand{\Xc}{\ensuremath{\mathcal{X}}}
\newcommand{\eC}{\mathbb{C}}
\newcommand{\eD}{\mathbb{D}}
\newcommand{\eE}{\mathbb{E}}
\newcommand{\eI}{\mathbb{I}}
\newcommand{\eN}{\mathbb{N}}
\newcommand{\eR}{\mathbb{R}}
\newcommand{\eS}{\mathbb{S}}

\theoremstyle{plain}{\theorembodyfont{\rmfamily}%
}

\title[Non-convex optimization for DDE self-calibration]{
Non-convex optimization for self-calibration of direction-dependent effects in radio interferometric imaging
}

\author[A. Repetti et al.]{
Audrey Repetti\thanks{E-mail: a.repetti@hw.ac.uk},
Jasleen Birdi,
Arwa Dabbech,
and Yves Wiaux
\\
Heriot Watt University, Institute of Sensors, Signals and Systems, Edinburgh EH14 4AS, UK\
}

\date{Accepted XXX. Received YYY; in original form ZZZ}

\pubyear{2017}

\begin{document}
\label{firstpage}
\pagerange{\pageref{firstpage}--\pageref{lastpage}}
\maketitle

\begin{abstract}
Radio interferometric imaging aims to estimate an unknown sky intensity image from degraded observations, acquired through an antenna array. 
In the theoretical case of a perfectly calibrated array, it has been shown that solving the corresponding imaging problem by iterative algorithms based on convex optimization and compressive sensing theory can be competitive with classical algorithms such as CLEAN. 
However, in practice, antenna-based gains are unknown and have to be calibrated. Future radio telescopes, such as the SKA, aim at improving imaging resolution and sensitivity by orders of magnitude. At this precision level, the direction-dependency of the gains must be accounted for, and radio interferometric imaging can be understood as a blind deconvolution problem. 
In this context, the underlying minimization problem is non-convex, and adapted techniques have to be designed. 
In this work, leveraging recent developments in non-convex optimization, we propose the first joint calibration and imaging method in radio interferometry, with proven convergence guarantees. Our approach, based on a block-coordinate forward-backward algorithm, jointly accounts for visibilities and suitable priors on both the image and the direction-dependent effects (DDEs). As demonstrated in recent works, sparsity remains the prior of choice for the image, while DDEs are modelled as smooth functions of the sky, i.e. spatially band-limited.
Finally, we show through simulations the efficiency of our method, for the reconstruction of both images of point sources and complex extended sources. 
MATLAB code is available on GitHub.

\end{abstract}




\section{Introduction}

Radio interferometry (RI) aims to observe an area of the sky at high angular resolution through an array of antennas. More precisely, each measurement is acquired by a pair of antennas, in the Fourier domain of the intensity image of interest. These measurements consist of complex noisy visibilities, and are related to an undersampled selection of the Fourier coefficients of the intensity image degraded by antenna gains \citep{Thompson_2001}. 

When the antenna gains are perfectly calibrated, the imaging problem relies on estimating the observed sky intensity from the acquired visibilities. 
To this aim, several efficient methods have been developed since the early 1970's, starting with the celebrated CLEAN algorithm \citep{Hogbom_1974, Schwarz_1978, Thompson_2001}. 
Essentially, this algorithm is a non-linear deconvolution method based on local iterative beam removal.
At each iteration, it involves computing the residual dirty image and removing the dirty beam at the point with maximum absolute value in the residual image. The beam removal step is dependent on a loop gain factor, controlling the fraction of the value of the brightest point to be removed. This whole process implicitly assumes sparsity of the sought image. 
In recent years, several variants of CLEAN have been proposed (e.g. Multi-Scale CLEAN \citep{Cornwell_2008}, Adaptive Scale Pixel- CLEAN \citep{Bhatnagar_2004}). 
Furthermore, it is interesting to note that it has been shown recently that the CLEAN methods can be described within the framework of optimization algorithms \citep{Cornwell_2008, Wiaux_2009}. In particular, it has similarities with the matching pursuit algorithm \citep{Mallat_1993}, and it can be seen as a gradient descent algorithm regularized with sparsity constraint on the image \citep{Rau_2009,Carrillo_2014,Onose_2016}. The latter comes from the fact that the computation of the residual image is analogous to a gradient step, while the local beam removal by the loop gain factor resembles the thresholding operation to impose sparsity.

Recently, new imaging approaches have been developed to tackle the RI imaging problem leveraging optimization methods \citep{Rockafellar70, Urruty_Conv_an, Boyd2004_book}, and compressive sensing theory \citep{Candes_2006a, Candes_2006b, Candes_2006, Donoho2006, Baraniuk_2007}. 
The additive noise on the visibilities being assumed to be independent and identically distributed (i.i.d.) Gaussian, these approaches rely on solving a regularized least squares (LS) minimization problem. 
The regularization term is then chosen to incorporate prior information on the target intensity image. In particular, 
in \citep{Wiaux_2009}, the authors 
exploit the fact that many images are sparse in some dictionary (e.g. image domain, possibly redundant wavelet basis \citep{Mallat_book}, gradient domain \citep{Rudin_1992_total_variation, Gilboa_2008}, etc.) 
by using an $\ell_1$ regularization term. 
Note that this approach has subsequently been investigated in several works \citep{Wiaux_2009b, Wenger_2010, Li_2011, McEwen_2011, Carrillo_2012, Dabbech_2015, Onose_2017}. 
Furthermore, in the field of RI imaging, the resulting minimization problem involves large dimensional variables, especially for the big data problems encountered in the era of modern radio telescopes such as the Square Kilometer Array (SKA)\footnote{http://www.skatelescope.org/} and LOw Frequency ARray (LOFAR)\footnote{http://www.lofar.org/}. 
In this context, it has been shown that proximal methods \citep{Combettes_Book_10, PB2013, Komodakis_N_2014_playing_dor} are particularly well-suited, since they are designed to solve large scale problems. 
Notably, a first proximal method for RI imaging has been designed by \cite{Carrillo_2014} and developed in the \texttt{PURIFY} package \citep{pratley_2016}, using an simultaneous direction method of multipliers (SDMM) based algorithm \citep{Setzer_2010}. This method has been further improved by \cite{Onose_2016}, using the alternating direction method of multipliers (ADMM) algorithm \citep{BPCPE11} and a primal-dual forward-backward-based algorithm \citep{Condat_2011_primal_dual, V2013, Pesquet2014}. 
It is worth noting that the image reconstruction quality obtained by these methods was shown to have the potential to supersede CLEAN for recovering extended emission \citep{Onose_2016}.

In practice, however, antenna-based gains are unknown. These gains, related to each antenna of the interferometer, are classified as direction-independent and dependent effects (DIEs and DDEs, respectively). On the one hand, DDEs are time-variable complex gains, different for each antenna and varying across the field of view. They correspond to a multiplication in the image domain, and hence a convolution in the Fourier domain. On the other hand, DIEs stand for a particular case of the DDEs, when the spatial dependency can be ignored at each instant of the integration (i.e. only a scalar complex gain factor is considered for each antenna).
Traditionally, the calibration process, i.e. estimation of DIEs/DDEs, focusses on calibration of DIEs only. However, the new generation radio telescopes are being designed to produce orders of magnitude of improvement in both resolution and dynamic range. In order to match the imaging capabilities of these instruments, estimating only the DIEs is no longer sufficient and the DDEs must be incorporated in the calibration process.
In general, the calibration problem, either to calibrate for DIEs or independently for DDEs, can be formulated as a non-linear LS minimization problem \citep{Mitchell_2008, Salvini_2014, Smirnov_2015}. In particular, traditional DIEs self-calibration methods solve this problem 
using gradient-based techniques such as Levenberg-Marquardt (LM) algorithm \citep{Levenberg1944, Marquardt1963}. However, \cite{Mitchell_2008, Salvini_2014} pointed out its high computational cost and proposed a fast solver, namely StEFCal, based on an alternating direction implicit (ADI) method. 
Nevertheless, these methods are only designed to solve the DIE calibration problem, and with an upsurge in the number of unknown parameters, the calibration of DDEs is a highly challenging task. 
Keeping this in mind, few methods have been developed recently for DDE calibration. 
On the one hand, 
\cite{Intema_2009} have proposed a DDE calibration method that attempts to iteratively solve and correct for ionospheric phase errors, namely the \emph{Source Peeling and Atmospheric Modeling} (SPAM) method. Basically, SPAM solves for complex gains towards a number of bright sources, then approximates the DDEs by a sum of low-order Zernike polynomials, fitting the phase component of the solutions.
On the other hand, traditional methods for calibration have been improved by the SAGE algorithm in \citep{Yatawatta2009, Kazemi2011}, while another general framework has been developed by 
\cite{Smirnov_2015} solving for non-linear LS, considering the complex Jacobian formalism. Furthermore, a different approach based on non-linear Kalman filter is adopted by \cite{Tasse_2014}.
Finally, a new calibration scheme based on facet calibration, has been developed in the last years (see e.g. \cite{Tasse_2014b, Smirnov_2015, Weeren_2016}). 
To apply this method, it is assumed that the DDEs are piece-wise constant in the image domain. Therefore, the sky is partitioned into facets, with the facet center determined by the brightest source or the approximate center of a source group. At each step, the calibration solutions are obtained for the facet center, which is then applied over the whole facet. 
It is worth mentioning that, to the best of our knowledge, this method has only been applied to point sources models of the sky, which at most include few extended sources. 
Also note that, when this DDE calibration method is combined with an imaging method, the obtained global reconstruction algorithm does not benefit from any global convergence guarantee.

In this paper, we develop a method estimating jointly the unknown intensity image and the DDEs, 
reconstructing not only point sources but also sophisticated extended sources, 
and considering a smooth DDE screen applied to all the sources across the image, preventing the need to select calibrator directions.
Importantly, it is to our knowledge the first joint imaging and DDE calibration algorithm with convergence guarantees. 
Moreover, it is important to emphasize that the proposed method can be applied to joint DIE calibration and imaging problem as a particular case, providing a robust convergent self-calibration (selfcal) method.
Our approach is inspired by both the imaging techniques using optimization and compressive sensing theories, and the alternating calibration method StEFCal. It aims to minimize a regularized non-linear LS criterion, with respect to both the image and the DDEs.
To regularize the problem, in addition to an $\ell_1$ regularization term applied to the image (to promote the sparsity of the image, or its sparsity in a given dictionary), we make use of the assumption that the brightest sources of the image are known. Moreover, concerning the DDEs, we assume that they are smooth functions across the field of view, i.e. spatially band-limited.
Therefore, we design an algorithm based on recent non-convex optimization techniques, which benefits from convergence guarantees deduced from \cite{Bolte_2014, Frankel2014, Chouzenoux_2016}.
It involves alternating between the estimation of the image and the DDEs, relying on an iterative structure based on the same optimization toolbox for both image and DDEs estimation, consisting of a gradient step followed by a projection step.
In addition to the convergence guarantees, the key point of our algorithm is that it works globally on the whole image and in an automatic manner, without any requirement of partitioning the sky into different directions. 
Furthermore, it is adapted not only to images of point sources, but also to reconstruct images with extended and complex structures. Finally, the results of our simulations suggest that using our method to jointly estimate the DDEs and the image 
leads to large improvement of the dynamic range, which is orders of magnitude higher when compared to accounting for DIEs only.

The remainder of the paper is organised as follows. 
A simple description of our method is provided in Section~\ref{Sec:Idea}, followed by a complete description given in Sections~\ref{Sec:Imaging}, \ref{Sec:Calib} and \ref{Sec:algo}. 
More precisely, in Section~\ref{Sec:Imaging}, we describe the imaging inverse problem and the associated minimization problem appearing in the case of a perfectly calibrated telescope array. 
In Section~\ref{Sec:Calib}, we introduce the calibration problem to estimate the DDEs from a known sky intensity image. 
The joint imaging and DDE calibration problem and the proposed alternating method are given in Section~\ref{Sec:algo}. Moreover, an alternative DDE calibration method based on the StEFCal algorithm is described in Section~\ref{Sec:stefcal}.
The simulations performed and the results obtained thereby using the proposed approach are discussed in Section~\ref{Sec:Sim}. 
Finally, in Section~\ref{Sec:Conc}, we briefly summarize the main contributions presented in this paper and we describe envisaged future work directions. 
Note that even though the main formal explanations concerning the proposed approach are given in Sections~\ref{Sec:Imaging}, \ref{Sec:Calib} and \ref{Sec:algo}, the reader should be able to understand the main results of Section~\ref{Sec:Sim} after the intuitive description of our method given in Section~\ref{Sec:Idea}.

\section{Block-coordinate approach in a nutshell}
\label{Sec:Idea}

The joint imaging and DDE calibration method proposed in this article is based on a block-coordinate technique. The mathematical details of this approach are given in Sections~\ref{Sec:Imaging}, \ref{Sec:Calib} and \ref{Sec:algo}, while Section~\ref{Sec:stefcal} gives details of an alternative StEFCal-based method. Nevertheless, the current section is dedicated to an intuitive description that should enable readers to jump to Section~\ref{Sec:Sim} and understand our results and conclusions.
Moreover, tables of notations are provided in Appendix~\ref{App:notations}.

The proposed approach is in the spirit of the current imaging techniques based on optimization and compressive sensing theories, and can be seen as a generalization of an existing DIE calibration method, namely StEFCal \citep{Salvini_2014}.

\subsection{Imaging problem}
\label{Ssec:Idea:image}

The imaging problem for RI corresponds to an inverse problem \citep{Rau_2009, Carrillo_2014, Onose_2016}. In this context, the objective is to estimate an original unknown intensity image ${\xb}$ from the complex and noisy visibilities $\yb$ given by
\begin{equation}	\label{eq:simple:pb_inv}
\yb = {\Gbs} \Fbs {\xb} + \bb,
\end{equation}
where $\Fbs$ is the Fourier matrix, ${\Gbs}$ is the matrix containing on each row the antenna-based gain related to each antenna pair acquiring the complex visibilities, and $\bb$ is a realization of a complex i.i.d. Gaussian additive noise. 
In the case of a perfectly calibrated antenna array (i.e. $\Gbs$ is known), problem~\eqref{eq:simple:pb_inv} is linear and can be solved efficiently using convex optimization methods \citep{Onose_2016}. 
Basically, these approaches define the estimate of the image as a solution to a regularized LS minimization problem (see Section~\ref{Sec:Imaging} for more details). 
In particular, compressive sensing theory ensuring that combining an LS data-fidelity with a regularization term promoting sparsity leads to good reconstruction results \citep{Candes_2006}, $\ell_1$-based regularization terms have been extensively used in the last years for RI imaging. 
Therefore, the underlying minimization problem consists of a sum of an $\ell_2$ term and an $\ell_1$ term. A simple optimization algorithm to solve this problem is the forward-backward algorithm \citep{Combettes_Wajs_2005}. Basically, it is an iterative method, alternating at each iteration between a gradient step (forward step) on the $\ell_2$ function, and a proximity step (backward step) on the $\ell_1$ function. Note that in the context of the $\ell_1$ norm, the backward step corresponds to a soft-thresholding operation, obliging the small values of the signal to be set to zero, thus promoting sparsity of this signal.
Furthermore, as pointed out by \cite{Rau_2009, Carrillo_2014, Onose_2016}, the forward-backward algorithm is very similar to the CLEAN methods. Indeed, the forward step computes the residual dirty image, while the backward step is used to promote sparsity. Thus, similarities can be seen between these two steps and the major and the minor cycles of CLEAN methods (see e.g. \cite{Onose_2016} for details). 
Nevertheless, it is worth emphasizing that the forward-backward algorithm, and in general optimization algorithms \citep{Combettes_Book_10, Komodakis_N_2014_playing_dor}, can be used to introduce more sophisticated regularization terms. 
Moreover, it has been shown in the last years that using optimization and compressive sensing theories leads to very competitive results with respect to traditional radio interferometric methods such as CLEAN \citep{Onose_2016, Onose_2017}, yet comparisons have mostly been performed on smaller images, and the computational costs so far have not been competitive.

\subsection{Calibration problem}
\label{Ssec:idea:calib}

In practice, the antenna-based gains contained in ${\Gbs}$ are also unknown and have to be estimated. 
In the case when they reduce to DIEs and the image is known, \cite{Salvini_2014} have proposed an efficient method solving for the calibration problem, namely the StEFCal method. 
In the Fourier domain, DIEs consist of only one complex non-zero coefficient centred at the zero spatial frequency.
In this context, problem~\eqref{eq:simple:pb_inv} can be rewritten using a matrix formulation:
\begin{equation}	\label{eq:simple:matrix_pb_inv}
\Ybs = \Cbs \Xbs \Cbs^* + \Bbs,
\end{equation}
where $\Cbs$ is the diagonal matrix containing the non-zero elements of the direction-independent Fourier coefficients, $(.)^*$ denotes the complex conjugate of its argument, $\Xbs$ is the matrix containing the Fourier coefficients of the image ${\xb}$ associated with the frequencies acquired by the antenna pairs, and $\Bbs$ is the matrix formulation of the additive noise. 
In order to have a bi-linear inverse problem, \cite{Salvini_2014} propose to introduce in~eq.~\eqref{eq:simple:matrix_pb_inv} matrices $\Cbs_1 = \Cbs_2 = \Cbs$, and to solve the non-linear LS problem associated with~eq.~\eqref{eq:simple:matrix_pb_inv} using an iterative method based on an ADI algorithm. This method alternates at each iteration between the estimation of ${\Cbs}_1$ and ${\Cbs}_2$, assuming they are independent. 
More precisely, the update of ${\Cbs}_1$ is taken to be the exact minimizer of the LS objective function, i.e.
\begin{equation}	\label{eq:stef_simple:i}
\Cbs_1^{+} = \argmin{\Cbs_1}  \| \Cbs_1 \Xbs \Cbs_2^* - \Ybs \|_2^2,
\end{equation}
while the update of ${\Cbs}_2$ is taken to be exactly equal to $\Cbs_1$, i.e. $\Cbs_2^{+} = \Cbs_1^{+}$. 
As explained by \cite{Salvini_2014}, this method converges when the original image is known exactly. However, it has been shown that in practice, it works when only the bright sources of the original image are known. In this case, \cite{Salvini_2014} suggest to combine the StEFCal calibration method with an imaging algorithm in order to estimate more accurately the image once we have an estimation of the DIEs, and to iterate this process. 
However, this combined DIEs calibration and imaging approach does not benefit from the StEFCal convergence guarantees, and is not adapted to DDE calibration.

\subsection{Proposed joint DDE calibration and imaging method}

Taking advantage of the above-mentioned methods, we propose a joint imaging and DDE calibration algorithm. 
In this context, we generalize the inverse problem~\eqref{eq:simple:matrix_pb_inv} in order to take into account the DDEs in ${\Cbs}_1$ and ${\Cbs}_2$. 
To this aim, we assume that the DDEs are smooth functions across the field of view, i.e. they are spatially band-limited (see Figure~\ref{Fig:DDEs_image} in Section~\ref{Sec:Sim}). 
Therefore, we propose to reduce drastically the dimensionality of the problem by estimating only the non-zero Fourier coefficients of the DDEs in ${\Cbs}_1$ and ${\Cbs}_2$, now represented by matrices $\Ubs_1$ and $\Ubs_2$ (see Section~\ref{Sec:Calib} for more details). 
Moreover, we make use of the prior information on the bright sources of the original image. 
Note that this assumption is common in the context of DDE calibration methods \citep{Yatawatta2009, Weeren_2016}, and is useful to reduce the ambiguity problems appearing between the image and the DDEs \citep{Smirnov_2011b}. 
More precisely, we assume that the original image $\xb$ can be split as a sum of two images $\xb_o$ and $\epsilonb$, where $\xb_o$ is assumed to be known exactly, while $\epsilonb$ has to be estimated (see Figure~\ref{Fig:images} in Section~\ref{Sec:Sim} for an example in the case of point sources). 
Then, instead of using two different algorithms to estimate the DDEs and the image, respectively, we design a joint framework. In essence, this joint method involves alternating between the estimation of $\Ubs_1$, $\Ubs_2$ and the faint sources $\epsilonb$ contained in ${\xb}$, using the same algorithmic structure, based on the forward-backward iterations. 
To do so, we propose to 
\begin{equation}		\label{min_prob_glob_simp}
\minimize{\epsilonb, \Ubs_1, \Ubs_2}
h(\epsilonb, \Ubs_{1}, \Ubs_{2}) +  r(\epsilonb) +  p(\Ubs_{1}, \Ubs_{2}),
\end{equation}
where $h$ is the data fidelity term corresponding to a least squares criterion and depending on both the image and the DDEs, $r$ is the regularization function for the image, and $p$ is the regularization function for the DDEs. 
In particular, $r$ is chosen to constrain the image to be positive and to promote sparsity either directly in the image domain or in a given dictionary. Concerning the DDEs, $p$ is chosen to incorporate constraints on the direction-dependent Fourier coefficients, and to control the similarities between $\Ubs_1$ and $\Ubs_2$ (see Section~\ref{Sec:Calib} for more details). 
Note that problem~\eqref{min_prob_glob_simp} is non-convex with respect to the concatenation of the variables $(\epsilonb, \Ubs_1, \Ubs_2)$, but is convex with respect to each of them. In other words, keeping $(\Ubs_1, \Ubs_2)$ (resp. $(\epsilonb, \Ubs_2)$ and $(\epsilonb, \Ubs_1)$) fixed, problem~\eqref{min_prob_glob_simp} is convex with respect to the variable $\epsilonb$ (resp. $\Ubs_1$ and $\Ubs_2$).

The proposed iterative method is based on a block-coordinate forward-backward approach, solving the regularized LS problem~\eqref{min_prob_glob_simp}, and benefits from the convergence guarantees in \cite{Chouzenoux_2016}. 
More precisely, at each iteration, we first estimate approximately the DDEs computing a fixed number of forward-backward iterations, and then estimate approximately the image again using forward-backward iterations. 
It is important to emphasize that, as a particular case, the proposed approach can be applied to solve the joint DIE calibration and imaging problem.

\begin{figure*}
\includegraphics[width=12cm, angle=-90]{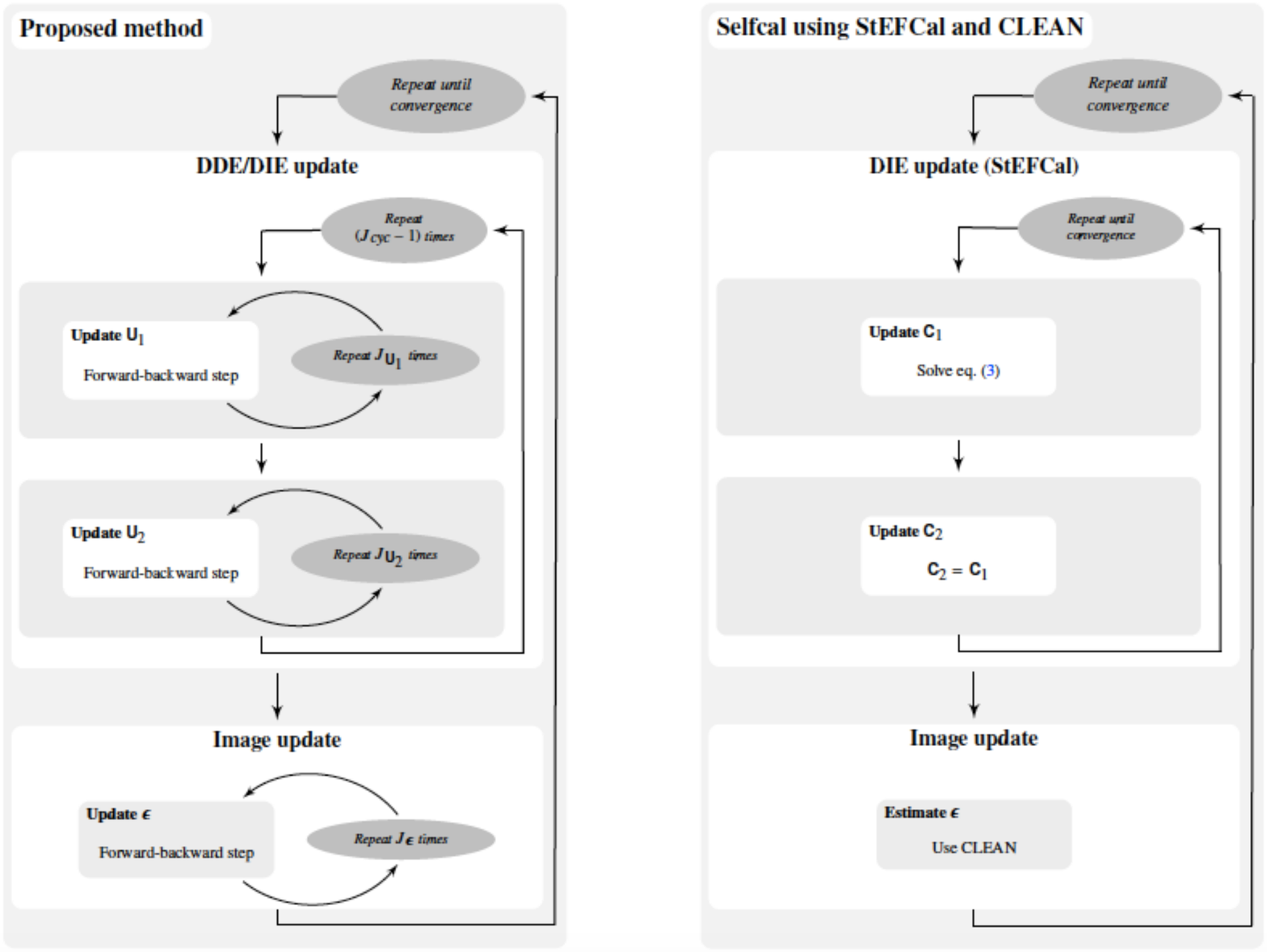}
\caption{\label{Graph:algo_BCFB}
The diagrams of the proposed method (left) and the traditional selfcal method using StEFCal and CLEAN (right). 
In the proposed method, only a finite number of iterations are performed to estimate $\Ubs_1$, $\Ubs_2$, and $\epsilonb$, given by $J_{\Ubs_1}\in \eN$, $J_{\Ubs_2}\in \eN$, and $J_{\epsilonb}\in \eN$ respectively. The parameter $J_{\text{cyc}} \in \eN$ gives the number of sub-iterations performed on the DDEs before estimating the image. 
The structure of the two methods is very similar, it consists in alternating between the estimation of the DIEs (and possibly the DDEs for the proposed method) and the estimation of the image. 
The main differences are: (i) our method use the same forward-backward based steps to estimate both the DDEs and the image, while the selfcal method uses two independent techniques to estimate them, and (ii) unlike the selfcal method, our method computes only a finite number of sub-iterations in each inner-loop. Note that these two differences are crucial to ensure the convergence of the global proposed method. }
\end{figure*}

It is interesting to note that the global structure of the proposed algorithm is very similar to the traditional selfcal method, since they both aim to alternate between the estimation of the gains and the estimation of the image. 
To illustrate the similarities and differences between the two methods, diagrams depicting the relevant steps of the proposed method and the traditional selfcal method are given in Figure~\ref{Graph:algo_BCFB} (left) and (right), respectively\footnote{The selfcal method presented in this diagram consists in alternating between the StEFCal algorithm described in Section~\ref{Ssec:idea:calib} and an imaging method, namely CLEAN. However, note that StEFCal can be coupled with other imaging methods, as described in Section~\ref{Ssec:Idea:image}.}.
One can observe that the global structure of the two diagrams are very similar. 
However, the structure of the inner-loops are different. 
In particular, in our method, $J_{\text{cyc}}-1$ inner-loops are performed to estimate \emph{approximately} the DDEs $(\Ubs_1, \Ubs_2)$. 
Within each of these inner-loops, $J_{\Ubs_1}$ forward-backward steps are performed to estimate $\Ubs_1$, followed by $J_{\Ubs_2}$ forward-backward steps to estimate $\Ubs_2$. 
Note that $J_{\Ubs_1}$ and $J_{\Ubs_2}$ are finite so as to have \emph{approximate} estimations of $\Ubs_1$ and $ \Ubs_2$, respectively. 
Then, to complete one global iteration (i.e. a \emph{cycle}), $J_{\epsilonb}$ forward-backward steps are performed to estimate \emph{approximately} the image $\epsilonb$. 
Note that computing only \emph{approximated} estimates is very important in practice to ensure the convergence of the overall algorithm. Intuitively, one can understand this point as follows: when the algorithm is initialized with a very poor estimation of the image, it is obvious that estimating \emph{completely} the DDEs from this incorrect image can be inefficient. Therefore, it is important to control the accuracy of the estimates at each iteration in order to make the overall algorithm converge step by step for the image and the DDEs together. 
This adopted methodology for approximate estimates is different from the traditional selfcal method, where the problem of estimating the gains (restricted to DIEs) is solved \emph{completely} before solving the imaging problem, and \emph{vice versa}. As explained above, 
this strategy can lead to poor reconstruction results. 
Furthermore, while the imaging update of our method is very similar to the forward-backward approach described in Section~\ref{Ssec:Idea:image}, the calibration part is different from the StEFCal approach given in Section~\ref{Ssec:idea:calib}. 
Firstly, using a forward-backward based algorithm allows us to introduce constraints on the DDEs. 
Moreover, StEFCal is an \emph{implicit} method, stating directly that the update of $\Cbs_2$ is equal to $\Cbs_1$, whereas our method updates independently the variables $\Ubs_1$ and $\Ubs_2$. Thus, in order to constrain them to be equal at convergence, the distance between $\Ubs_1 $ and $ \Ubs_2$ is \emph{explicitly} controlled in the minimization problem. 
Finally, StEFCal is only designed for DIE calibration, while our method jointly corrects for DDEs and estimates the image.

Therefore our method can be seen as a generalization of selfcal method, with theoretical convergence guarantees. 
Note that traditional selfcal method does not benefit from the convergence guarantees of the proposed method since both rely on different algorithmic structures. 
More details of the proposed algorithm are given in Section~\ref{Sec:algo}.

For the sake of completeness, a naive generalization of the StEFCal method for DDE calibration is also proposed in Section~\ref{Sec:stefcal}. Note that in contrast with the StEFCal method for DIE calibration, in the DDE case, it is not enough to consider the LS data-fidelity term, and the minimization problem needs to be regularized. 
As for the basic StEFCal method for DIE calibration, we propose to combine this naive generalization with an imaging technique. The obtained method consists in alternating between a DDE calibration algorithm (i.e. the naive generalization of StEFCal) and an imaging algorithm (not necessarily based on the CLEAN method).
Nevertheless, the obtained method is not guaranteed to converge, fully justifying the development of a more sophisticated approach, described in Sections~\ref{Sec:Imaging}, \ref{Sec:Calib} and \ref{Sec:algo}.

In Section~\ref{Sec:Sim}, we describe the simulations performed considering both point source images and an image of M31. In both cases, we show the results obtained using our method, reconstructing the DDEs and the image $\epsilonb$, and the results obtained with the StEFCal algorithm solving only for the DIEs, and combined with an imaging method based on the forward-backward iterations. 
In particular, in Section~\ref{Ssec:sim:tests1} we focus on point source images, where the original image $\xb$ is the sum of three point source images: the known image $\xb_o$,  and the unknown images ${\epsilonb}_1$ and ${\epsilonb}_2$ (i.e. $\epsilonb = \epsilonb_1+\epsilonb_2$, see Figure~\ref{Fig:images}). In our simulations, ${\epsilonb}_2$ is considered to be an astrophysical noise of very low intensity, which can be seen as the confusion noise induced by the resolution limitation of the radio telescope. Thus, the main objective is to estimate ${\epsilonb}_1$. Our experiments in this context aim to study the behaviour of the proposed approach varying the dynamic range of the image (see Figures~\ref{Fig:test1:curves} and \ref{Fig:test1:images}), the size of the DDEs (see Figures~\ref{Fig:test2:curves} and \ref{Fig:test2:images}), the number of antennas considered, and the time dependency of the DDEs (see Figures~\ref{Fig:test3:curves} and \ref{Fig:test3:images}). 
In Section~\ref{Ssec:sim:tests2}, we describe the results obtained by considering an image of M31 (see Figure~\ref{Fig:M31}). In this case, we assume that the brightest part of the image $\xb_o$ is known, and the objective is to estimate the unknown sources ${\epsilonb}$. In particular, we perform three experiments, changing the prior information on the bright sources. Figures~\ref{Fig:testM31:images_delta01}, \ref{Fig:testM31:images_delta05} and \ref{Fig:testM31:images_delta1} show the results obtained for these simulations, giving in the first row the bright sources contained in $\xb_o$ on the left, and the unknown image ${\epsilonb}$ on the right.
It is worth noting that all the results presented in Section~\ref{Sec:Sim} suggest that using our method to jointly estimate the DDEs and the image enables to improve the dynamic range by orders of magnitude compared to accounting for DIEs only. 
Moreover, in order to present a study as complete as possible, we compared our approach with the naive StEFCal method for DDE calibration presented in Section~\ref{Sec:stefcal}, and we show that our method leads to better reconstruction results.

The following three sections give more details on the proposed method. In particular, Section~\ref{Sec:Imaging} is dedicated to the imaging part, Section~\ref{Sec:Calib} to the calibration part, and the complete algorithm is explained in detail in Section~\ref{Sec:algo}. 
However, at this stage the reader might directly be interested in the simulations and results given in Section~\ref{Sec:Sim}.

\section{Imaging problem formulation}
\label{Sec:Imaging}

\subsection{Observation model}

A radio interferometer consists of an array of $n_a$ antennas, measuring the radio emission from a given area of the sky.
The complex data, namely the visibilities, acquired at instant $t \in \{1, \ldots, T\}$, are determined by the relative position between each antenna pair indexed by $(\alpha, \beta) \in \{1, \ldots, n_a\}^2$, with $\alpha < \beta$. 
The baseline components, denoted by $\big( u_{t,\alpha,\beta}, v_{t,\alpha,\beta}, w_{t,\alpha,\beta} \big)$, are measured in units of the observation wavelength. 
On the one hand, the components in the plane perpendicular to the line of sight $\vb_{t,\alpha,\beta} = \big( u_{t,\alpha,\beta}, v_{t,\alpha,\beta} \big)$ specify the coordinates of the projected the baseline, $\vb_{t,\alpha,\beta} = \vb_{t,\alpha} - \vb_{t,\beta}$, $\vb_{t,\alpha}$ (resp. $\vb_{t,\beta}$) being the position of antenna $\alpha$ (resp. $\beta$) in units of the observation wavelength.
On the other hand, the third component $w_{t,\alpha,\beta}$ identifies the coordinate in the line of sight. 
The sky brightness distribution $\overline{x}$ is described in the same coordinate system, with components $l$, $m$, $n$ and with $\lb = (l,m)$ and $n(\lb) = \sqrt{1 - l^2- m^2}$, $l^2+m^2 \le 1$. 
Then, the general measurement equation of non-polarized monochromatic RI imaging, for the antenna pair $(\alpha,\beta)$ at instant $t$, is given by
\begin{equation}	\label{Pb:inter_cont}
y_{t,\alpha,\beta} = \int \overline{d}_{t,\alpha}(\lb) \overline{d}_{t,\beta}(\lb)^*  \overline{x}(\lb) \es^{-2\is \pi \vb_{t,\alpha,\beta} \cdot \lb } \ds\lb,
\end{equation}
where 
$\overline{\db}_{t,\alpha} $ represents a DDE in the image domain related to antenna $\alpha$ at instant $t$.

To recover the original unknown image from incomplete visibility measurements, 
a discretized version of problem~\eqref{Pb:inter_cont} is considered,
which corresponds to a sampled version of the continuous sky brightness distribution $\overline{x}$. 
We denote this sampled original intensity signal by the vector $\overline{\xb} = (\overline{x}(n))_{-N/2 \le n \le N/2-1} \in \eR_+^N$, and the complex visibilities are represented by the vector $\yb \in \eC^{TM}$. 
More precisely, at each instant $t\in \{1,\ldots,T\}$, an interferometer with $n_a$ antennas acquires $M = n_a (n_a-1)/2$ different measurements in the Fourier domain of the image of interest. 
Therefore, the visibility $y_{t, \alpha,\beta} \in \eC$ measured by the antenna pair $(\alpha,\beta)$ at instant $t$ at the discrete spatial frequency $k_{t,\alpha,\beta} = k_{t,\alpha} - k_{t,\beta}$ 
can be modelled as the following inverse problem:
\begin{equation} \label{eq:obs_alpha_beta}
y_{t, \alpha,\beta} = 
\sum_{n=-N/2}^{N/2-1} \overline{d}_{t,\alpha}(n) \overline{d}_{t,\beta}(n)^* \overline{x}(n) \es^{-2\is \pi k_{t,\alpha,\beta} \frac{n}{ N}} \\
+ 
b_{t,\alpha,\beta},
\end{equation}
where 
$\overline{\db}_{t,\alpha} = (\overline{d}_{t,\alpha}(n))_{-N/2 \le n \le N/2-1} \in \eC^N$ is the sampled DDE related to antenna $\alpha$, and $\bb = (b_{t, \alpha, \beta})_{\underset{ 1 \le \alpha < \beta \le n_a}{1\le t \le T}} \in \eC^{TM}$ is a realization of a complex i.i.d. Gaussian additive noise. 
Note that, using these notations, the DIEs can be seen as a special case of the DDEs where $\overline{\db}_{t,\alpha} = \delta_{t,\alpha} \unb_N$, with $\delta_{t,\alpha} \in \eC$ and $\unb_N$ being the unitary vector of dimension $N$.

When the antenna array is perfectly calibrated, i.e. when the DDEs/DIEs are perfectly known, the observations~ described in eq.~\eqref{eq:obs_alpha_beta} can be rewritten using a matrix formulation
\begin{equation}	\label{pb:inv_pb_image}
\yb = \overline{\Gbs} \Fbs \overline{\xb} + \bb,
\end{equation}
where $\Fbs \in \eC^{N \times N}$ represents the Fourier matrix, and $\overline{\Gbs} \in \eC^{TM \times N}$ is the matrix containing on each row the antenna-based gain for the pair $(\alpha,\beta)$. 
More precisely, each row of $\overline{\Gbs}$, indexed by  $\big( t, \alpha, \beta \big)$, corresponds to the convolution of the Fourier transforms $\widehat{\overline{\db}}_{t,\alpha}$ and $\widehat{\overline{\db}^*}_{t,\beta}$ of the gains $\overline{\db}_{t,\alpha}$ and $\overline{\db}^*_{t,\beta}$ respectively, centred at the spatial frequency $k_{t,\alpha,\beta}$. 

Note that the relevant variables and dimensions associated with the RI inverse problem are summarized in Table~\ref{Tab:not:problem}, given in Appendix~\ref{App:notations}.

\subsection{Minimization problem}
\label{Ssec:Min_x}

Considering the inverse problem~\eqref{pb:inv_pb_image}, 
new methods based on both convex optimization and compressive sensing theory have been developed recently to find an estimate $\xb^\star$ of $\overline{\xb}$ from the observations $\yb$ \citep{Wiaux_2009}. 
In this context, the estimated image is defined as the minimizer of a sum of two functions: the data fidelity term related to the observation model, and the regularization term incorporating \emph{a priori} information on the target image. 
In the context of RI imaging, the additive noise being assumed i.i.d. Gaussian, 
a usual choice for the data fidelity term is the least squares criterion:
\begin{equation}
(\forall \xb \in \eR^N)\quad
\widetilde{h} (\xb) = \frac12 \| \overline{\Gbs} \Fbs \xb - \yb \|_2^2,
\end{equation}
where $\| \cdot \|_2$ denotes the standard Euclidean norm. Then, 
the estimated image can be defined as a solution to: 
\begin{equation}	\label{pb:minx}
\minimize{\xb \in \eR^N} \widetilde{h} (\xb) + \eta \, \widetilde{r}(\xb),
\end{equation}
where $\widetilde{r} \colon \eR^N \to ]-\infty , + \infty]$ is the regularization function, and $\eta>0$ is the regularization parameter chosen to balance the data fidelity and the regularization terms.

Note that in previous works \citep{Wiaux_2009, Carrillo_2014, Onose_2016}, the authors have chosen to solve a constrained version of problem~\eqref{pb:minx}:
\begin{equation}	\label{eq:pb:contr}
\minimize{\xb \in \eR^N} \widetilde{r}(\xb)
\text{ subject to }
\| \overline{\Gbs} \Fbs \xb - \yb \|_2 \le \theta,
\end{equation}
where $\theta>0$ is a parameter to be fixed by the user, associated with the upper bound of the norm of the additive noise in eq.~\eqref{pb:inv_pb_image}. Historically, this formulation has been preferred over problem~\eqref{pb:minx} since in theory, $\theta$ can be determined as a function of the additive noise in~eq.~\eqref{pb:inv_pb_image}, whereas $\eta$ is a free parameter. However, due to technical assumptions related to the algorithm proposed in the current work, we will only focus on problem~\eqref{pb:minx} for the remaining of this article \citep{Chouzenoux_2016}. 
Moreover, the constrained problem given in eq.~\eqref{eq:pb:contr} might be unstable in the context of joint imaging and calibration, since, in this case, the bound $\theta$ cannot be properly fixed.

Compressive sensing theory makes use of the assumption that the image $\overline{\xb}$ has a sparse representation $ \Psib^\dagger \overline{\xb} \in \eC^D$ in a given dictionary $\Psib \in \eC^{N \times D}$ \citep{Fornasier_book_2011}, where $\dagger$ is the transpose conjugate operator. 
This framework has shown to give good reconstruction results in several application fields such as astronomical remote sensing \citep{Bobin_2008}, optical interferometric imaging \citep{Auria_2014, Birdi2016}, 
and RI imaging \citep{Wiaux_2009, Li_2011, Garsden2015, Dabbech_2015}. 
In this context, the regularization function $\widetilde{r}$ in eq.~\eqref{pb:minx} is chosen in order to promote sparsity. Intuitively, the $\ell_0$ pseudo-norm, counting the non-zero elements of a signal, is the best choice to promote sparsity \citep{Donoho1995}. More generally $\ell_p$ norms, with $0 \le p <1$, can be efficiently used \citep{Bouman1996}. However, due to their non-convexity and non-differentiability, these functions are difficult to optimize in practice, and often an $\ell_1$ convex relaxation is used \citep{BLANCFERAUD_ECCV2004_l1, Donoho2006, Figueiredo_2007}. Note that (re)weighted $\ell_1$ regularization \citep{Candes2008, Carrillo_2012} can also be used, in order to approximate the $\ell_0$ regularization function. The weighted $\ell_1$ regularization is defined by
\begin{equation}
(\forall \xb \in \eR^N)\quad
\widetilde{r}(\xb) =  \| \Wbs \Psib^\dagger \xb \|_1,
\end{equation}
where $\Wbs \in \eR^{D \times D}$ is a weighting matrix, often chosen to be diagonal (if $\Wbs$ is the identity matrix, the usual $\ell_1$ regularization term is recovered).
Concerning $\Psib$, different dictionaries can be chosen, depending on the nature of the target image. 
On the one hand, if the image is assumed to have only few non-zero coefficients (e.g. point sources), one can choose $\Psib = \Ibs_N$ to be the identity operator in $\eR^N$. 
On the other hand, natural images are not necessarily sparse, but can have a sparse representation in other domains. 
For instance, piece-wise constant images are sparse on the gradient domain, and thus total variation based regularizations can be used for such images \citep{Rudin_1992_total_variation, Gilboa_2008, Wiaux_2010}. Another choice for $\Psib$ is to promote sparsity in the wavelet domain, using, e.g., isotropic undecimated wavelet (IUW) transforms \citep{Starck_1994, Li_2011, Dabbech_2015, Garsden2015}, or a concatenation of wavelet transforms \citep{Vannier_2010, Carrillo_2012}. 

Note that constraints can also be taken into account in the formulation given in eq.~\eqref{pb:minx}. To this aim, we introduce the indicator function of a closed non-empty subset $\eE$ of $\eR^N$, penalizing all the coefficients of the image which are not in $\eE$.
More formally, we define the indicator function as follows:
\begin{equation}
\iota_{\eE}(\xb) = 
\begin{cases}
0, 	&	\text{if } \xb \in \eE,	\\
+\infty,	&	\text{otherwise.}
\end{cases}
\end{equation}
In problem~\eqref{pb:minx}, introducing this indicator function in the regularization term is equivalent to constrain $\xb$ to belong to $\eE$, since the objective function cannot be minimized as long as $\xb \not\in \eE$. 
Constraints, and subsequently indicator functions, can be used to impose positivity of the image by defining $\eE = [0, +\infty[^N$, or more generally to constrain the amplitude of a signal to stay in a box. They can be used as well to impose sparsity in the case when the sparsity degree $\kappa \in \eN^*$ is known, by choosing $\eE = \menge{\xb \in \eR^N}{\| \xb \|_1 \le \kappa}$.

\section{Calibration problem formulation}
\label{Sec:Calib}

\subsection{Observation model}
\label{Ssec:calib:model}

In practice, in the context of RI, for every $t\in \{1,\ldots,T\}$, the antenna-based gains $(\overline{\db}_{t,\alpha})_{1 \le \alpha \le n_a}$ described in eq.~\eqref{eq:obs_alpha_beta} are unknown and have to be calibrated. 
To this aim, the visibilities $\yb$ can be reordered, and a matrix formulation can be used.
Let $\overline{\Dbs}_{t,1} \in \eC^{n_a \times N}$ 
be the matrix containing on each row $\alpha$ the reordered kernel 
$\big(\widehat{\overline{d}}_{t,\alpha}(-k)\big)_k$
centred at $k_{t,\alpha}$, and 
$ \overline{\Dbs}_{t,2} \in \eC^{n_a \times N}$ be the matrix containing on each row $\alpha$ the conjugate $\widehat{\overline{\db}}^{\,*}_{t,\alpha}$ of $\widehat{\overline{\db}}_{t,\alpha}$ centred at $-k_{t,\alpha}$. 
Moreover, let $\Ybs_t = \big( \Ys_{t,(\alpha,\beta)} \big)_{1 \le \alpha, \beta \le n_a} \in \eC^{n_a \times n_a}$ be the matrix containing the visibilities, with zero coefficients on its diagonal and such that $\Ys_{t,(\alpha,\beta)} = y_{t, \alpha, \beta}	$ if $\alpha < \beta$, and $\Ys_{t,(\alpha,\beta)} = y_{t,\beta,\alpha}^*$ if $\alpha > \beta$. 
Then, the inverse problem described in~eq.~\eqref{eq:obs_alpha_beta} can be equivalently reformulated as follows
\begin{equation}	\label{eq:pb_inv_matrix}
(\forall t \in \{1,\ldots, T\})\quad
\Ybs_t = \Sc \big( \overline{\Dbs}_{t,1} \,\overline{\Xbs}   \,\overline{\Dbs}_{t,2}^\top \big) + \Bbs_t,
\end{equation}
where $(.)^\top$ denotes the transpose operation of its argument, $\Sc \colon \eC^{n_a \times n_a} \to \eC^{n_a \times n_a}$ is the operator setting to zero the coefficients on the diagonal of its argument, $\Bbs_t$ is the matrix formulation associated with the realization of the additive noise in~eq.~\eqref{eq:obs_alpha_beta}, and $\overline{\Xbs} \in \eC^{N \times N}$  contains on each row/column a shifted version of the Fourier transform of the original image $\widehat{\overline{\xb}}$. More formally, for every $(k,k') \in \{1, \ldots,N\}$, we have $ \overline{\Xs}(k,k') = \widehat{\overline{x}}(k+k'-1) $ if $k+k'-1 \le N$, and $\overline{\Xs}(k,k') = 0$ otherwise. This matrix is used to model the convolution operation in the matrix formulation to define $\Ybs_t$. 
Note that the matrix formulation given by eq.~\eqref{eq:pb_inv_matrix} symmetrizes the data, contrary to the inverse problem given in eq.~\eqref{pb:inv_pb_image}. Thus, each measurement appears twice in $\Ybs_t$ (itself and its conjugate). However, we use this formulation in order to simplify the notation in the remaining of the section.
Another remark concerns the fact that, in our formulation, we assume that the DDEs are time-dependent. Therefore, when the image is known, $T$ inverse problems of the form of eq.~\eqref{eq:pb_inv_matrix} have to be independently solved.

In the particular case when only DIEs are considered, the inverse problem given by~eq.~\eqref{eq:pb_inv_matrix} can be simplified, and the formulation given by \cite{Salvini_2014} can be used. Indeed, in this case, the Fourier transform $\widehat{\overline{\db}}_{t,\alpha}$ of $\overline{\db}_{t,\alpha}$ has only one non-zero coefficient centred at the zero spatial frequency, denoted by $\widehat{\overline{d}}_{t,\alpha}(0)$. Thus, each row of $\overline{\Dbs}_{t,1}$ and $\overline{\Dbs}_{t,2}$ contains only one non-zero element, centred in $k_{t,\alpha}$ and $-k_{t,\alpha}$, respectively, and eq.~\eqref{eq:pb_inv_matrix} reduces to 
\begin{equation}	\label{eq:pb_inv_mat_dies}
(\forall t \in \{1,\ldots, T\})\quad
\Ybs_t = \Sc \big( \overline{\Cbs}_{t} \, \widetilde{\Xbs}   \,\overline{\Cbs}_{t}^* \big) + \Bbs_t,
\end{equation} 
where $\overline{\Cbs}_t \in \eC^{n_a \times n_a}$ is the diagonal matrix containing the elements $\big( \widehat{\overline{d}}_{t,\alpha}(0) \big)_{1 \le \alpha \le n_a}$ on its diagonal. 
In this case, $\widetilde{\Xbs}$ is a matrix of size $n_a \times n_a$ with entries corresponding to the coefficients of $\overline{\Xbs}$ associated with the frequencies selected by the antenna pairs (i.e. the rows associated with the spatial frequency $k_{t,\alpha}$, and the columns associated with the spatial frequency $-k_{t,\beta}$, for every $(\alpha, \beta) \in \{1, \ldots, n_a\}^2$, with $\beta\neq \alpha$).

\subsection{Minimization problem}

As described in Section~\ref{Ssec:Min_x}, for the imaging problem, we assume that the additive noise in the model described by eq.~\eqref{eq:pb_inv_matrix} is an i.i.d. Gaussian noise. In this context, several works have proposed to choose the data fidelity term associated with the DIEs or DDEs to be the LS criterion \citep{Salvini_2014,Smirnov_2015}. 
In particular, in the case when only DIEs are considered, 
according to~eq.~\eqref{eq:pb_inv_mat_dies}, the estimate of $\overline{\Cbs}_t$ can be defined as a solution to
\begin{equation}	\label{eq:min_dies_sym}
\minimize{(\Cbs_{t})_{1 \le t \le T} \in ( \eC^{n_a \times n_a} )^{T}  } 
\sum_{t=1}^T \frac12 \|  \Sc \big( \Cbs_{t} \, \widetilde{\Xbs} \, \Cbs_{t}^* \big) - \Ybs_t \|_2^2.
\end{equation}
Note that this minimization problem is non-convex.
Then, to solve it, \cite{Salvini_2014} proposed to introduce two intermediary variables $\Cbs_{t,1} $ and $\Cbs_{t,2}$ such that $\Cbs_t = \Cbs_{t,1} = \Cbs_{t,2}$. In this case, the overall minimization problem remains non-convex, but it becomes convex with respect to $\Cbs_{t,1}$ when $\Cbs_{t,2}$ is fixed, and convex with respect to $\Cbs_{t,2}$ when $\Cbs_{t,1}$ is fixed.

More generally, as described by \cite{Smirnov_2015}, the calibration problem for DDEs can also be written as a LS minimization problem. 
In particular, by analogy with eq.~\eqref{eq:min_dies_sym}, we can define estimates of the DDEs, stored in the matrices $(\overline{\Dbs}_{t,1}, \overline{\Dbs}_{t,2})$, as a solution to a LS minimization problem. 
However, only the non-zero coefficients of the Fourier transforms of the DIEs appear in eq.~\eqref{eq:pb_inv_mat_dies}, while $\overline{\Dbs}_{t,1}$ and $ \overline{\Dbs}_{t,2}$ both contain $n_a \times N$ unknown coefficients. 
Therefore, the DDE calibration minimization problem is not only non-convex, but also highly under-determined. 
Furthermore, the non-convexity becomes even more problematic when the image is also unknown, and the associated minimization problem tends to be more difficult to solve. 
In order to overcome this difficulty, we make use of the assumption that DDEs are smooth functions of the sky, i.e. spatially band-limited, and we propose to regularize the LS minimization problem accordingly.
More precisely, we assume that, for every $t\in \{1, \ldots,T\}$ and $\alpha\in \{1, \ldots,n_a\}$, the Fourier transform $\widehat{\overline{\db}}_{t,\alpha}$ of ${\overline{\db}}_{t,\alpha}$ has a known bounded support, we denote by $\eS$ the largest of these supports (i.e. for every $k \not\in \eS$, $\widehat{\overline{d}}_{t,\alpha}(k) = 0$), and by $S$ its cardinal. 
In addition, we assume that the amplitude of the coefficients of $\widehat{\overline{\db}}_{t,\alpha}$ are bounded, and the amplitude of the central coefficient $\widehat{\overline{d}}_{t,\alpha} (0)$ is much larger than the amplitude of the other coefficients. More formally, we assume that there exist $\vartheta_1 \gg \vartheta_2 >0$ such that 
\begin{equation}		\label{ass:ddes:ii:1}
\big| \real\big( \widehat{\overline{d}}_{t,\alpha} (0) \big) \big| \le \vartheta_1,	
\;
\big| \imag\big( \widehat{\overline{d}}_{t,\alpha} (0) \big) \big| \le \vartheta_1,
\end{equation}
and
\begin{equation}	\label{ass:ddes:ii:2}
(\forall k \in \eS \setminus \{0\}) \quad
\big| \real\big( \widehat{\overline{d}}_{t,\alpha} (k) \big) \big| \le \vartheta_2,	
\;
\big| \imag\big( \widehat{\overline{d}}_{t,\alpha} (k) \big) \big| \le \vartheta_2	.
\end{equation}

Under this assumption, the DDE calibration problem can be reformulated in order to estimate only the non-zero coefficients of $\overline{\Dbs}_{t,1}$ and $\overline{\Dbs}_{t,2}$. 
To this aim, we introduce the matrix $\overline{\Ubs}_{t,1} \in \eC^{n_a \times S}$ (resp. $\overline{\Ubs}_{t,2} \in \eC^{n_a \times S}$) containing on each row $\alpha$ only the non-zero coefficients of $\overline{\Dbs}_{t,1}$ (resp. $\overline{\Dbs}_{t,2}$). 
More formally, the $\alpha$-th row of $\overline{\Ubs}_{t,1}$, denoted by $\overline{\ub}_{t,\alpha,1} \in \eC^S$, contains $\big( \widehat{\overline{d}}_{t,\alpha}(-k) \big)_{-k \in \eS}^\top \in \eC^{1 \times S}$, and the $\alpha$-th row of $\overline{\Ubs}_{t,2}$, denoted by $\overline{\ub}_{t,\alpha,2} \in \eC^S$, contains 
$\big( \widehat{\overline{d}}_{t,\alpha}(k)^* \big)_{k \in \eS}^\top \in \eC^{1 \times S}$. 
Moreover, let $\Dc_{t,1}\colon \eC^{n_a \times S} \to \eC^{n_a \times N}$ and $\Dc_{t,2}\colon \eC^{n_a \times S} \to \eC^{n_a \times N}$ be the functions building respectively matrices $\overline{\Dbs}_{t,1}$ and $\overline{\Dbs}_{t,2}$ from the non-zero coefficient matrices $\overline{\Ubs}_{t,1}$ and $\overline{\Ubs}_{t,2}$. Thus, we have $\overline{\Dbs}_{t,1} = \Dc_{t,1}\big( \overline{\Ubs}_{t,1} \big) $ and $\overline{\Dbs}_{t,2} = \Dc_{t,2}\big( \overline{\Ubs}_{t,2} \big) $. 
Then, we propose to reformulate the minimization problem associated with the DDE calibration as 
\begin{equation}	\label{eq:min_ddes_constraints_C}
\minimize{(\Ubs_{1}, \Ubs_{2}) \in (\eC^{T n_a \times S})^{2}} 
 \overline{h}(\Ubs_{1}, \Ubs_{2})  + \nu \, {p}(\Ubs_{1}, \Ubs_{2}),
\end{equation}
where  
$\Ubs_1$ (resp. $\Ubs_2$), for all $t\in \{1,\ldots,T\}$, is the concatenation of matrices $\Ubs_{t,1}$ (resp. $\Ubs_{t,2}$), 
the function $\overline{h} \colon \eC^{T n_a \times S} \times \eC^{T n_a \times S} \to \eR$ is the data fidelity term chosen to be the least squares criterion, and ${p} \colon \eC^{T n_a \times S} \times \eC^{T n_a \times S} \to ]-\infty, +\infty]$ is the regularization term enforcing similarity between $\Ubs_{1}$ and $\Ubs_{2}$, and constraining the direction-dependent Fourier coefficients to satisfy conditions~\eqref{ass:ddes:ii:1}-\eqref{ass:ddes:ii:2}.

In addition to the dimensionality reduction, working only on the non-zero coefficients $\Ubs_1$ and $\Ubs_2$ allows to handle independently the DDE associated with each of the antennas. In particular, in this context the data-fidelity term becomes additively separable in terms of antenna specific terms and each time instant. This separability will be a key feature in Section~\ref{Sec:algo} to design a highly parallelizable algorithm.  
More precisely, we have
\begin{align}	
\overline{h}( \Ubs_{1}, \Ubs_{2} )  	
& 	=\sum_{t=1}^T \sum_{\alpha = 1}^{n_a} 
\frac12 \|  \Hbs_{t,\alpha,1}  \ub_{t, \alpha,1} - \Ybs_{t,\alpha} \|_2^2		\label{eq:data_fid_d_C1}	\\
& 	=\sum_{t=1}^T \sum_{\alpha = 1}^{n_a} 
\frac12 \|  \Hbs_{t,\alpha,2}  \ub_{t, \alpha,2} - \Ybs_{t,\alpha} \|_2^2	 ,	\label{eq:data_fid_d_C2}
\end{align}
where $\Ybs_{t,\alpha} = \big( \Ys_{t,(\alpha,\beta)} \big)_{\underset{\footnotesize \beta \neq \alpha}{\footnotesize \beta\in \{1 \ldots, n_a\}}}$, 
and, 
in eq.~\eqref{eq:data_fid_d_C1}, 
$\Hbs_{t,\alpha,1} = \Dc_{t,\alpha,2} \big( \Ubs_{t,2} \big)  \Lc_{t,\alpha,1} \big( \overline{\Xbs} \big)$ with 
$\Dc_{t,\alpha,2} \big( \Ubs_{t,2} \big)  \in \eC^{(n_a-1) \times N} $
corresponds to the matrix $\Dbs_{t,2} = \Dc_{t,2} \big( \Ubs_{t,2} \big)$ without the row $\alpha$, 
and 
$\Lc_{t,\alpha,1} \colon \eC^{N \times N} \to \eC^{N \times S}$ is the operator selecting only the columns of $\overline{\Xbs}$ associated with the frequencies corresponding to the support $\eS$ centred in $k_{t,\alpha}$. 
Similarly, in eq.~\eqref{eq:data_fid_d_C2}, $\Hbs_{t,\alpha,1} = \Dc_{t,\alpha,1} \big( \Ubs_{t,1} \big)  \Lc_{t,\alpha,2} \big( \overline{\Xbs} \big) $ where 
$\Dc_{t,\alpha,1} \big( \Ubs_{t,1} \big)  \in \eC^{(n_a-1) \times N} $
corresponding to the matrix $\Dbs_{t,1} = \Dc_{t,1} \big( \Ubs_{t,1} \big)$ without the row $\alpha$, 
and
$\Lc_{t,\alpha,2} \colon \eC^{N \times N} \to \eC^{N \times S}$ being the operator selecting only the columns of $\overline{\Xbs}$ associated with the frequencies $k\in \{-N/2, \ldots, N/2-1\}$ corresponding to the support $\eS$ centred in $-k_{t,\alpha}$. 
Note that the data fidelity term $\overline{h}$ defined by~eq.~\eqref{eq:data_fid_d_C1}-\eqref{eq:data_fid_d_C2} is non-convex in $(\Ubs_1, \Ubs_2)$. However, \eqref{eq:data_fid_d_C1} is convex with respect to $\Ubs_1$ when $\Ubs_2$ is fixed, and \eqref{eq:data_fid_d_C2} is convex with respect to $\Ubs_2$ when $\Ubs_1$ is fixed.

On the other hand, the regularization function $p$ in~eq.~\eqref{eq:min_ddes_constraints_C} is given by
\begin{multline}	\label{eq:reg_DD}
p(\Ubs_{1}, \Ubs_{2}) = 
\sum_{t = 1}^T  \sum_{\alpha = 1}^{n_a} \frac12  \| \ub_{t,\alpha,2} - \sigma(\ub_{t,\alpha,1}) \|_2^2		\\
+ \iota_{\eD}(\ub_{t,\alpha,1}) + \iota_{\eD}(\ub_{t, \alpha,2}) 
\end{multline}
where $\sigma \colon \eC^{S} \to \eC^{S}$ is the bijection 
defined as $\sigma(\ub_{t,\alpha,1}) = \big(u_{t,\alpha,1}(-s)^*\big)_{1 \le s \le S} $, flipping the elements of $\ub_{t,\alpha,1}$ to obtain the vector $ \big(u_{t,\alpha,1}(-s)\big)_{1 \le s \le S} $, and then taking the complex conjugate of each of its element $u_{t,\alpha,1}(-s)^*$, for every $s \in \{1, \ldots, S\}$. Note that we have $\sigma = \sigma^{-1}$. 
This bijection is used in order to impose the kernels contained in $\Ubs_1$ to be equal to the complex conjugate of the kernels contained in $\Ubs_2$, up to a reorganization of its coefficients.
The set $\eD \subset \eR^{2S}$ is chosen in order to constrain the DDEs amplitudes to satisfy conditions~\eqref{ass:ddes:ii:1}-\eqref{ass:ddes:ii:2}. Formally, $\eD$ is defined by
\begin{align}
\eD = \Big\{ \ub  \in \eC^{S}
&	\Big| \;  | \real\big( u (0) \big) | \le \vartheta_1, \, | \imag( u (0) \big) | \le \vartheta_1,  \nonumber\\
& \text{and } (\forall s \in \{-S/2, \ldots, S/2-1\}\setminus \{0\}) \nonumber\\
&	| \real\big( u (s) \big) | \le \vartheta_2, \, 
| \imag\big( u (s) \big) | \le \vartheta_2  \Big\}.	\label{eq:set_D}
\end{align}

\section{Proposed joint imaging and DDE calibration}
\label{Sec:algo}

\subsection{Proposed blind approach}
\label{Ssec:Blind_problem}

In this work, we develop a new method for the joint estimation of the original image $\overline{\xb}$, and the DDEs $\overline{\db}_{t,\alpha}$ when both are unknown. To this aim, we consider a global minimization problem, mixing eq.~\eqref{pb:minx} and eq.~\eqref{eq:min_ddes_constraints_C}. Moreover, we assume that the original image $\overline{\xb}$ is the sum of two images, $\xb_o \in \eR^N$ and $\overline{\epsilonb} \in \eR^N$. The first image contains the brightest sources of the original image, and is assumed to be known as prior information, while the second image contains the other unknown sources and has to be estimated. 
Note that this assumption is a usual assumption in the context of DDE calibration methods \citep{Yatawatta2009, Weeren_2016}, and is useful to reduce the ambiguity problems. In particular, the joint calibration and imaging problem being a blind deconvolution problem, any arbitrary modulation of the intensity image can be absorbed in the DDEs without affecting the data in eq.~\eqref{pb:inv_pb_image}. The data are also blind to a common translation of the DDE kernels in the Fourier domain. These two points are mitigated by the assumption that both the intensity and the position of the brightest sources in the sky are known.
Then, the global minimization problem can be written as follows:
\begin{equation}	\label{pb:min_glob}
\minimize{\underset{\small  (\Ubs_{1}, \Ubs_{2})\in (\eC^{Tn_a \times S})^2 }{\small \epsilonb \in \eR^N}}
h(\epsilonb, \Ubs_{1}, \Ubs_{2}) + \eta \, r(\epsilonb) + \nu \, p(\Ubs_{1}, \Ubs_{2}),
\end{equation}
where $h\colon \eR^N\times \eC^{T n_a \times S}\times \eC^{T n_a \times S} \to \eR$ is the data fidelity term corresponding to a least squares criterion and depending on both the image and the DDEs, $r$ is the regularization function for the image (see Section~\ref{Ssec:Min_x}) and $p$ is the regularization function for the DDEs given by eq.~\eqref{eq:reg_DD}. 

According to Sections~\ref{Sec:Imaging} and \ref{Sec:Calib}, the data fidelity term has three different formulations, respectively to the image, to the DDEs contained in $\Ubs_1$, and those contained in $\Ubs_2$. More precisely, we have
\begin{align}	
&	h(\epsilonb, \Ubs_1, \Ubs_2) 	\nonumber\\
&	\! =  \| \Gc(\Ubs_1, \Ubs_2) \Fbs (\xb_o + \epsilonb) - \yb \|_2^2 	\label{eq:h_x_final}\\
&	\! = 
\!\!\!\! \sum_{\underset{1 \le \alpha \le n_a}{1 \le t \le T}} \!
\frac12 \|  \Dc_{t,\alpha,2} \big( \Ubs_{t,2} \big)  \Xc_{t,\alpha,1} \big( \Fbs (\xb_o + \epsilonb) \big) \ub_{t, \alpha,1} - \Ybs_{t,\alpha} \|_2^2
\label{eq:h_u1_final} 	\\
&	\! = 
\!\!\!\! \sum_{\underset{1 \le \alpha \le n_a}{1 \le t \le T}} \!
\frac12 \|  \Dc_{t,\alpha,1} \big( \Ubs_{t,1} \big)  \Xc_{t,\alpha,2} \big( \Fbs (\xb_o + \epsilonb) \big) \ub_{t, \alpha,2} - \Ybs_{t,\alpha} \|_2^2 .	
\label{eq:h_u2_final}
\end{align}
In the formulation~\eqref{eq:h_x_final}, 
$\Gc\colon \eC^{Tn_a \times S} \times \eC^{Tn_a \times S} \to \eC^{TM \times N}$ is the function defined such that $\Gc(\overline{\Ubs}_1, \overline{\Ubs}_2) =\overline{\Gbs}$, $\overline{\Gbs}$ being the matrix described in eq.~\eqref{pb:inv_pb_image}, where each row indexed by $\big(t, \alpha, \beta\big)$ relates to the convolution between the kernels $\ub_{t,\alpha,1}$ and $\ub_{t, \beta,2}$. 
This data fidelity term is used to estimate the image, and we can notice that eq.~\eqref{eq:h_x_final} is convex with respect to $\epsilonb$ when $\Ubs_1$ and $\Ubs_2$ are fixed. 
In the formulation given by eq.~\eqref{eq:h_u1_final} (resp. eq.~\eqref{eq:h_u2_final}), 
$\Xc_{t,\alpha,1} \colon \eC^N \to \eC^{N \times S}$ (resp. $\Xc_{t,\alpha,2} \colon \eC^N \to \eC^{N \times S}$) is the function defined such that $\Xc_{t,\alpha,1}(\Fbs \overline{\xb}) = \Lc_{t,\alpha,1} \big( \overline{\Xbs} \big)$ (resp. $\Xc_{t,\alpha,2}(\Fbs \overline{\xb}) = \Lc_{t,\alpha,2} \big( \overline{\Xbs} \big)$), where $\overline{\Xbs}$ is the matrix defined in eq.~\eqref{eq:pb_inv_matrix}. 
The data fidelity term given in~eq.~\eqref{eq:h_u1_final} (resp. eq.~\eqref{eq:h_u2_final}) is convex with respect to $\Ubs_1$ (resp. $\Ubs_2$) when the other variables are fixed. 
Therefore, even if the function $h$ is non-convex with respect to $(\epsilonb, \Ubs_1, \Ubs_2) $, it is convex when each of the three variables is taken independently and the two others are fixed.

Note that the relevant variables and dimensions associated with the joint calibration and imaging minimization problem are summarized in Table~\ref{Tab:not:min}, and the variables used in the proposed algorithm are summarized in Table~\ref{Tab:not:algo}, given in Appendix~\ref{App:notations}.

\subsection{Optimization tools}

In order to describe our optimization method to solve the joint calibration and imaging problem given in Section~\ref{Ssec:Blind_problem}, we first present in this section the basics of optimization. Note that the definitions given below stand for non-necessarily convex functions. 
We refer the reader to \cite{Rockafellar70, Bauschke_H_2011_book_con_amo} for an overview in convex optimization, and to \cite{Rockafellar97, Mordukhovich06} for non-convex optimization.

A function $\psi \colon \eR^Q \to ]-\infty, +\infty]$ is proper if its domain, denoted by $\dom \psi$, is non-empty. 

Let $\psi$ be a proper, lower-semicontinuous function, bounded from below by an affine function. Its proximity operator \citep{Moreau_J_1965} defined at $\zb \in \eR^Q$, denoted by $\prox_\psi(\zb)$, corresponds to the set of minimizers of $\psi+\frac12 \| \cdot - \zb \|_2^2$, i.e.
\begin{equation}	\label{eq:def_prox}
\prox_\psi(\zb) = \argmin{\ub \in \eR^Q} \psi(\ub) +\frac12 \| \ub - \zb \|_2^2.
\end{equation}
Note that, in the case when $\psi$ is convex, $\prox_\psi(\zb)$ reduces to a singleton. 
Proximity operators are used in optimization algorithms to deal with non-smooth functions, whereas differentiable functions are usually handled by computing gradient steps.

A particular case of proximity operator is the projection operator onto a closed non-empty set $\eE$, denoted by $\Pi_{\eE}$, obtained when $\psi$ is the indicator function of $\eE$.
Another well known proximity operator, often used in signal processing, is the soft-thresholding operator \citep{Chaux_CPW_2007} corresponding to the case when $\psi$ is the $\ell_1$ norm. The $n$-th component of this operator, when applied to a signal $\zb$, is given by
\begin{equation}	\label{eq:prox_l1}
\left[ \prox_{\eta \ell_1}(\zb) \right] (n) = 
\begin{cases}
-z(n) +\eta,	&	\text{if } z(n) < -\eta,	\\
0	,				&	\text{if } -\eta \le z(n) \le \eta,	\\
z(n) - \eta,		&	\text{otherwise},
\end{cases}
\end{equation}
where $\eta>0$ is the threshold parameter. Basically, this operator imposes small values of a signal $\zb$ to be equal to $0$ (more precisely, values $z(n)$ such that $| z(n) | \le \eta$), while the other values are reduced accordingly to $\eta$. 
Note that this operator is a convex relaxation of the hard-thresholding operator \citep{Blumensath_2008} corresponding to the case when $\psi$ is the $\ell_0$ pseudo-norm.

\subsection{Proposed alternated forward-backward algorithm}
\label{Ssec:algo}

To solve the joint calibration and imaging problem described in Section~\ref{Ssec:Blind_problem}, we propose to exploit the block-variable structure of problem~\eqref{pb:min_glob}. To this aim, we design an iterative alternated minimization algorithm, alternating between the estimation of the unknown image $\overline{\epsilonb}$ and the estimation of the DDEs represented by the matrices $\overline{\Ubs}_1$ and $\overline{\Ubs}_2$. 
Furthermore, as explained in Sections~\ref{Sec:Imaging} and \ref{Sec:Calib}, due to the constraints and the $\ell_1$ regularization term, the global objective function is non-smooth. More precisely, it corresponds to a sum of smooth and non-smooth functions, and we propose to solve it using an alternated forward-backward algorithm \citep{Chouzenoux_2016}. This approach combines, at each iteration, a gradient step (forward step) on the Lipschitz-differentiable functions with a proximity step (backward step) on the non-smooth functions.
 
In the context of problem~\eqref{pb:min_glob}, on the one hand, for the image estimation, at each iteration, a gradient step is performed on the differentiable data fidelity term given by eq.~\eqref{eq:h_x_final} while a proximity step is computed to deal with the non-smooth regularization function $r$. On the other hand, for the DDEs estimation, at each iteration the forward step involves not only the data fidelity term given by eq.~\eqref{eq:h_u1_final} (resp. eq.~\eqref{eq:h_u2_final}) but also the differentiable part of the function $p$ given by eq.~\eqref{eq:reg_DD}, i.e. the squared norm of the difference between $\Ubs_1$ and $\Ubs_2$. Then a projection step is required to ensure that the updates of the DDEs satisfy conditions~\eqref{ass:ddes:ii:1}-\eqref{ass:ddes:ii:2}.

The iterations of this method to solve problem~\eqref{pb:min_glob} are given in Algorithm~\ref{algo:global}, and the details are given below.

\begin{algorithm}
\caption{Alternated forward-backward algorithm}\label{algo:global}
\begin{algorithmic}[1] 

\vspace*{0.1cm}

\State 
\fbox{\fbox{\textbf{Initialization}}} 
Let $\epsilonb^{(0)} \in \dom r$ and, for every $t \in \{1, \ldots,T\}$ and $\alpha \in \{1, \ldots, n_a\}$, $(\ub_{t,\alpha,1}^{(0)}, \ub_{t,\alpha,2}^{(0)}) \in \eD^2$. 
Let, for every $i \in \eN$, 
$(J^{(i)}_{\epsilonb}, J^{(i)}_{\Ubs_1}, J^{(i)}_{\Ubs_2}) \in \eN^3$,  
$\gamma^{(i)}_{\epsilonb} \in [0,+\infty[$, 
and, for every $(t,\alpha) \in \{1, \ldots,T\} \times \{1, \ldots, n_a\}$, $ \big( \gamma^{(i)}_{\ub_{t,\alpha,1}} , \gamma^{(i)}_{\ub_{t,\alpha,2}} \big) \in [0,+\infty[^2$.

\vspace*{0.2cm}

\State 
\fbox{\fbox{\textbf{Iterations}}}

\vspace*{0.2cm}

\State 
\textbf{For} $i = 0, 1, \ldots$

\vspace*{0.2cm}

\hspace*{-0.6cm}\fbox{DDEs updating steps: $\Ubs_1$ and $\Ubs_2$}

\vspace*{0.2cm}

\State		\label{algo:steps:Fourier_image_for_DDEs}
\quad	
$\displaystyle \widehat{\xb}^{(i)} =  \Fbs(\xb_o + \epsilonb^{(i)})  $


\vspace*{0.2cm}

\hspace*{-0.2cm}\fbox{Updating steps: $\Ubs_1$}

\vspace*{0.2cm}

\State 	\label{algo:steps:start_V1}
\quad \textbf{for } $t \in \{1, \ldots, T\}$ \textbf{ do in parallel} 

\vspace*{0.1cm}

\State
\quad\quad \textbf{for } $\alpha \in \{1, \ldots, n_a\}$ \textbf{ do in parallel}

\vspace*{0.1cm}

\State
\quad\quad\quad 
$\displaystyle \widetilde{\ub}_{t,\alpha,1}^{(i,0)} = \ub_{t,\alpha,1}^{(i)}$	

\State	\label{algo:step:H1}
\quad\quad\quad 
$\displaystyle  \Hbs_{t,\alpha,1}^{(i)} = \Dc_{t,\alpha,2} \big( \Ubs_{t,2}^{(i)} \big) \Xc_{t,\alpha,1} \big( \widehat{\xb}^{(i)} \big) $

\vspace*{0.1cm}

\State 	\label{algo:step:star_U1}
\quad\quad\quad 
 \textbf{for} $j = 0, \ldots, J^{(i)}_{\Ubs_1}-1$

\State	\label{algo:step:grad_u1}
\quad\quad\quad\quad 
\hspace*{-0.1cm}$\displaystyle \begin{array}{ll}
\wb_{t,\alpha,1}^{(i,j)} = 
&	2 \Hbs_{t,\alpha,1}^{(i) \,\,\, \dagger} ( \Hbs_{t,\alpha,1}^{(i)} \widetilde{\ub}_{t,\alpha,1}^{(i,j)}- \Ybs_{t,\alpha} )	\\
&	\displaystyle + \nu \big( \widetilde{\ub}_{t,\alpha,1}^{(i,j)} - \sigma(\ub_{t,\alpha,2}^{(i)}) \big)
\end{array}$

\State	\label{algo:step:proj_u1}
\quad\quad\quad\quad 
$\displaystyle \widetilde{\ub}_{t,\alpha,1}^{(i,j+1)} = \Pi_{\eD} \left( \widetilde{\ub}_{t,\alpha,1}^{(i,j)} - \gamma^{(i)}_{\ub_{t,\alpha,1}} \wb_{t,\alpha,1}^{(i,j)} \right)$

\State
\quad\quad\quad 
 \textbf{end for}

\State
\quad\quad\quad
$\displaystyle \ub_{t,\alpha,1}^{(i+1)} = \widetilde{\ub}_{t,\alpha,1}^{(i,J^{(i)}_{\Ubs_1})}$	

\State 	\label{algo:step:end_U1}
\quad\quad
 \textbf{end for}

\State	\label{algo:steps:stop_V1}
\quad
 \textbf{end for}


\vspace*{0.2cm}

\hspace*{-0.2cm}\fbox{Updating steps: $\Ubs_2$}

\vspace*{0.2cm}

\State \label{algo:steps:start_V2}
\quad \textbf{for } $t \in \{1, \ldots, T\}$ \textbf{ do in parallel} 

\vspace*{0.1cm}

\State
\quad\quad \textbf{for } $\alpha \in \{1, \ldots, n_a\}$ \textbf{ do in parallel}

\vspace*{0.1cm}

\State
\quad\quad\quad
$\displaystyle \widetilde{\ub}_{t,\alpha,2}^{(i,0)} = \ub_{t,\alpha,2}^{(i)}$	

\State	\label{algo:step:H2}
\quad\quad\quad
$\displaystyle  \Hbs_{t,\alpha,2}^{(i)} = \Dc_{t,\alpha,1} \big( \Ubs_{t,1}^{(i+1)} \big) \Xc_{t,\alpha,2} \big( \widehat{\xb}^{(i)} \big)$

\vspace*{0.1cm}

\State 	\label{algo:step:star_U2}
\quad\quad\quad
 \textbf{for} $j = 0, \ldots, J^{(i)}_{\Ubs_2}-1$

\State	\label{algo:step:grad_u2}
\quad\quad\quad\quad
\hspace*{-0.1cm}$\displaystyle \begin{array}{ll}
\wb_{t,\alpha,2}^{(i,j)} = 
&	2 \Hbs_{t,\alpha,2}^{(i) \,\,\, \dagger} ( \Hbs_{t,\alpha,2}^{(i)} \widetilde{\ub}_{t,\alpha,2}^{(i,j)} - \Ybs_{t,\alpha} )	\\
&	\displaystyle + \nu \big( \widetilde{\ub}_{t,\alpha,2}^{(i,j)} - \sigma(\ub_{t,\alpha,1}^{(i+1)}) \big)
\end{array}$

\State	\label{algo:step:proj_u2}
\quad\quad\quad\quad
$\displaystyle \widetilde{\ub}_{t,\alpha,2}^{(i,j+1)} = \Pi_{\eD} \left( \widetilde{\ub}_{t,\alpha,2}^{(i,j)} - \gamma^{(i)}_{\ub_{t,\alpha,2}} \wb_{t,\alpha,2}^{(i,j)} \right)$

\State
\quad\quad\quad 
 \textbf{end for}

\State
\quad\quad\quad
$\displaystyle \ub_{t,\alpha,2}^{(i+1)} = \widetilde{\ub}_{t,\alpha,2}^{(i,J^{(i)}_{\Ubs_2})}$	

\State 	\label{algo:step:end_U2}
\quad\quad
 \textbf{end for}

\State \label{algo:steps:stop_V2}
\quad
 \textbf{end for}


\vspace*{0.2cm}

\hspace*{-0.6cm}\fbox{Image updating steps: $\epsilonb$}
 
 \vspace*{0.2cm}

\State \label{algo:steps:start_im}
\quad $\displaystyle \widetilde{\epsilonb}^{(i,0)} = \epsilonb^{(i)}$

\State 	\label{algo:step:G}
\quad $\Gbs^{(i)} = \Gc( \Ubs_1^{(i)}, \Ubs_2^{(i)})$

\vspace*{0.1cm}

\State 	\label{algo:step:star_eps}
\quad \textbf{for} $j = 0, \ldots, J^{(i)}_{\epsilonb}-1$

\State 	\label{algo:step:grad_x}
\quad\quad 
$\displaystyle \zb^{(i,j)} =  2 \Fbs^\dagger \Gbs^{(i)}\phantom{\big|}^\dagger \Big(\Gbs^{(i)} \Fbs (\xb_o+ \widetilde{\epsilonb}^{(i,j)}) - \yb \Big)$

\State 	\label{algo:step:prox_x}
\quad\quad 
$\displaystyle \widetilde{\epsilonb}^{(i,j+1)} = \prox_{\gamma^{(i)}_{\epsilonb} \eta r} \Big( \widetilde{\epsilonb}^{(i,j)} - \gamma^{(i)}_{\epsilonb} \zb^{(i,j)}  \Big) $

\State  	\label{algo:step:end_eps}
\quad \textbf{end for}

\State \label{algo:steps:stop_im}
\quad 
$\displaystyle \epsilonb^{(i+1)} = \widetilde{\epsilonb}^{(i,J^{(i)}_{\epsilonb})}$
 
 \vspace*{0.2cm}

\State
\textbf{end for}
 
 \vspace*{0.1cm}

\end{algorithmic}
\end{algorithm}

\subsubsection{Number of sub-iterations}
\label{Sssec:algo:sub-it}

At each global iteration $i\in \eN$ of the algorithm, for both variables $\Ubs_1$ and $\Ubs_2$, a given number of sub-iterations $J_{\Ubs_1}^{(i)}$ (resp. $J_{\Ubs_2}^{(i)}$) are performed in order to obtain an update $\Ubs_1^{(i+1)}$ (resp. $\Ubs_2^{(i+1)}$) as accurate as desired by the user. 
Similarly, at iteration $i$, a fixed number of sub-iterations $J_{\epsilonb}^{(i)}$ are performed in order to obtain the update $\epsilonb^{(i+1)}$ of the image. 

Basically the user can choose any values for $J_{\Ubs_1}^{(i)}$, $J_{\Ubs_2}^{(i)}$ and $J_{\epsilonb}^{(i)}$ provided they satisfy the conditions given in Section~\ref{Ssec:algo:conv} (i.e. they are finite values, and each of the variables $\Ubs_1$, $\Ubs_2$ and $\epsilonb$ need to be estimated at least once during the iteration process). 
However, to simplify the explanation of our method, we propose to estimate at each iteration $i$ either the DDEs or the image. 
To this purpose, 
note that for a given iteration $i$, when $J_{\Ubs_1}^{(i)} = J_{\Ubs_2}^{(i)} = 0$ and $J_{\epsilonb}^{(i)} > 0$, only the image is updated while the DDEs are kept fixed, i.e. $\Ubs_1^{(i+1)} = \Ubs_1^{(i)}$ and $\Ubs_2^{(i+1)} = \Ubs_2^{(i)}$. 
In contrast, if we choose $J_{\Ubs_1}^{(i)}, J_{\Ubs_2}^{(i)} > 0$ and $J_{\epsilonb}^{(i)} = 0$, then only the DDEs are updated and the image is kept fixed at iteration $i$, i.e. $\epsilonb^{(i+1)} = \epsilonb^{(i)}$.

As mentioned in Section~\ref{Sec:Idea}, Figure~\ref{Graph:algo_BCFB} (left) shows a diagram giving the relevant steps of the proposed method. It resumes the global structure of the algorithm and describes the number of sub-iterations performed on $\Ubs_1$, $\Ubs_2$ and $\epsilonb$, respectively. 
In particular, in our method, we define a \emph{cycle} to be the minimum iterations of Algorithm~\ref{algo:global} needed to estimate at least once the DDEs and once the image, and we denote by $J_{\text{cyc}}\ge 2$ the global number of iterations of Algorithm~\ref{algo:global} to perform a cycle. 
More precisely, to estimate the DDEs, $(J_{\text{cyc}}-1)$ iterations of Algorithm~\ref{algo:global} are performed to estimate $(\Ubs_1, \Ubs_2)$. For each of these iterations, $J_{\Ubs_1}^{(i)}$ forward-backward steps are computed to estimate $\Ubs_1$, followed by $J_{\Ubs_2}^{(i)}$ forward-backward steps to estimate $\Ubs_2$. Then, to complete one global cycle, $J_{\epsilonb}^{(i)}$ forward-backward steps are performed to estimate the image $\epsilonb$.

The parameters $(J_{\Ubs_1}^{(i)}, J_{\Ubs_2}^{(i)}, J_{\epsilonb}^{(i)}, J_{\text{cyc}})$ are used in order to alternate more than once between $\Ubs_1$ and $\Ubs_2$ before updating the image $\epsilonb$. 
This flexibility is important in practice since it can accelerate the convergence speed of the global algorithm. 
Furthermore, the reconstruction quality is not highly dependent on the numbers of sub-iterations, provided that they are finite to ensure the convergence of the algorithm (see Section~\ref{Ssec:algo:conv}). 
In fact, if these parameters are chosen too large, the number of iterations performed on the DDEs and on the image may not be balanced, and the algorithm may reach local minima.

Note that to simplify the readability of Algorithm~\ref{algo:global}, the parameter $ J_{\text{cyc}}$ does not appear explicitly. Then to take it into account, we propose to choose $(J_{\Ubs_1}^{(i)}, J_{\Ubs_2}^{(i)})$ such that 
\begin{enumerate}
\item
For every $(i +1)\neq 0 \Big[\text{mod } J_{\text{cyc}}\Big]$, we have $J_{\epsilonb}^{(i)} = 0$ and $J_{\Ubs_1}^{(i)}, J_{\Ubs_2}^{(i)} > 0$, 
\item
For every $(i+1) = 0 \Big[\text{mod } J_{\text{cyc}}\Big]$, we have $J_{\Ubs_1}^{(i)} = J_{\Ubs_2}^{(i)} = 0$ and $J_{\epsilonb}^{(i)} > 0$. 
\end{enumerate}

For instance, if $J_{\text{cyc}} = 2$, at iterations $i=0, 2, 4, \ldots$ of the algorithm, only the DDEs $\Ubs_1^{(i)}$ and $\Ubs_2^{(i)}$ are updated, computing respectively $J_{\Ubs_1}^{(i)}>0$ and $ J_{\Ubs_2}^{(i)}>0$ forward-backward steps, while at iterations $i=1, 3, 5, \ldots$ of the algorithm only the image $\epsilonb^{(i)}$ is updated, computing $J_{\epsilonb}^{(i)}>0$ forward-backward steps. 
If we want to update 2 times more the DDEs, we choose $J_{\text{cyc}} = 3$. Then, at iterations $i = 0, 1, 3, 4, 6, 7, \ldots$ of the algorithm, only the DDEs are updated, while the image is only updated at iterations $i = 2, 5, 8, \ldots$. Consequently, increasing $J_{\text{cyc}}$ allows to alternate more often between $\Ubs_1^{(i)}$ and $\Ubs_2^{(i)}$ before estimating $\epsilonb^{(i)}$. However, this number has to be finite in accordance with the convergence conditions given in Section~\ref{Ssec:algo:conv}.

\subsubsection{Estimation of DDEs}
\label{Sssec:param_DDE_estim}

Algorithm~\ref{algo:global} alternates between the estimation of the non-zero coefficients of the Fourier transform of the DDEs (steps~\ref{algo:steps:start_V1} to \ref{algo:steps:stop_V2}) and the estimation of the image (steps~\ref{algo:steps:start_im} to \ref{algo:steps:stop_im}). 
Concerning the estimation of the DDEs, 
the proposed algorithm updates alternately $\Ubs_1$ in steps~\ref{algo:steps:start_V1}-\ref{algo:steps:stop_V1} and $\Ubs_2$ in steps~\ref{algo:steps:start_V2}-\ref{algo:steps:stop_V2}. 
It is worth noting that these steps update in parallel the DDEs for all the antennas and at all time instants. 
In the following, several additional remarks are made on these steps.

\paragraph*{Gradient steps}

Steps~\ref{algo:step:grad_u1} and \ref{algo:step:grad_u2} correspond to the gradient steps performed on the variables $\Ubs_1$ and $\Ubs_2$, respectively. 
These steps being symmetric, we focus on the description of step~\ref{algo:step:grad_u1}. 

In problem~\eqref{pb:min_glob}, there are two differentiable functions corresponding to the data fidelity term given by eq.~\eqref{eq:h_x_final} and the smooth regularization function controlling the distance between $\Ubs_1$ and $\Ubs_2$. These two functions are complex valued, and as described in \citep{Smirnov_2015}, to deal directly with these functions, adapted tools from complex analysis have to be used. 
However, one can note that, for every $\ub \in \eC^S$ and for every $\Hbs \in \eC^{(n_a-1) \times S}$, we have 
$\Hbs \ub  = [ \Hbs \, | \, \is \, \Hbs ] 
\left[\begin{matrix}
\real(\ub)	\\
\imag(\ub)
\end{matrix}\right]$. 
This relation implies that working with complex valued variables of dimension $S$ is equivalent to deal with real valued variables of dimension $2S$. Thus, to leverage recent results from non-convex optimization, the gradient given in step~\ref{algo:step:grad_u1} corresponds to the gradient of the smooth functions involved in eq.~\eqref{pb:min_glob}, rewritten as real-valued functions. 

\paragraph*{Projection steps}

Steps~\ref{algo:step:proj_u1} and \ref{algo:step:proj_u2} correspond to projection steps on $\eD$, where $\eD$ is defined by eq.~\eqref{eq:set_D}, for each row of variables $\Ubs_1$ and $\Ubs_2$, respectively. 
Similarly to the gradient steps, these steps are symmetric. Thus we focus on the description of step~\ref{algo:step:proj_u1}.

Let $\ub = \big(u(s)\big)_{-S/2 \le s \le S/2-1} \in \eC^S$ be a complex vector which can be decomposed as $\ub = \real(\ub) + \is \, \imag(\ub)$. Then the projection of $\ub$ in $\eD$, is given by $ \Pi_{\eD}(\ub) = \big( v(s) \big)_{-S/2 \le s \le S/2-1} $, where 
\begin{equation}
\begin{cases}
\real(v(0)) = \min\{ \max \{ \real(u(0)), -\vartheta_1 \}, \vartheta_1 \},	\\
\imag(v(0)) = \min\{ \max \{ \imag(u(0)), -\vartheta_1 \}, \vartheta_1 \},
\end{cases}
\end{equation}
and, for every $s \neq 0$,
\begin{equation}
\begin{cases}
\real(v(s)) = \min\{ \max \{ \real(u(s)), -\vartheta_2 \}, \vartheta_2 \},	\\
\imag(v(s)) = \min\{ \max \{ \imag(u(s)), -\vartheta_2 \}, \vartheta_2 \}.	
\end{cases}
\end{equation}
This operation corresponds to a shrinkage of the highest/lowest amplitude values of the Fourier coefficients of the DDEs, in order to satisfy conditions~\eqref{ass:ddes:ii:1}-\eqref{ass:ddes:ii:2}.

\subsubsection{Estimation of the image}

The proposed Algorithm~\ref{algo:global} has the same structure for the estimation of the DDEs and the image, in the sense that forward-backward iterations are used in both cases.

\paragraph*{Gradient step}

Step~\ref{algo:step:grad_x} corresponds to the gradient step performed on the image. For the imaging problem, this step involves only the least-squares criterion given by eq.~\eqref{eq:h_x_final}. Note that, the gradient of this function, computed at the current sub-iterate $\widetilde{\epsilonb}^{(i,j)}$ relies on computing the residual image associated with the full current image $ (\xb_o+ \widetilde{\epsilonb}^{(i,j)}) $.

\paragraph*{Proximity step}

Step~\ref{algo:step:prox_x} involves the computation of the proximity operator of the function $r$, corresponding to the regularization term associated with the image. 
According to remarks made in Section~\ref{Ssec:Min_x}, we propose to promote sparsity of the global image $\xb_o + \epsilonb$ using the $\ell_1$ norm and to penalize its negative coefficients thanks to the indicator function of the positive orthant $[0,+\infty[^N$. 

In the case when the original image $\overline{\xb}$ is assumed to be sparse, we choose
\begin{equation}  \label{eq:def_reg_x_sparse}
r(\epsilonb) = \| \epsilonb \|_1 + \iota_{\eE}(\epsilonb),
\end{equation} 
where $\eE $ is an $N$ dimensional box. 
More precisely, we know that the original global image $\overline{\xb}$ corresponds to the sum of $\xb_o$ and $\overline{\epsilonb}$. Let $\eI\subset \{-N/2, \ldots, N/2-1\}$ be the set of indices giving the exact position of the known bright sources contained in the image $\xb_o$. In other words, $\eI$ is the support of $\xb_o$. Then, we have, for every $n \in \eI$, $x_o(n) = \overline{x}(n)$. In this context, we oblige the coefficients $\epsilon(n)$ such that $n \in \eI$ to be equal to $0$, while we constrain the other coefficients to be positive. More formally, we define
\begin{equation}\label{eq:def_C}
\eE = \Menge{\epsilonb \in \eR^N}{ (\forall n \in \eI) \, \epsilon(n) = 0, \text{ and } (\forall n \not\in \eI) \, \epsilon(n) \ge 0}.
\end{equation}
In this context, the proximity operator $\prox_{\gamma_{\epsilonb}^{(i)} \eta r }$ of $\gamma_{\epsilonb}^{(i)}\eta r$ at $\zb \in \eR^N$, for $\gamma_{\epsilonb}^{(i)}>0$ and $\eta>0$, corresponds to the projection onto $\eE$ composed with a soft-thresholding operator (see eq.~\eqref{eq:prox_l1}), i.e.
\begin{equation}
\prox_{\gamma_{\epsilonb}^{(i)} \eta r }(\zb) =
\Pi_{\eE} \Big( \prox_{\gamma_{\epsilonb}^{(i)} \eta \| \cdot \|_1}(\zb )  \Big).
\end{equation}

As described in Section~\ref{Ssec:Min_x}, when the unknown original image $\overline{\xb}$ is not sparse in its domain, we can decompose it in a given dictionary $\Psib$ in order to promote sparsity of its coefficients $\Psib^\dagger \overline{\xb}$. In this case, the regularization term is chosen to be
\begin{equation} \label{eq:def_reg_x_dict}
r(\epsilonb) = \| \Psib^\dagger (\xb_o + \epsilonb) \|_1 + \iota_{\eE}(\epsilonb).
\end{equation}
The proximity operator of the above function does not have an explicit formula, and it has to be computed approximately with sub-iterations (e.g. using the Dual forward-backward algorithm \citep{Combettes_Vu_2011}). 

More generally, note that the regularization function $r$ can be non-convex but, for convergence purposes, needs to be semi-algebraic\footnote{Semi-algebraicity is a property satisfied by a wide class of functions, which means that their graph is a finite union of sets defined by a finite number of polynomial inequalities. In particular, it is satisfied for the functions mentioned in~eq.~\eqref{eq:def_reg_x_sparse} and~eq.~\eqref{eq:def_reg_x_dict}.}.

\subsection{Convergence results}
\label{Ssec:algo:conv}

The proposed method is based on a block-coordinate forward-backward algorithm developed by \cite{Chouzenoux_2016} (other versions can be found in \cite{Frankel2014, Bolte_2014}). In particular, we can deduce from this paper the convergence guarantees of Algorithm~\ref{algo:global} given below.\footnote{Note that the algorithms, and the associated convergence results, presented in \cite{Chouzenoux_2016, Frankel2014, Bolte_2014} are designed to minimize real valued functions, and thus are not directly applicable to solve problem~\eqref{pb:min_glob}. 
However, to design Algorithm~\ref{algo:global}, problem~\eqref{pb:min_glob} has been reformulated to deal with real valued functions.}

Assume that, for each iteration $i\in \eN$ of Algorithm~\ref{algo:global}, the step-size $\gamma_{\epsilonb}^{(i)}$, associated with the image update, and, for every $(t,\alpha) \in \{1 , \ldots, T\}\times \{1, \ldots, n_a\}$ the step-sizes $ \gamma_{\ub_{t,\alpha,1}}^{(i)} $, $ \gamma_{\ub_{t,\alpha,2}}^{(i)} $, associated with the DDEs updates, satisfy
\begin{align}
&	0 < \gamma_{\ub_{t,\alpha,1}}^{(i)} < 2 / \big(\nu + 2\| \Hbs_{t,\alpha,1}^{(i)} \|_{S} \big)	,	\\
&	0 < \gamma_{\ub_{t,\alpha,2}}^{(i)} < 2 / \big(\nu + 2\| \Hbs_{t,\alpha,2}^{(i)} \|_{S} \big)	,	\\
&	0 < \gamma_{\epsilonb}^{(i)} <   \| \Gbs^{(i)} \|_{S} 	,	
\end{align}
where $\| . \|_S$ denotes the spectral norm, and $\Hbs_{t,\alpha,1}^{(i)}$, $\Hbs_{t,\alpha,2}^{(i)}$, and $\Gbs^{(i)}$ are the matrices built in steps~\ref{algo:step:H1}, \ref{algo:step:H2} and \ref{algo:step:G}, respectively. 
Moreover, assume that the number of sub-iterations determined by $J_{\Ubs_1}^{(i)}$, $J_{\Ubs_2}^{(i)}$, and $J_{\epsilonb}^{(i)}$ are finite and that there exists $(i',i'',i''') \in \eN^3$ such that $J_{\Ubs_1}^{(i')}>0$, $J_{\Ubs_2}^{(i'')}>0$ and $J_{\epsilonb}^{(i''')}>0$. This latter condition ensures that all the different variables (i.e. the DDEs and the image) will be updated at least once during the iterative process. 
Then, 
the sequence of iterates $\Big( \epsilonb^{(i)}, \Ubs_1^{(i)}, \Ubs_2^{(i)} \Big)_{i \in \eN}$ generated by Algorithm~\ref{algo:global} converges to a critical point $ \Big( \epsilonb^{\star}, \Ubs_1^{\star}, \Ubs_2^{\star} \Big) $ of the objective function in eq.~\eqref{pb:min_glob}. Moreover, the objective function value is decreasing along the iterations.

Finally, \citet[Cor. 3.1]{Chouzenoux_2016} ensures that Algorithm~\ref{algo:global} converges globally to a solution to problem~\eqref{pb:min_glob} if it is initialized not too far from this solution. This result shows the importance of the initialization of algorithms solving non-convex problems. 
Therefore, the initialization of Algorithm~\ref{algo:global} is explained in detail in the following section.

\subsection{Initialization}
\label{Ssec:Algo:Init}

As explained above, problem~\eqref{pb:min_glob} being non-convex, the final solution of the DDEs and the image obtained by Algorithm~\ref{algo:global} is likely to be sensitive to the initialization used for these unknowns. 
Therefore, we propose an empirical approach to initialize the proposed algorithm, leading in practice to a stable global method, as attested by the simulations presented in Section~\ref{Sec:Sim}. 
This initialization is composed of two steps described below.

\paragraph*{Step 1.}

The initialization of the image and the DDEs is done during the same process. It relies on the assumption that the DDEs are band-limited, and that the zero spatial frequency coefficients (the DIEs) of the DDEs $\widehat{\overline{\db}}_{t,\alpha}$ are much larger than the other coefficients. 
Thus, we propose to first jointly estimate the image $\overline{\epsilonb}$ and only the DIEs $\widehat{\overline{d}}_{t,\alpha}(0)$ using Algorithm~\ref{algo:global}. 
This involves solving problem~\eqref{pb:min_glob} considering a support of size $ 1\times 1$ for the DDEs, ignoring the other Fourier coefficients $\big( \widehat{\overline{d}}_{t,\alpha}(k) \big)_{k \in \Sc\setminus \{0\}}$ modelled in the inverse problem~\eqref{eq:obs_alpha_beta}. 
In other words, even if the inverse problem of interest includes DDEs, the initialization step we propose ignores this fact and solves the problem taking into account only DIEs in the algorithm.

For this first run of Algorithm~\ref{algo:global}, DIEs $\widehat{\overline{d}}_{t,\alpha}(0)$ are initialized randomly, and the image $\overline{\epsilonb}$ is chosen to be the zero image.

At the end of this first step, we obtain an estimation of the DIEs, denoted by $\check{\Ubs}_1$ and $\check{\Ubs}_2$, and an estimation of the unknown image $\overline{\epsilonb}$, denoted by $\check{\epsilonb}$.

\paragraph*{Step 2.}

Once we have a first estimation of the DIEs from \textbf{Step 1}, we use it to initialize the variables $\widehat{{\db}}_{t,\alpha}^{(0)}$ to estimate the full DDEs. 
More precisely, in Algorithm~\ref{algo:global}, $\widehat{{d}}_{t,\alpha}^{(0)}(0)$ is taken to be the DIE estimated in \textbf{Step 1}, while the other coefficients are chosen randomly.

Concerning the initialization of the image, preliminary simulations have shown that the first estimation obtained from \textbf{Step 1} is not necessarily good. In some cases, the estimation only contains artefacts and will slow down the convergence of the Algorithm, but in worse cases, it can even lead to poor reconstruction results. 
This can be explained by the fact that in \textbf{Step 1} only the DIEs are estimated and not the DDEs, which is not enough to reconstruct high dynamic range images.

Therefore we examine the information content of the reconstructed image from \textbf{Step 1} by comparing it with the considered prior image $\xb_o$.

Intuitively, if the amplitudes of the unknown sources belonging to $\overline{\epsilonb}$ are close to the amplitudes of the known sources in $\xb_o$, considering DIEs may lead to a first good estimation of $\overline{\epsilonb}$. However, in the case when the amplitudes of the unknown sources are much lower than the amplitudes of the known sources, DIEs will not lead to an accurate estimation. 
For this purpose, we design an automatic selective method, based on the comparison of the energy contained in both images, in the Fourier domains. 

Let $\check{\Gbs} = \Gc(\check{\Ubs}_1, \check{\Ubs}_2)$ be the convolution matrix associated with the DIEs estimated in \textbf{Step 1} ($\Gc$ being defined in eq.~\eqref{eq:h_x_final}). 
Then, we compare the energy, i.e. the $\ell_2$ norm, of $\check{\Gbs} \Fbs \check{\epsilonb}$ with that of $\check{\Gbs} \Fbs \xb_o$, corresponding respectively to the energies of images $\check{\epsilonb}$ and $\xb_o$ taking into account only the measured frequencies from their Fourier coefficients. 
In the context of radio interferometry imaging, $u-v$ coverages tend to measure more frequently low frequencies in the Fourier domain, containing mainly the information on the approximations of the image. Then, comparing the energy of $\check{\Gbs} \Fbs \check{\epsilonb}$ with the energy of $\check{\Gbs} \Fbs \xb_o$ gives a weighted information, without taking into account details of the image which are known to be not recovered by DIEs. 

Empirically, we propose to distinguish three different cases:
\begin{enumerate}
\item
If $\ell_2( \check{\Gbs} \Fbs \check{\epsilonb} ) < \underline{\tau} \ell_2( \check{\Gbs} \Fbs \xb_o)$, where $\underline{\tau}\ge 0$, we assume that the unknown sources have an amplitude much smaller than the amplitude of the known bright sources, and thus cannot be reconstructed correctly only by considering DIEs. Then we consider that the image estimated with \textbf{Step 1} cannot be trusted. In this case, we discard this estimate, and we initialize Algorithm~\ref{algo:global} for joint imaging with DDE calibration considering $\epsilonb^{(0)}$ to be a zero image.

\item
If $ \underline{\tau} \ell_2( \check{\Gbs} \Fbs \xb_o) \le \ell_2( \check{\Gbs} \Fbs \check{\epsilonb} ) \le \overline{\tau} \ell_2( \check{\Gbs} \Fbs \xb_o)$, with $\overline{\tau} \ge \underline{\tau}$, we assume that the unknown sources and the known bright sources have similar amplitudes, and thus $\overline{\epsilonb}$ could be reconstructed approximately only considering DIEs. Then we consider that the image estimated in \textbf{Step 1} can be trusted. 
In this case, we use $\check{\epsilonb}$ to initialize $\epsilonb^{(0)}$ in Algorithm~\ref{algo:global} for joint imaging with DDE calibration.

\item
If $ \overline{\tau} \ell_2( \check{\Gbs} \Fbs \xb_o) \le \ell_2( \check{\Gbs} \Fbs \check{\epsilonb} )$, 
we consider that the image estimated with \textbf{Step 1} can be trusted but may contain some artefacts due to DIEs. 
Thus, we threshold the estimate $\check{\epsilonb}$, until having $ \overline{\tau} \ell_2( \check{\Gbs} \Fbs \xb_o)  \ge \ell_2( \check{\Gbs} \Fbs \check{\epsilonb} )$.
Then, $\epsilonb^{(0)}$ in Algorithm~\ref{algo:global} for joint imaging with DDE calibration is taken to be the thresholded image.
\end{enumerate}

\paragraph*{Remarks.}

Note that the proposed initialization process is totally automatic. Moreover, we observed during preliminary simulations that it leads to stable results, in the sense that considering, for the same experiment, different random initializations both for DIEs in \textbf{Step 1} and DDEs in \textbf{Step 2} leads to similar reconstruction results. 
In addition, it is interesting to emphasize that \textbf{Step 1} of the proposed initialization is very similar to perform the StEFCal method developed by~\cite{Salvini_2014}, coupled with an imaging method based on the forward-backward algorithm \citep{Combettes_Wajs_2005}. In particular, we observed numerically that these two methods give very similar results. However, from an algorithmic view point, the calibration part of these methods are different since \textbf{Step 1} relies on an alternated forward-backward iterations, while StEFCal is based on an alternating direction implicit method. Moreover, unlike \textbf{Step 1} which can be seen as a joint DIE calibration and imaging method, StEFCal is not an imaging method.
Finally, note that if only DIEs are considered in the observation model, then \textbf{Step 1} can be seen as a selfcal method, with convergence guarantees (and is enough to jointly estimate the DIEs and the image).

\subsection{Additional stopping criteria}
\label{Ssec:algo:stop_crit}

In order to avoid useless gradient and proximal steps computation, 
we introduce additional stopping criteria in Algorithm~\ref{algo:global}. 
Firstly, as explained earlier, the algorithm performs $J_{\text{cyc}}-1$ updates of the DDEs for a fixed image, where $J_{\text{cyc}}$ is chosen in order to avoid a complete convergence of the DDEs when the image is only known approximately. 
However, in practice, when the global algorithm has already performed a large number of iterations, the DDE updates can stabilise in less than $J_{\text{cyc}}-1$ iterations. Therefore, we introduce parameter $ \xi_{\Ubs_{\text{tot}}}>0$, corresponding to an additional stopping criterion such that, for every $i \in \eN$ satisfying $(i+1) \neq 0 \;  \Big[\text{mod } J_{\text{cyc}}\Big]$, 
\begin{multline}	\label{crit_stop:Utot}
\max \left\{ 
\max_{t,\alpha} \Big\{
\dfrac{\| \ub_{t,\alpha,1}^{(i+1)} - \ub_{t,\alpha,1}^{(i)} \|_2}{\| \ub_{t,\alpha,1}^{(i+1)} \|_2} \Big\},  
\right.	\\
\left.
\max_{t,\alpha} \Big\{
\dfrac{\| \ub_{t,\alpha,2}^{(i+1)} - \ub_{t,\alpha,2}^{(i)} \|_2}{\| \ub_{t,\alpha,2}^{(i+1)} \|_2} \Big\}
\right\} < \xi_{\Ubs_{\text{tot}}}.
\end{multline}
Then, if this criterion is satisfied, the algorithm stops estimating the DDEs during the current cycle, and estimates directly the image.

Similarly, we introduce parameter $\xi_{\epsilonb}>0$, corresponding to an additional stopping criterion to avoid useless updates of the image in the inner-loop given by steps \ref{algo:step:star_eps}-\ref{algo:step:end_eps} in Algorithm~\ref{algo:global}.
Then, for every $i \in \eN$ satisfying $(i+1) = 0 \;  \Big[\text{mod } J_{\text{cyc}}\Big]$, we consider 
\begin{equation}	\label{crit_stop:im}
(\forall j \in \{0, \ldots, J_{\epsilonb}^{(i)}-1\})\quad
\dfrac{\| \widetilde{\epsilonb}^{(i,j+1)} - \widetilde{\epsilonb}^{(i,j)} \|_2}{\| \widetilde{\epsilonb}^{(i,j+1)} \|_2} < \xi_{\epsilonb}.
\end{equation}
Then, if this criterion is satisfied, the algorithm stops estimating the image and starts a new cycle, estimating again the DDEs.

Finally, to stop the global algorithm at convergence, we use a stopping criterion depending on the objective function. More precisely, we introduce $\zeta>0$, and we define, for every $i\in \eN$,
\begin{equation}	\label{crit_stop:obj}
| \varphi^{(i+1)}  - \varphi^{(i)} | / \varphi^{(i+1)} < \zeta,
\end{equation}
where $\varphi^{(i+1)}$ is the value of the objective function to be minimized in~eq.~\eqref{pb:min_glob} evaluated at the current update $(\epsilonb^{(i+1)}, \Ubs_1^{(i+1)}, \Ubs_2^{(i+1)})$. 

Note that these stopping criteria are used in the algorithm to avoid useless calculus when the iterates are almost constant along the iterations. They can be seen as additional optional parameters to $J_{\text{cyc}}$, $J_{\Ubs_1}$, $J_{\Ubs_2}$, and $J_{\epsilonb}$ to ensure the global convergence of Algorithm~\ref{algo:global}.

\subsection{Computational cost}
\label{Ssec:algo:cost}

The objective of this section is to provide information on the computational cost of the proposed algorithm. 
As described in Section~\ref{Sssec:algo:sub-it}, at each iteration $i\in \eN$, either the DDEs or the image are updated, and one global cycle of the algorithm performs $J_{\text{cyc}}-1$ iterations updating only the DDEs and one iteration updating only the image. 
Therefore, we describe below the behaviour of the first cycle of Algorithm~\ref{algo:global}, which has to be repeated until convergence, in order to provide rough information on the computational cost of the proposed method.

The first $J_{\text{cyc}}-1$ iterations of the cycle are used to update the DDEs. 
To this aim, a discrete Fourier transform of the global image $(\xb_o + \epsilonb)$ is performed in step~\ref{algo:steps:Fourier_image_for_DDEs}. 
Then, for every $i \in \{0, \ldots, (J_{\text{cyc}}-1)-1\}$, 
$\Ubs_1$ and $\Ubs_2$ are updated sequentially. Firstly, $\Ubs_1$ is updated by implementing $T  n_a $ parallel steps consisting in (i) the construction of the observation matrices of dimension $(n_a-1) \times S$ (described in step~\ref{algo:step:H1}), and (ii) $J_{\Ubs_1}^{(i)}$ forward-backward steps on a variable of dimension $S$ (described in steps~\ref{algo:step:grad_u1}-\ref{algo:step:proj_u1}). Then $\Ubs_2$ is updated similarly (using steps~\ref{algo:step:H2}, \ref{algo:step:grad_u2}, and \ref{algo:step:proj_u2}). 
Therefore, in total for the update of the DDEs, $T  n_a $ parallel steps are performed, each computing $2 (J_{\text{cyc}}-1)$ times (i), and 
$\sum_{i=0}^{J_{\text{cyc}}-2} (J_{\Ubs_1}^{(i)} + J_{\Ubs_2}^{(i)}) $
times (ii).

The last iteration $i = J_{\text{cyc}}-1$ of the first cycle is dedicated to the update of the image. In this case, the convolution matrix is built in step~\ref{algo:step:G}, followed by $J_{\epsilonb}^{(i)}$ forward-backward steps on a variable of dimension $N$. These forward-backward steps are described in steps~\ref{algo:step:grad_x}-\ref{algo:step:prox_x}. Note that they are the heaviest steps of the algorithm. 
Indeed, the forward step consists in performing a discrete Fourier transform on the global image, followed by an inverse discrete Fourier transform. 
Concerning the backward step, if the regularisation described in eq.~\eqref{eq:def_reg_x_sparse} is considered, then step~\ref{algo:step:prox_x} reduces to a soft-thresholding operation. However, in the case when eq.~\eqref{eq:def_reg_x_dict} is considered, then step~\ref{algo:step:prox_x} is computed using to sub-iterations.

Note that the description of the algorithm given here does not take into account the stopping criteria described in Section~\ref{Ssec:algo:stop_crit}. In particular, the condition given in eq.~\eqref{crit_stop:Utot} is used to perform at maximum $J_{\text{cyc}}-1$ iterations estimating the DDEs, while the condition given in eq.~\eqref{crit_stop:im} is used to perform at maximum $J_{\epsilonb}^{(i)}$ sub-iterations estimating the image. 
Examples of numbers of sub-iterations $J_{\epsilonb}^{(i)}$, $J_{\Ubs_1}^{(i)}$, $J_{\Ubs_2}^{(i)}$ and $J_{\text{cyc}}$ and of stopping criteria values are given in the simulation sections, in Tables~\ref{Tab:param_point_sources} and \ref{Tab:param_M31}.

Finally, it is important to emphasize that for the sake of simplicity the current implementation uses fast Fourier transform (FFT) for the Fourier transforms and do not take into account the gridding and degridding steps needed to perform non-uniform Fourier transforms \citep{Fessler_2003}. These steps need to be incorporated in the convolution matrices built in steps~\ref{algo:step:H1}-\ref{algo:step:H2} (for the DDEs estimation) and~\ref{algo:step:G} (for the image estimation).

\section{StEFCal for DDE calibration}
\label{Sec:stefcal}

For the sake of completeness, we propose in this section a naive generalization of the StEFCal method developed by~\cite{Salvini_2014} for DDE calibration. However, the obtained method is not guaranteed to converge, fully justifying the development of a more sophisticated approach, described in the previous sections.

In the particular case when only DIEs appear in the inverse problem~\eqref{pb:inv_pb_image} (or its equivalent matrix formulation), the calibration problem can be solved efficiently using the StEFCal algorithm. This method consists in estimating the zero spatial frequency elements $\big( \widehat{\overline{d}}_{t,\alpha}(0) \big)_{1 \le \alpha \le n_a}$ of the DIEs, by solving problem~\eqref{eq:min_dies_sym} using an alternating direction implicit (ADI) method. 
Note that this approach is a DIE calibration method, which can be combined with an imaging algorithm, and is not adapted for the DDE calibration problem.

The StEFCal approach used in the previous sections can be naively generalized to solve the DDE calibration problem described in~eq.~\eqref{eq:min_ddes_constraints_C}, with $\nu = 0$. In this case, the ADI algorithm defines at each iteration the update $\Ubs_1^{+}$ of $\Ubs_1$ by setting, for every $t\in \{1, \ldots,T\}$ and $\alpha \in \{1, \ldots,n_a\}$,
\begin{multline}
\ub_{t,\alpha,1}^{+} \\
= \argmin{\ub} 
\|  \Dc_{t,\alpha,2} \big( \Ubs_{t,2} \big)  \Xc_{t,\alpha,1} \big( \Fbs (\xb_o+\epsilonb^\star) \big) \ub - \Ybs_{t,\alpha} \|_2^2,
\end{multline}
where $\Dc_{t,\alpha,2} \big( \Ubs_{t,2} \big)$ and $ \Ybs_{t,\alpha} $ are defined in~eq.~\eqref{eq:data_fid_d_C1}, $\Xc_{t,\alpha,1}$ is the function defined in eq.~\eqref{eq:h_u1_final}, and $\epsilonb^\star$ is the current estimation of the image. 
Then, $\Ubs_2$ is updated by taking $\ub_{t,\alpha,2}^{+} = \sigma \big( \ub_{t,\alpha,1}^{+} \big)$, where $\sigma$ is the bijection defined in~eq.~\eqref{eq:reg_DD}. 
Note that this approach requires the inversion of the matrix $\Big( \Dc_{t,\alpha,2} \big( \Ubs_{t,2} \big)  \Xc_{t,\alpha,1} \big( \Fbs (\xb_o+\epsilonb^\star) \big) \Big)^\dagger \Dc_{t,\alpha,2} \big( \Ubs_{t,2} \big)  \Xc_{t,\alpha,1} \big( \Fbs (\xb_o+\epsilonb^\star) \big)$ at each iteration. Unfortunately, this matrix is not necessarily invertible, and the StEFCal method cannot be directly applied to the DDE calibration problem. 

We propose to modify the minimization problem solved by~\cite{Salvini_2014} by introducing a regularization term. More precisely, we propose to solve problem~\eqref{eq:min_ddes_constraints_C} where $p$ is given by:
\begin{equation}	 
p(\Ubs_{1}, \Ubs_{2}) = 
\sum_{t = 1}^T  \sum_{\alpha = 1}^{n_a}  \| \ub_{t,\alpha,1} - \sigma(\ub_{t,\alpha,2}) \|_2^2		.
\end{equation}
In this context, the update of $\Ubs_1$ is given by, for every $t\in \{1, \ldots,T\}$ and $\alpha \in \{1, \ldots,n_a\}$,
\begin{multline}	\label{eq:update:ADI_DD}
\ub_{t,\alpha,1}^{+} 
= \argmin{\ub} 
\|  \Dc_{t,\alpha,2} \big( \Ubs_{t,2} \big)  \Xc_{t,\alpha,1} \big( \Fbs (\xb_o+\epsilonb^\star) \big) \ub - \Ybs_{t,\alpha} \|_2^2 \\
+ \nu \| \ub - \sigma(\ub_{t,\alpha,2}) \|_2^2  ,
\end{multline}
which admits an explicit solution. In eq.~\eqref{eq:update:ADI_DD}, the parameter $\nu$ not only is a regularization parameter, but also allows to have a more stable minimization problem to be solved at each iteration. 
Moreover, as mentioned earlier, to estimate both the DDEs and the image, one can combine a calibration method to an imaging algorithm. 
Therefore, the proposed generalized StEFCal based method repeats the following two steps until convergence:
\begin{enumerate}
\item	\label{StEFCal:DD:step1}
Estimate the DDEs from the current estimate $\epsilonb^\star$ of the original image: 
Update iteratively $\Ubs_1$ using eq.~\eqref{eq:update:ADI_DD}, from the current estimation of the image, and $\Ubs_2$ such that, for every $t\in \{1, \ldots,T\}$ and $\alpha \in \{1, \ldots,n_a\}$, $\ub_{t,\alpha,2}^{+} = \sigma \big( \ub_{t,\alpha,1}^{+} \big)$. 
Stop at convergence.
\item \label{StEFCal:DD:step2}
Estimate the unknown faint sources $\overline{\epsilonb}$ of the image from the current estimate $(\Ubs_1^\star, \Ubs_2^\star)$ of the DDEs : 
Minimize $h(\epsilonb, \Ubs_1^\star, \Ubs_2^\star) + \eta r(\epsilon)$, where $h$ is given by eq.~\eqref{eq:h_x_final}, and $r$ is defined by eq.~\eqref{eq:def_reg_x_sparse} or eq.~\eqref{eq:def_reg_x_dict}.
\end{enumerate}
In step~\ref{StEFCal:DD:step2}, a simple method to minimize $h(\epsilonb, \Ubs_1^\star, \Ubs_2^\star) + \eta r(\epsilon)$ is to use the forward-backward algorithm \citep{Combettes_Wajs_2005}. It corresponds to the steps~\ref{algo:step:grad_x}-\ref{algo:step:prox_x} in Algorithm~\ref{algo:global}, when $J_{\epsilonb}^{(i)} \to +\infty$.

The main differences between this StEFCal-based approach and our method described in Section~\ref{Sec:algo} are the following:
\begin{enumerate}
\item
Our method allows to take into account constraints in the amplitude of the Fourier coefficients of the DDEs, thanks to the projection steps \ref{algo:step:proj_u1} and \ref{algo:step:proj_u2} in Algorithm~\ref{algo:global}. Note that these constraints could be taken into account also using the ADI algorithm by using sub-iterations, which may slow down drastically the algorithm.

\item
The ADI method is an \emph{implicit} algorithm, meaning that it takes advantage from the symmetric formulation of eq.~\eqref{eq:min_ddes_constraints_C} by obliging the second variable $\Ubs_2$ to be exactly equal to the first variable $\Ubs_1$, up to a transformation. 
This symmetry is not directly used in the structure of our algorithm, since in Algorithm~\ref{algo:global} both $\Ubs_1$ and $\Ubs_2$ are updated alternately using forward-backward iterations.

\item
While the StEFCal algorithm is initially only a calibration method, which can be combined with an imaging algorithm, the proposed method described in Section~\ref{Sec:algo} is directly designed to estimate \emph{jointly} the image and the DDEs. Therefore, at each iteration, our method does not require a prefect estimation of both DDEs and the image. The accuracy of these estimates is controlled, at each iteration $i\in \eN$ of Algorithm~\ref{algo:global}, by the parameters $J_{\Ubs_1}^{(i)}$, $J_{\Ubs_2}^{(i)}$, $J_{\text{cyc}}$, and $J_{\epsilonb}^{(i)}$. Thus, in our method, the global estimations are obtained at the convergence of the algorithm, as described in Section~\ref{Ssec:algo:conv}, while the proposed naive generalization of StEFCal does not have any convergence guarantee. In particular, we will show in Section~\ref{Ssec:sim:test1-4} that the results obtained with the StEFCal based method are less accurate than those obtained using our algorithm.
\end{enumerate}

\section{Simulations and results}
\label{Sec:Sim}

\subsection{Simulation setting}
\label{Ssec:sim:setting}

To show the performance of the proposed method, we computed numerical experiments, implemented in MATLAB\footnote{The MATLAB code is available at \texttt{https://basp-group.github.io/joint-img-dde-calibration/}.}. 
We consider $n_a$ randomly distributed antennas, where each antenna pair acquires $T$ measurements, $n_a$ and $T$ being specified for the different simulations performed in this section. 
Note that Earth rotation is incorporated to track the $u-v$ positions of each resulting baseline by considering a time interval of 10 hours\footnote{The $u-v$ tracks are simulated using the code available at \texttt{http://www.astro.umd.edu/$\sim$cychen/MATLAB/ASTR410/uvAnd} \texttt{Beams.html}}.
Moreover, in order to simplify the experiments, we use discrete versions of the associated $u-v$ coverages. 
This is done by considering the nearest discrete $u-v$ position of each antenna. This approximation is adopted in order to avoid to introduce gridding and degridding  steps to model the non-uniform FFT in this first presentation of the proposed DDE selfcal algorithm.

\begin{figure}
\begin{center}
\begin{tabular}{c@{}c@{}}
	\includegraphics[width=4.3cm]{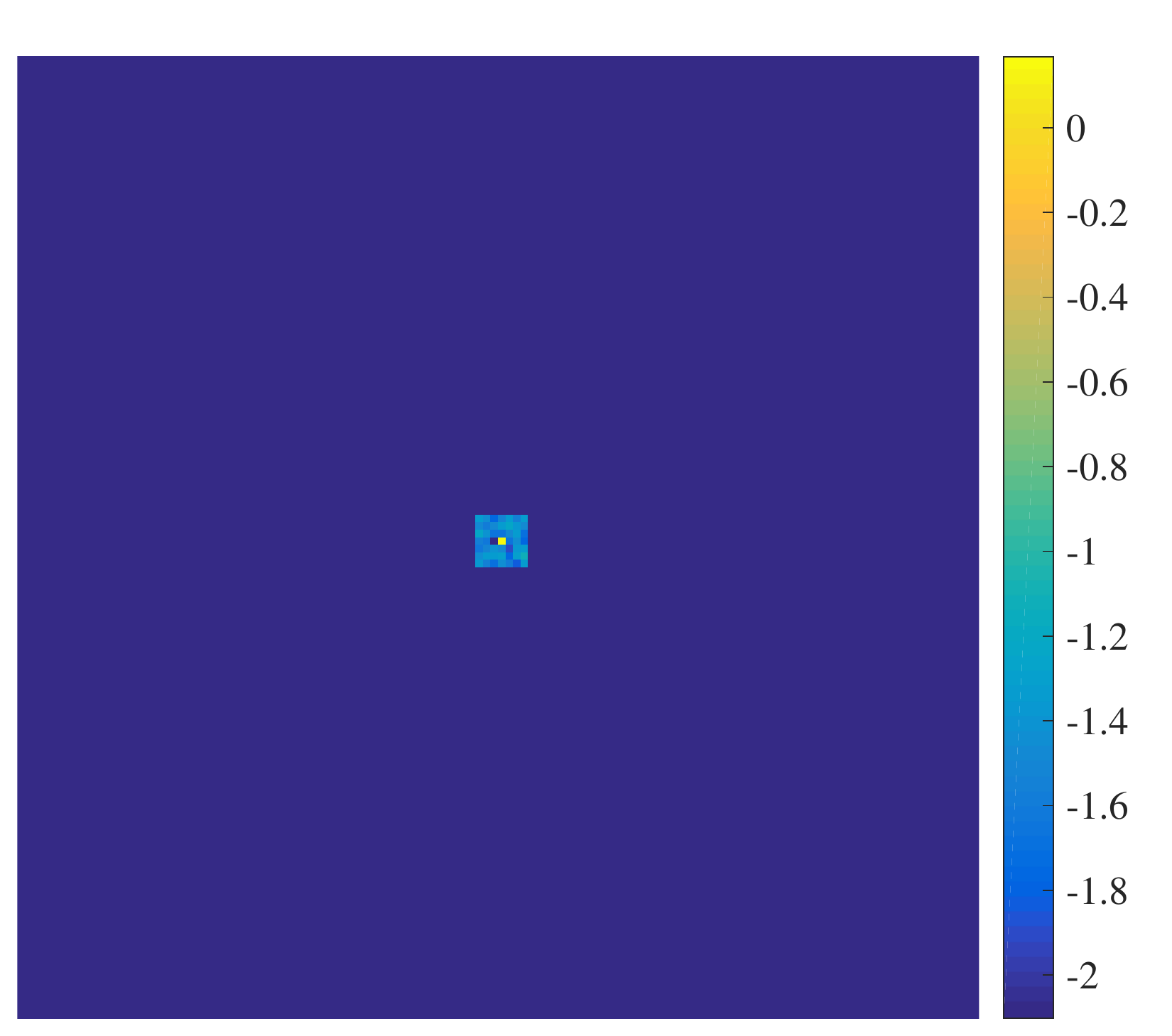}	
&	\includegraphics[width=4.3cm]{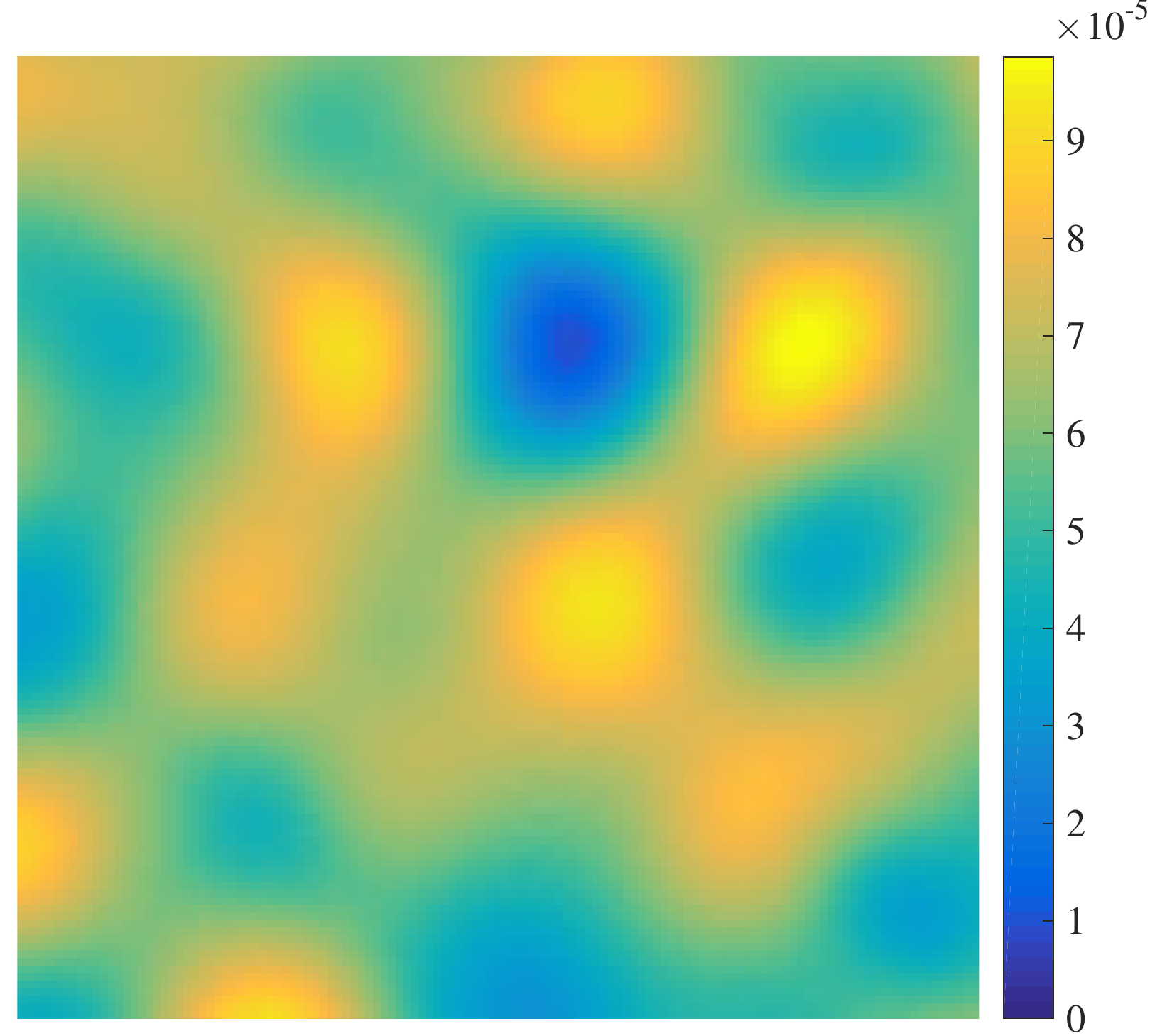}	
\end{tabular}
\end{center}
\caption{\label{Fig:DDEs_image}
Example of considered DDEs, with $S=7 \times 7$ non-zero Fourier coefficients. 
Modulus of (left) the Fourier transform of the DDE $\widehat{\overline{\db}}_{t,\alpha}$ in log scale, and (right) the DDE $\overline{\db}_{t,\alpha}$ in image space in linear scale.}
\end{figure}

For every $\alpha \in \{1, \ldots, n_a\}$, the DDEs kernel $\widehat{\overline{\db}}_{t,\alpha}$ associated with antenna $\alpha$ at instant $t\in \{1, \ldots,T\}$ is simulated randomly in the Fourier domain 
and the support $\eS$ of the direction-dependent Fourier coefficients is assumed to be known exactly. 
In our approach, we do not assume that calibration transfer has been performed but that both DIEs and DDEs belong to some complex valued intervals centred in $0$. 
In practice, in the proposed algorithm, these intervals are handled by considering the set $\eD$ defined by~eq.~\eqref{eq:set_D} and parametrized by $(\vartheta_1,\vartheta_2)$. 
In particular, we choose $\vartheta_1 = 1.5+\theta$ and $\vartheta_2 = \theta$, where $\theta>0$ is a parameter depending on the standard deviation $\upsilon>0$ used to generate the DDEs. 
In other words, we impose to our algorithm to estimate DDEs such that the real and imaginary parts of their zero spatial frequency coefficients (i.e. the DIEs) belong to $ [ -1.5-\theta , + 1.5+\theta ] $, while the other coefficients belong to $[-\theta, + \theta]$.
Note that it is reasonable to consider small values for the higher order spatial frequencies since they represent direction-dependent variations in the gain across the field of view with respect to the mean gain. 
Therefore, in our simulations, we generate the DIEs such that both their real and imaginary parts have values around $\pm 1$, while we generate the other coefficients of the DDEs such that their real and imaginary parts have values in the neighbourhood of $0$ (in both cases with standard deviation $\upsilon$). 
In the context when calibrator transfer has been performed, normalized DIEs are obtained. Then, in our method we can introduce stronger constraints obliging the reconstructed DIEs to belong to a complex neighbourhood of $1+0\mathsf{i}$. 
More formally, these constraints can be taken into account by our algorithm by imposing that the real part of the DIEs belongs to $[ 1-\theta, 1+\theta]$ and their imaginary part belongs to $[-\theta, +\theta]$.
An example of one considered DDE is shown in Figure~\ref{Fig:DDEs_image}, in image and Fourier spaces. 
For instance, in the context of the SPAM method \citep{Intema_2009}, such DDEs can be seen as ionospheric phase screen.

\begin{figure*}
\begin{center}\footnotesize
\begin{tabular}{c@{}c@{}c@{}c@{}}
\hspace*{-0.2cm} \includegraphics[width=4.3cm]{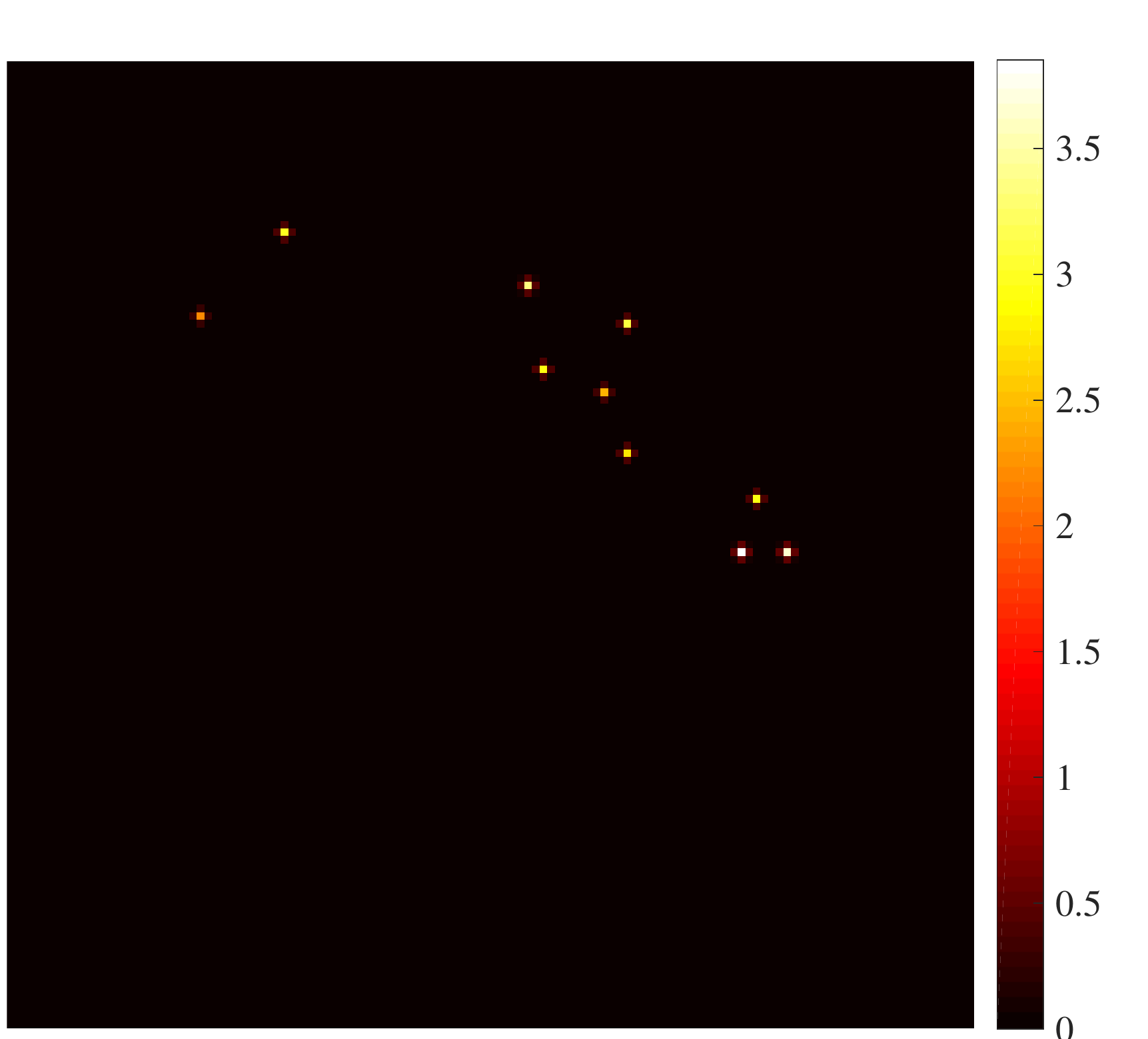}
&\includegraphics[width=4.3cm]{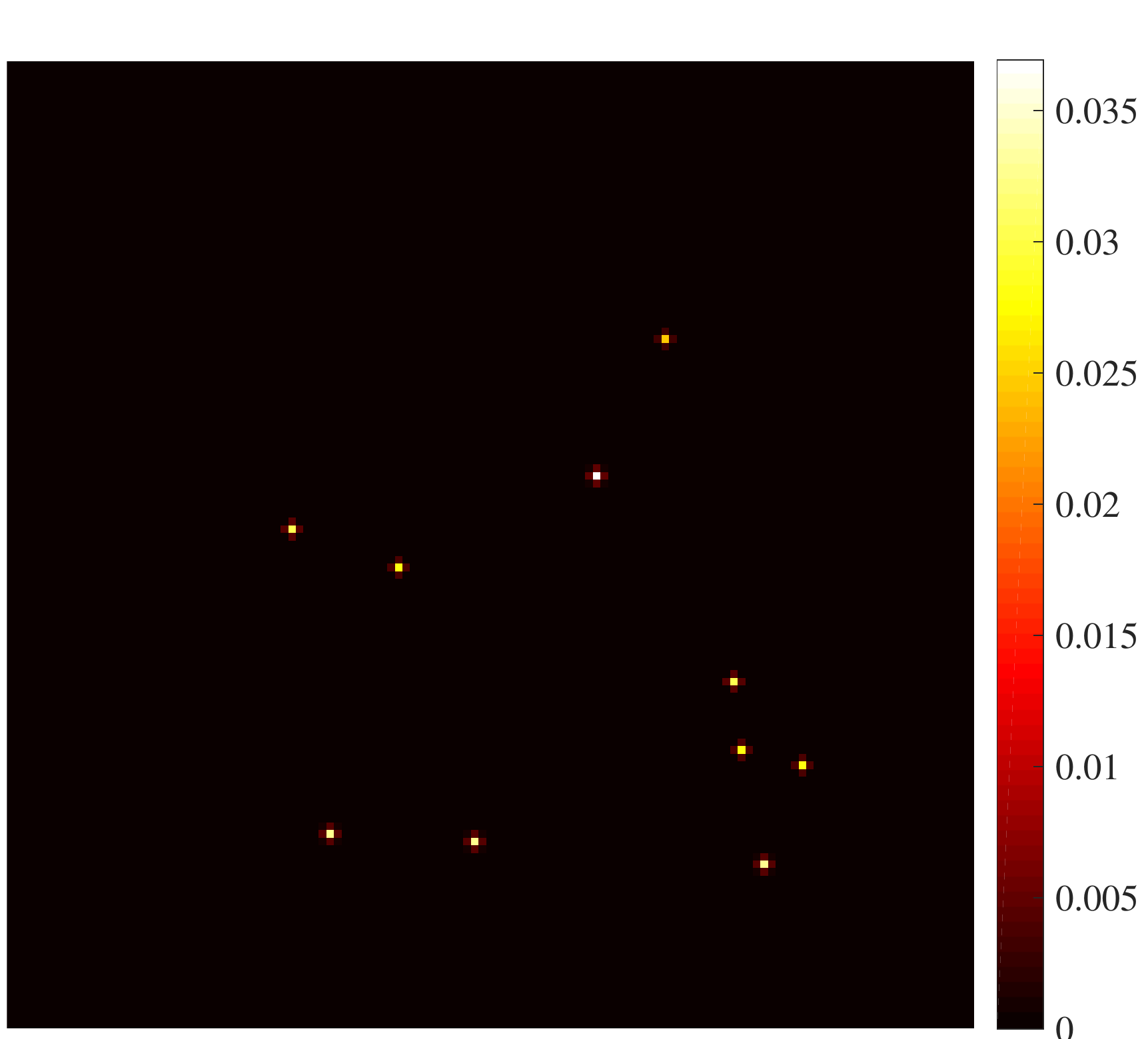}
&\includegraphics[width=4.3cm]{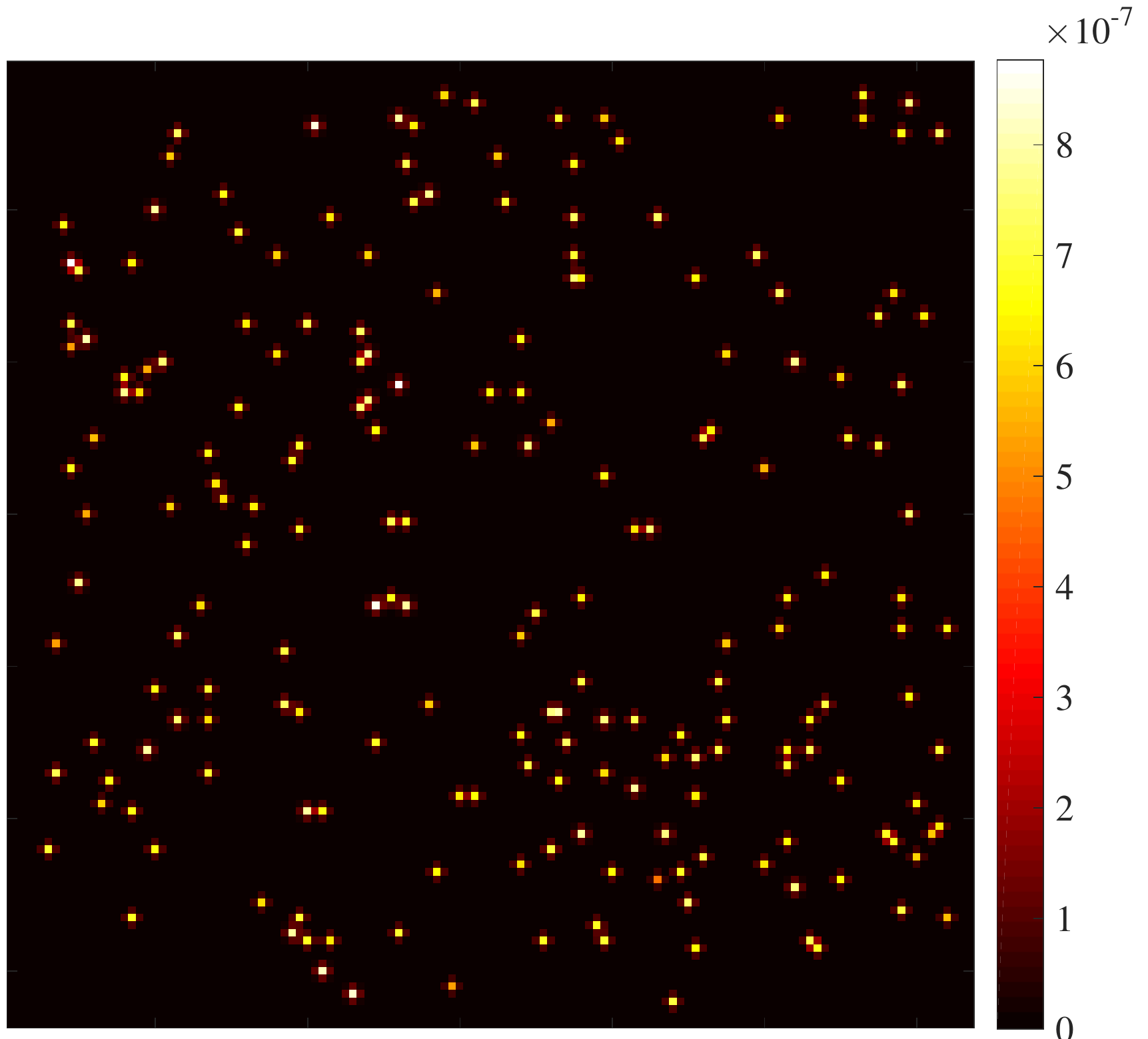}
&\includegraphics[width=4.3cm]{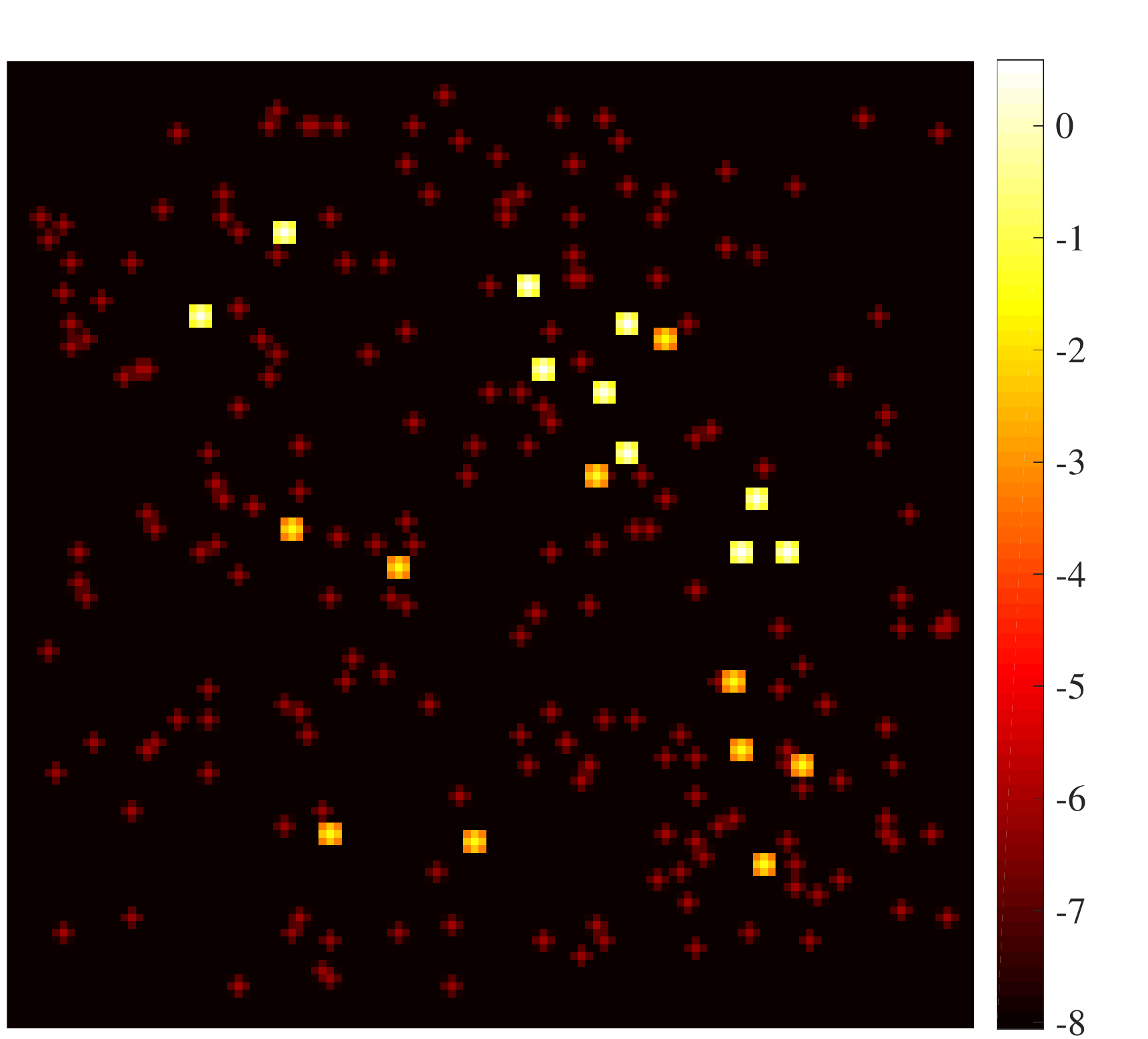}	
\end{tabular}
\vspace*{-0.4cm}
\end{center}
\caption{\label{Fig:images}
Example of point sources images considered in the simulations of Section~\ref{Ssec:sim:tests1}, where $\overline{\xb} = \xb_o + \overline{\epsilonb}$, and $\overline{\epsilonb}$ is a sum of two images $\overline{\epsilonb}_1$ and $\overline{\epsilonb}_2$. 
From left to right: 
The known image $\xb_o$ with 10 sources and corresponding energy $\Es(\xb_o) = 10$. 
The unknown first intensity level image $\overline{\epsilonb}_1$ with 10 sources and corresponding energy $\Es(\overline{\epsilonb}_1) = 0.1$.
The unknown second intensity level image $\overline{\epsilonb}_2$ with 200 sources and corresponding energy $\Es(\overline{\epsilonb}_2) = 10^{-5}$.
The resulting global original image $\overline{\xb} = \xb_o + \overline{\epsilonb}_1 + \overline{\epsilonb}_2$ in log scale. Note that each source being defined as a 2D Gaussian kernel of size $3 \times 3$, the sources of the first two levels appear as squares in the last image due to the log scale.}
\end{figure*}

An extensive study has been performed by considering a wide variety of cases, both varying parameters for the calibration and the imaging part. 
In Section~\ref{Ssec:sim:tests1}, we consider a simulated sky of point sources, and we investigate the impact of the range of the image, the size of the support of the Fourier direction-dependent kernels along with the associated standard deviation $\upsilon$, and the time dependency. 
Finally, in Section~\ref{Ssec:sim:tests2} we show that our method can be used as well to reconstruct complex structured images such as the image of M31. 
For all these cases, results will be compared in terms of different metrics. In particular, we will consider the signal to noise ratio (SNR) of the reconstructed image $\epsilonb^\star$ with respect to the original $\overline{\epsilonb}$, which is expressed as
\begin{equation}
\text{SNR} = 20 \log_{10} \Big( \dfrac{\| \overline{\epsilonb} \|_2}{\| \epsilonb^\star - \overline{\epsilonb} \|_2} \Big),
\end{equation}
and the $\ell_2$ norm of the residual images, obtained either considering the estimated DDEs or the true DDEs. More precisely, we will consider the weighted $\ell_2$ norm of the residual images $\| \Fbs^\dagger \Gbs^{\star \, \dagger} \big( \Gbs^{\star} \Fbs \xb^\star - \yb \big) \|_2 /\sqrt{N}$, where $\xb^\star = \xb_o + \epsilonb^\star$ corresponds to the global estimated image and $\Gbs^\star$ corresponds to the estimated DDEs, i.e. $\Gbs^\star= \Gc(\Ubs_1^\star, \Ubs_2^\star)$ ($\Gc$ being defined in eq.~\eqref{eq:h_x_final}). 
Similarly, we will consider the weighted $\ell_2$ norm of the residual images $\| \Fbs^\dagger\overline{\Gbs}^{ \dagger} \big( \overline{\Gbs}\Fbs \xb^\star - \yb \big) \|_2 /\sqrt{N}$, where the matrix $\overline{\Gbs}$ introduced in~eq.~\eqref{pb:inv_pb_image} corresponds to the original DDEs. 
Note that the latter metric is used in order to point out the ambiguity problems which can appear between $\xb^\star$ and $\Gbs^\star$, i.e. due to the inverse problem formulation given in eq.~\eqref{pb:inv_pb_image}, parts of $\xb^\star$ can be absorbed in $\Gbs^\star$, and \emph{vice versa}.

In the two simulation settings, we show results obtained using both the proposed method and the StEFCal algorithm \citep{Salvini_2014}. Note that the latter has been designed to solve only for DIEs, not for DDEs, and without estimating the image. 
Thus, the results shown in our experiments are obtained combining the StEFCal algorithm, correcting for DIEs, with an imaging algorithm based on the forward-backward algorithm \citep{Combettes_Wajs_2005}, considering the same regularization terms as used in our method, and the same associated parameters. 
We alternate between the DIEs estimation and the image estimation until stabilization of the overall process. 
In the following, we will refer to this method using the notation StEFCal-FB. 
It is interesting to note that in our experiments, the StEFCal-FB method gives results of the same quality as those obtained after the first step of the initialization of our algorithm, where only the DIEs are considered, as described in Section~\ref{Ssec:Algo:Init}. 

Finally, as proposed in Section~\ref{Sec:stefcal}, we perform simulations on images of point sources using the proposed naive generalization of the StEFCal algorithm to correct for the DDEs. Discarding for one instant the important problem of the lack of convergence guarantee, we show that this generalization of StEFCal combined with a forward-backward algorithm already gives good reconstruction results, although they are not as good as those obtained using our method.

\subsection{Sparse images with point sources}
\label{Ssec:sim:tests1}

In this first part, we consider simulated sky images of size $128 \times 128$, 
consisting of point sources, where each source corresponds to a small 2D Gaussian kernel of size $3\times 3$. 
As described in Section~\ref{Ssec:Blind_problem}, we consider that the original image can be decomposed as $\overline{\xb} = \xb_o + \overline{\epsilonb}$, where $\xb_o$ is known, while $\overline{\epsilonb}$ is unknown and has to be estimated jointly with the DDEs. 
More precisely, in all the experiments of this section, we assume that $\xb_o$ corresponds to 10 bright sources generated randomly such that their total energy\footnote{The energy $\Es(\xb)$ of a vector $\xb\in \eC^N$ is defined as \newline $\Es(\xb) = ( \sum_{n=1}^N |x(n)|^2 )^{1/2}$.} is $\Es(\xb_o) = 10$. An example of $\xb_o$ is shown in the first image in Figure~\ref{Fig:images}. 
Moreover, the sources in $\overline{\epsilonb}$ are generated randomly with two intensity levels, i.e. $\overline{\epsilonb} = \overline{\epsilonb}_1 + \overline{\epsilonb}_2$. 
In all the experiments of this section, we consider $\overline{\epsilonb}_2$ having an energy $\Es(\overline{\epsilonb}_2) = 10^{-5}$ and to be composed of $200$ sources (see the third image of Figure~\ref{Fig:images} for an example), while, depending on the experiments, we vary the energy and the number of sources belonging to $\overline{\epsilonb}_1$. In particular, an example of $\overline{\epsilonb}_1$ containing $10$ sources with $\Es(\overline{\epsilonb}_1) = 0.1$ is shown in the second image of Figure~\ref{Fig:images}, and the global image $\overline{\xb} = \xb_o + \overline{\epsilonb}_1 + \overline{\epsilonb}_2$ in this case is given in the right image of Figure~\ref{Fig:images} (in log scale). In this context, $\overline{\epsilonb}_2$ has sources with intensity $\sim 10^{7}$ weaker than the sources belonging to $\xb_o$. Thus, $\overline{\epsilonb}_2$ is considered in our simulations as astrophysical noise, and we do not focus on its estimation. Therefore, our main objective is to find an estimate of $\overline{\epsilonb}_1$, reconstructing its faint sources.

\begin{table}
\begin{center}
\begin{tabular}{| l | l | l | l |}
  \hline
\multicolumn{2}{l|}{Parameters} & Initialization  & Algorithm~\ref{algo:global} \\
\hline\hline
\multicolumn{2}{l|}{$J_{\text{cyc}}$}	&	101	&	11	\\
\hline
$J_{\Ubs_1}^{(i)} = J_{\Ubs_2}^{(i)}$	& {{if $(i+1) = 0 \, \Big[\text{mod } J_{\text{cyc}}\Big]$}} 	&	0	&	0	\\
& {{otherwise}} 	&	2 & 5	\\ 
\hline
$J_{\epsilonb}^{(i)}$& {{if $(i+1) = 0 \, \Big[\text{mod } J_{\text{cyc}}\Big]$}} 	&	1000 & 1000	\\ 
& {{otherwise}} 	&	0 & 0	\\ 
\hline
\multicolumn{2}{l|}{$\xi_{\Ubs_{\text{tot}}}$}	&	$10^{-6}$	&		\\
\hline
\multicolumn{2}{l|}{$\xi_{\epsilonb}$}	&	$10^{-5}$	&	$10^{-5}$	\\
\hline
\multicolumn{2}{l|}{$\zeta$}	&	$2\times10^{-2}$	&	$2\times10^{-2}$	\\
\hline
\multicolumn{2}{l|}{$\underline{\tau}$}	&	0.6	&		\\
\hline
\multicolumn{2}{l|}{$\overline{\tau}$}	&	0.9	&		\\
\hline
\end{tabular}
\end{center}
\caption{\label{Tab:param_point_sources}
Parameters considered in Section~\ref{Ssec:sim:tests1} for Algorithm~\ref{algo:global} and its initialization.}
\end{table}

In this section, the original image being sparse in its domain, we use~eq.~\eqref{eq:def_reg_x_sparse} as regularization term for the image, and we fix the regularization parameters $\eta = 10^{-3}$ for the image, $\nu = 2000$ for the initialization estimating the zero spatial frequency coefficients of the DDEs (i.e. the DIEs), and $\nu=1000$ for the main algorithm estimating DDEs. Moreover, the different parameters for Algorithm~\ref{algo:global} are given in Table~\ref{Tab:param_point_sources}. 
As described in Section~\ref{Ssec:algo}, on the one hand, at each iteration $i\in \eN$ such that $(i+1) \neq 0 \, \Big[\text{mod } J_{\text{cyc}}\Big]$, the algorithm alternates between the estimation of $\Ubs_1^{(i)}$ and $\Ubs_2^{(i)}$ computing $J_{\Ubs_1}^{(i)}$ (resp. $J_{\Ubs_2}^{(i)}$) forward-backward sub-iterations, while the image is not updated. On the other hand, if $(i+1) = 0 \, \Big[\text{mod } J_{\text{cyc}}\Big]$, then the algorithm does not update the DDEs but computes $J_{\epsilonb}^{(i)}$ sub-iterations to update the image $\epsilonb^{(i)}$. 
Moreover, $ \xi_{\Ubs_{\text{tot}}}$, $\xi_{\epsilonb}$, and $\zeta$ are the stopping criteria defined in Section~\ref{Ssec:algo:stop_crit}.

Finally, in this section the considered images consisting of point sources, in addition to the SNR and the $\ell_2$ norm of the residual images, we will consider a success rate counting the positions of the sources in $\overline{\epsilonb}_1$ which are recovered successfully. More precisely, we compare the positions of the brightest sources found in the estimate ${\epsilonb}^\star$ with the true positions of sources belonging to $\overline{\epsilonb}_1$. Then, we deduce from this comparison a percentage of success for the position estimation of the faint sources belonging to $\overline{\epsilonb}_1$.

\subsubsection{Dynamic range of the image}
\label{Ssec:sim:test1-1}

\paragraph*{Simulation settings.}

One of the main advantages of taking DDEs into account, and not only DIEs, is to be able to reconstruct more accurately images with high dynamic range \citep{Smirnov_2011b, Smirnov_2015}. 
To illustrate this assertion, and in order to show the performance and the limits of our method, in this first test, we vary the energy of $\overline{\epsilonb}_1$ and the number of sources it contains, while all other parameters are kept fixed. 
More precisely,  we consider the measurements made by $n_a = 200$ antennas for a single time interval $T = 1$, and considering DDEs Fourier kernels with support size $S = 7 \times 7$, and standard deviation 
set to be $\upsilon = 0.05$ (see Section~\ref{Ssec:sim:setting}). 
Then, we perform simulations for 10, 50 and 90 sources belonging to $\overline{\epsilonb}_1$, and for each of these values, we consider $\Es(\overline{\epsilonb}_1) \in \{ 10, 1, 10^{-1}, 10^{-2}, 10^{-3}\}$. 
Note that the only case when the intensity of sources belonging to $\xb_o$ is of the same order as those belonging to $\overline{\epsilonb}_1$ appears if $\overline{\epsilonb}_1$ has 10 sources with global energy $\Es(\overline{\epsilonb}_1)=10$.

\begin{figure*}
\begin{tabular}{cccc}
 \hspace*{-0.2cm}	\includegraphics[height=4.0cm]{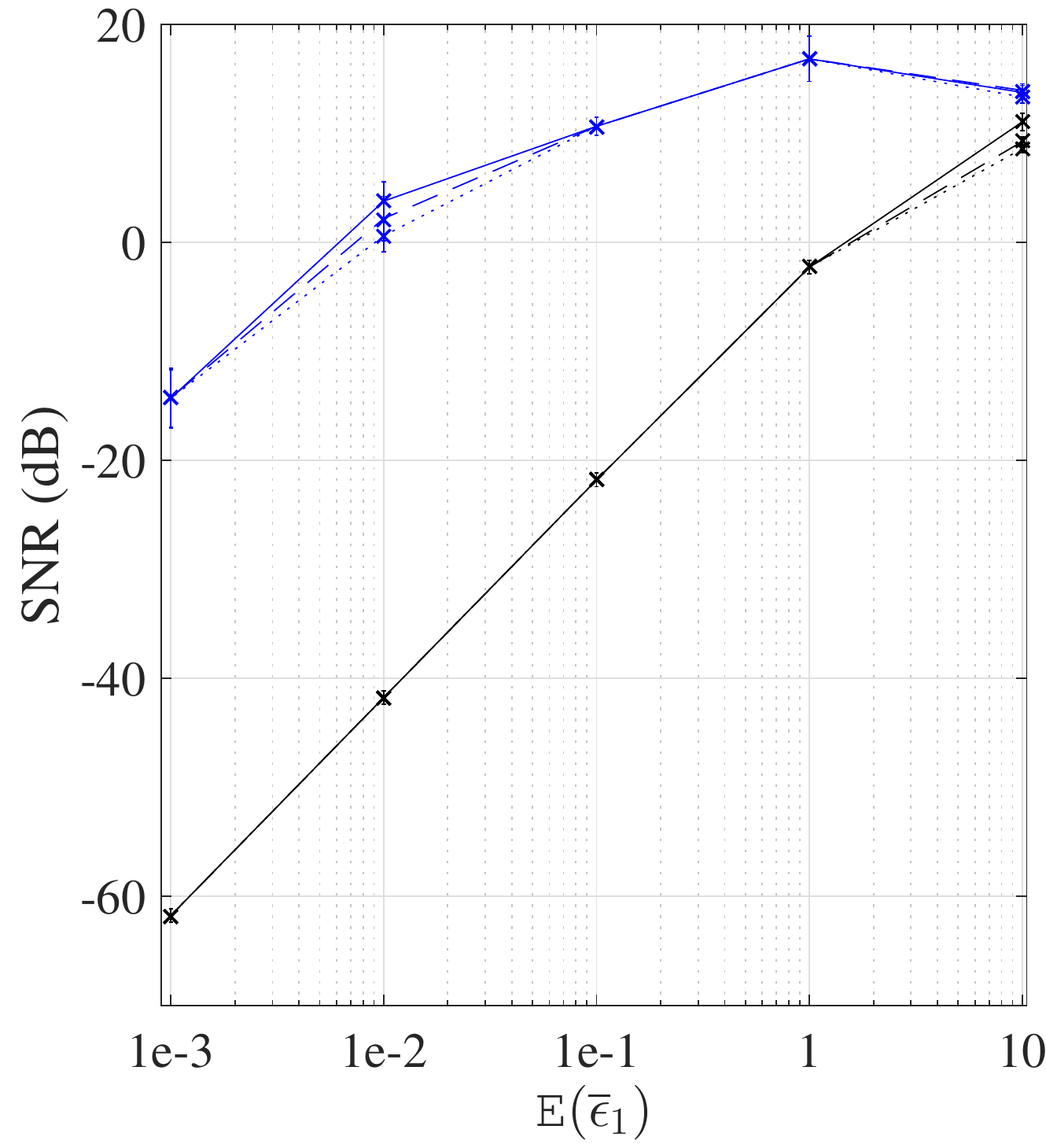}
&\hspace*{-0.2cm}	\includegraphics[height=4.0cm]{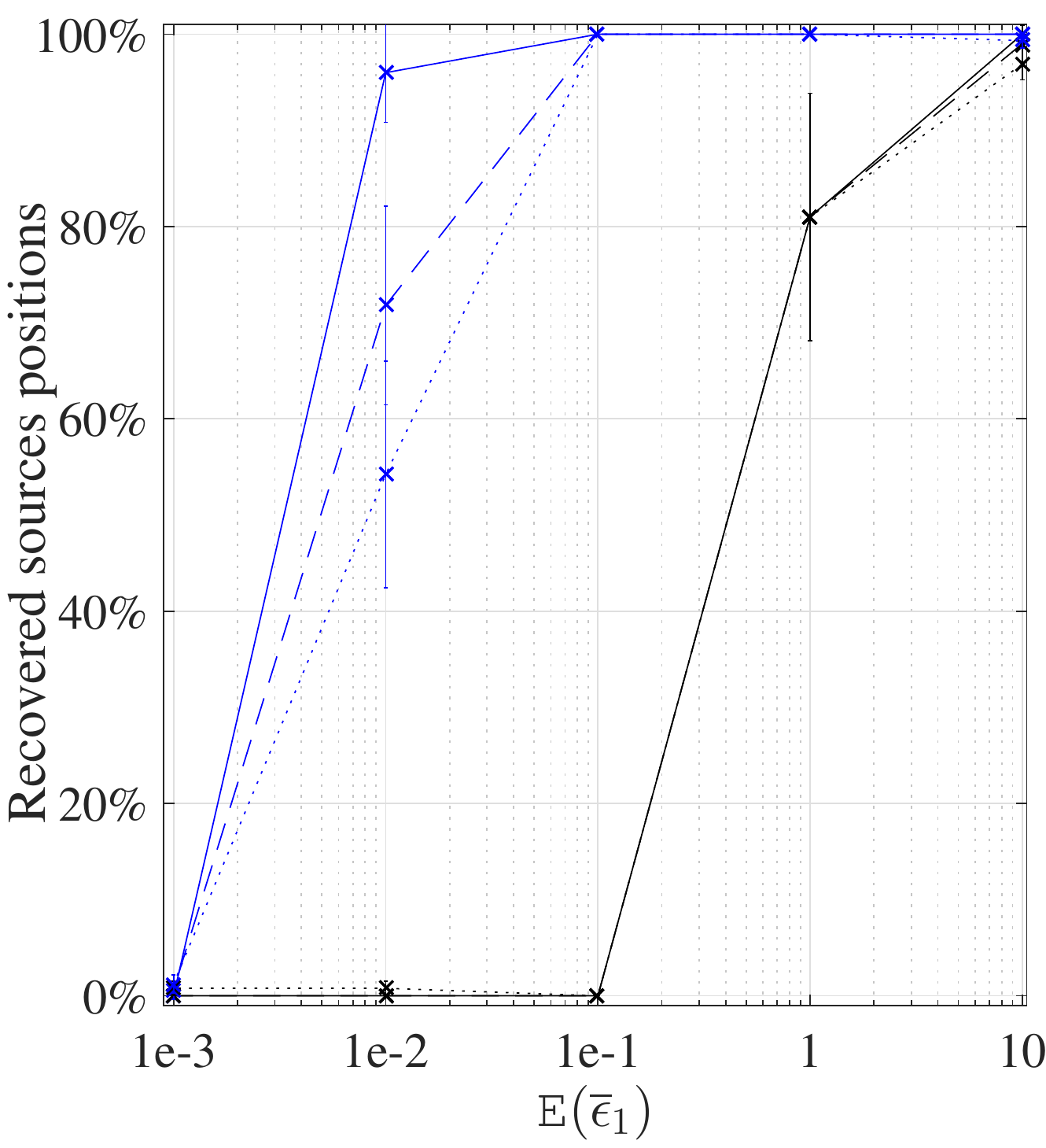}
&\hspace*{-0.2cm}	\includegraphics[height=4.0cm]{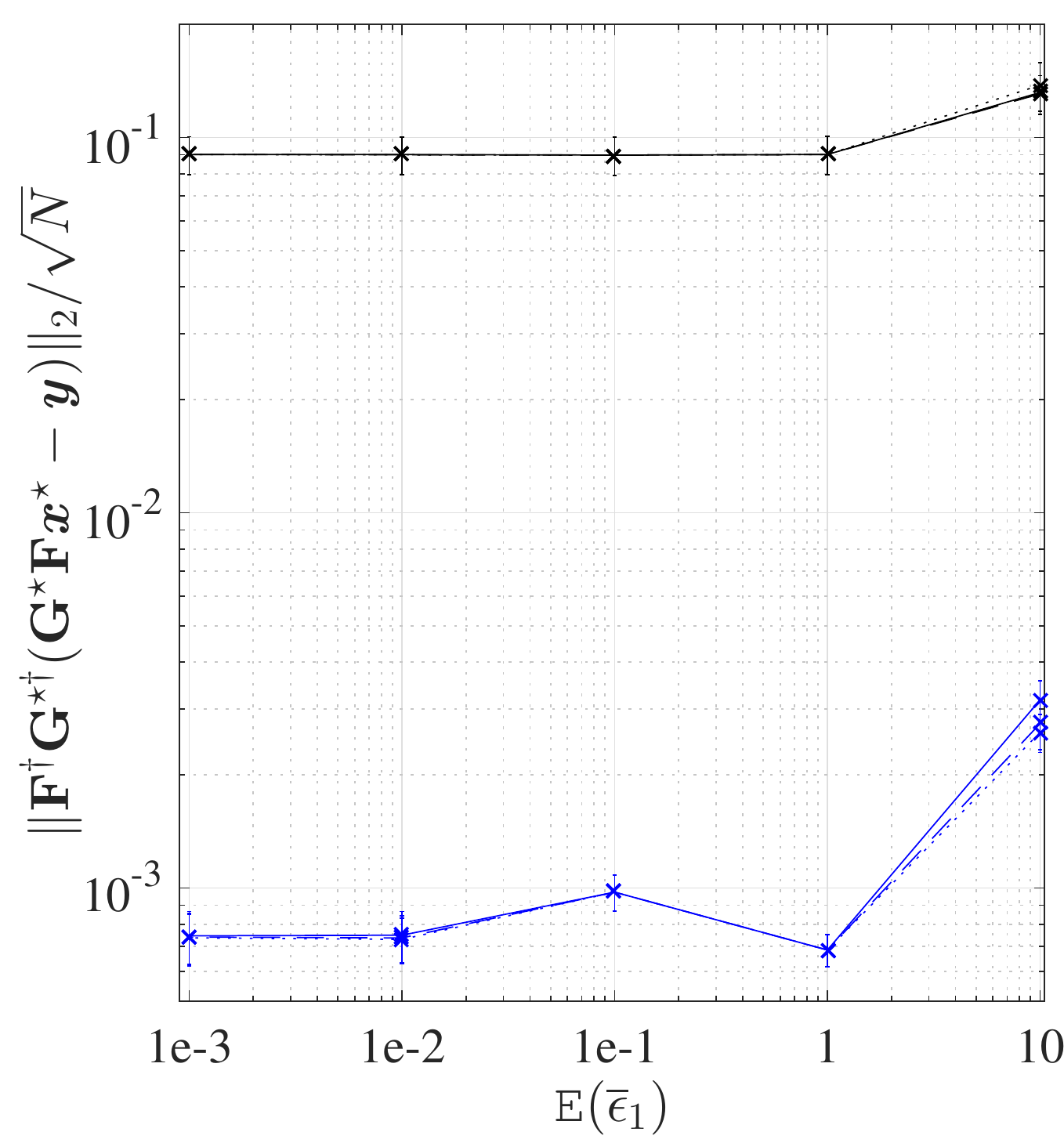}
&\hspace*{-0.2cm}	\includegraphics[height=4.0cm]{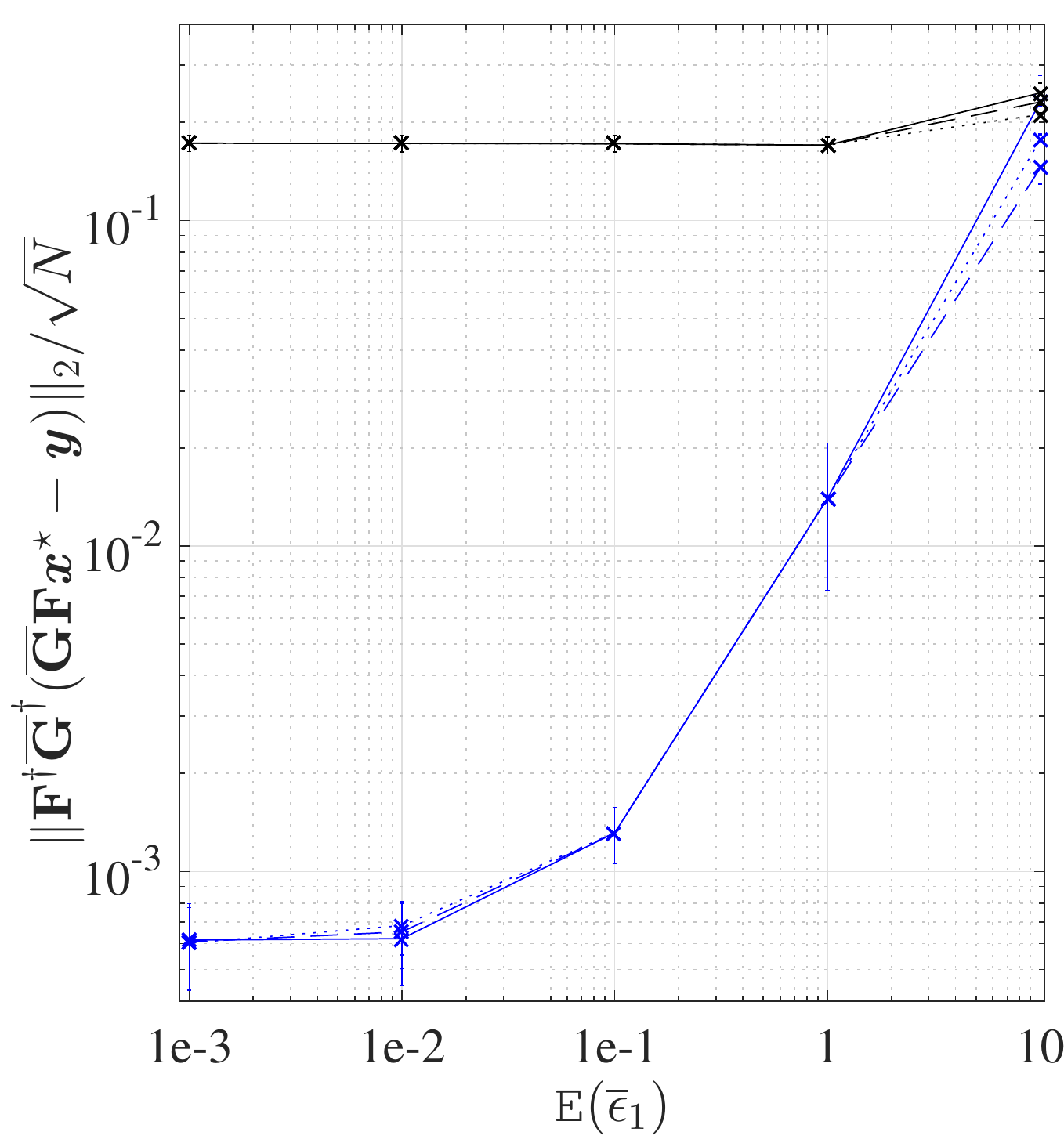}
\end{tabular}
\vspace*{-0.2cm}
\caption{\label{Fig:test1:curves}
Results obtained for simulations of Section~\ref{Ssec:sim:test1-1} using the proposed method (blue lines) and estimating only the DIEs with StEFCal-FB (black lines), considering 10, 50, and 90 sources in $\overline{\epsilonb}_1$ (resp. solid lines, dashed lines, and dotted lines), varying $\Es(\overline{\epsilonb}_1) \in \{10^{-3},10^{-2},10^{-1},1,10\}$, while $\Es(\xb_o) = 10$. 
From left to right: SNR of the reconstructed $\epsilonb^\star$ with respect to $\overline{\epsilonb}$;
Success rate determining the percentage of recovered sources positions from $\overline{\epsilonb}_1$;
$\ell_2$ norm of the residual image $\| \Fbs^\dagger \Gbs^{\star \, \dagger} \big( \Gbs^\star \Fbs \xb^\star - \yb \big) \|_2 / \sqrt{N} $ considering $\Gbs^\star$ obtained with the estimated DDEs;
$\ell_2$ norm of the residual image $\| \Fbs^\dagger \overline{\Gbs}^{\dagger} \big( \overline{\Gbs} \Fbs \xb^\star - \yb \big) \|_2 / \sqrt{N} $ considering $\overline{\Gbs}$ obtained with the true DDEs.
Results are given for an average over 10 realizations varying the antenna distribution, the random images, and the DDEs.}
\vspace*{-0.3cm}
\end{figure*}
\begin{figure*}
\begin{tabular}{c@{}c@{}c@{}c@{}}
\hspace*{-0.4cm}	\includegraphics[width=4.2cm]{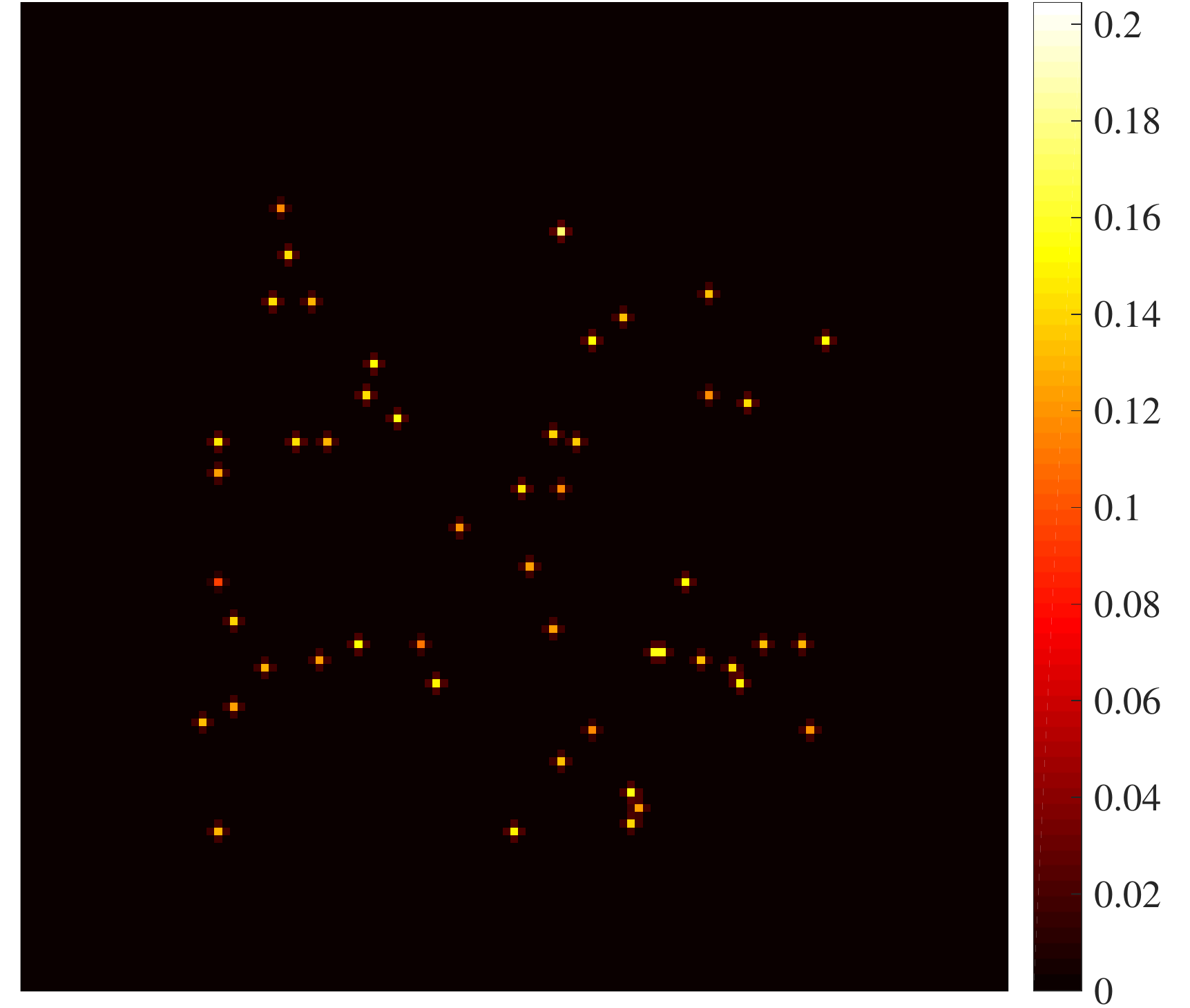}
&	\hspace*{0.1cm}\includegraphics[width=4.2cm]{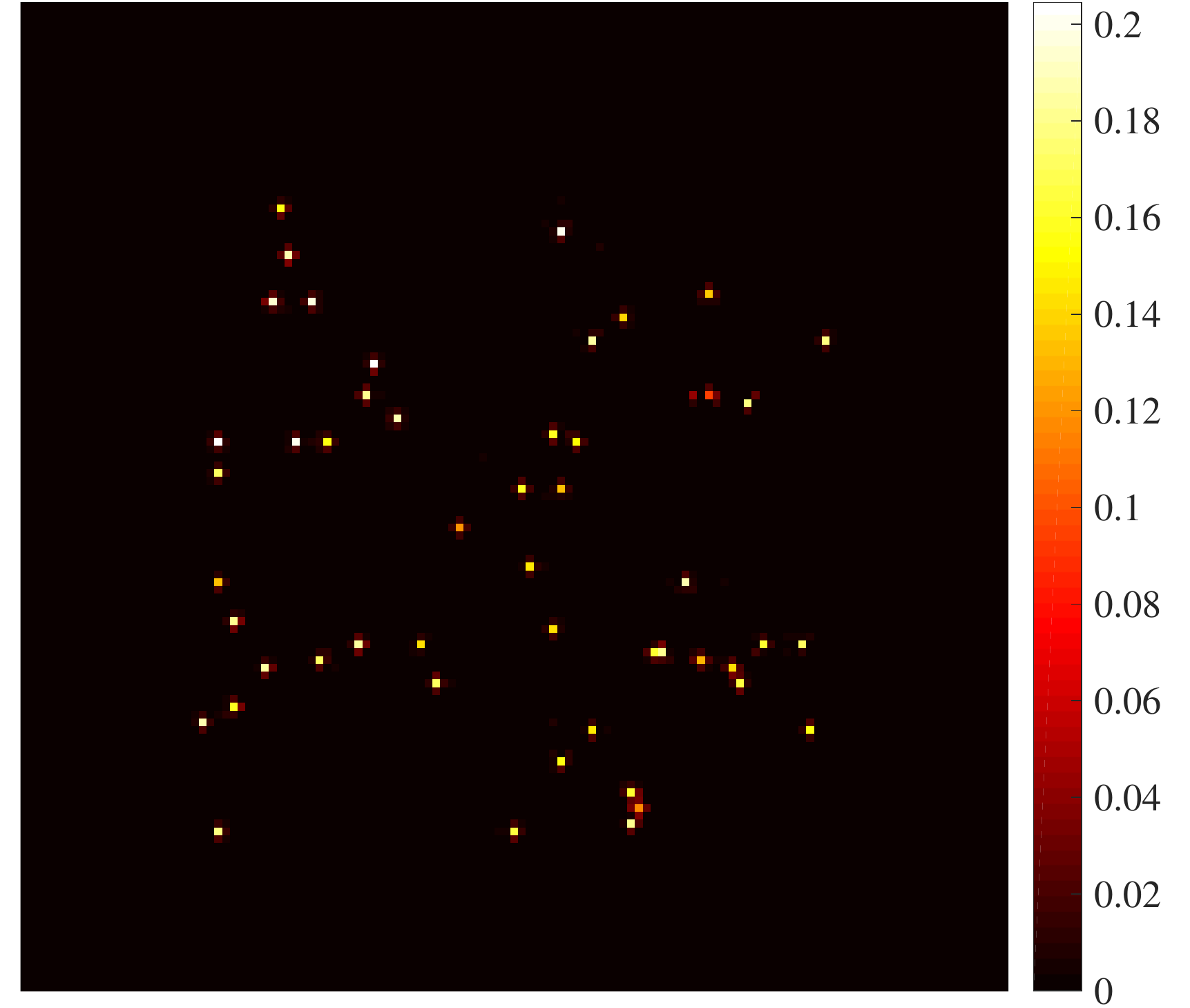}
&	\hspace*{0.1cm}\includegraphics[width=4.2cm]{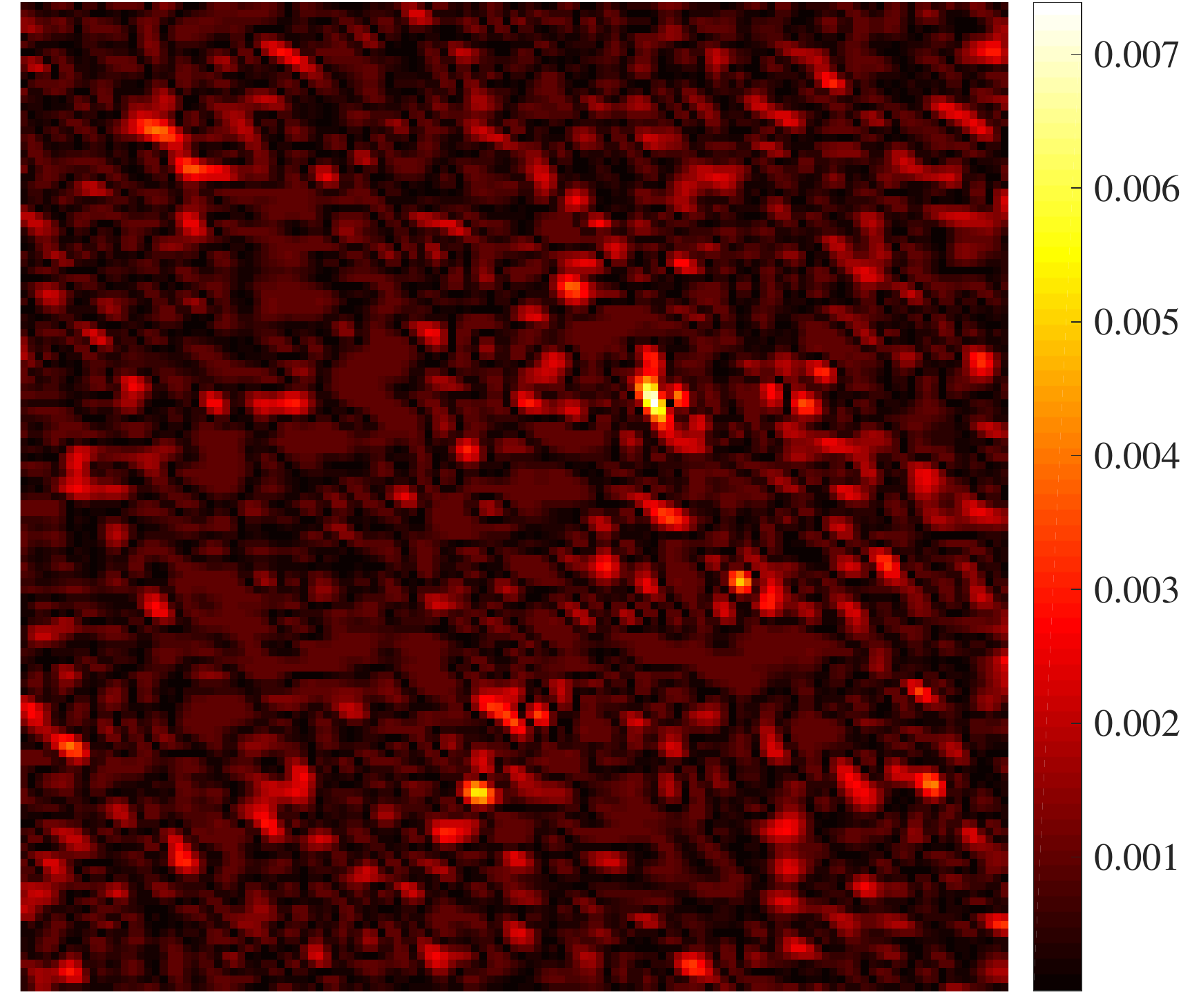}
&	\hspace*{0.1cm}\includegraphics[width=4.2cm]{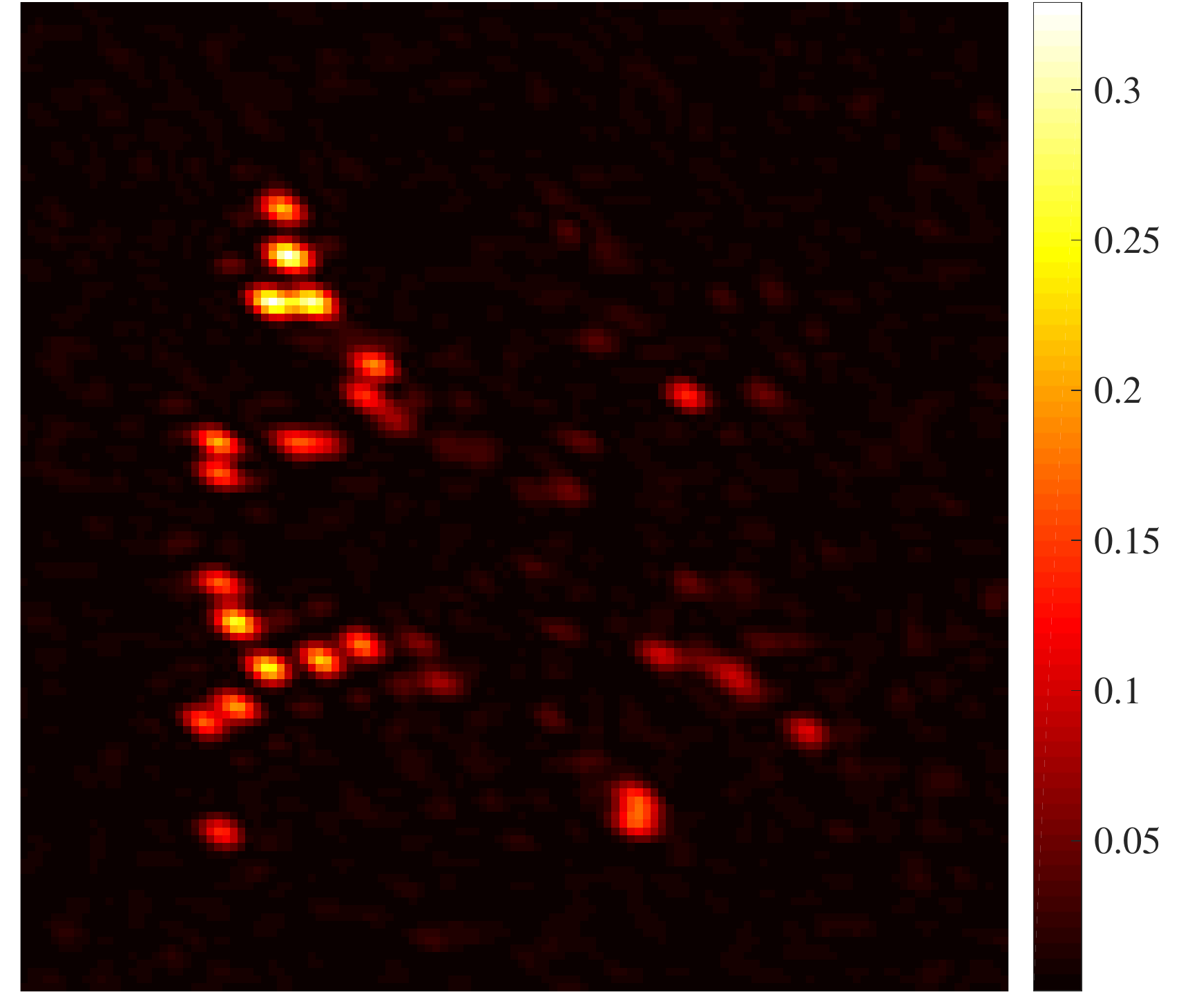}	\\
\hspace*{-0.4cm}	
&	\hspace*{0.1cm}\includegraphics[width=4.2cm]{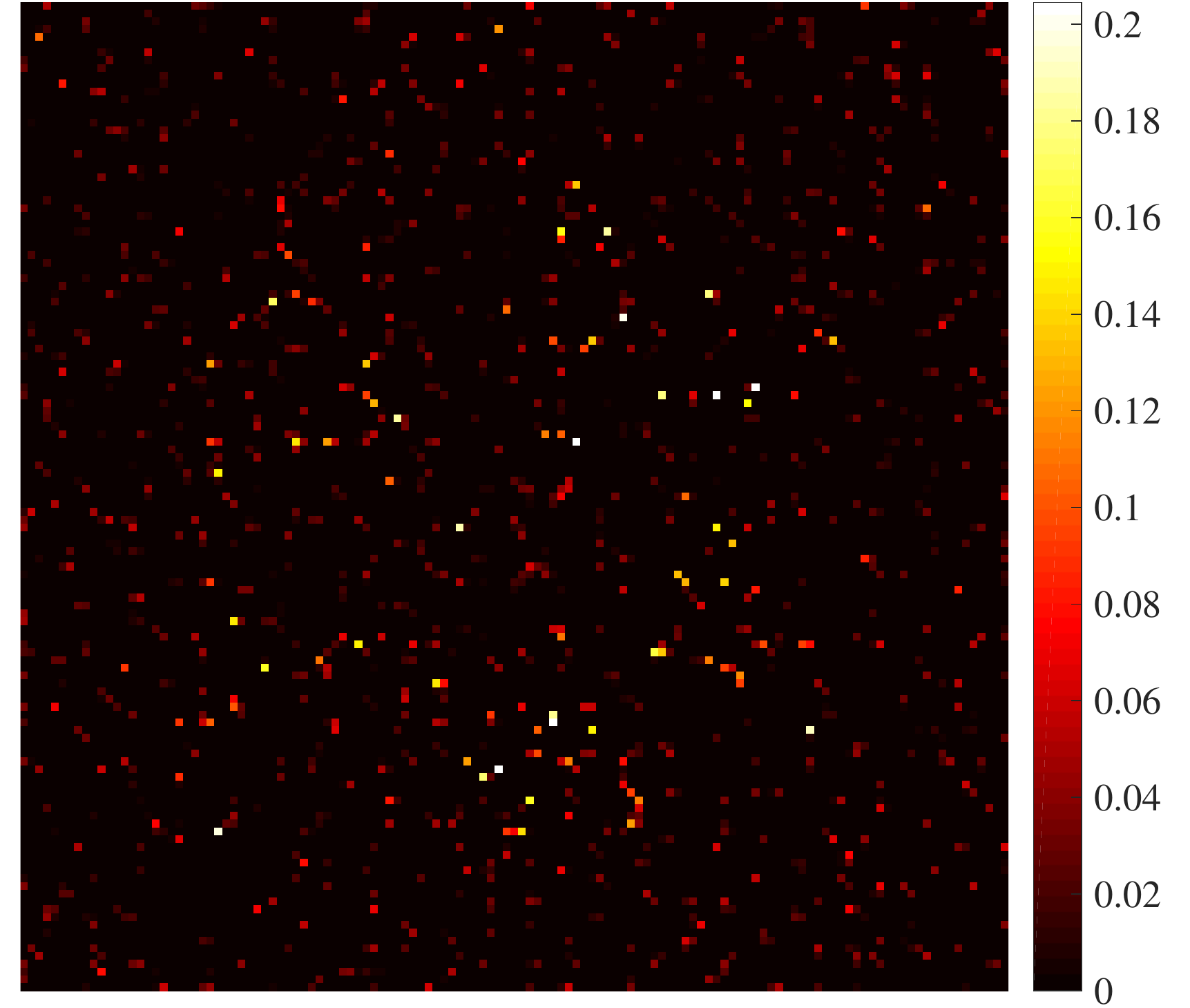}
&	\hspace*{0.1cm}\includegraphics[width=4.2cm]{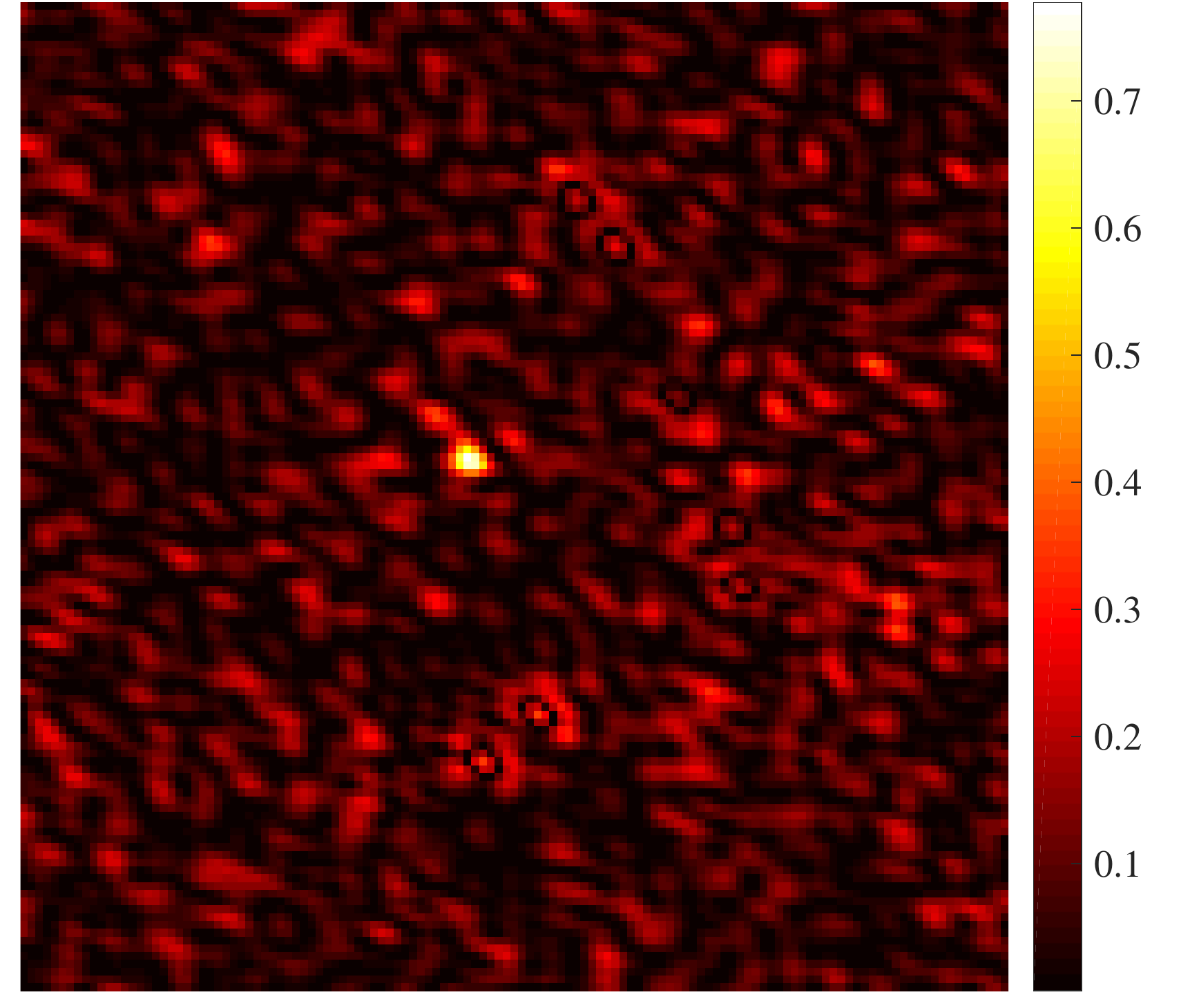}
&	\hspace*{0.1cm}\includegraphics[width=4.2cm]{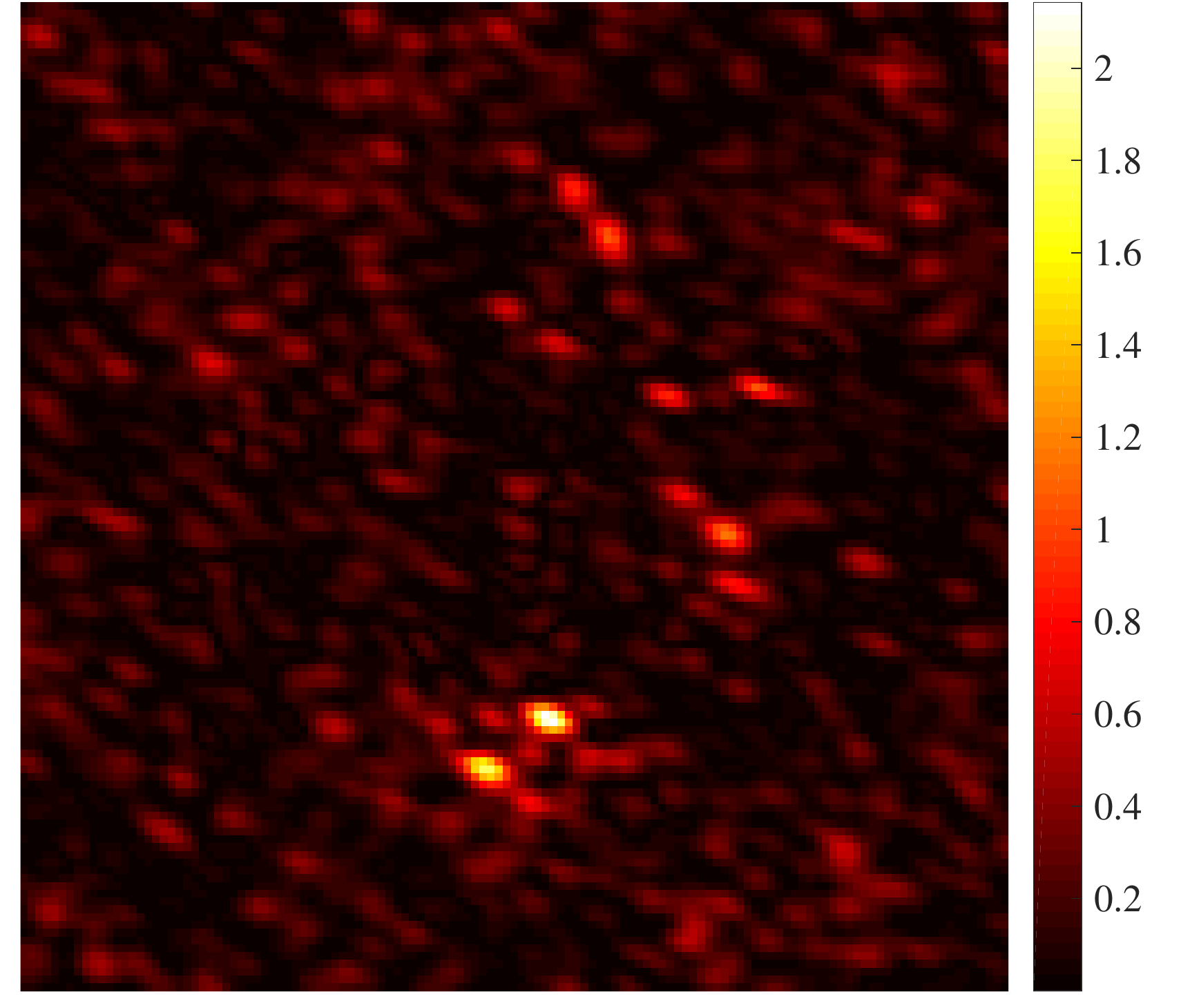}	\\
\hspace*{-0.4cm}	\includegraphics[width=4.2cm]{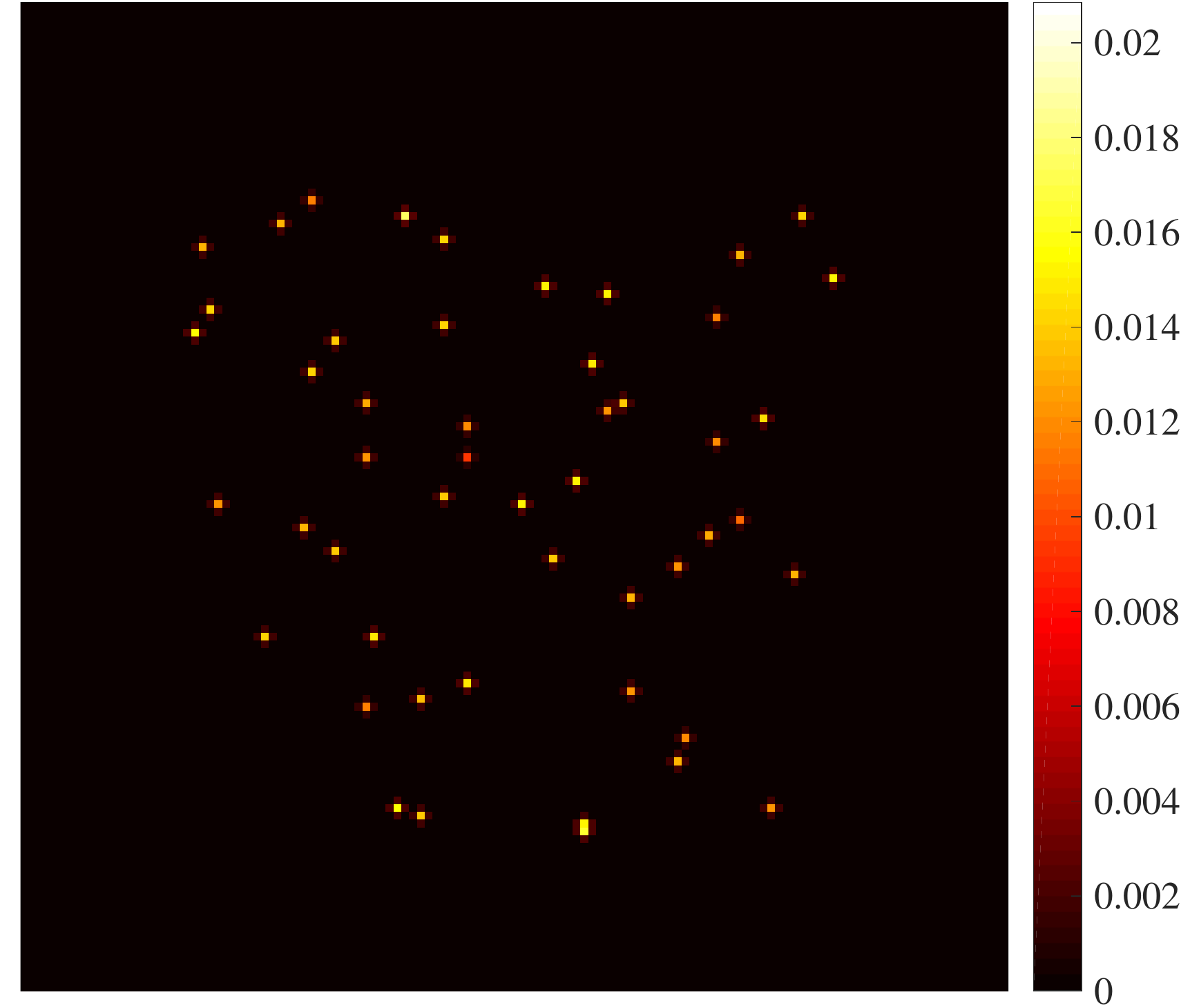}
&	\hspace*{0.1cm}\includegraphics[width=4.2cm]{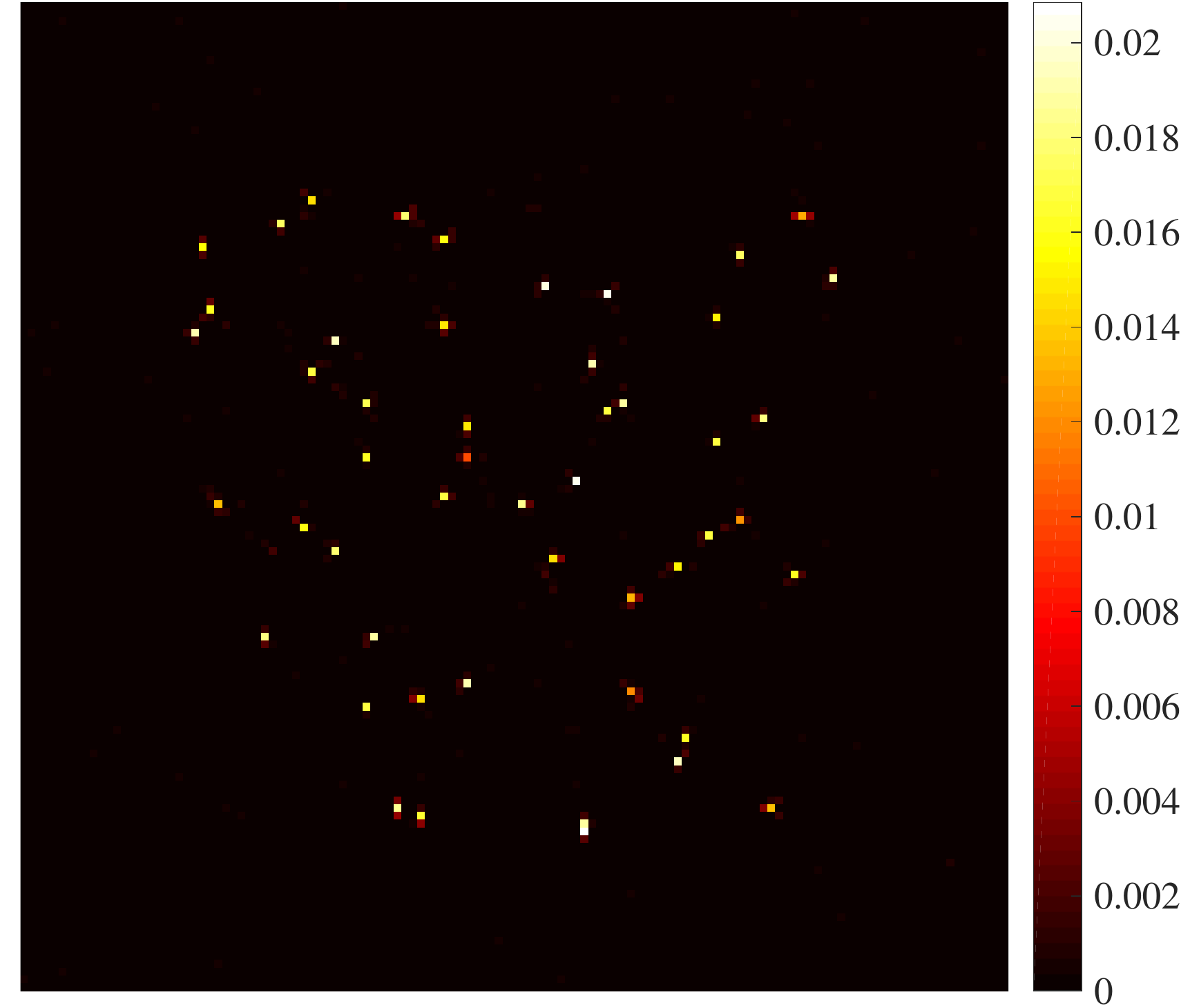}
&	\hspace*{0.1cm}\includegraphics[width=4.2cm]{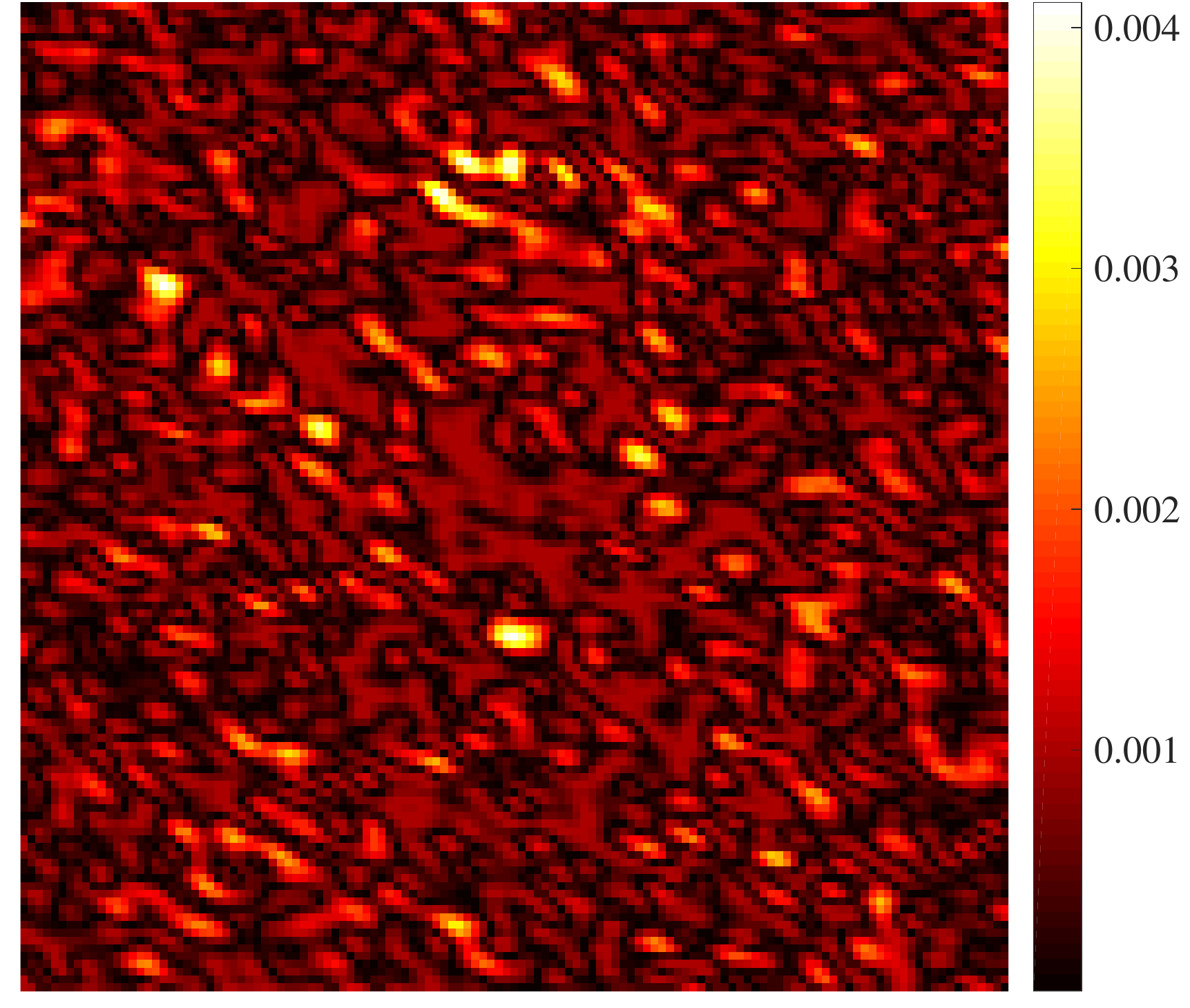}
&	\hspace*{0.1cm}\includegraphics[width=4.2cm]{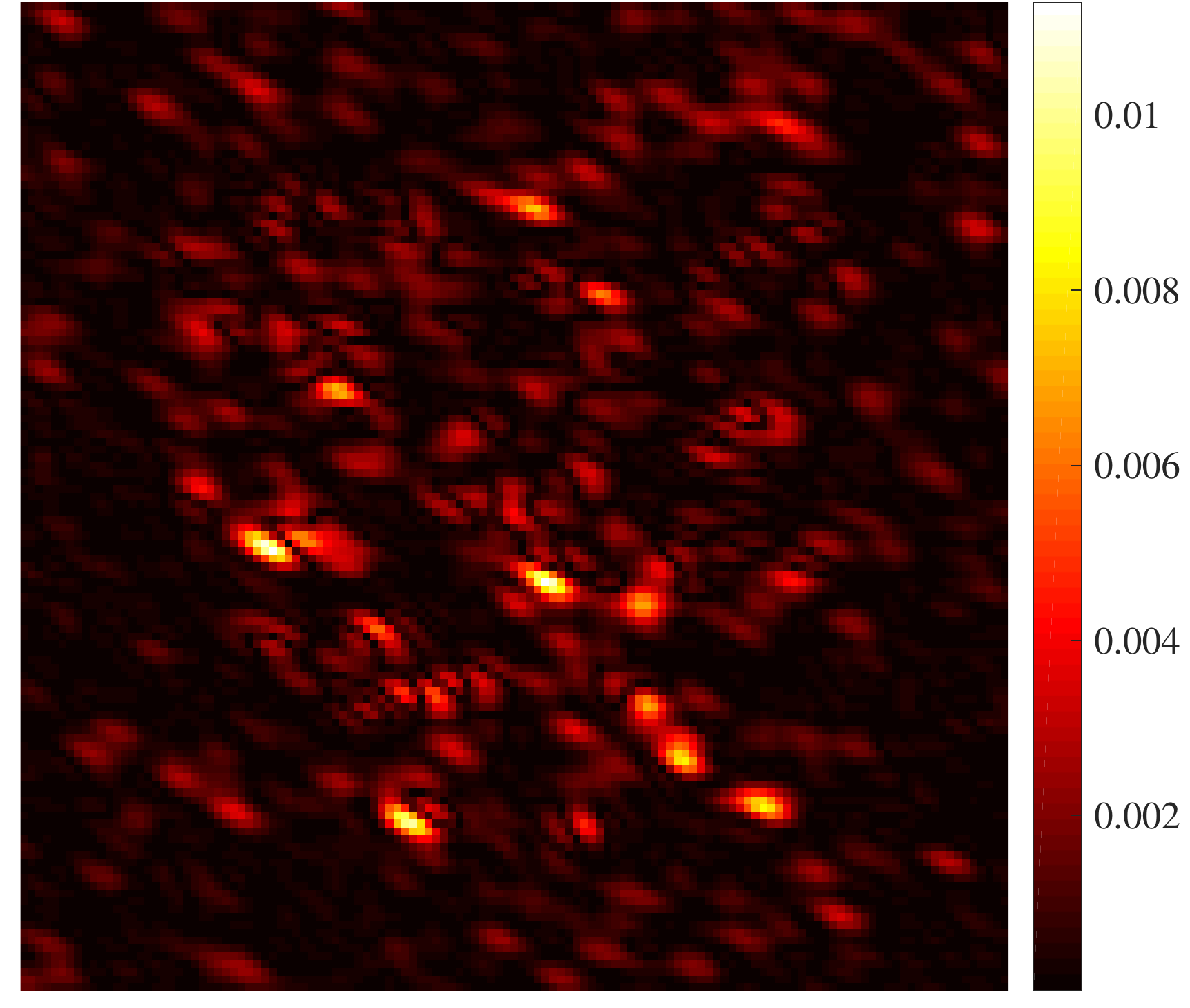}	\\
\hspace*{-0.4cm}	\includegraphics[width=4.2cm]{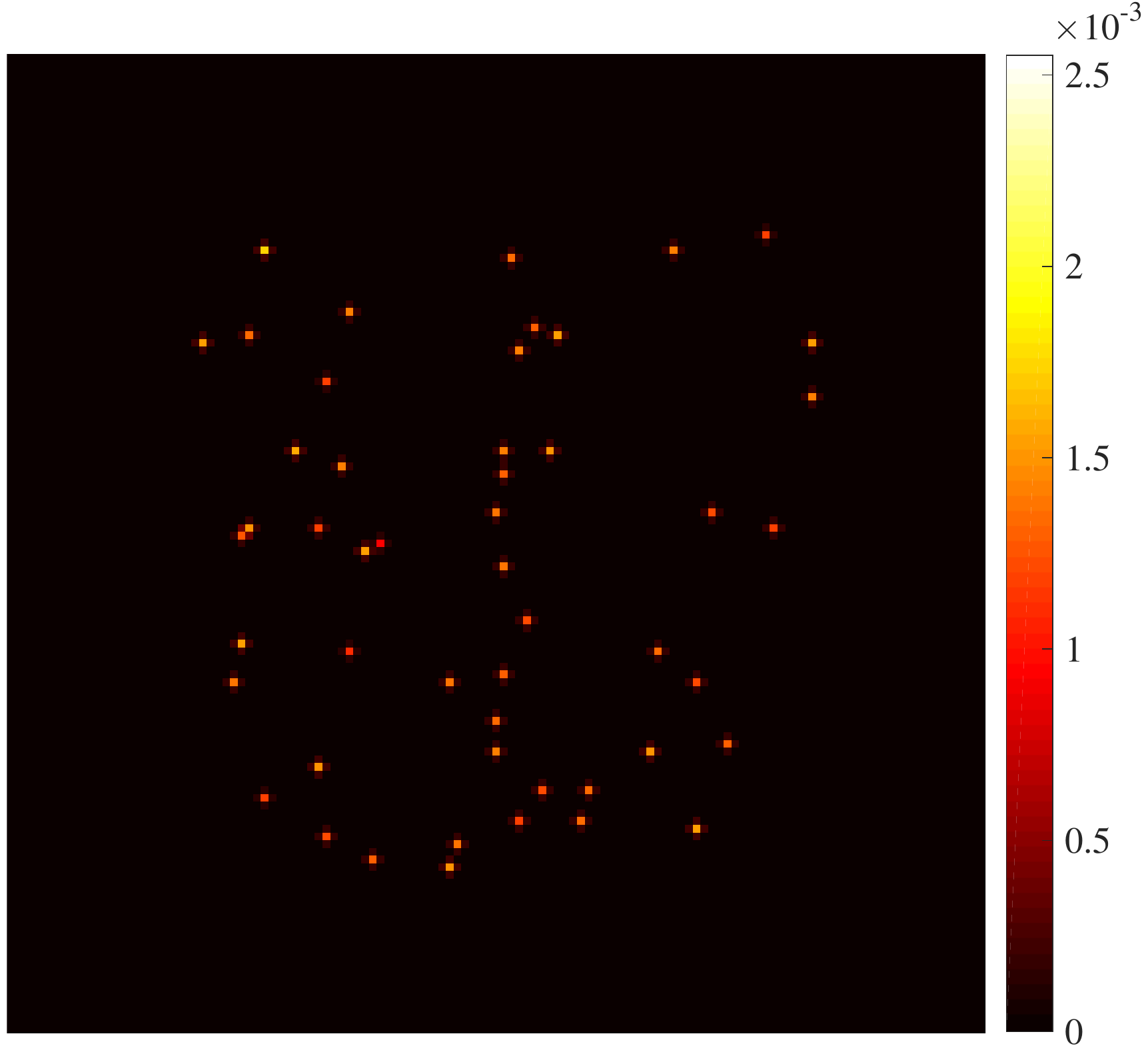}
&	\hspace*{0.1cm}\includegraphics[width=4.2cm]{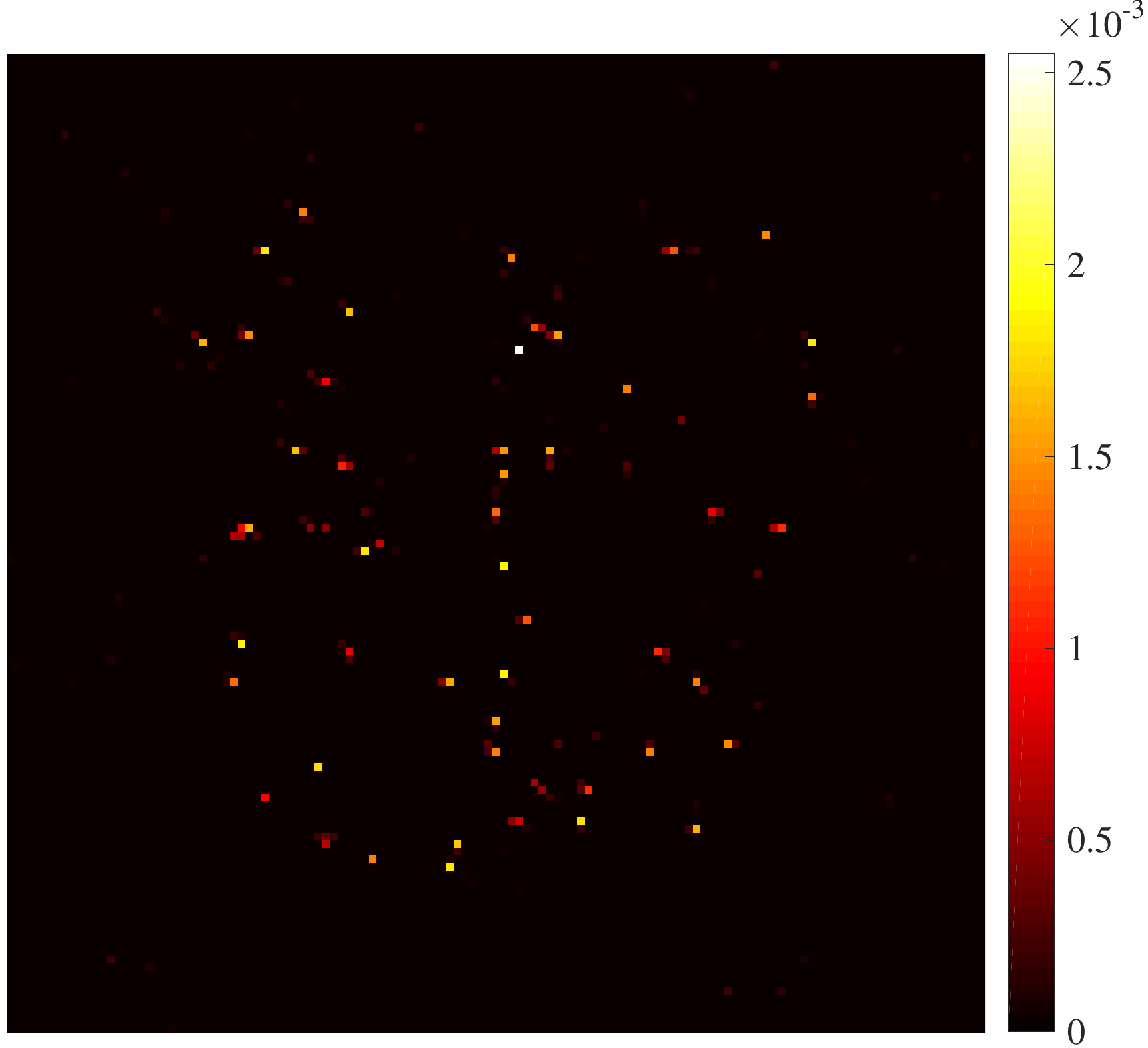}
&	\hspace*{0.1cm}\includegraphics[width=4.2cm]{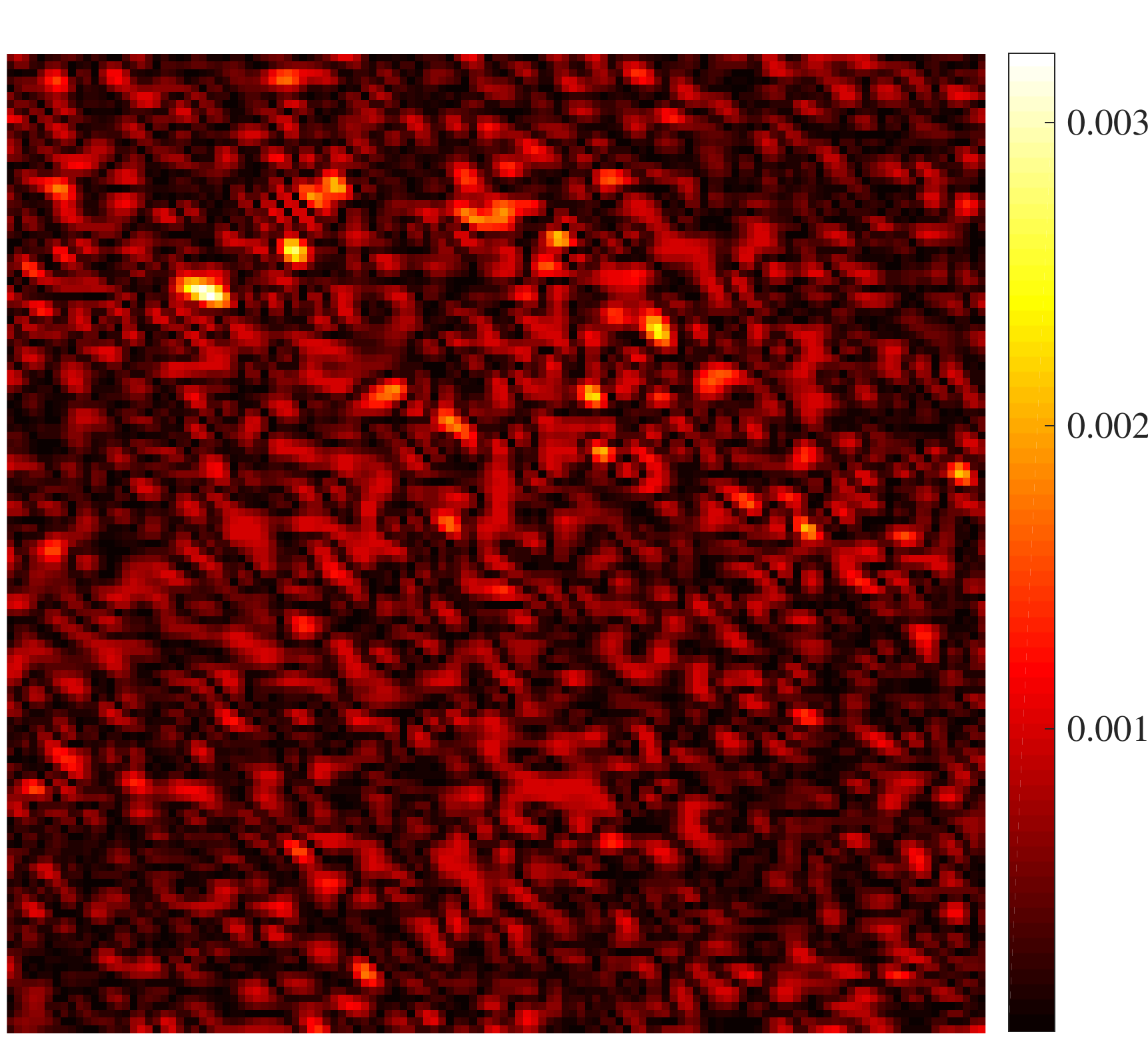}
&	\hspace*{0.1cm}\includegraphics[width=4.2cm]{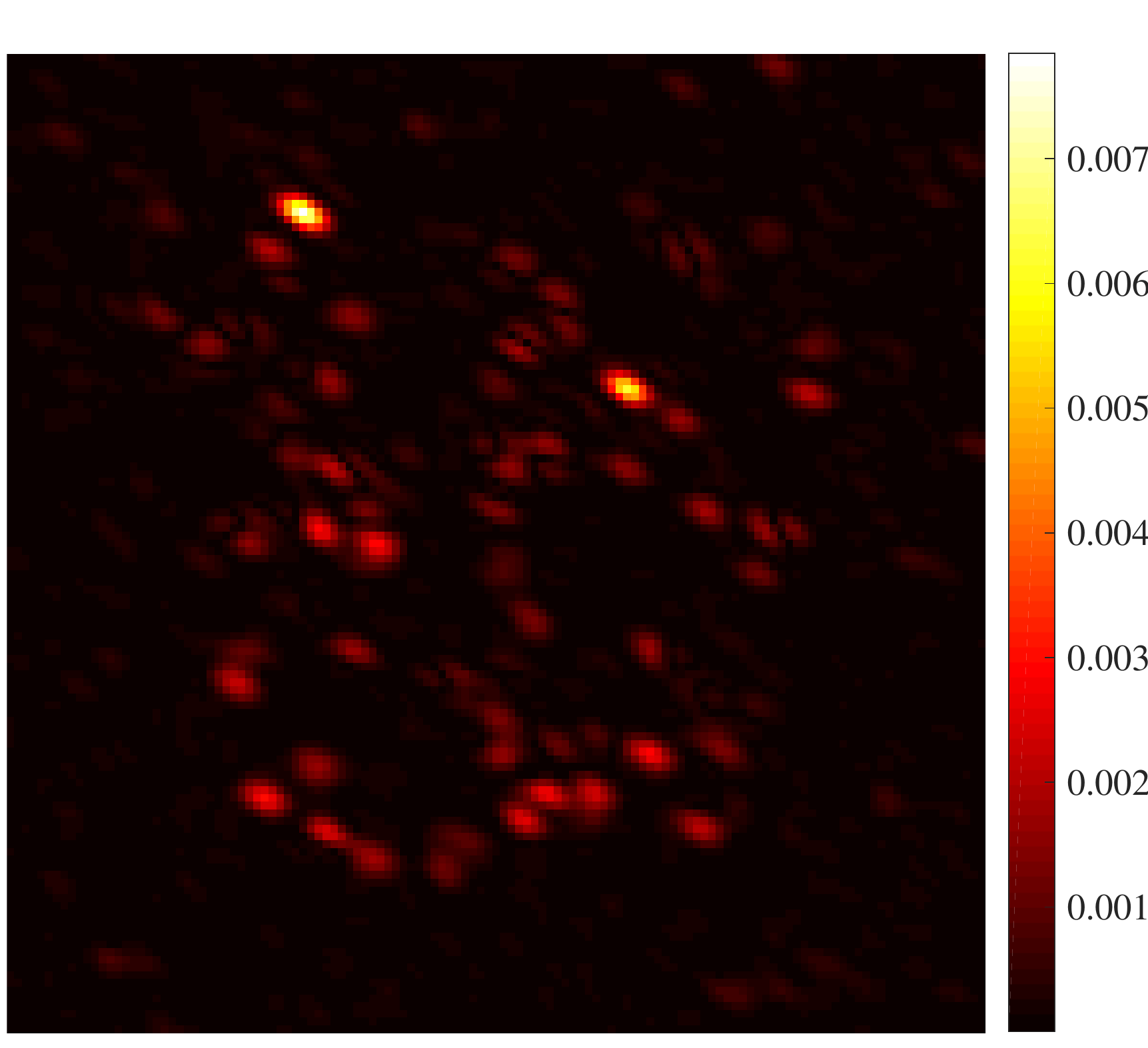}	
\end{tabular}
\vspace*{-0.2cm}
\caption{\label{Fig:test1:images}
Images corresponding to simulations performed in Section~\ref{Ssec:sim:test1-1} with 50 sources in $\overline{\epsilonb}_1$. 
The first column corresponds to the original unknown images $\overline{\epsilonb}$; the second column gives the associated reconstructions; the third column shows the residual images considering $\Gbs^\star$ and $\epsilonb^\star$ obtained either with the proposed method or with StEFCal-FB; and the fourth column corresponds to the residual images considering the true DDEs, and $\epsilonb^\star$ obtained either with the proposed method or with StEFCal-FB. 
The first and second rows correspond to the case when $\Es(\overline{\epsilonb}_1) = 1$. In the first row the results are obtained using our method, and in the second row using StEFCal-FB. 
The third row corresponds to the case when $\Es(\overline{\epsilonb}_1) = 0.1$ and results are obtained using our method. 
The fourth row corresponds to the case when $\Es(\overline{\epsilonb}_1) = 0.01$ and results are obtained using our method. 
}
\end{figure*}

\paragraph*{Simulation results.}

Figure~\ref{Fig:test1:curves} gives the results, as a function of $\Es(\overline{\epsilonb}_1)$ using different metrics, obtained for an average over 10 realizations varying the antenna distribution, the random images, and the DDEs coefficients.
The first graph shows the SNR of the reconstructed image $\epsilonb^\star$ with respect to the original image $\overline{\epsilonb}$.
In the second graph, we consider a success rate counting the positions of the sources in $\overline{\epsilonb}_1$ which are recovered successfully.
These first two graphs show that, whatever the number of considered sources in $\overline{\epsilonb}_1$ and the value of $\Es(\overline{\epsilonb}_1)$, our method outperforms the StEFCal-FB method, showing the importance of reconstructing the DDEs. 
More precisely, on the one hand, our method is able to recover $100\%$ of sources' positions for $\Es(\overline{\epsilonb}_1) \in \{0.1, 1, 10\}$, with respective SNR values of $10.7$, $16.9$ and $13.4$ dBs, independently of the number of sources belonging to $\overline{\epsilonb}_1$. 
It is important to emphasize that, in these three cases, the reconstruction quality depends mainly on the energy of the image and not on its intensity level. Indeed, if $\overline{\epsilonb}_1$ contains $90$ sources with global energy equals to $\Es(\overline{\epsilonb}_1) =10$, the intensity of sources belonging to $\overline{\epsilonb}_1$ is almost two times lower than the intensity of the sources in $\xb_o$. 
On the other hand, the StEFCal-FB obtains approximately $100\%$ only in the case $\Es(\overline{\epsilonb}_1) = 10$ with SNR value of $11$ dB (resp. $9.1$ dB and $8.6$ dB) when 10 (resp. 50 and 90) sources belong to $\overline{\epsilonb}_1$. Note that this latter corresponds to the case when the sources in $\overline{\epsilonb}_1$ are almost of same intensity as the sources in $\xb_o$. 
Concerning our method, the only case when the number of sources considered in $\overline{\epsilonb}_1$ gives different reconstruction results is when $\Es(\overline{\epsilonb}_1) = 10^{-2}$. Therefore, this case can be seen as a limit case, since for $\Es(\overline{\epsilonb}_1) = 10^{-3}$ our method has a success rate for position recovery of $0\%$. 
It suggests that using the proposed method we are able to improve the dynamic range by at least three orders of magnitude compared to accounting for DIEs only.
The third graph of Figure~\ref{Fig:test1:curves} shows the weighted $\ell_2$ norm of the residual images $\| \Fbs^\dagger \Gbs^{\star \, \dagger} \big( \Gbs^{\star} \Fbs \xb^\star - \yb \big) \|_2 /\sqrt{N}$. 
Here, $\xb^\star = \xb_o + \epsilonb^\star$ corresponds to the global estimated image (obtained either using our method or StEFCal-FB), 
and $\Gbs^\star$ corresponds either to the DDEs estimated with our method, i.e. $\Gbs^\star= \Gc(\Ubs_1^\star, \Ubs_2^\star)$ ($\Gc$ being defined in eq.~\eqref{eq:h_x_final}), or to the DIEs obtained using the StEFCal-FB method. 
Similarly, the last graph in Figure~\ref{Fig:test1:curves} shows the weighted $\ell_2$ norm of the residual images $\| \Fbs^\dagger\overline{\Gbs}^{ \dagger} \big( \overline{\Gbs}\Fbs \xb^\star - \yb \big) \|_2 /\sqrt{N}$, where the matrix $\overline{\Gbs}$ introduced in~eq.~\eqref{pb:inv_pb_image} corresponds to the original DDEs. 
These two last graphs show again the advantage of reconstructing the full DDEs instead of considering only DIEs. 
In particular, the third graph shows that our method gives a smaller norm of the residual images (of order $10^{-3}$ vs. $10^{-1}$ with StEFCal-FB), taking into account both the estimated image and the estimated DDEs. However, the fourth graph suggests that there is an ambiguity error between $\Gbs^\star $ and $\xb^\star$, mainly when $\Es(\overline{\epsilonb}_1) \in \{1, 10\}$ (i.e. corresponding to the cases when the sources in $\overline{\epsilonb}$ are of similar range as the sources in $\xb_o$) leading to important errors in the residual images when the true DDEs are considered. 

The above global observations can be observed as well by Figure~\ref{Fig:test1:images} showing images obtained considering 50 sources in $\overline{\epsilonb}_1$, with energy $\Es(\overline{\epsilonb}_1) \in \{0.01, 0.1, 1\}$.
The first two rows consider the case when $\Es(\overline{\epsilonb}_1) = 1$, showing the original image $\overline{\epsilonb}$ in the first column. The second column gives the reconstructed images obtained using the proposed method (first row) and the StEFCal-FB method (second row). For this realization, our method finds $100\%$ of the positions of the sources in $\overline{\epsilonb}_1$ and the estimate has an SNR equal to $18.9$ dB, while the StEFCal-FB method finds $90\%$ of the sources with a reconstructed image of SNR $=-3.22$ dB. This difference of SNR in the reconstruction can be understood by noticing that the background of the estimated image with StEFCal-FB is very noisy due to the non-estimation of the DDEs, which is not the case using our method. 
The third and the fourth columns correspond respectively to the residual images $| \Fbs^\dagger \Gbs^{\star \, \dagger} \big( \Gbs^{\star} \Fbs \xb^\star - \yb \big) |$ and $ |\Fbs^\dagger \overline{\Gbs}^{ \dagger} \big( \overline{\Gbs} \Fbs \xb^\star - \yb \big) |$ obtained with our method (first row) and the StEFCal-FB method (second row). 
In both cases, one can notice the errors of the residual images obtained using our method and considering $\Gbs^\star$ (resp. $\overline{\Gbs}$) are $100$ (resp. $10$) times smaller than using the StEFCal-FB method. However, as already noticed in Figure~\ref{Fig:test1:curves}, there is obviously an ambiguity error between $\Gbs^\star$ and $\xb^\star$ leading to larger errors when the true observation matrix $\overline{\Gbs}$ is considered. 
The third and fourth rows show similar results, in the cases when $\Es(\overline{\epsilonb}_1) = 0.1$ and $\Es(\overline{\epsilonb}_1) = 0.01$, respectively. However, since in these cases the StEFCal-FB method has a success rate of $0\%$ for finding the positions of faint sources, only the results obtained using our method are shown. 
In the case $\Es(\overline{\epsilonb}_1) = 0.1$ presented in the third row, the proposed method recovers all the positions of the faint sources, and the reconstructed image has an SNR of $12.32$ dB. Similarly, the reconstructed image presented in the fourth row, in the case $\Es(\overline{\epsilonb}_1) = 0.01$, our method recovers $90\%$ of the faint sources positions, and the SNR of the estimate is equal to $6.5$ dB. 
Thus, when the SNR is low, one can still observe visually that our method leads to good reconstruction results. 
Furthermore, as observed in Figure~\ref{Fig:test1:curves}, the $\ell_2$ norm of the residual images decreases with the energy of $\overline{\epsilonb}_1$.

\paragraph*{Ambiguity.}

As mentioned in the previous paragraph, Figure~\ref{Fig:test1:curves} and Figure~\ref{Fig:test1:images} emphasize the ambiguity in the reconstruction of the DDEs and the reconstruction of the image. 
This ambiguity can be observed by comparing the $\ell_2$ errors of the third and fourth graphs of Figure~\ref{Fig:test1:curves}, and comparing the error images shown in the third and fourth graphs of Figure~\ref{Fig:test1:images}. 
Indeed, one can observe that the residual images obtained with both the estimated image and DDEs (i.e. $ \Fbs^\dagger \Gbs^{\star \, \dagger} \big( \Gbs^{\star} \Fbs \xb^\star - \yb \big) $) have much smaller $\ell_2$ norm and amplitude than the residual images obtained considering the true DDEs (i.e. $ \Fbs^\dagger \overline{\Gbs}^{ \dagger} \big( \overline{\Gbs} \Fbs \xb^\star - \yb \big) $). 
Note that in particular, in Figure~\ref{Fig:test1:images}, the most significant ambiguities are associated with the brightest sources positions. 
This behaviour has already been observed in previous works, even in the DIE calibration case, and is known as ``ghost sources'' (see e.g. \cite{Grobler_MNRAS_2014}).

\subsubsection{Size of direction-dependent Fourier kernels}
\label{Ssec:sim:test1-2}

\paragraph*{Simulation settings.}

A second set of simulations have been performed considering images with point sources, investigating the size $S$ of the support of the Fourier direction-dependent kernels, and the standard deviation $\upsilon$ defined in~ Section~\ref{Ssec:sim:setting}. 
In this section, we consider the measurements obtained with $n_a = 200$ antennas for a single time interval $T = 1$. Moreover, we generate images $\overline{\epsilonb}_1$ such that they contain $10$ sources, and we fix $\Es(\overline{\epsilonb}_1) = 0.1$. 
We perform simulations considering direction-dependent Fourier kernels with support size $S \in \{ 3 \times 3, 7\times7, 11 \times 11, 15 \times 15\}$, and, for each of these values, we consider the standard deviation 
to be $\upsilon \in \{0.005, 0.01, 0.05, 0.1\}$ (see Section~\ref{Ssec:sim:setting}). 
The objective of these tests is to analyse the estimation of the unknown image by the proposed method, depending on the importance of the direction-dependent Fourier kernels. Typically, for a given $\upsilon$, larger is the support size $S$, the more is the number of unknowns for DDEs. Similarly, for a given $S$, the DDEs will affect much more the observed sky image for larger values of $\upsilon$, leading to a more important need for DDE calibration. Hence, for both these cases, accurate estimation of both the image and the DDEs becomes difficult.

\begin{figure*}
\begin{tabular}{cccc}
 \hspace*{-0.2cm}	\includegraphics[height=4.0cm]{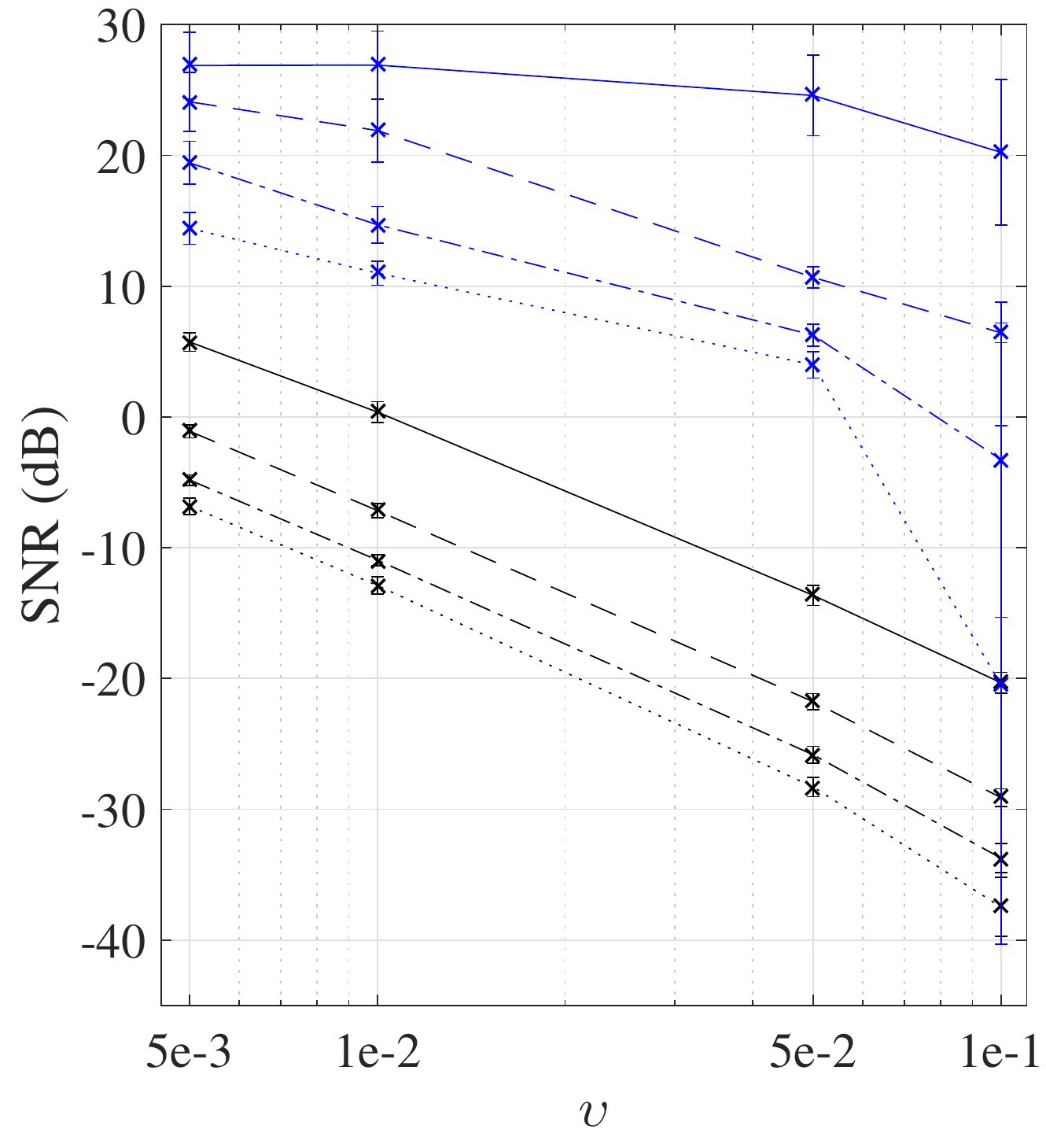}
&\hspace*{-0.2cm}	\includegraphics[height=4.0cm]{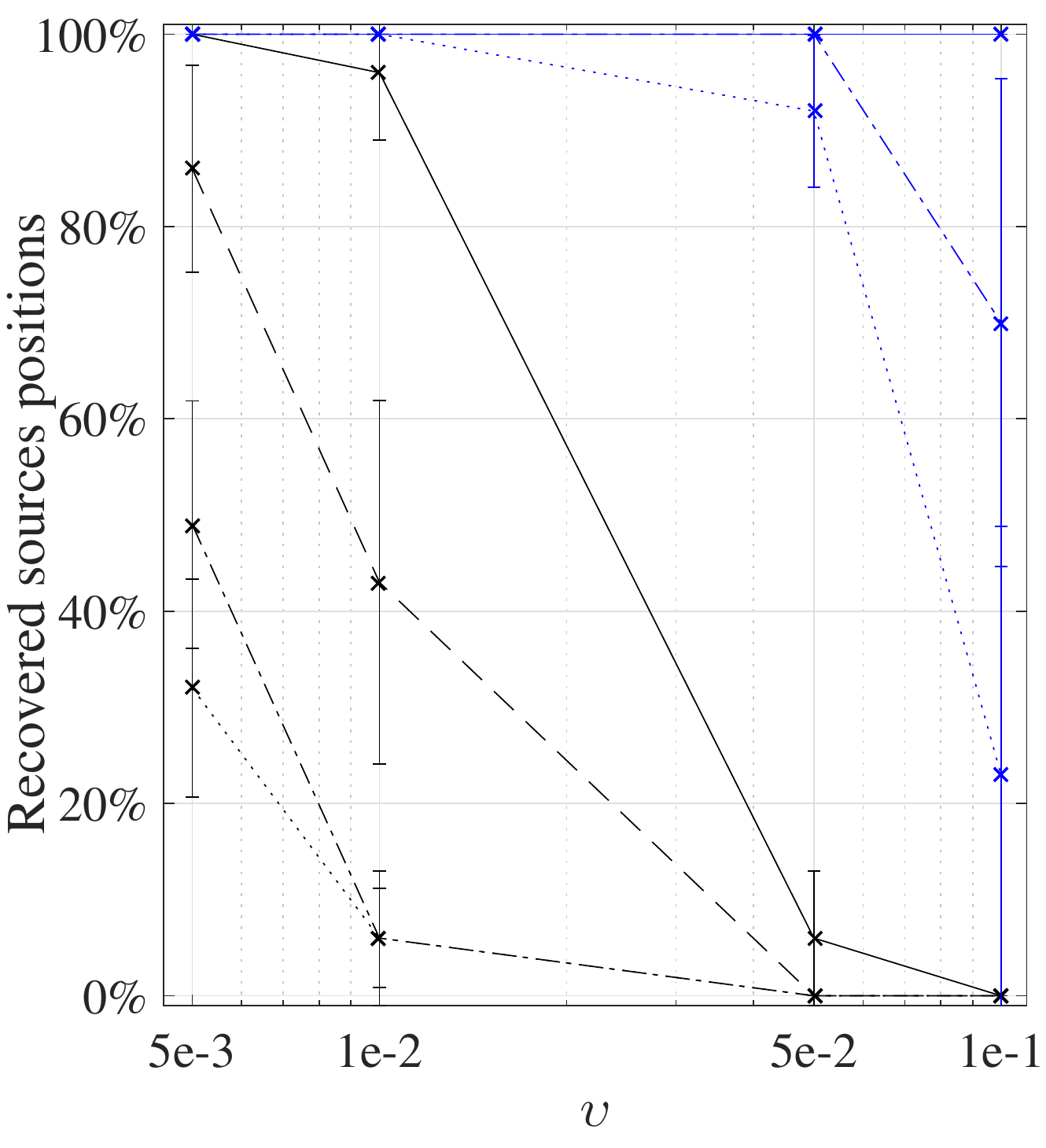}
&\hspace*{-0.2cm}	\includegraphics[height=4.0cm]{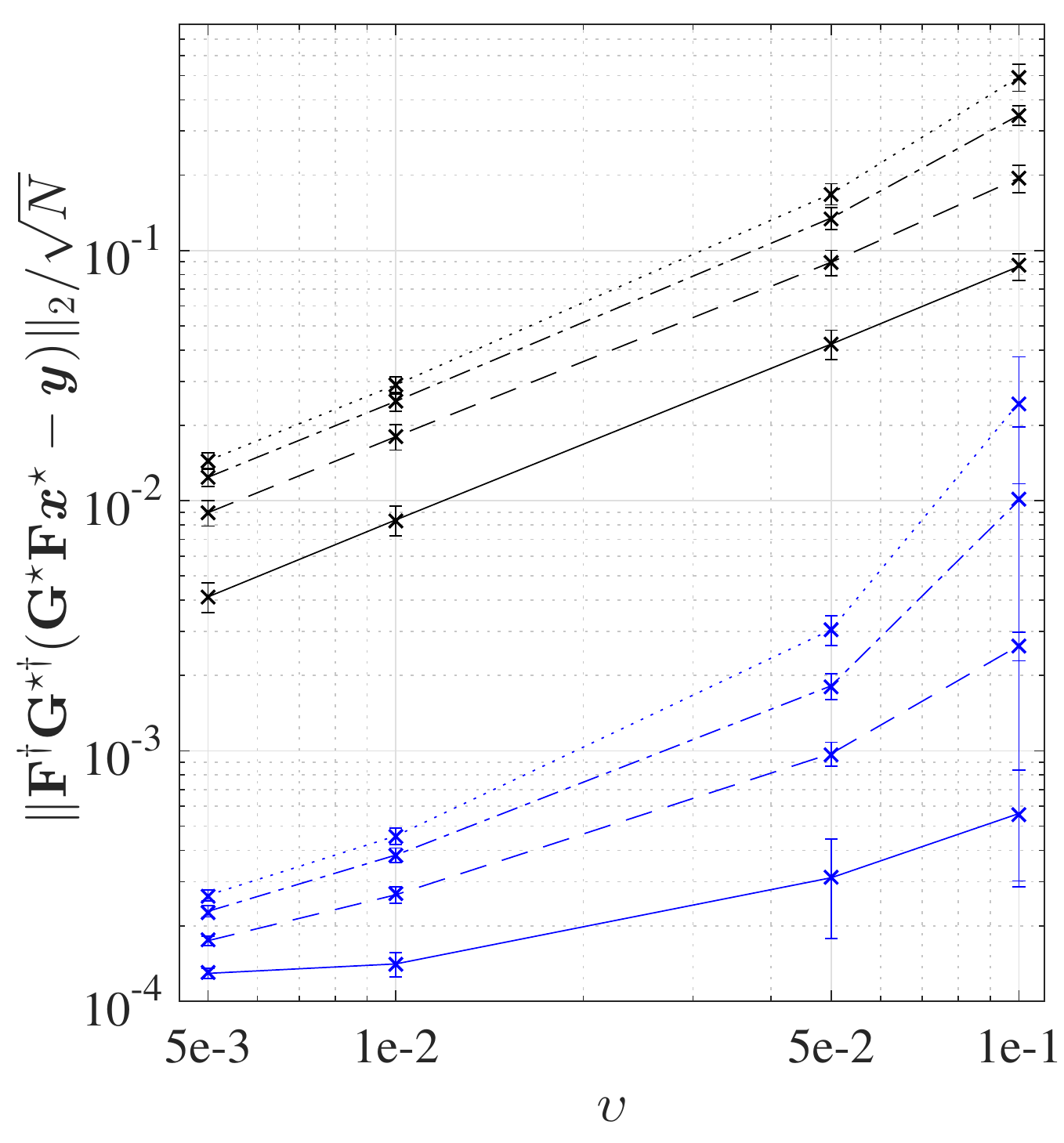}
&\hspace*{-0.2cm}	\includegraphics[height=4.0cm]{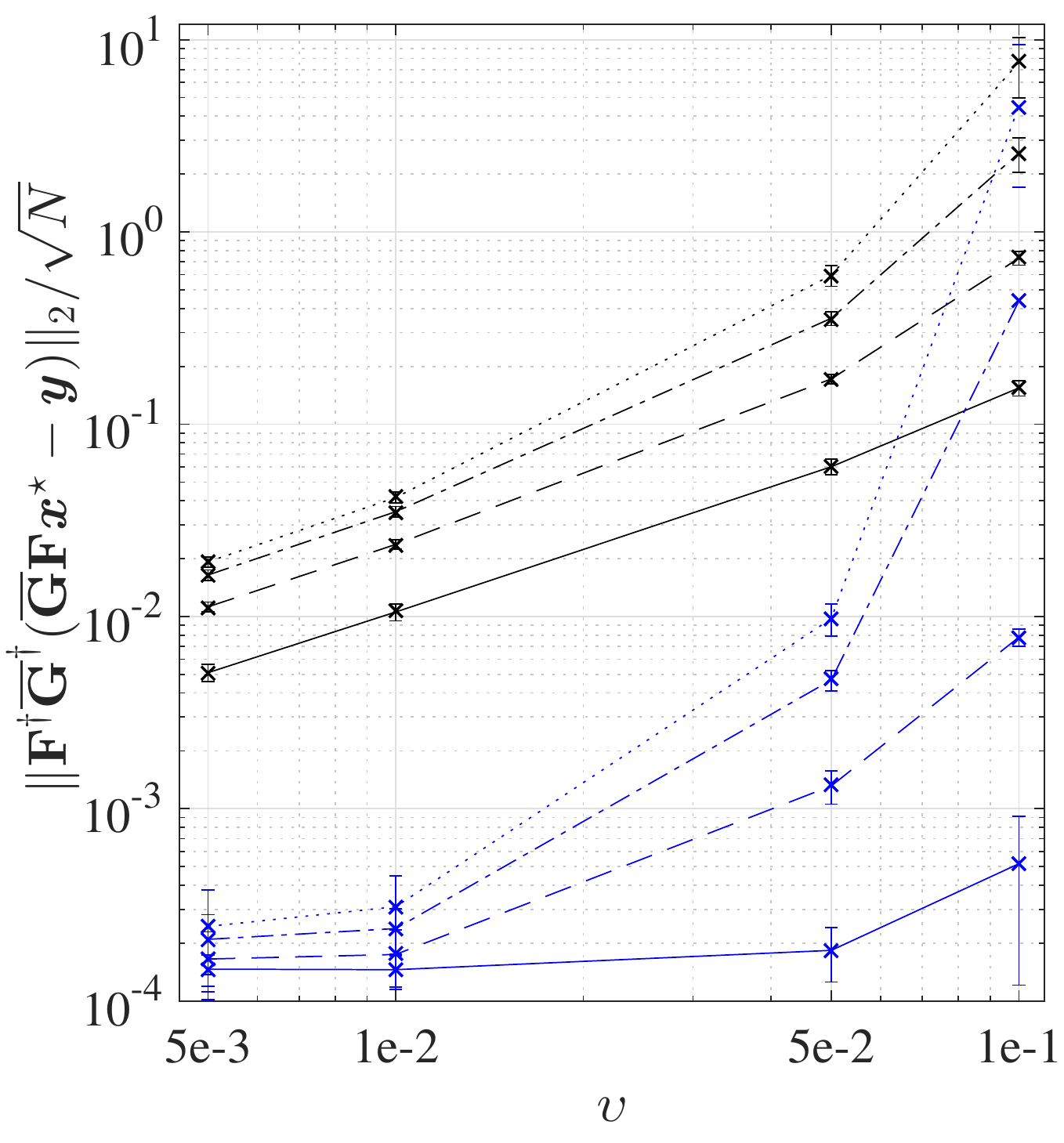}
\end{tabular}
\vspace*{-0.3cm}
\caption{\label{Fig:test2:curves}
Results obtained for simulations of Section~\ref{Ssec:sim:test1-2} using the proposed method (blue lines) and estimating only the DIEs with StEFCal-FB (black lines), considering the support size of Fourier direction-dependent kernels $S$ to be equal to $3\times3$, $7 \times 7$, $11 \times 11$ and $15 \times 15$ (resp. solid lines, dashed lines, dash-dotted lines and dotted lines), varying the standard deviation 
$\upsilon \in \{5\times 10^{-3},10^{-2},5\times 10^{-2}, 10^{-1}\}$. 
From left to right: SNR of the reconstructed $\epsilonb^\star$ with respect to $\overline{\epsilonb}$;
Success rate determining the percentage of recovered sources positions from $\overline{\epsilonb}_1$;
$\ell_2$ norm of the residual image $\| \Fbs^\dagger \Gbs^{\star \, \dagger} \big( \Gbs^\star \Fbs \xb^\star - \yb \big) \|_2 / \sqrt{N} $ considering $\Gbs^\star$ obtained with the estimated DDEs;
$\ell_2$ norm of the residual image $\| \Fbs^\dagger \overline{\Gbs}^{\dagger} \big( \overline{\Gbs} \Fbs \xb^\star - \yb \big) \|_2 / \sqrt{N} $ considering $\overline{\Gbs}$ obtained with the true DDEs.
Results are given for an average over 10 realizations varying the antenna distribution, the random images, and the DDEs.}
\vspace*{-0.6cm}
\end{figure*}
\begin{figure*}
\begin{tabular}{c@{}c@{}c@{}c@{}}
\hspace*{-0.4cm}	\includegraphics[width=4.2cm]{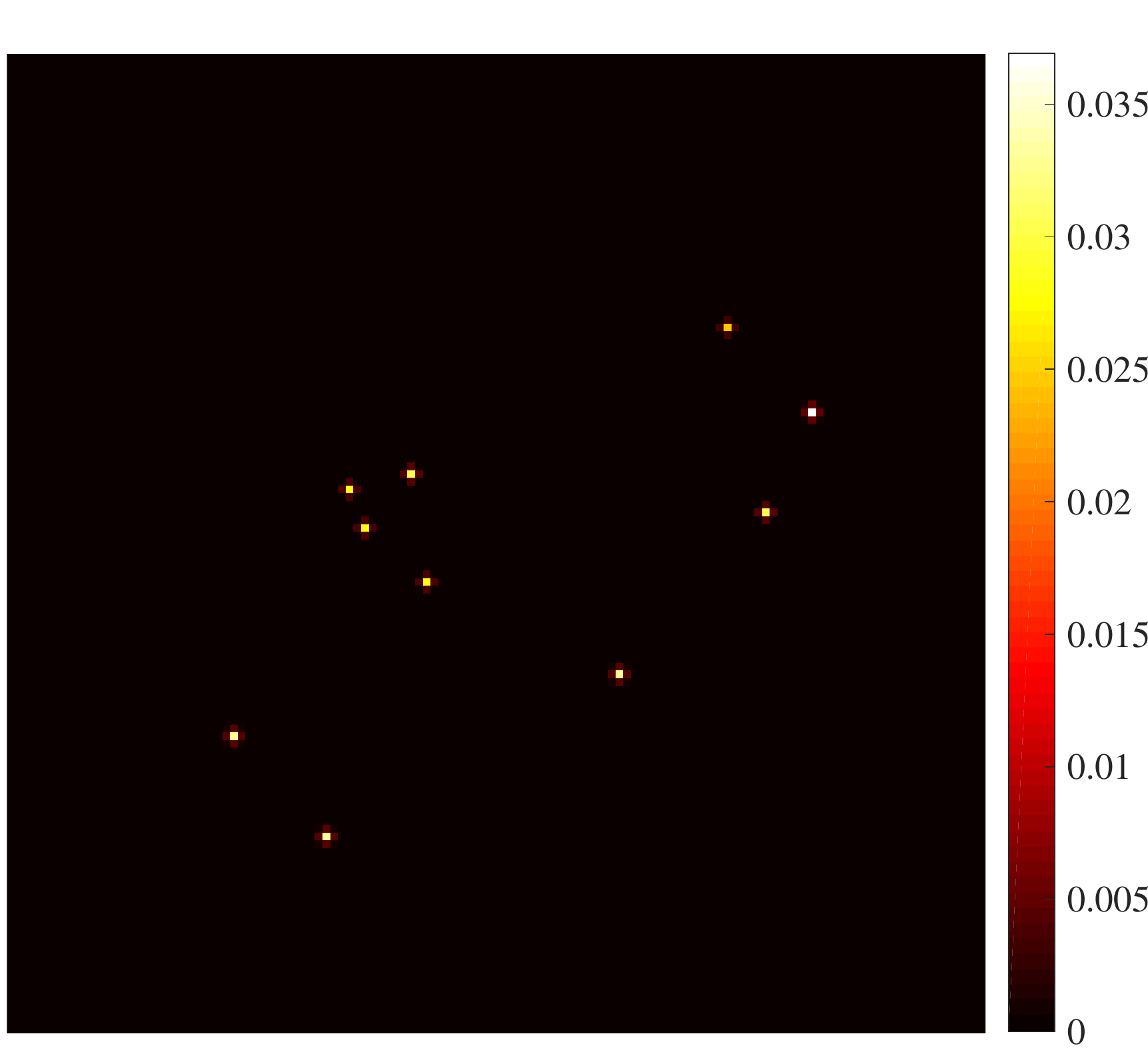}
&	\hspace*{0.1cm}\includegraphics[width=4.2cm]{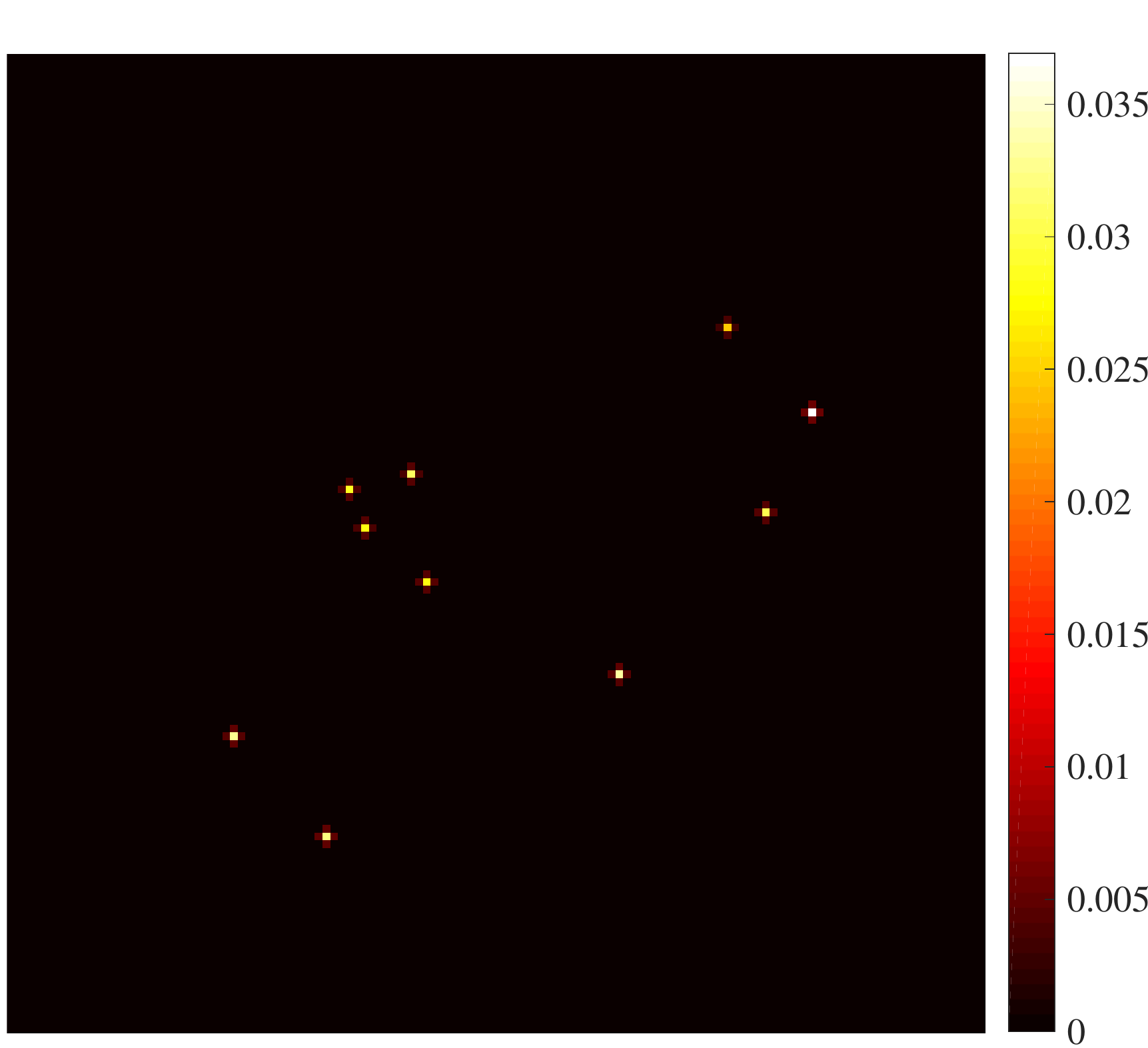}
&	\hspace*{0.1cm}\includegraphics[width=4.2cm]{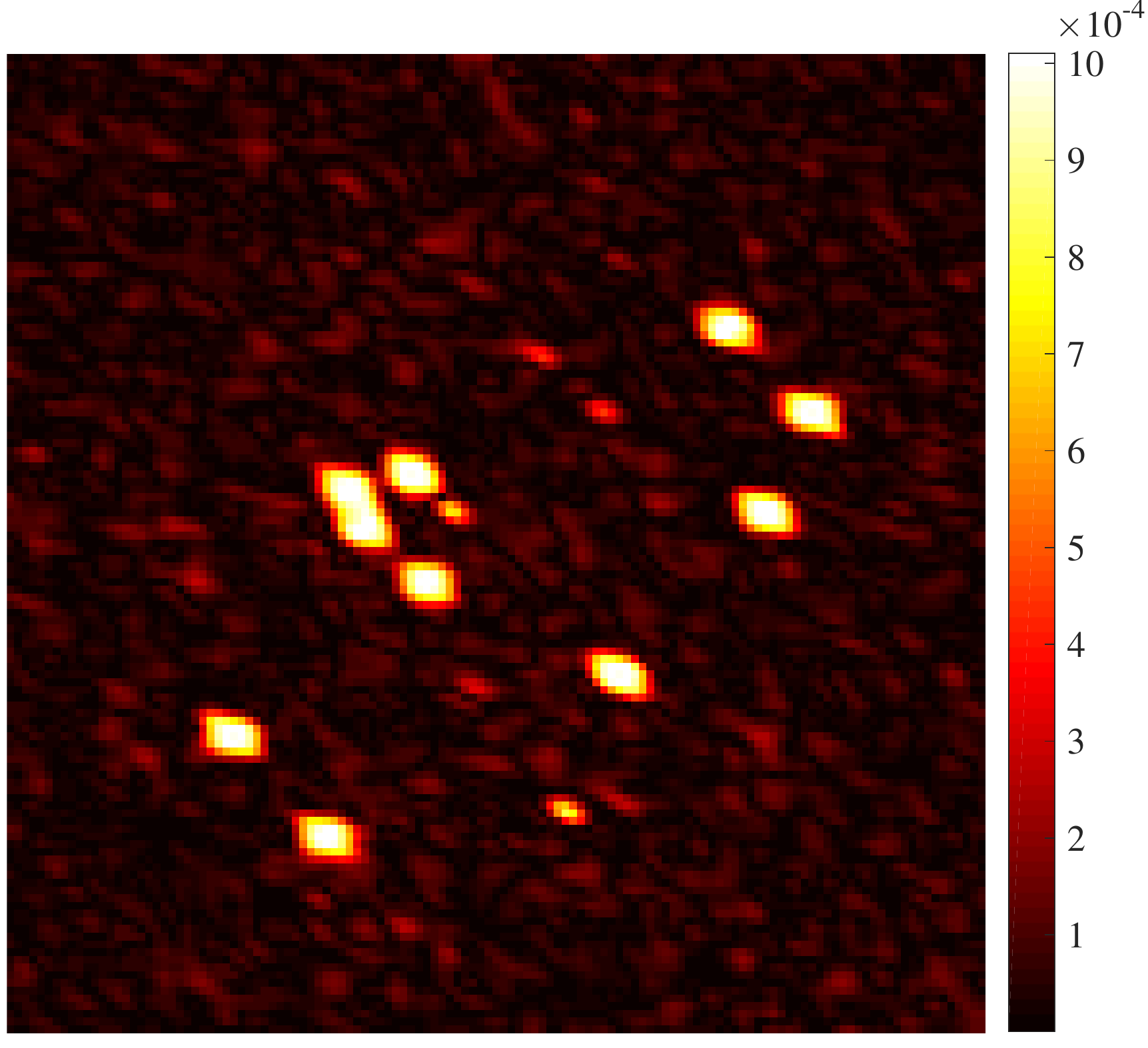}
&	\hspace*{0.1cm}\includegraphics[width=4.2cm]{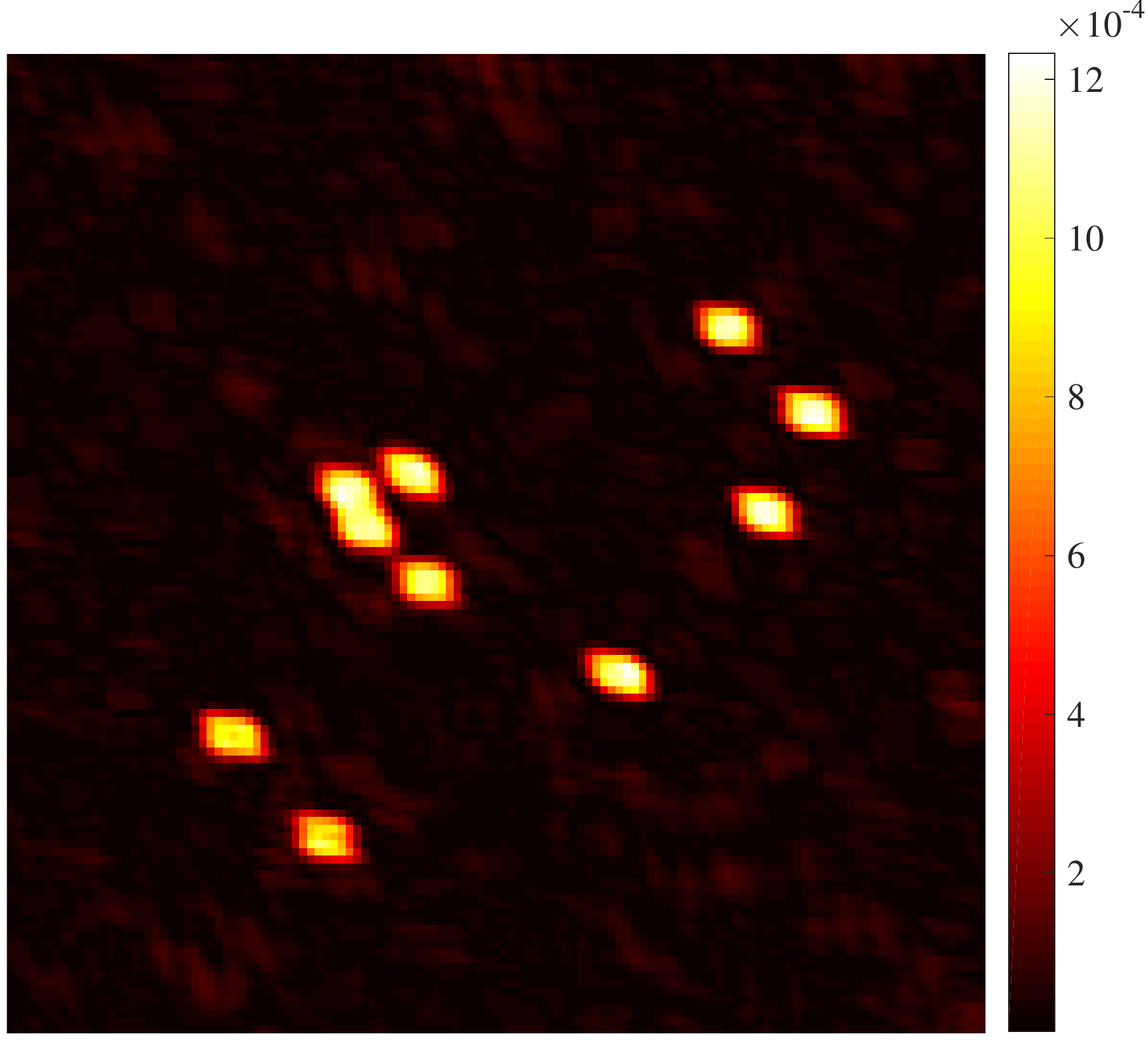}	\\[-0.3em]
\hspace*{-0.4cm}	
&	\hspace*{0.1cm}\includegraphics[width=4.2cm]{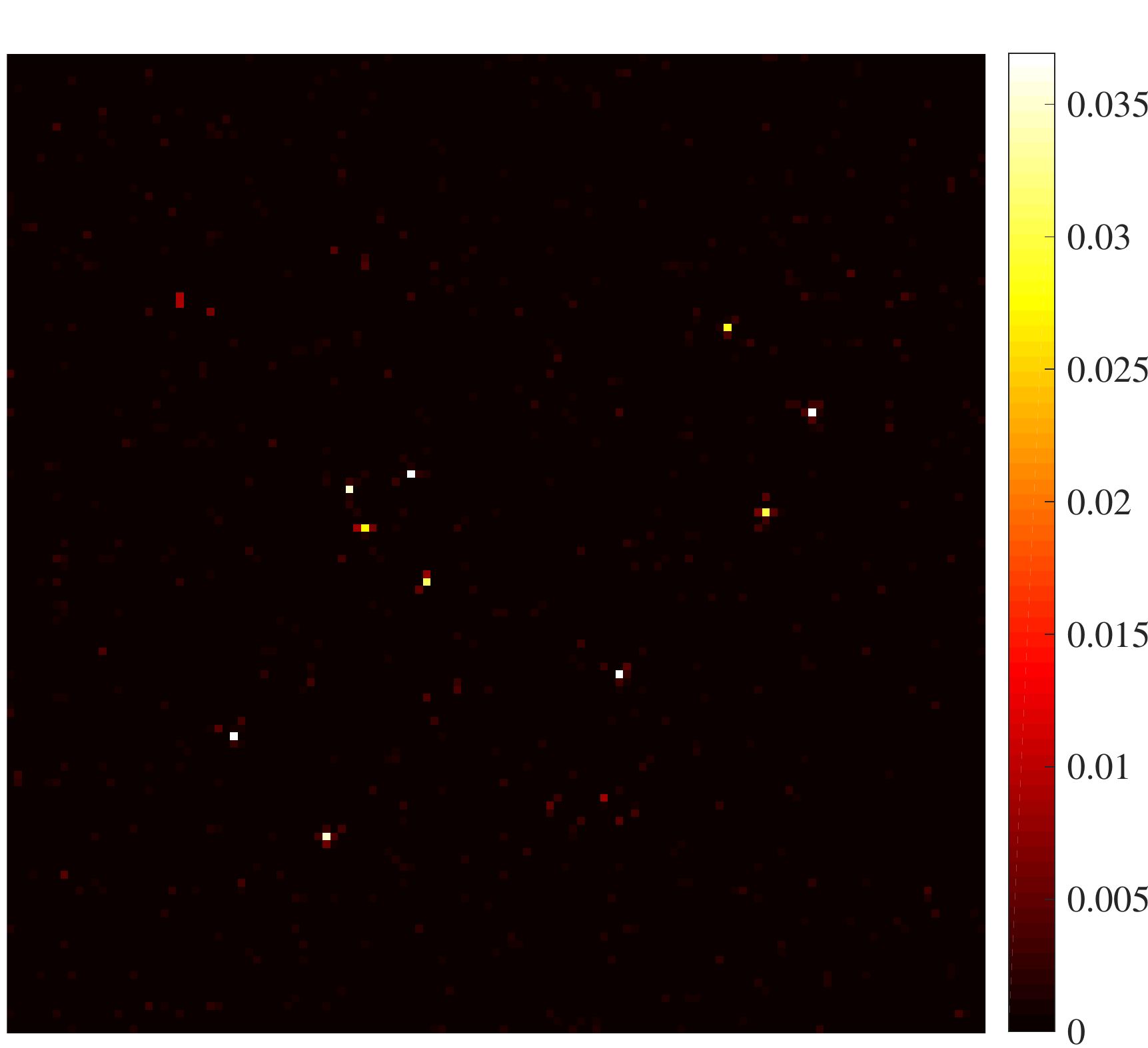}
&	\hspace*{0.1cm}\includegraphics[width=4.2cm]{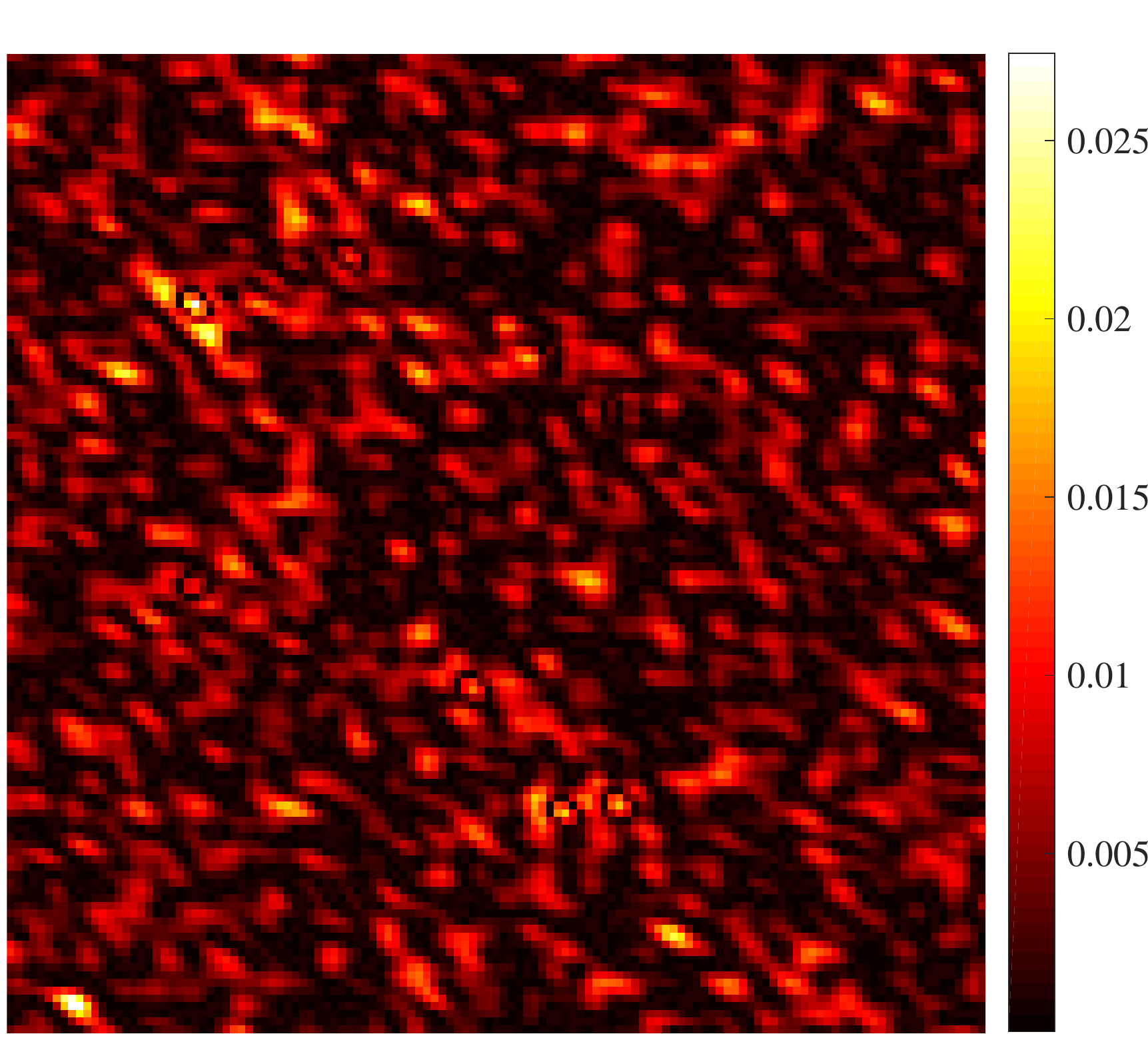}
&	\hspace*{0.1cm}\includegraphics[width=4.2cm]{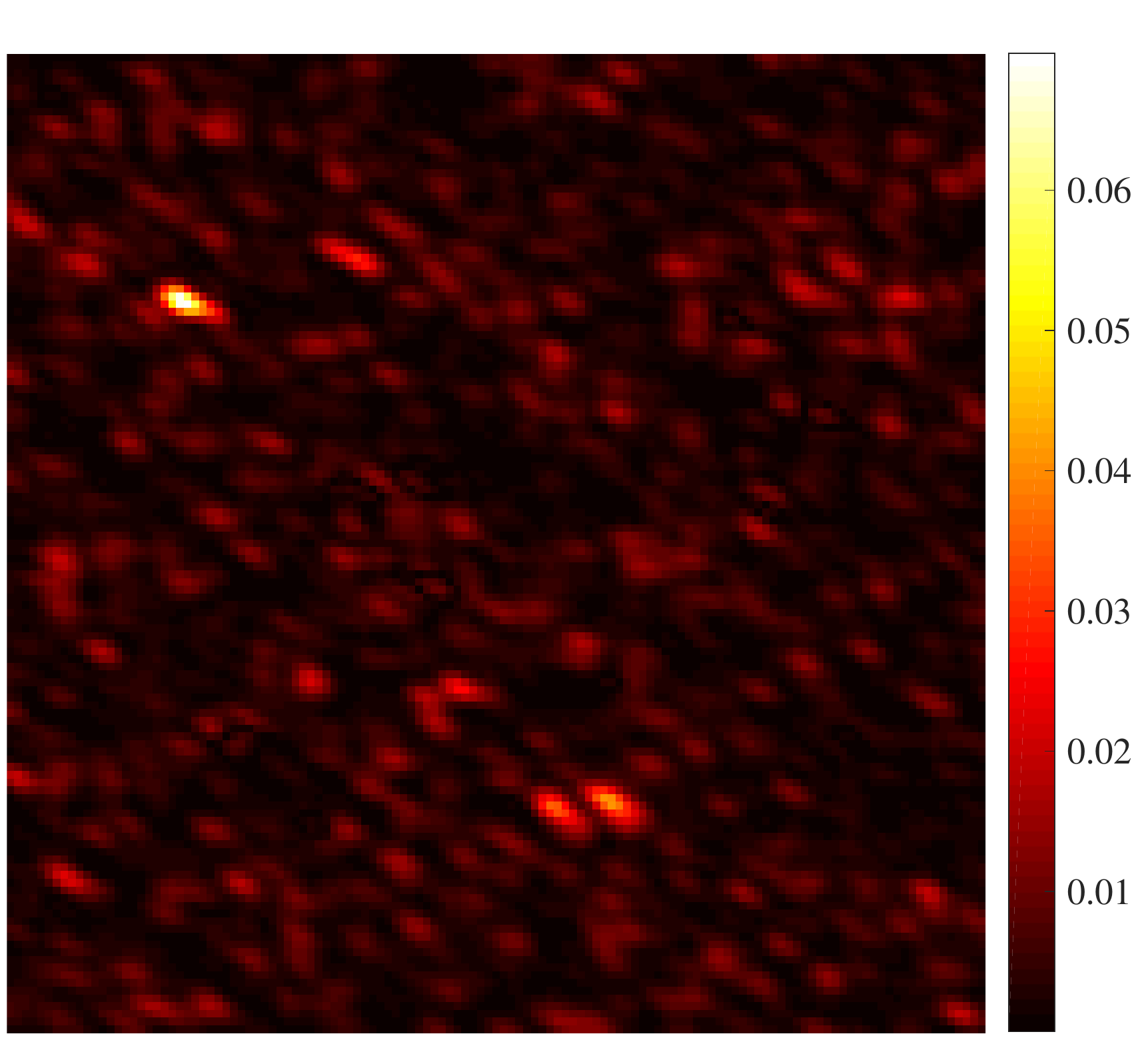}	\\[-0.3em]
\hspace*{-0.4cm}	\includegraphics[width=4.2cm]{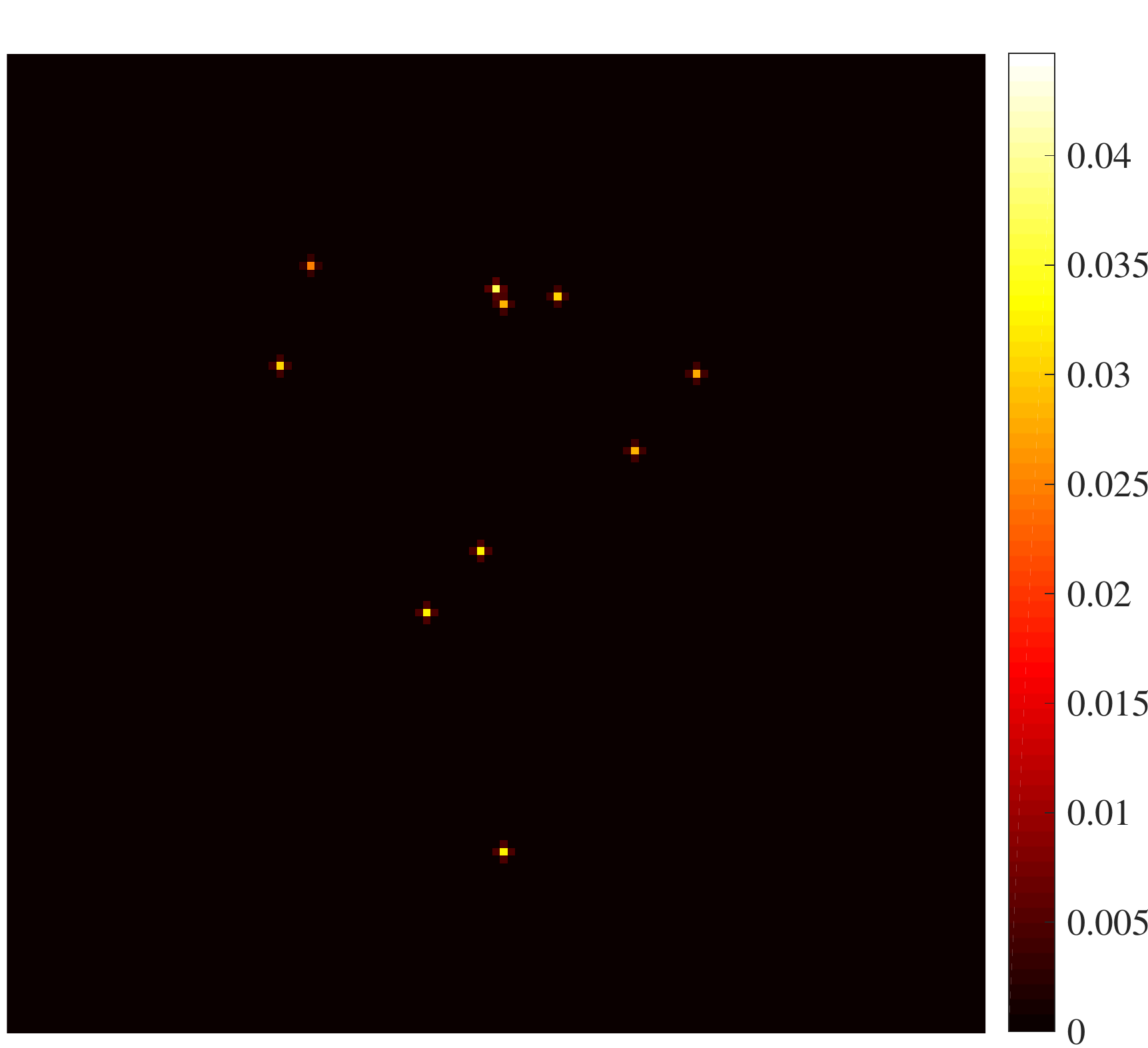}
&	\hspace*{0.1cm}\includegraphics[width=4.2cm]{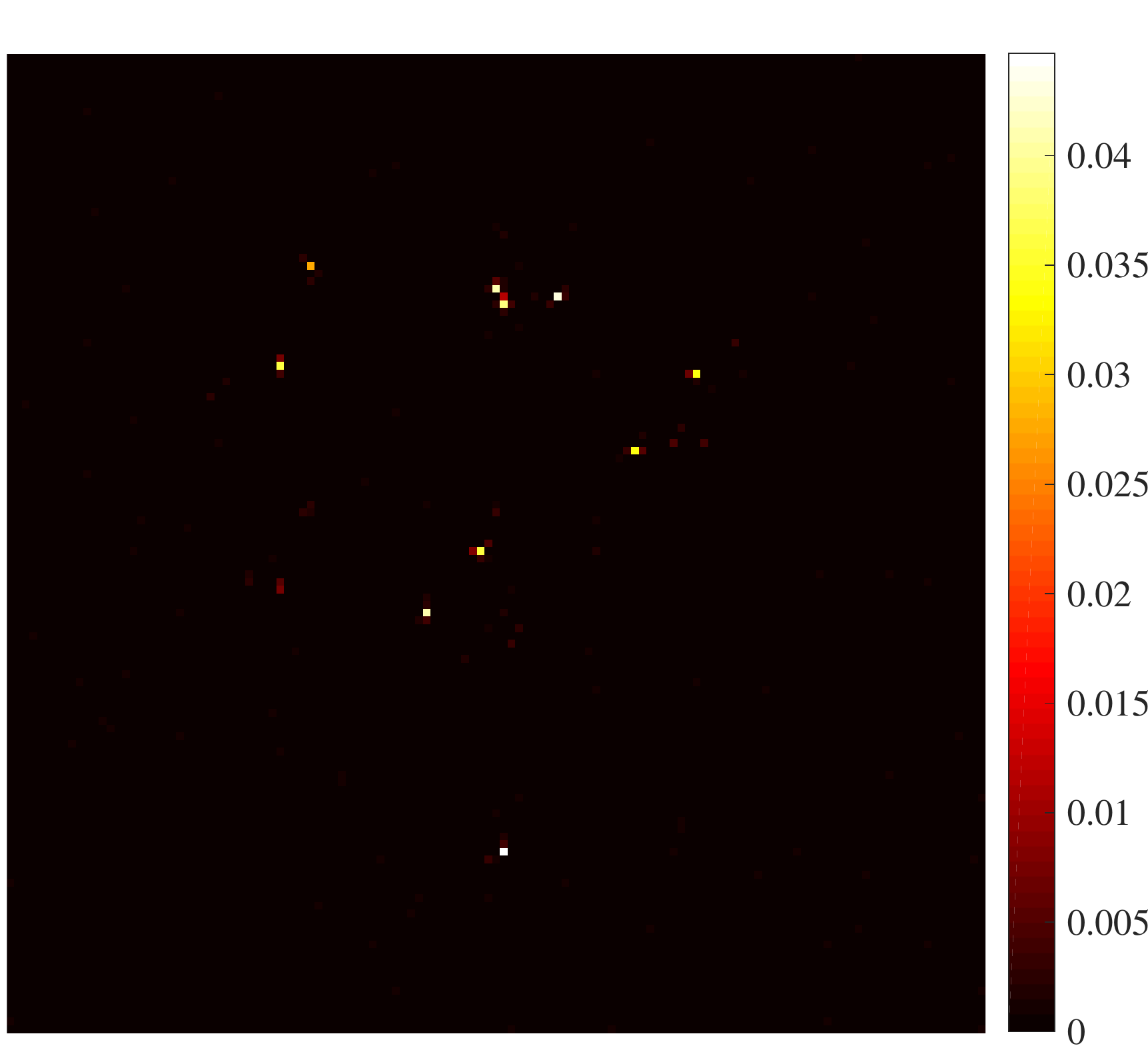}
&	\hspace*{0.1cm}\includegraphics[width=4.2cm]{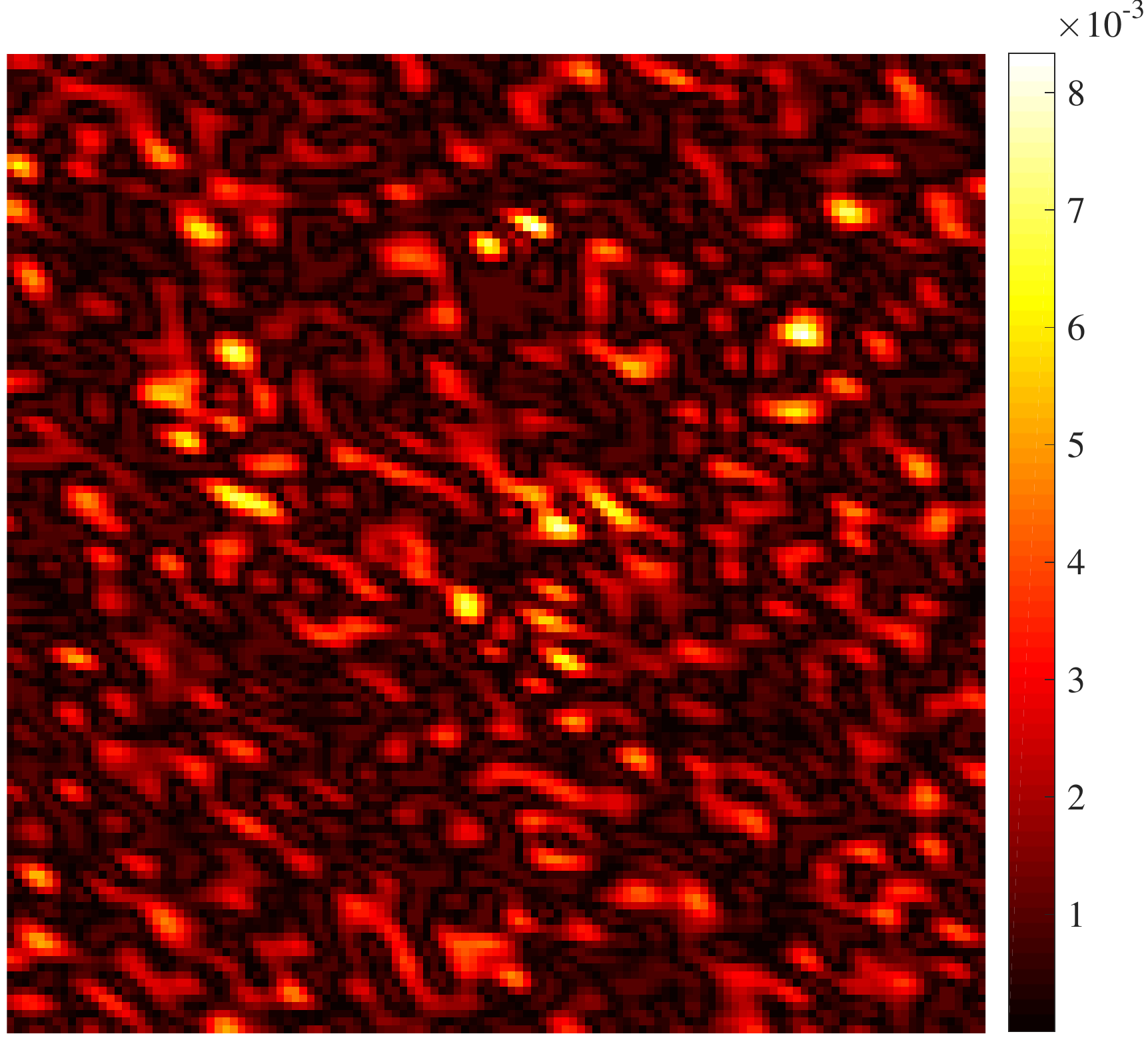}
&	\hspace*{0.1cm}\includegraphics[width=4.2cm]{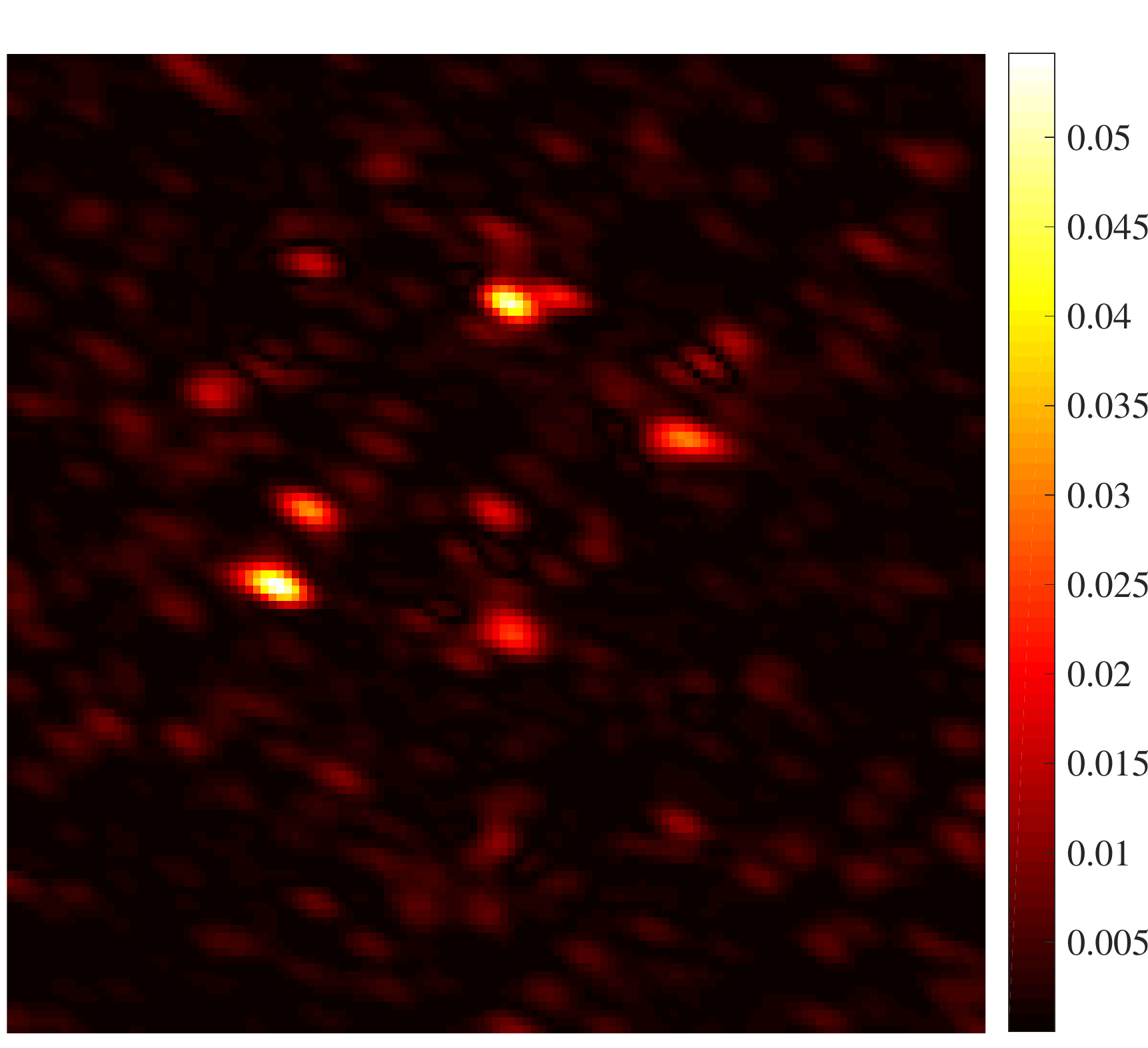}	\\[-0.3em]
\hspace*{-0.4cm}	\includegraphics[width=4.2cm]{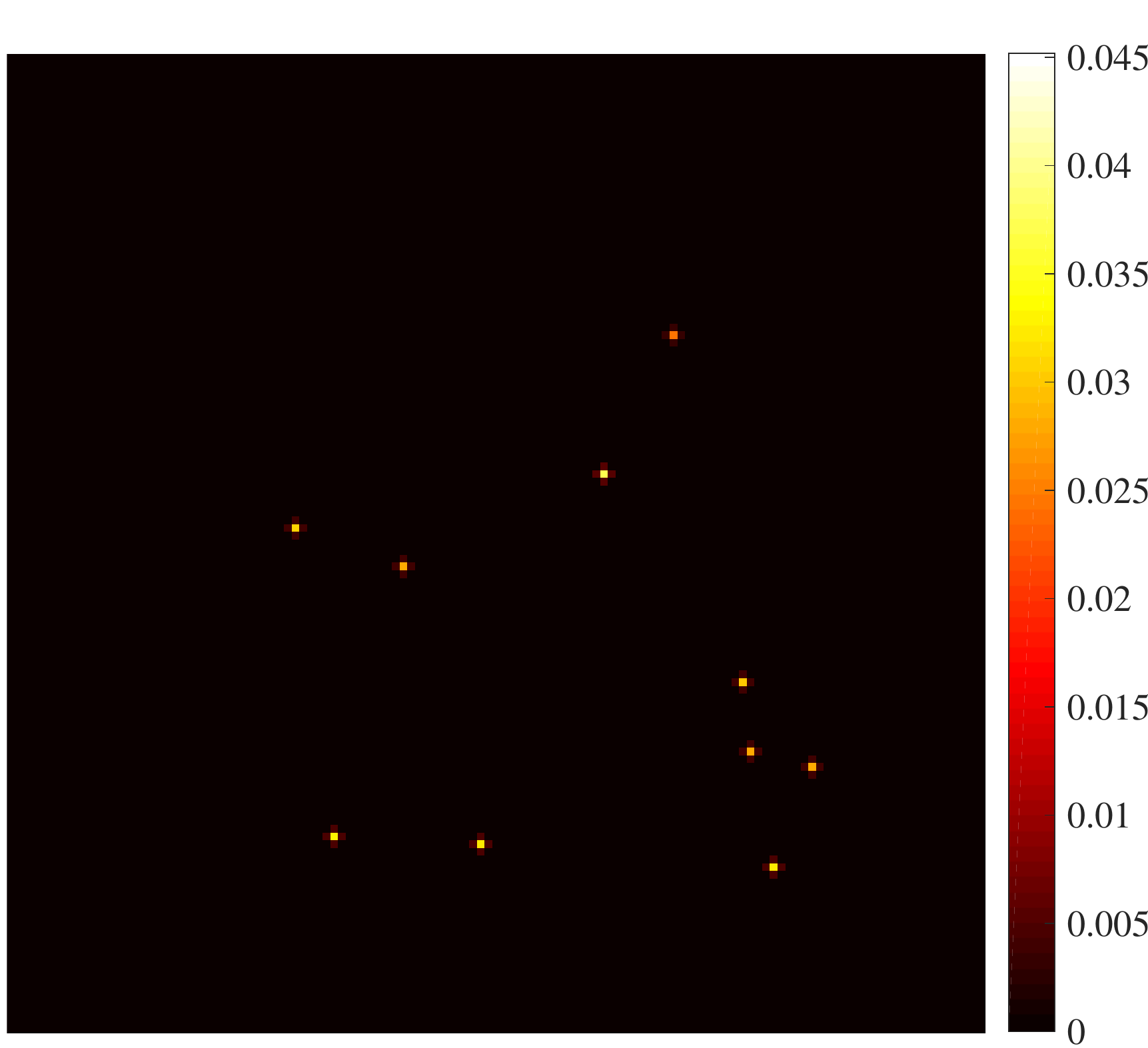}
&	\hspace*{0.1cm}\includegraphics[width=4.2cm]{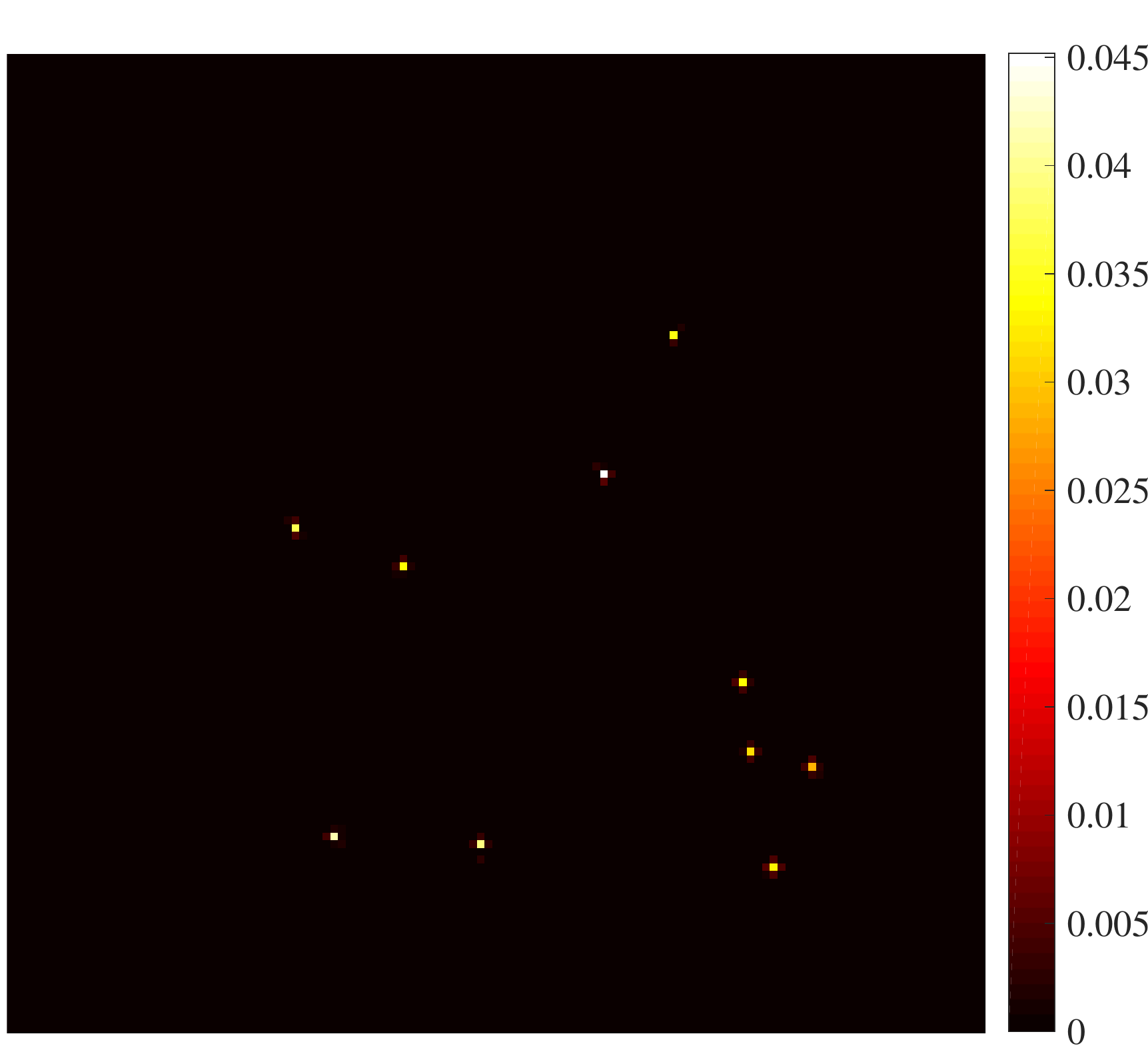}
&	\hspace*{0.1cm}\includegraphics[width=4.2cm]{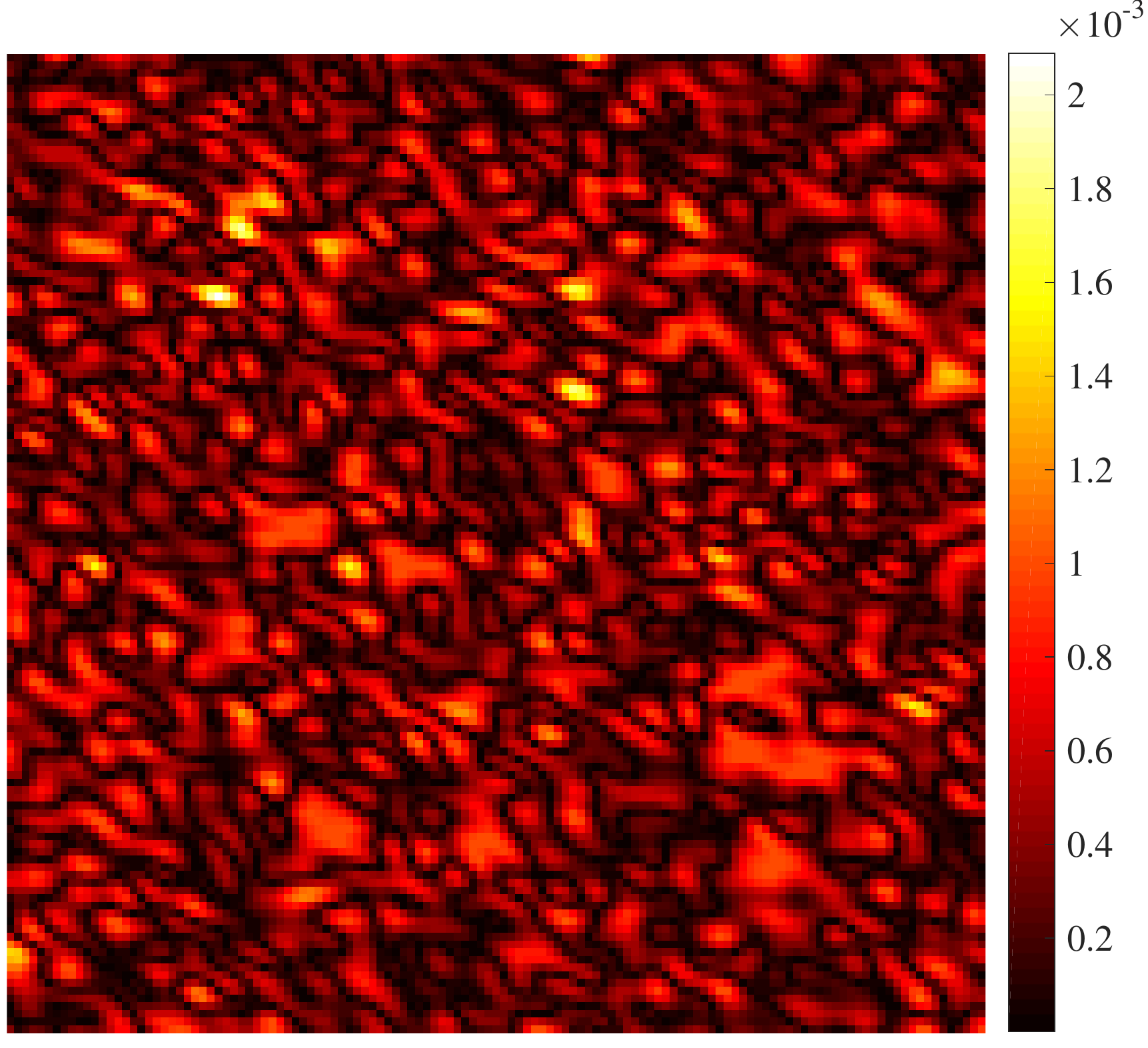}
&	\hspace*{0.1cm}\includegraphics[width=4.2cm]{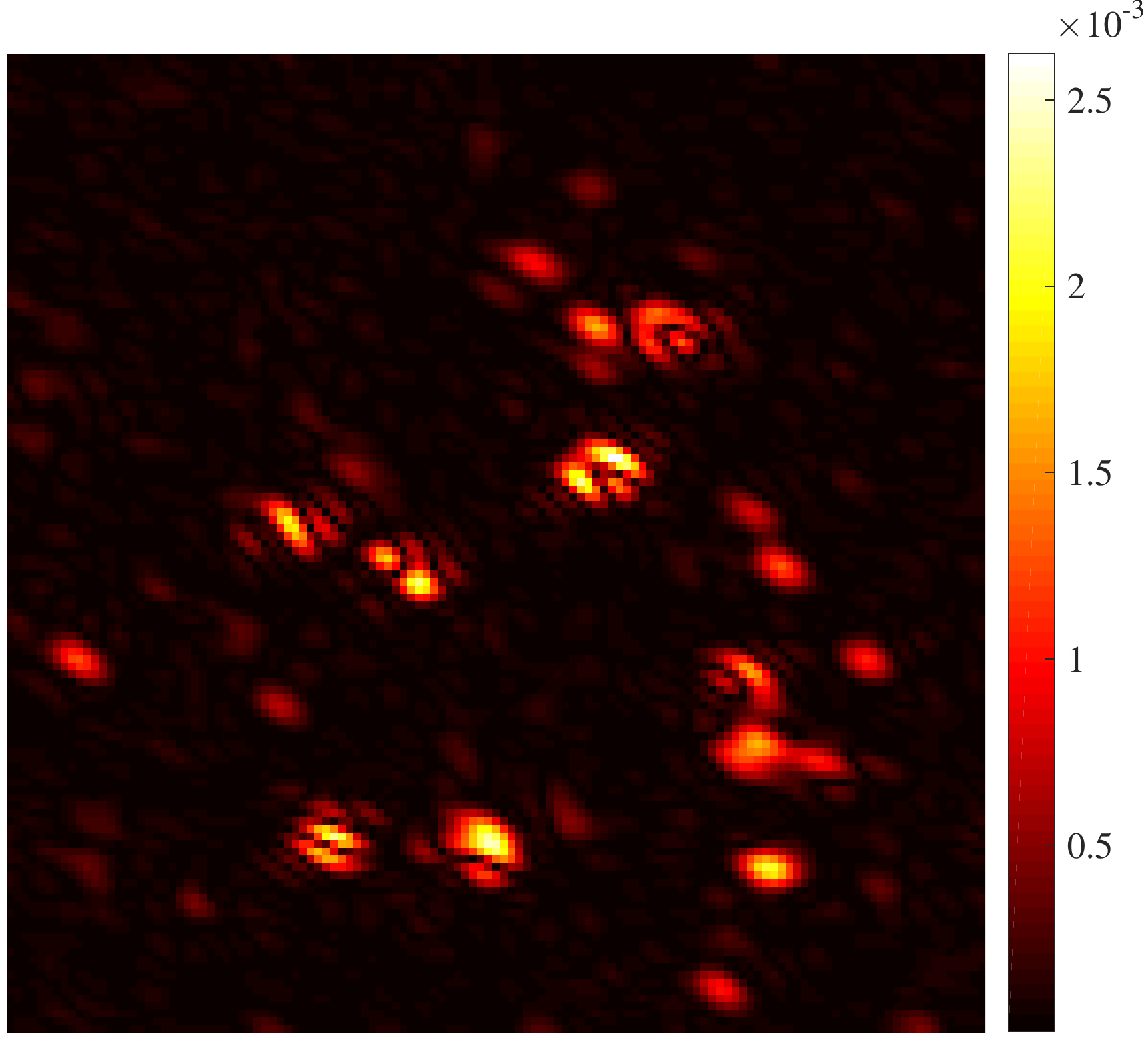}	
\end{tabular}
\vspace*{-0.2cm}
\caption{\label{Fig:test2:images}
Images corresponding to simulations in Section~\ref{Ssec:sim:test1-2}. 
The first column corresponds to the original unknown images $\overline{\epsilonb}$; the second column gives the associated reconstructions; the third column shows the residual images considering $\Gbs^\star$ and $\epsilonb^\star$ obtained either with the proposed method or with StEFCal-FB; and the fourth column shows the residual images considering the true DDEs, and $\epsilonb^\star$ obtained either with the proposed method or with StEFCal-FB. 
The first two rows correspond to the case when $S = 3\times3$ and $\upsilon=0.005$. In the first row are the results obtained using our method; in the second row the results obtained with StEFCal-FB. 
The third row corresponds to the case when $S = 11 \times 11$ and $\upsilon = 0.05$, where results are obtained using our method. 
The fourth row corresponds to the case when  $S = 15 \times 15$ and $\upsilon = 0.01$, where results are obtained using our method. 
}
\end{figure*}

\paragraph*{Simulation results.}

A comparison of the results obtained using the proposed approach and the StEFCal-FB method is presented in Figure~\ref{Fig:test2:curves}. As in Figure~\ref{Fig:test1:curves}, the four graphs represent respectively: the SNR of the estimated images $\epsilonb^\star$, the success rate determining the percentage of recovered sources positions in $\overline{\epsilonb}_1$, and the $\ell_2$ norm of the residual images obtained considering either the estimated DDEs (third graph) or the true DDEs (fourth graph). 
As expected, the difficulty of reconstructing accurately the image and the DDEs increases with $S$ and $\upsilon$. In particular, we can observe that the only case when StEFCal-FB estimates correctly $100\%$ of the sources' positions is for $S = 3 \times 3$ and $\upsilon = 0.005$. However, even in this case, our method leads to a better estimation of $\overline{\epsilonb}$ as attested by the SNR graph. Indeed, in this case the images reconstructed with our method have an averaged SNR of $26.9$ dB, while it is equal to $5.7$ dB with StEFCal-FB. In general, we can observe that for a support of Fourier direction-dependent kernels of sizes $S = 3 \times 3$ and $S = 7 \times 7$, our method recovers $100\%$ of the sources in $\overline{\epsilonb}_1$. Moreover, for $S = 3 \times 3$ the reconstructed images have a stable average SNR, reaching $20.2$ dB in the worst case when $\upsilon = 0.1$, and the residual images have small errors for all $\upsilon$ considered. 
This leads to the conclusion that even when direction-dependent Fourier kernels have very small amplitude, it is not enough to estimate only the DIEs, and our method improves significantly the quality of the reconstructed image. 
In the cases when $S = 11 \times 11$ and $S = 15 \times 15$, our method is able to detect correctly $100\%$ of the positions of sources in $\overline{\epsilonb}_1$ until $\upsilon = 0.01$ and $\upsilon = 0.05$, respectively. This shows that the proposed method is very stable even considering large Fourier direction-dependent kernels, impacting significantly the observations. However, it seems that there is a breaking point when $S = 15 \times 15$ and $\upsilon = 0.1$. In this case our algorithm is unstable and can lead to very poor reconstructions as attested by the four graphs.

In addition, images associated with these simulations are provided in Figure~\ref{Fig:test2:images}. 
The first two rows show the images associated with the simpler case when $S = 3 \times 3$ and $\upsilon = 0.005$, showing the original image $\overline{\epsilonb}$ in the first column. The other images in these rows consists of the results obtained using our method (first row) and the StEFCal-FB algorithm (second row). 
Visually, these three images are very similar and both methods recover the exact positions of the 10 sources. However, the quality of the image recovered with our method (first row, second column) is significantly better, since it has an SNR equals to $28.1$ dB, against $6.7$ dB for StEFCal-FB (second row, second column). This shows that even with very small DDEs, the details of the image are estimated more accurately with the proposed method, estimating the full DDEs. The third and fourth columns show the residual images obtained considering the estimated DDEs and the true DDEs, respectively. Note that in both cases, our method leads to residual images with two orders of magnitude lower than the residual images obtained using StEFCal-FB. 
Similarly, the third and fourth rows show the results obtained with our method for the cases $S = 11 \times 11$ and $\upsilon = 0.05$, and $S = 15 \times 15$ and $\upsilon = 0.01$, respectively. The third row considers a case with very large DDEs for both the support size and the standard deviation. For this simulation, the SNR is equal to $8.26$ dB, and we can observe few small artefacts in the reconstructed image. The fourth row gives results considering a very large support size $S$ but with a small standard deviation. In this case, the SNR of the reconstructed image is equal to $12.35$ dB, and we can observe that there are errors in the amplitude estimation but the sources positions are recovered exactly. 
Note that in both cases presented in third and fourth rows, the errors in the residual images are very small. 
Moreover, accordingly to the third and fourth graphs of Figure~\ref{Fig:test2:curves}, it seems that the ambiguity error is decreasing when both the standard deviation $\upsilon$ and the support size $S$ are decreasing.

\subsubsection{Time dependency and antennas distribution}
\label{Ssec:sim:test1-3}

\paragraph*{Simulation settings.}

\begin{figure}
\begin{tabular}{c@{}c@{}}
	\includegraphics[width=4.2cm]{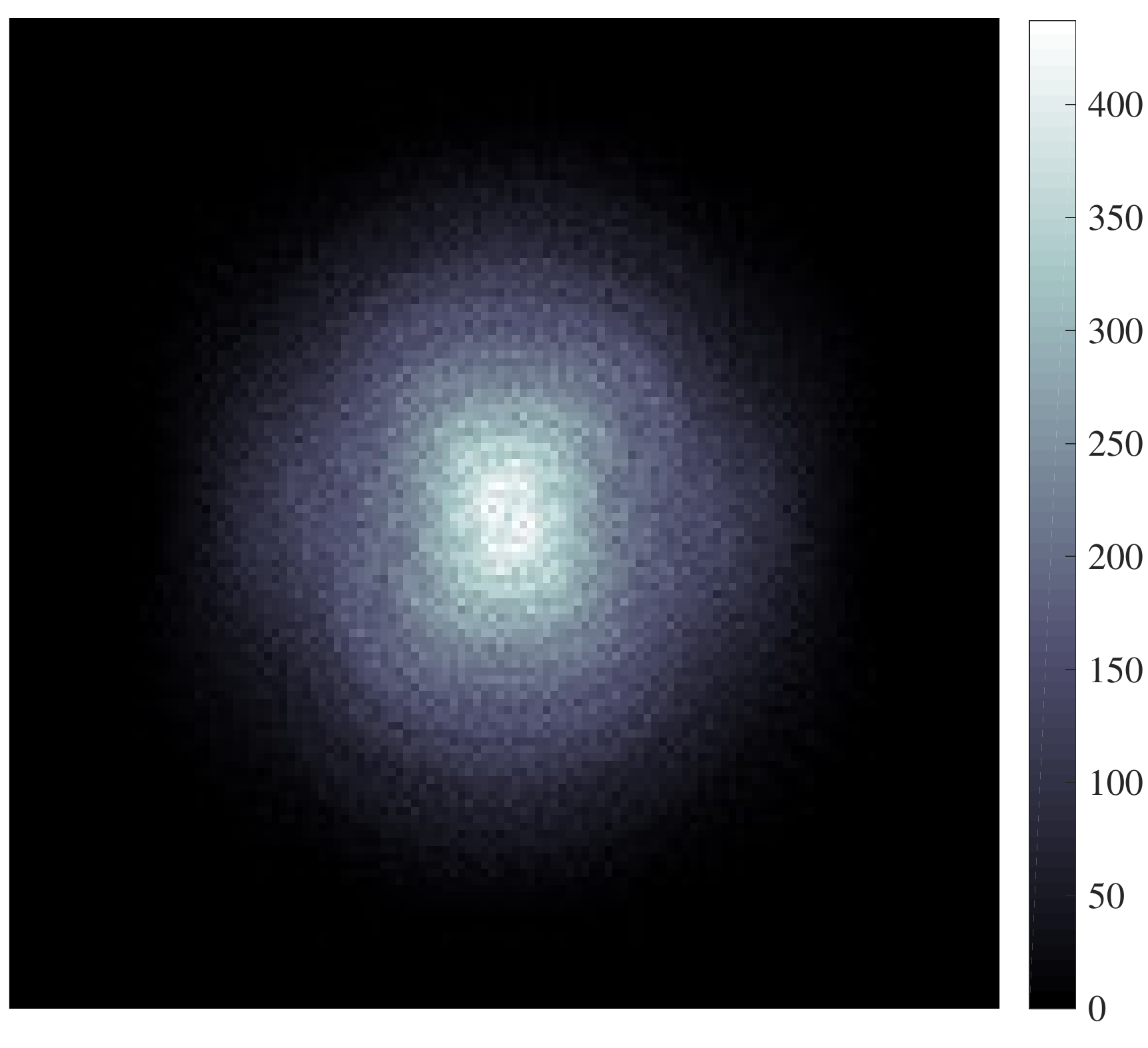}	
&	\includegraphics[width=4.2cm]{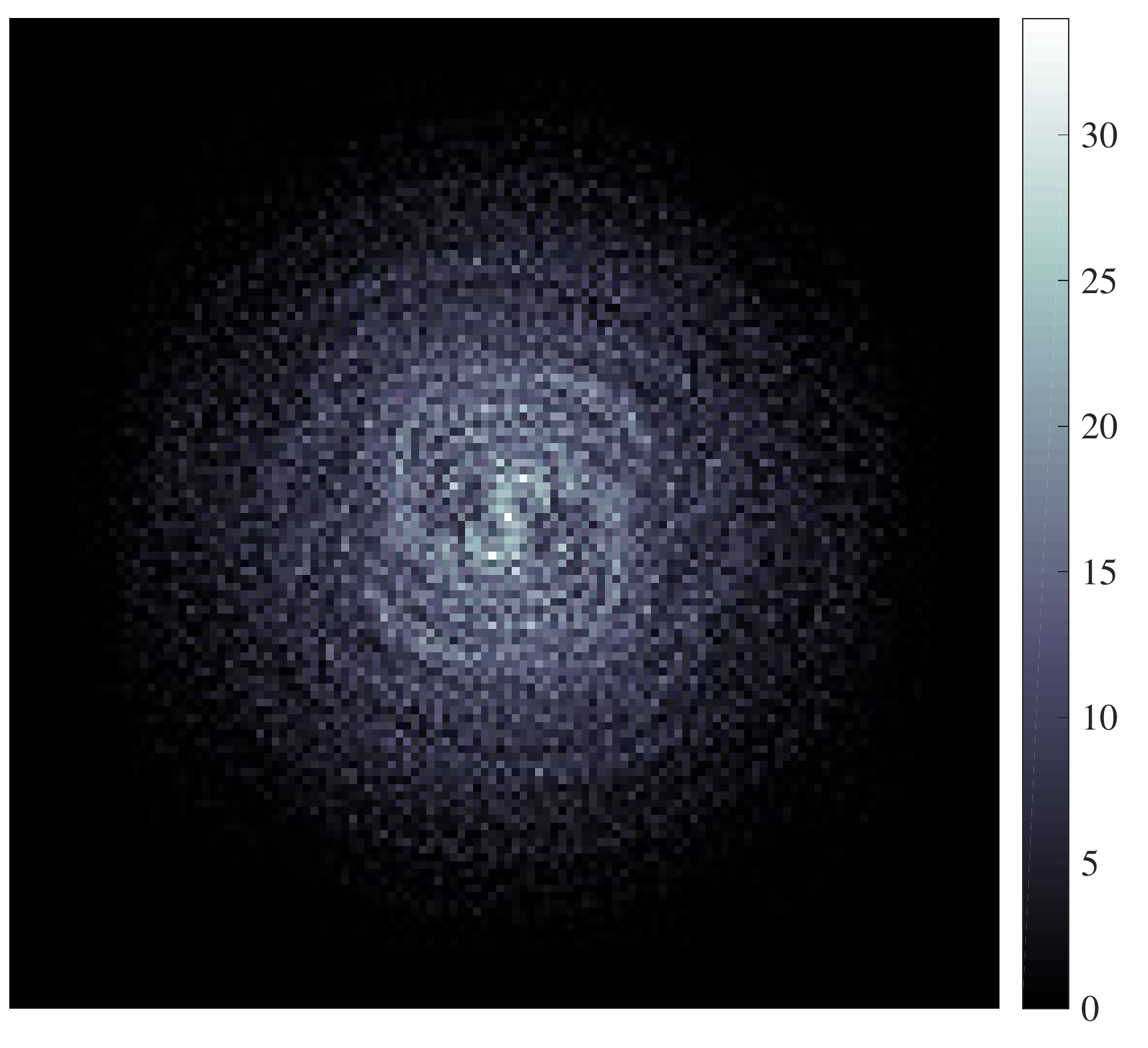}	
\end{tabular}
\vspace*{-0.2cm}
\caption{\label{Fig:test3:uv_cov}
Images corresponding to simulations in Section~\ref{Ssec:sim:test1-3}. 
Discrete symmetrized 2D masks selecting the measured frequencies in the Fourier domain, corresponding to the $u-v$ coverages in the case when (left) $n_a = 200$ and $T = 20$, and (right) $n_a = 50$ and $T = 20$. 
The colorbars indicate the number of measurements acquired for each spatial frequency.
}
\end{figure}

As explained in Section~\ref{Ssec:calib:model}, we consider a general model for DDEs where they can be time dependent. 
In the simulations presented in Sections~\ref{Ssec:sim:test1-1} and \ref{Ssec:sim:test1-2}, we have considered snapshot imaging, i.e. the measurements are acquired at a single instant corresponding to $T = 1$. However, in practice, the interferometric measurements can be acquired over a period of time at several instants, thereby increasing the sampling of the Fourier plane, and the number of different DDEs to be recovered. 
Moreover, at each instant, the sampling of the Fourier plane is determined by the number of antennas $n_a$ considered. 
Therefore, we implement our algorithm for different values of $T$ and $n_a$, fixing all the other parameters. 
In this experiment, we consider images with $10$ sources in $\overline{\epsilonb}_1$ and global energy $\Es(\overline{\epsilonb}_1) = 0.1$, while for the direction-dependent Fourier kernels we fix $S = 7 \times 7$ with $\upsilon = 0.05$. Then, we perform simulations with $n_a \in \{50, 100, 200\}$, considering $T \in \{1, 10, 20\}$ for $n_a \in \{100, 200\}$, and $T \in \{10, 20\}$ for $n_a = 50$. Note that we have not considered the case when $T=1$ and $n_a = 50$ since this configuration leads to a highly under-sampled $u-v$ coverage giving very poor results. 
Figure~\ref{Fig:test3:uv_cov} shows two examples of discrete versions of $u-v$ coverages obtained in the cases when (left) $n_a = 200$ and $T= 20$, and (right) $n_a = 50$ and $T=20$. They corresponds to 2D masks selecting the measured frequencies in the Fourier domain of the image of interest.

\begin{figure*}
\begin{tabular}{cccc}
 \hspace*{-0.2cm}	\includegraphics[height=4.0cm]{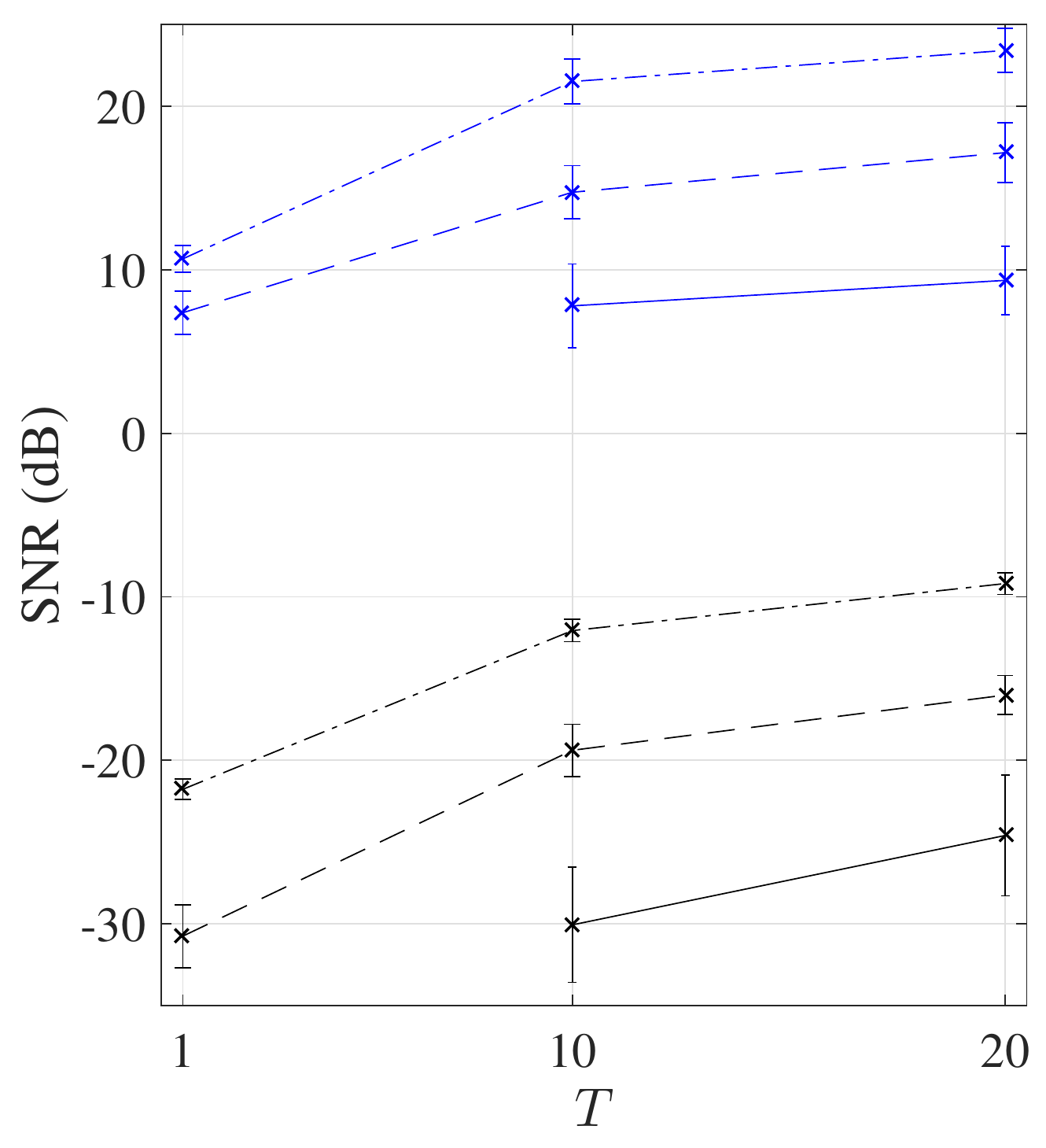}
&\hspace*{-0.2cm}	\includegraphics[height=4.0cm]{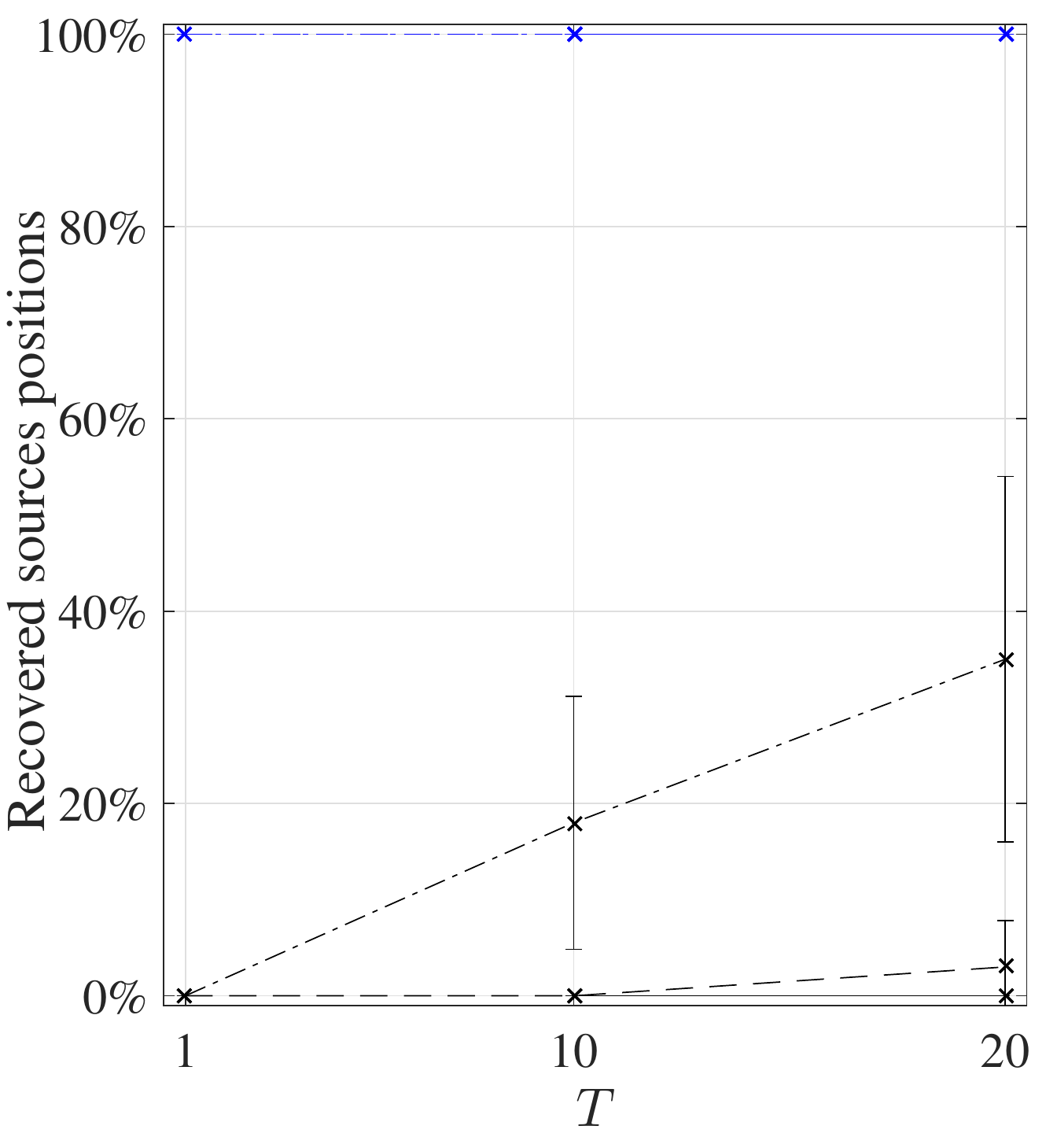}
&\hspace*{-0.2cm}	\includegraphics[height=4.0cm]{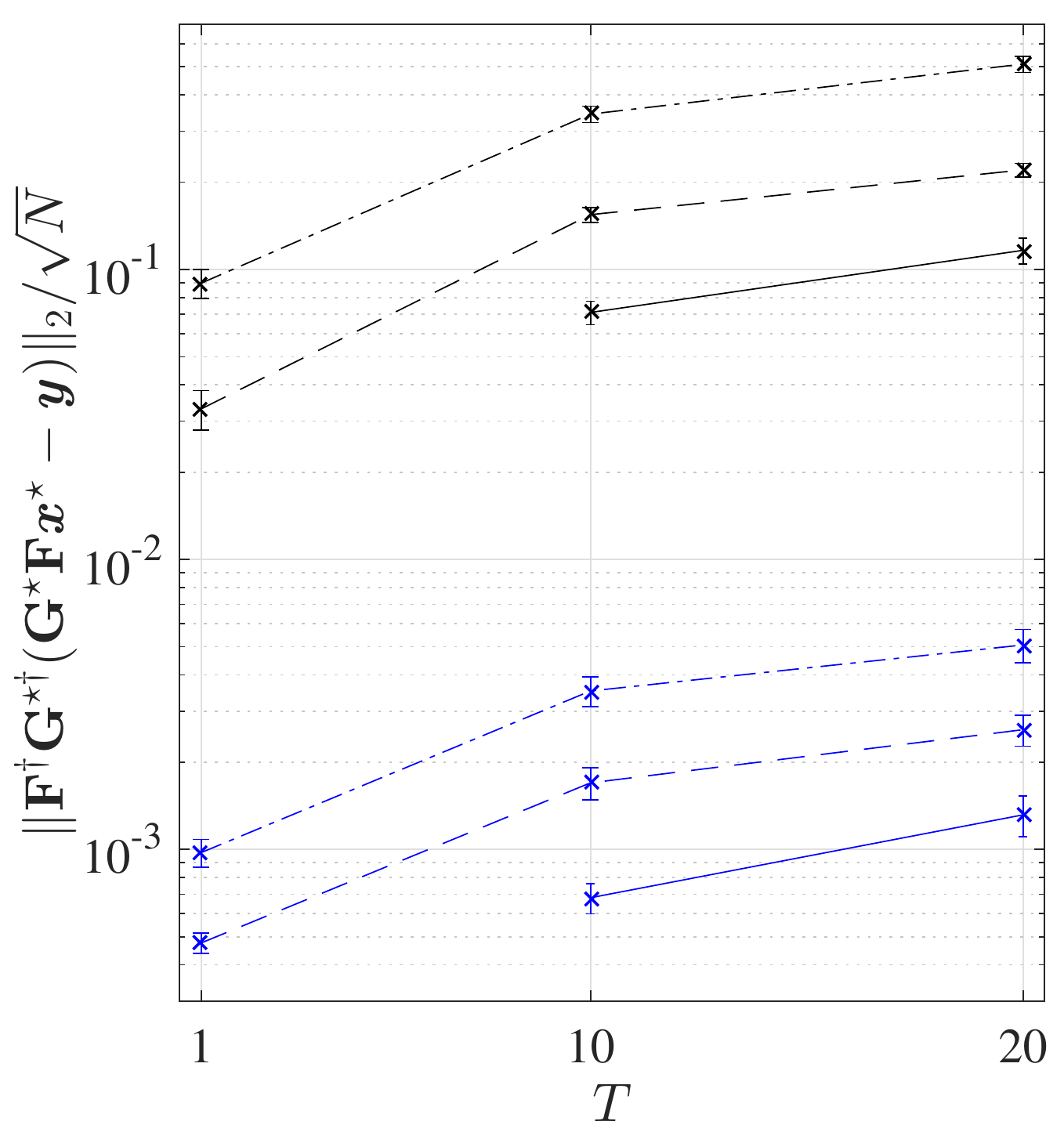}
&\hspace*{-0.2cm}	\includegraphics[height=4.0cm]{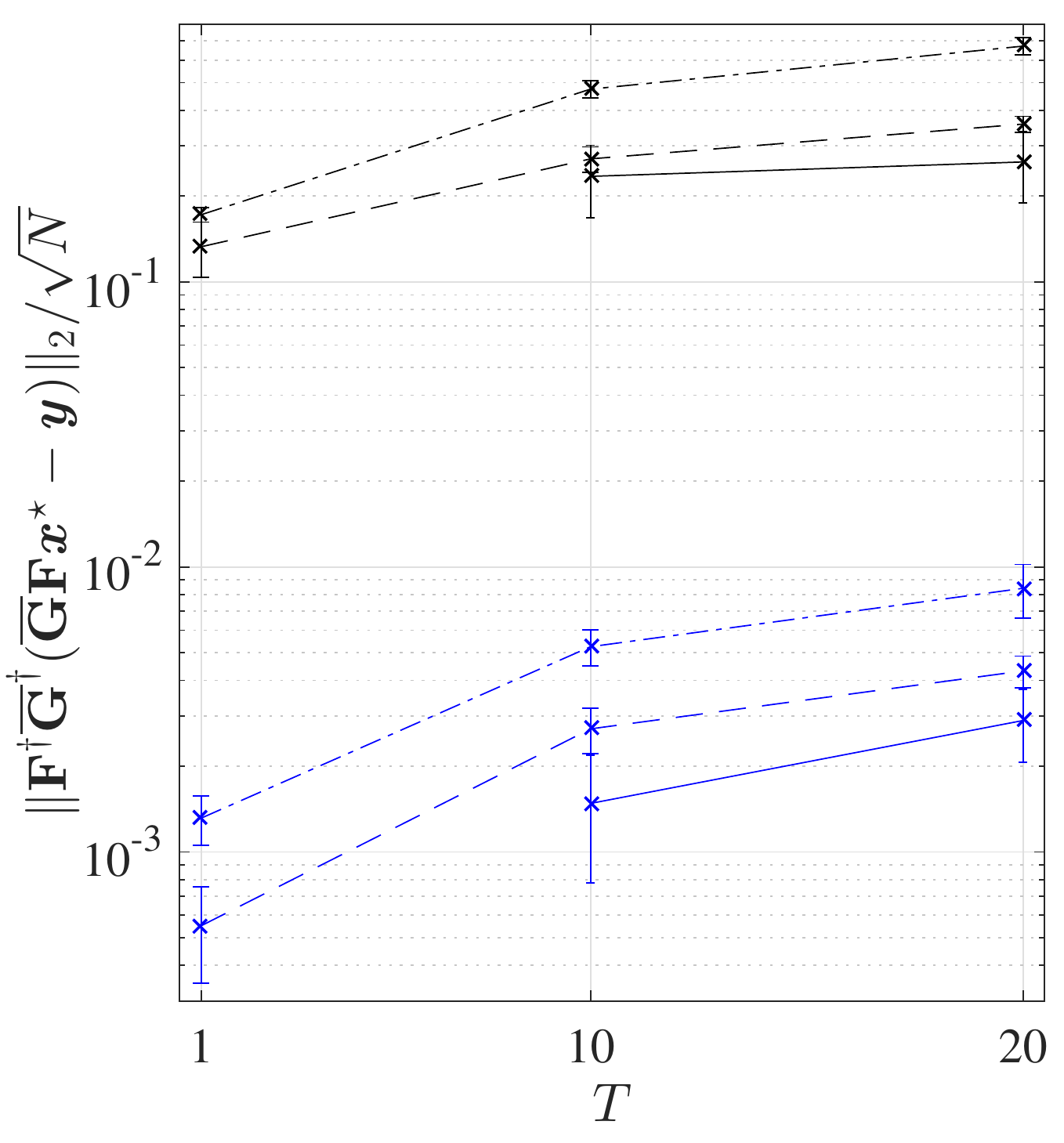}
\end{tabular}
\vspace*{-0.3cm}
\caption{\label{Fig:test3:curves}
Results obtained for simulations of Section~\ref{Ssec:sim:test1-3} using the proposed method (blue lines) and estimating only the DIEs with StEFCal-FB (black lines), considering the number of antennas to be $50$, $100$ and $200$ (resp. solid lines, dashed lines and dash-dotted lines), varying the number of integrations per antenna pair to be $T \in \{1, 10,20\}$. 
From left to right: SNR of the reconstructed $\epsilonb^\star$ with respect to $\overline{\epsilonb}$;
Success rate determining the percentage of recovered sources positions from $\overline{\epsilonb}_1$;
$\ell_2$ error of the residual image $\| \Fbs^\dagger \Gbs^{\star \, \dagger} \big( \Gbs^\star \Fbs \xb^\star - \yb \big) \|_2 / \sqrt{N} $ considering $\Gbs^\star$ obtained with the estimated DDEs;
$\ell_2$ error of the residual image $\| \Fbs^\dagger \overline{\Gbs}^{\dagger} \big( \overline{\Gbs} \Fbs \xb^\star - \yb \big) \|_2 / \sqrt{N} $ considering $\overline{\Gbs}$ obtained with the true DDEs.
Results are given for an average over 10 realizations varying the antenna distribution, the random images, and the DDEs.}
\vspace*{-0.6cm}
\end{figure*}
\begin{figure*}
\begin{tabular}{c@{}c@{}c@{}c@{}}
\hspace*{-0.4cm}	\includegraphics[width=4.2cm]{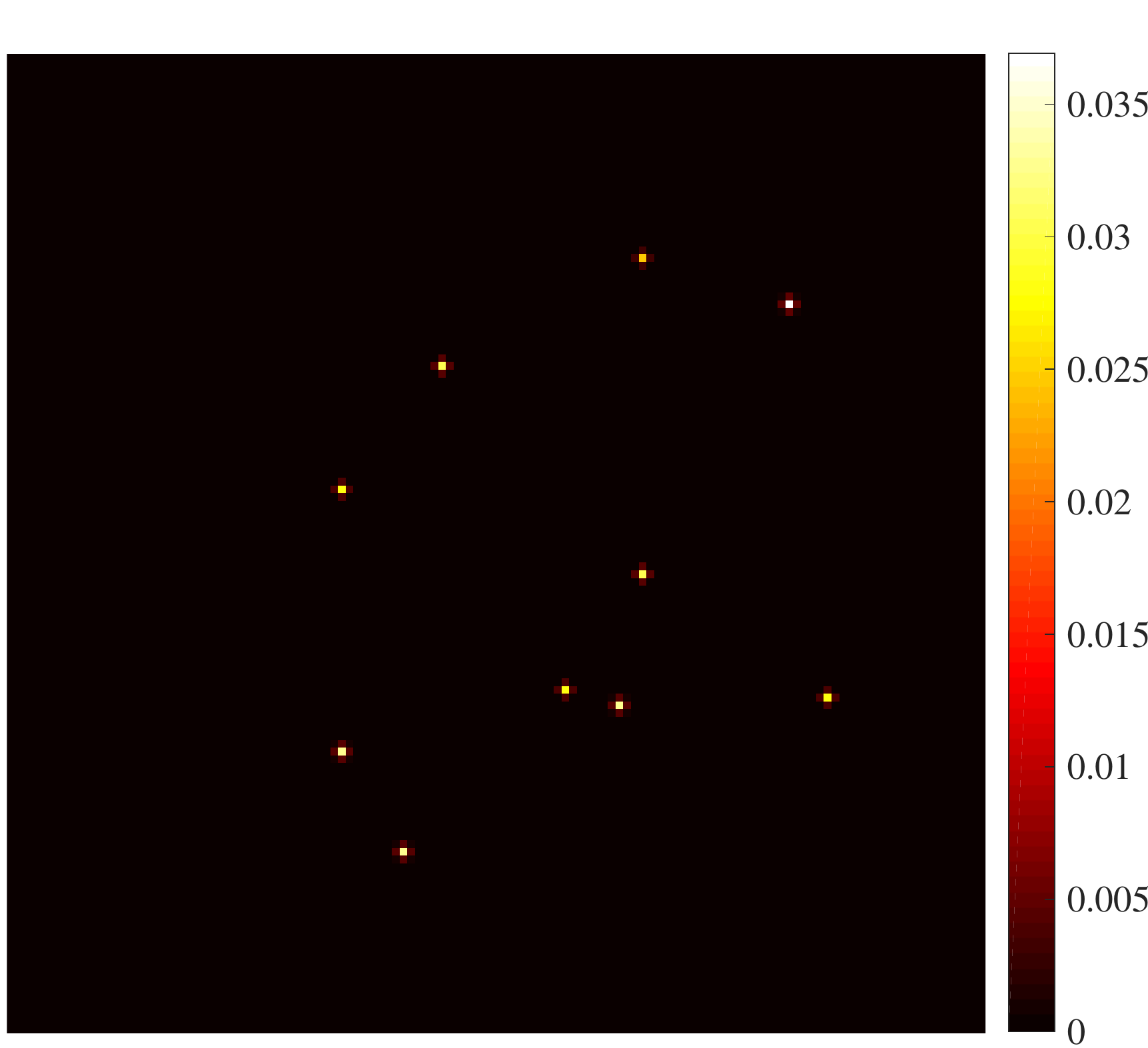}
&	\hspace*{0.1cm}\includegraphics[width=4.2cm]{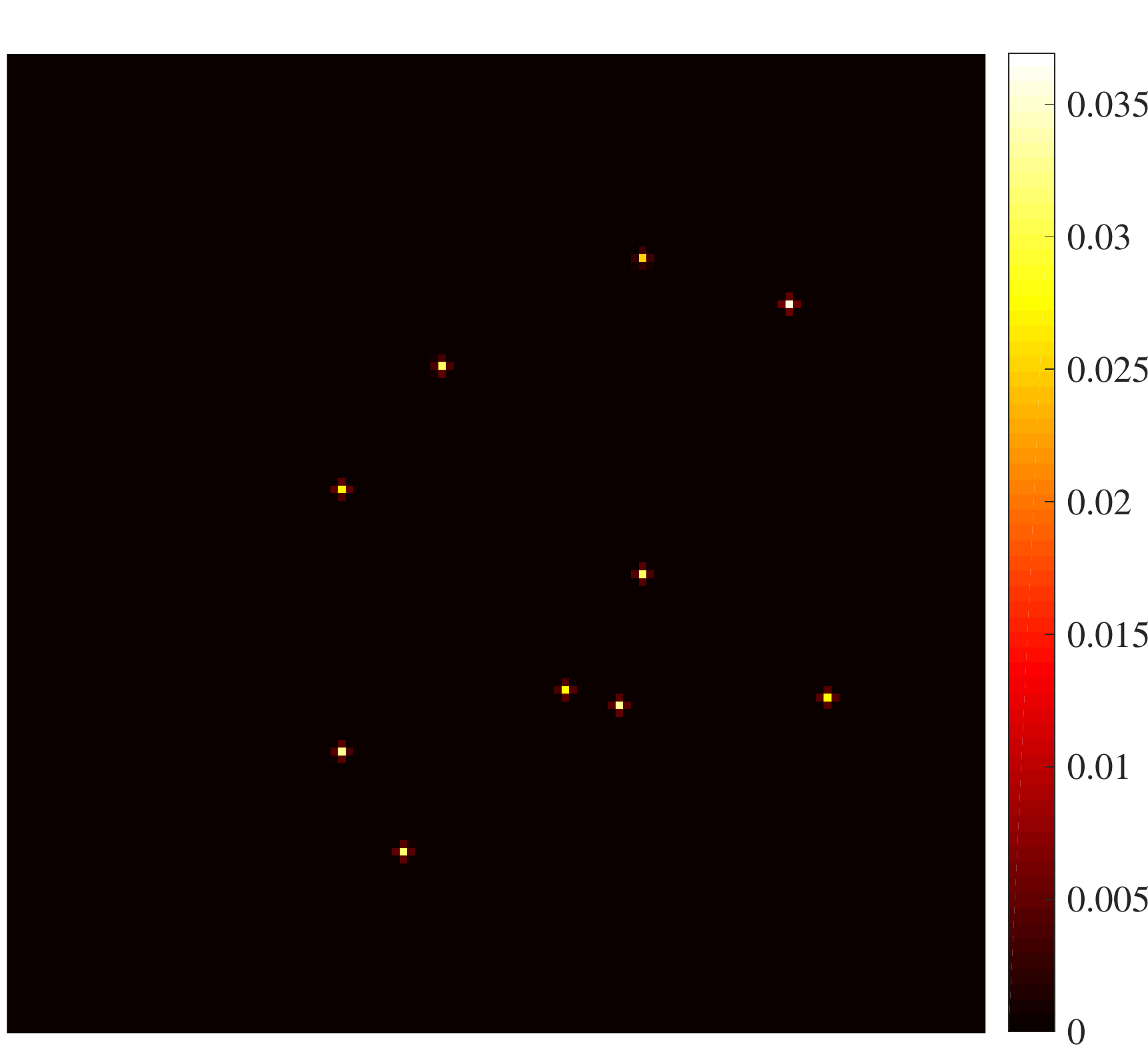}
&	\hspace*{0.1cm}\includegraphics[width=4.2cm]{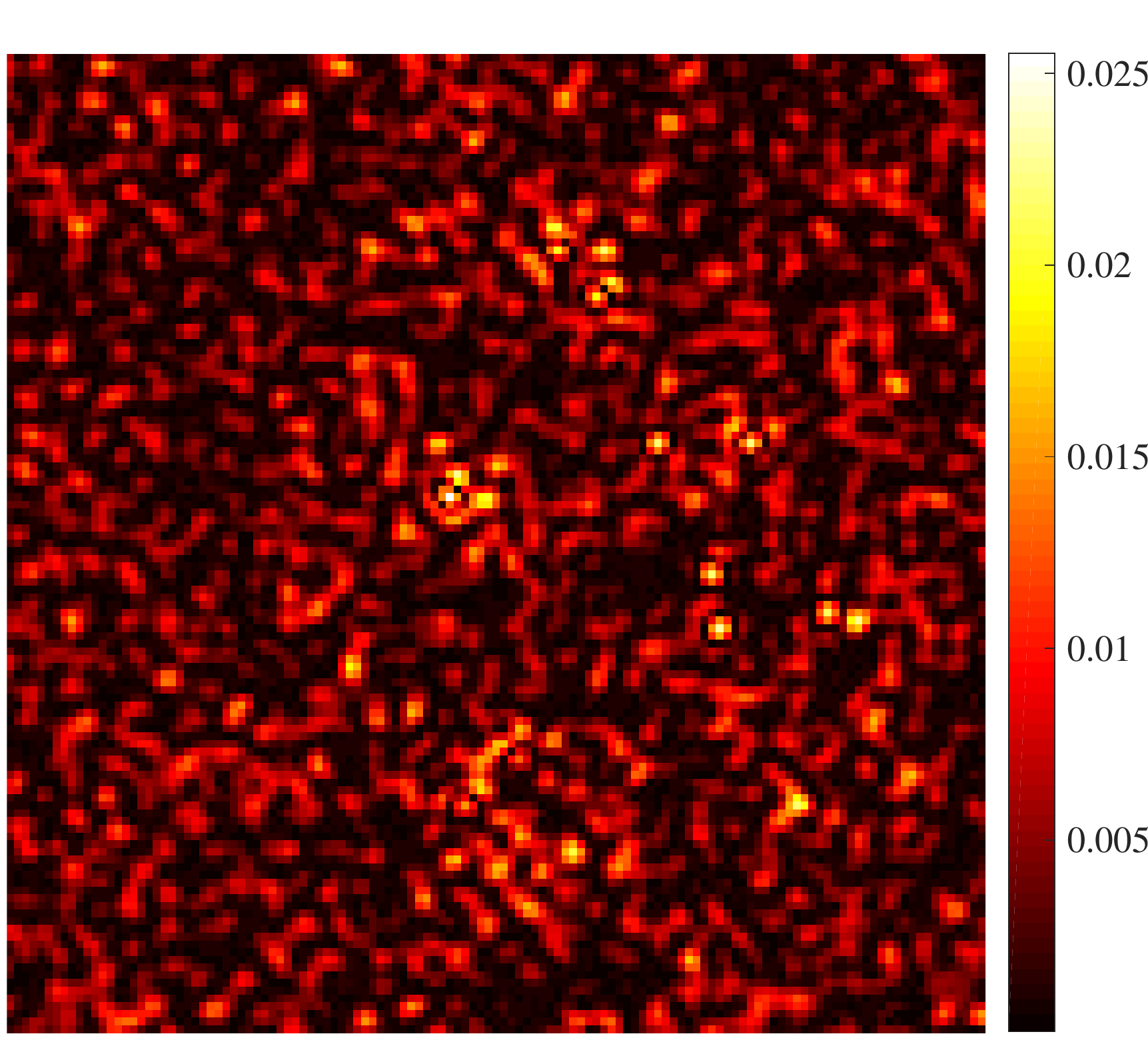}
&	\hspace*{0.1cm}\includegraphics[width=4.2cm]{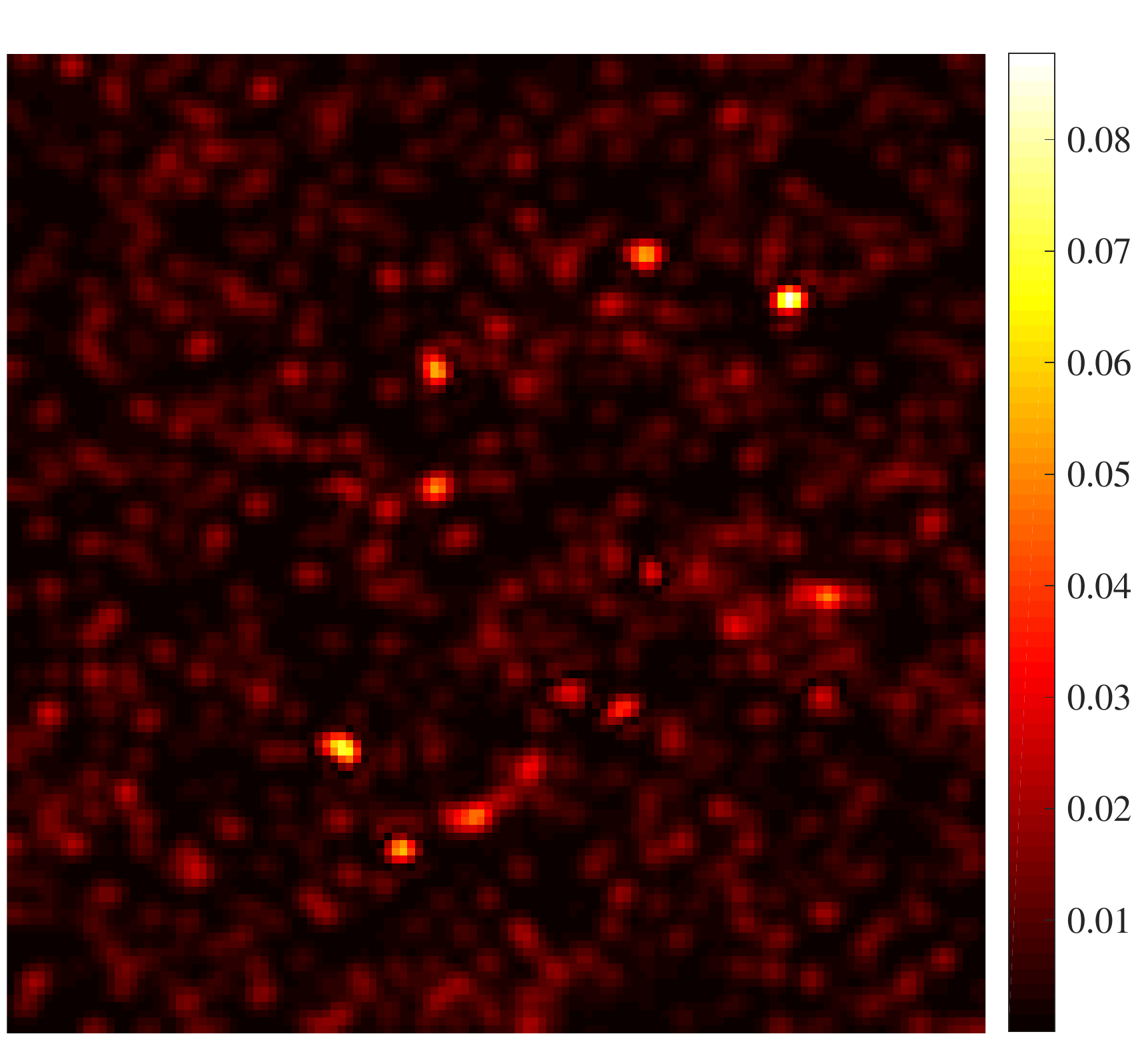}	\\[-0.1em]
\hspace*{-0.4cm}	\includegraphics[width=4.2cm]{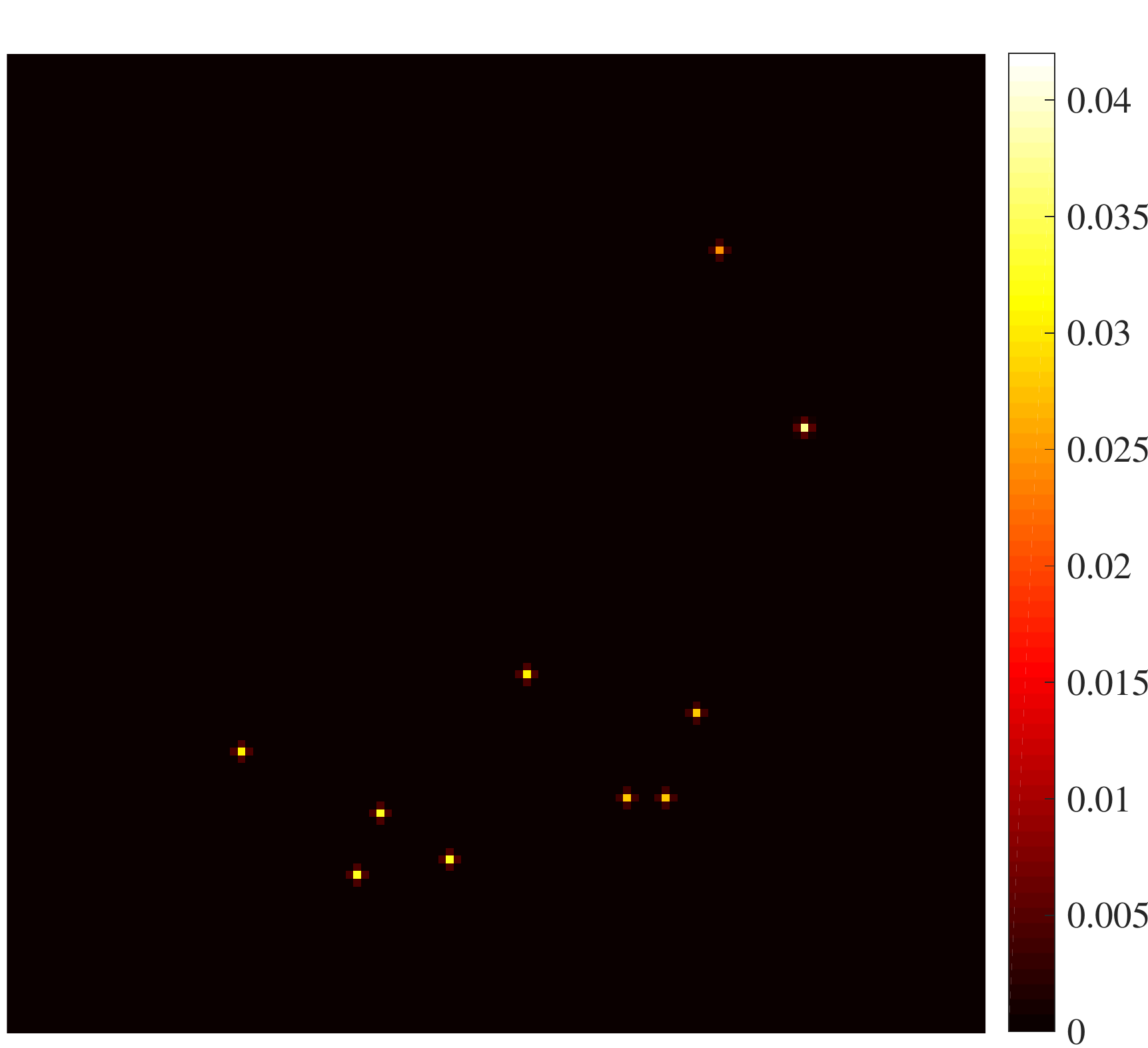}
&	\hspace*{0.1cm}\includegraphics[width=4.2cm]{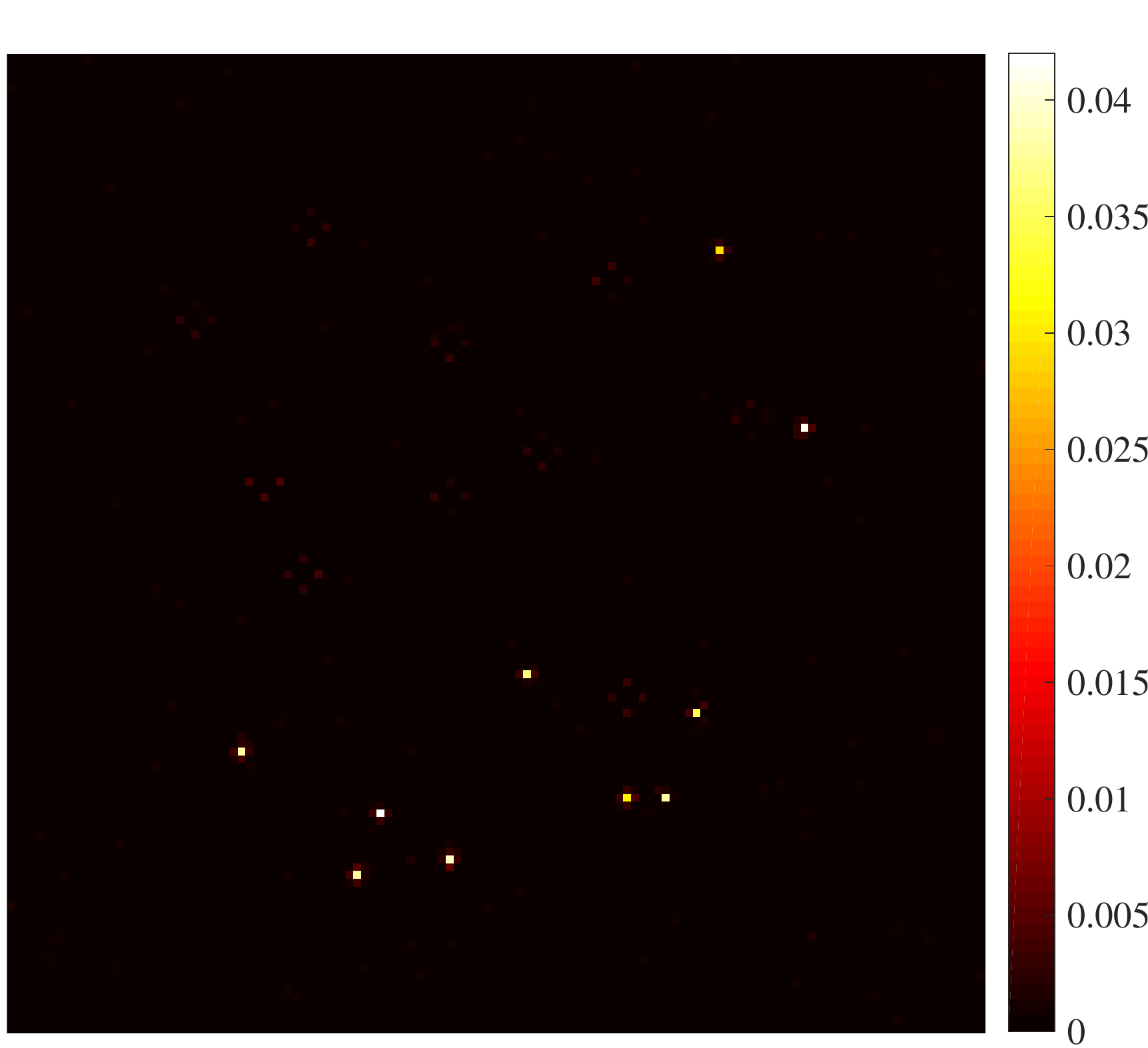}
&	\hspace*{0.1cm}\includegraphics[width=4.2cm]{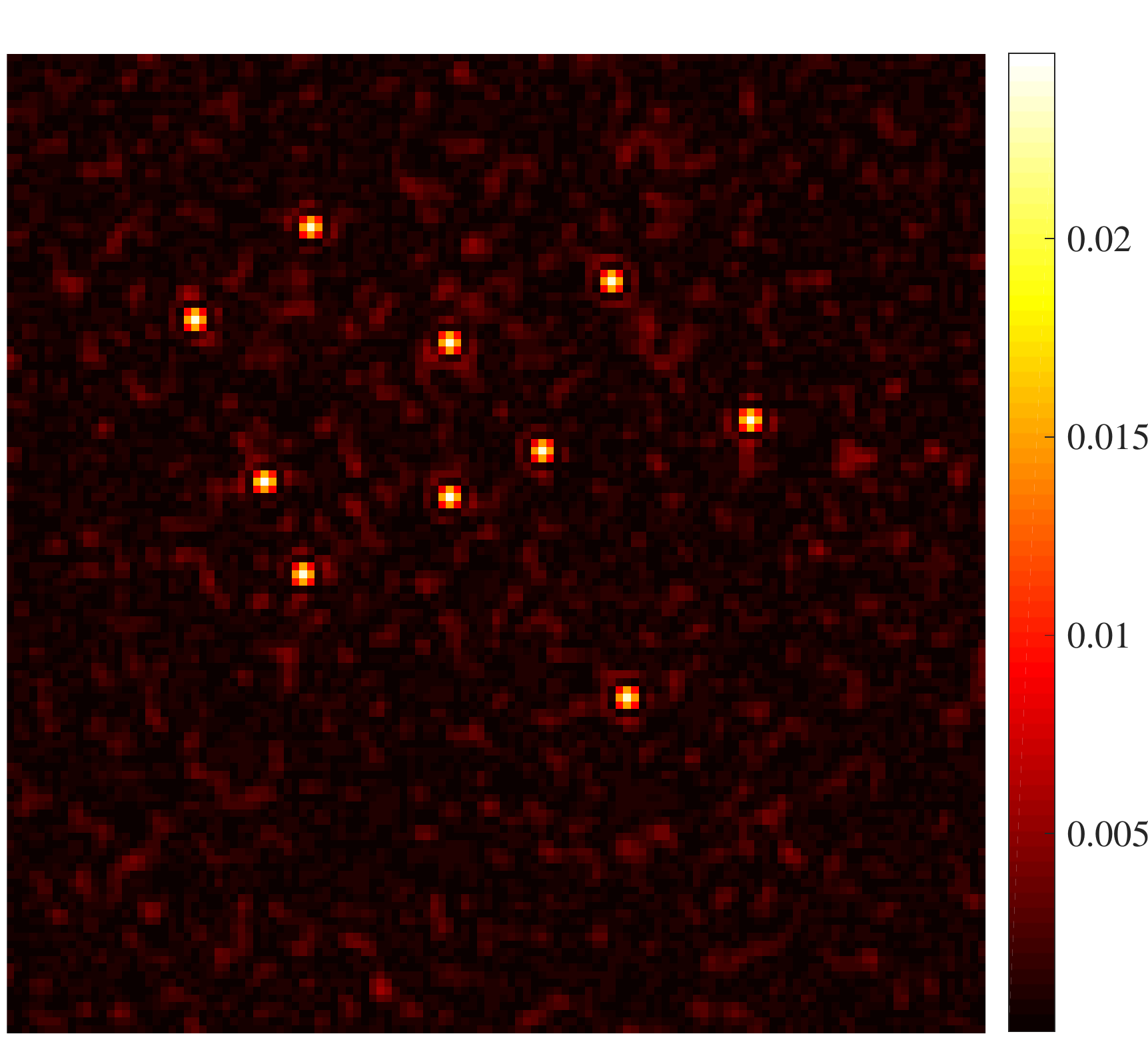}
&	\hspace*{0.1cm}\includegraphics[width=4.2cm]{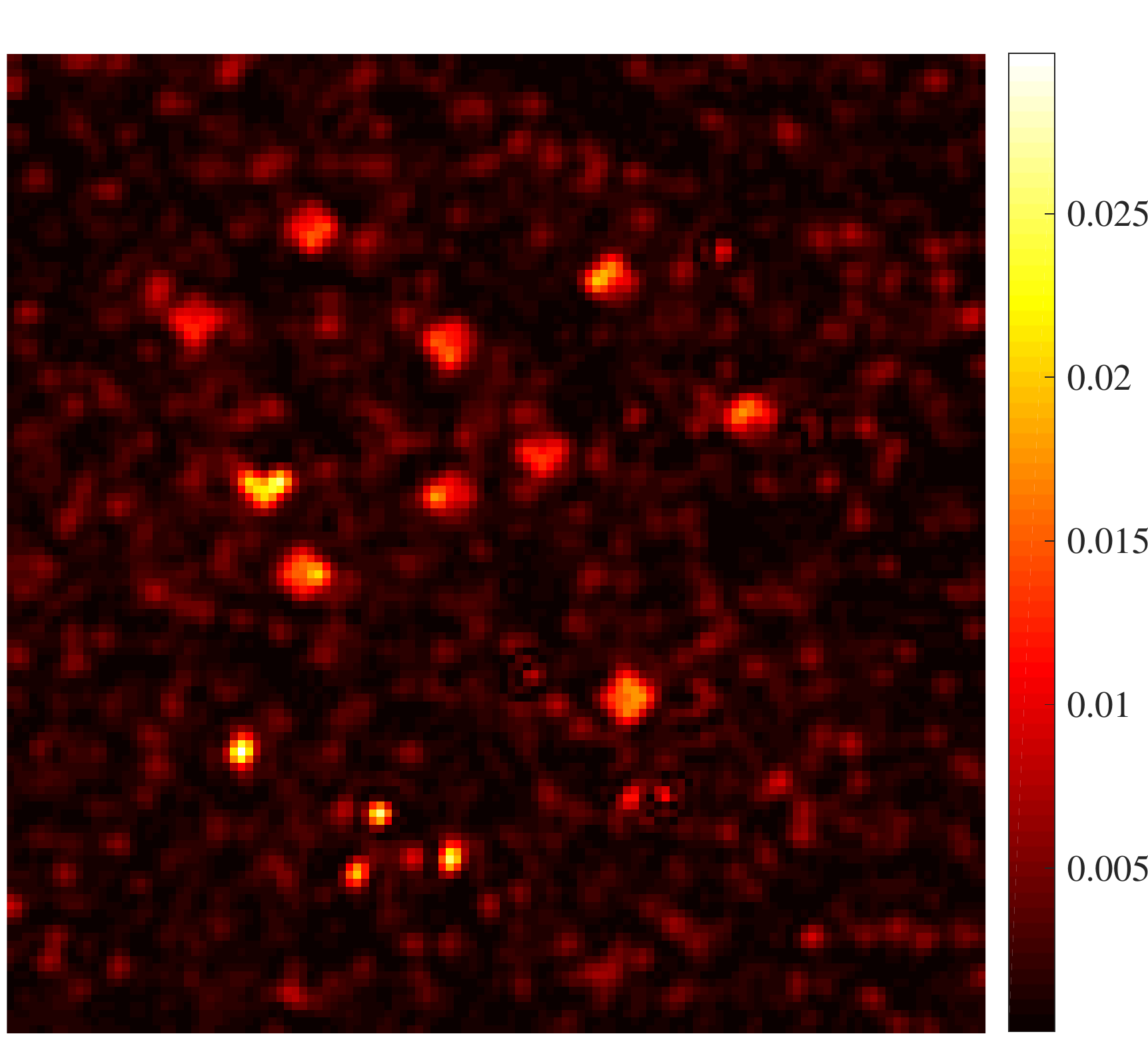}	
\end{tabular}
\vspace*{-0.2cm}
\caption{\label{Fig:test3:images}
Images corresponding to simulations in Section~\ref{Ssec:sim:test1-3}. 
The first row corresponds to the case when $n_a = 200$ and $T = 20$, while the second row corresponds to the case when $n_a = 50$ and $T = 20$.
In both cases, from left to right:  original unknown image $\overline{\epsilonb}$; associated reconstruction $\epsilonb^\star$ obtained with our method; residual image considering $\Gbs^\star$ and $\epsilonb^\star$ obtained with our method; residual image considering the true DDEs, and $\epsilonb^\star$ obtained with our method. 
}
\end{figure*}

\paragraph*{Simulation results.}

A comparison of the results obtained using the proposed approach and the StEFCal-FB method is presented in Figure~\ref{Fig:test3:curves}. As in the previous sections, the first graph represents the SNR of the estimated image $\epsilonb^\star$, the second graph gives the percentage of recovered sources positions in $\overline{\epsilonb}_1$, and the third and fourth graphs show the errors of the residual images obtained considering respectively the estimated DDEs and the true DDEs. 
Considering the reconstruction quality of $\epsilonb^\star$, we can observe that our method detects correctly $100\%$ of the sources positions for all the considered cases, which is not the case for StEFCal-FB. Note that increasing the number of integrations per antenna pair leads to better reconstruction results in terms of SNR of the image. However, it also increases the number of DDE unknowns which are uncorrelated due to time dependency. As a consequence, the error in the residual images increases with $T$.

These results are corroborated by the images shown in Figure~\ref{Fig:test3:images}. The first row corresponds to the case when $n_a =200$ and $T=20$, with images recovered considering the $u-v$ coverage associated with Figure~\ref{Fig:test3:uv_cov} (left). It gives, from left to right, the original unknown image $\overline{\epsilonb}$, its reconstruction obtained using our method with corresponding SNR equal to $25.58$ dB, and the residual images obtained using either the estimated DDEs, or using the true DDEs. The second row gives similar results for the case when $n_a =50$ and $T=20$, corresponding to the $u-v$ coverage associated with Figure~\ref{Fig:test3:uv_cov} (right). In this case, the SNR of the reconstructed image is equal to $10$ dB.

\subsubsection{StEFCal for DDE calibration}
\label{Ssec:sim:test1-4}

As described in Section~\ref{Sec:stefcal}, the StEFCal-FB method can be generalized to solve for the DDE calibration problem. However, this empirical method does not benefit from any convergence guarantees.

We have implemented the naive StEFCal-based method for DDE calibration described in Section~\ref{Sec:stefcal}, considering the measurements made by $n_a = 200$ antennas for a single time interval $T = 1$, and considering DDEs Fourier kernels with support size $S = 7 \times 7$, with standard deviation 
to be $\upsilon = 0.05$ (see Section~\ref{Ssec:sim:setting}). Moreover, we consider images with second level containing $10$ sources such that $\Es(\overline{\epsilonb}_1) = 0.1$. 

As already shown in the previous experiments, our method recover the exact position of $100\%$ of the sources in $\overline{\epsilonb}$ and the average SNR of the recovered image (over 10 experiments) is equal to $10.67$ dB. 
In comparison, the proposed generalized StEFCal based method recovers the position of $98\%$ of the faint sources on average, with an average SNR equal to $8$ dB. 
Note also that our method is on average 4 times faster since, as explained above, it does not require the DDEs estimation and image estimation steps to converge completely. 

The $\ell_2$ norm of the residual image $\| \Fbs^\dagger \Gbs^{\star \, \dagger} \big( \Gbs^\star \Fbs \xb^\star - \yb \big) \|_2 / \sqrt{N} $ is equal to $9.75 \times 10^{-4}$ (resp. $1.88 \times 10^{-4}$) with $\Gbs^\star $ and $\xb^\star$ obtained using our method (resp. using the proposed StEFCal based method). 
Moreover, the $\ell_2$ norm of the residual image $\| \Fbs^\dagger \overline{\Gbs}^{\dagger} \big( \overline{\Gbs} \Fbs \xb^\star - \yb \big) \|_2 / \sqrt{N} $ is equal to $1.31 \times 10^{-3}$ (resp. $5.78 \times 10^{-3}$) with $\xb^\star$ obtained using our method (resp. using the proposed StEFCal based method). These results show that the $\ell_2$ norm of the residual image with the estimated DDEs is smaller using the StEFCal based method than using our method. However, it suggests that the true $\ell_2$ norm of the residual image is smaller using our method. 

Therefore, according to the comments given above, we can conclude that, compared to the proposed naive StEFCal based method, our method based on the alternated forward-backward algorithm is faster, more stable and less sensitive to ambiguity problems.

\subsection{Reconstruction of images with an extended source}
\label{Ssec:sim:tests2}

\begin{figure}
\begin{center}\footnotesize
\begin{tabular}{c@{}c@{}}
\hspace*{-0.2cm} \includegraphics[width=4.3cm]{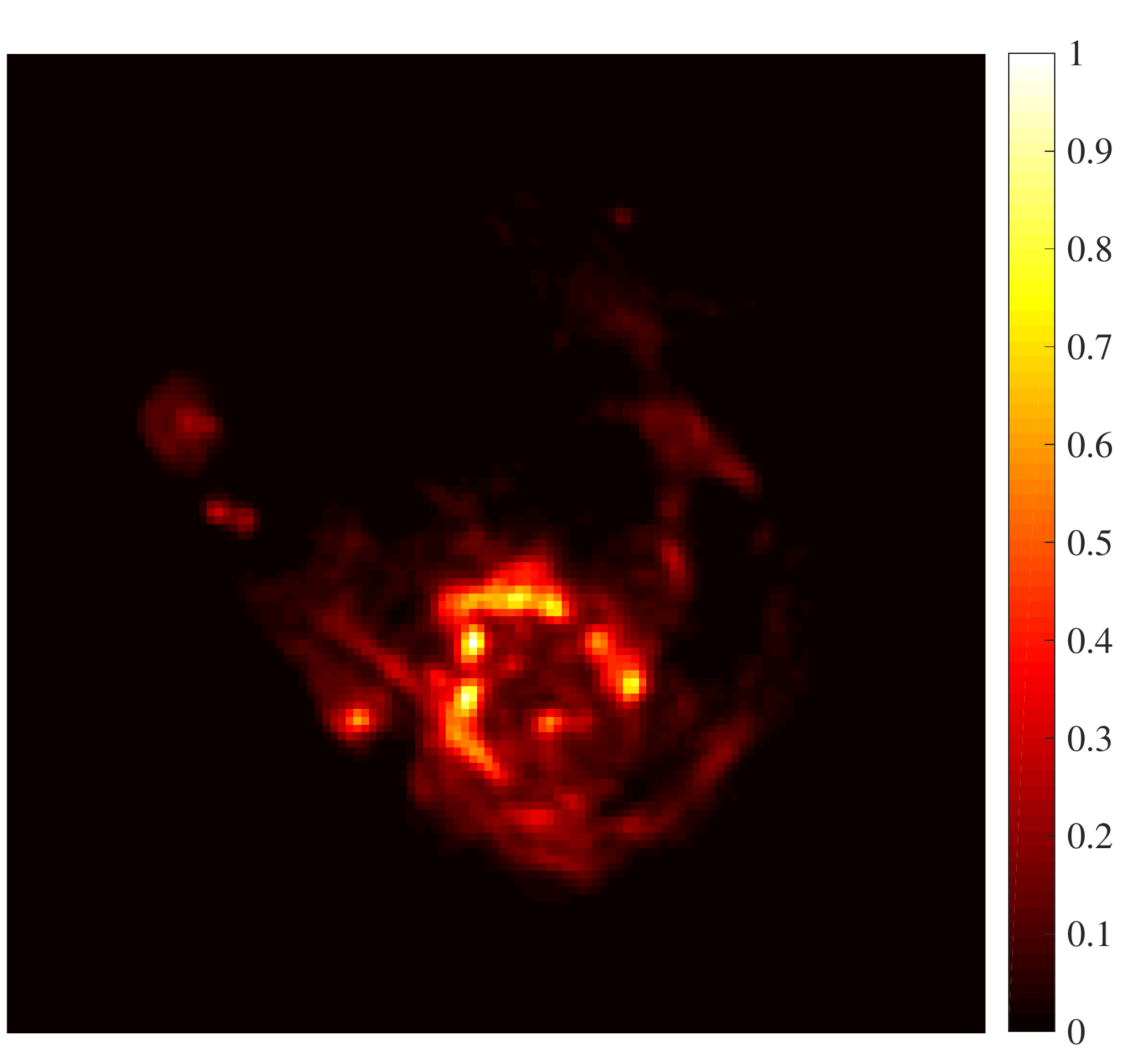}
&\includegraphics[width=4.3cm]{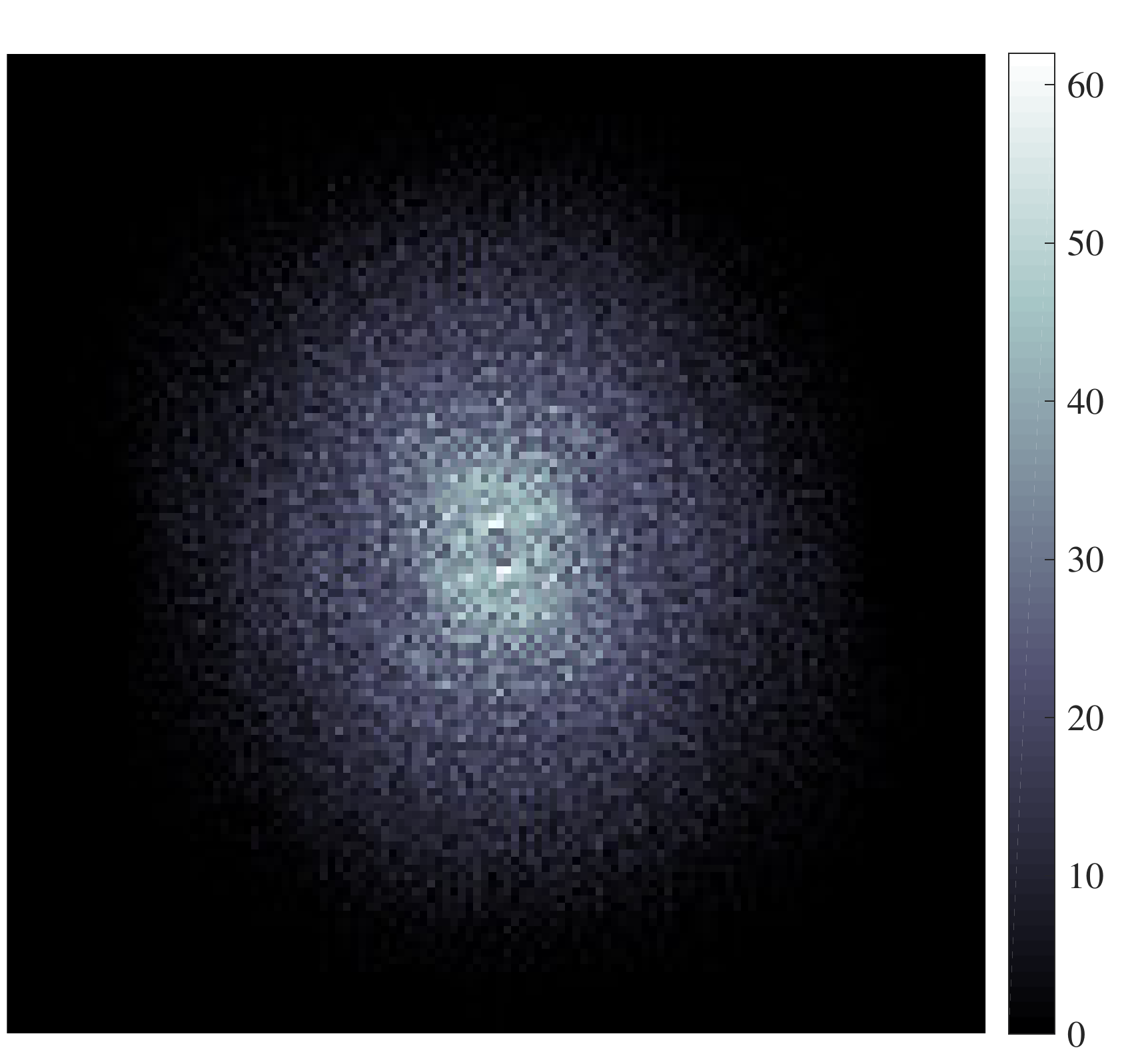}
\end{tabular}
\vspace*{-0.4cm}
\end{center}
\caption{\label{Fig:M31}
(left) Original global image of M31 of size $128 \times 128$ and (right) discrete symmetrized 2D mask selecting the measured frequencies in the Fourier domain, corresponding to the $u-v$ coverage in the case when $n_a = 100$ and $T=10$ (the colorbar indicates the number of measurements acquired for each spatial frequency). }
\end{figure}

In this second part, we consider an image of M31 of size $128 \times 128$ given in Figure~\ref{Fig:M31} (left). The objective of the simulations presented in this section is to show the behaviour of the proposed algorithm in the context of images involving extended sources. Therefore we consider only one simulation setting, consisting of measurements acquired by $n_a = 100$ antennas at $T=10$ instants. The corresponding discrete $u-v$ coverage is shown in Figure~\ref{Fig:M31} (right). Moreover, we fix the size of the support of the Fourier direction-dependent kernels to be $S=9\times 9$, with associated standard deviation, defined in Section~\ref{Ssec:sim:setting}, set to be $\upsilon = 0.05$. 
As described in Section~\ref{Ssec:Blind_problem}, we consider that the original image of M31 is decomposed as $\overline{\xb} = \xb_o + \overline{\epsilonb}$, where $\xb_o$ is known, while $\overline{\epsilonb}$ is unknown and has to be estimated jointly with the DDEs. 
In this case, we construct $\xb_o$ and $\overline{\epsilonb}$ such that
\begin{equation}
\Es( \overline{\epsilonb} ) = \kappa \Es( \xb_o ) ,
\end{equation}
where $\kappa>0$.
In the performed simulations, we investigate the reconstruction of the image $\overline{\epsilonb}$ for the cases when $\kappa\in \{0.1, 0.5, 1\}$.

\begin{table}
\begin{center}
\begin{tabular}{| l | l | l | l |}
  \hline
\multicolumn{2}{l|}{Parameters} & Initialization  & Algorithm~\ref{algo:global} \\
\hline\hline
\multicolumn{2}{l|}{$J_{\text{cyc}}$}	&	101	&	11	\\
\hline
$J_{\Ubs_1}^{(i)} = J_{\Ubs_2}^{(i)}$	& {{if $(i+1) = 0 \, \Big[\text{mod } J_{\text{cyc}}\Big]$}} 	&	0	&	0	\\
& {{otherwise}} 	&	2 & 5	\\ 
\hline
$J_{\epsilonb}^{(i)}$& {{if $(i+1) = 0 \, \Big[\text{mod } J_{\text{cyc}}\Big]$}} 	&	1000 & 1000	\\ 
& {{otherwise}} 	&	0 & 0	\\ 
\hline
\multicolumn{2}{l|}{$\xi_{\Ubs_{\text{tot}}}$}	&	$10^{-6}$	&		\\
\hline
\multicolumn{2}{l|}{$\xi_{\epsilonb}$}	&	$10^{-5}$	&	$10^{-5}$	\\
\hline
\multicolumn{2}{l|}{$\zeta$}	&	$10^{-2}$	&	$10^{-2}$	\\
\hline
\multicolumn{2}{l|}{$\underline{\tau}$}	&	0.2	&		\\
\hline
\multicolumn{2}{l|}{$\overline{\tau}$}	&	1.1	&		\\
\hline
\end{tabular}
\end{center}
\caption{\label{Tab:param_M31}
Parameters considered in Section~\ref{Ssec:sim:tests2} for Algorithm~\ref{algo:global} and its initialization.}
\end{table}

In this section, the original image is not sparse in its domain. Therefore, we propose to use the function given by eq.~\eqref{eq:def_reg_x_dict} as regularization term for the image, considering $\Psib$ to be the SARA collection of wavelets \citep{Carrillo_2012}. More precisely, $\Psib$ is chosen to be the concatenation of a Dirac basis with the first eight Daubechies wavelets \citep{Mallat_book}.
Moreover, we fix the regularization parameters in eq.~\eqref{pb:min_glob} to be $\eta = 10^{-2}$ and $\nu =  2000$ for the initialization estimating the zero spatial frequency coefficients of DDEs, and $\nu=1000$ for the main algorithm estimating DDEs. 
Finally, the different parameters for Algorithm~\ref{algo:global} are given in Table~\ref{Tab:param_M31}, including the stopping criteria defined in Section~\ref{Ssec:algo:stop_crit}.

\begin{figure}
\begin{tabular}{c@{}c@{}}
\hspace*{-0.3cm}
	\includegraphics[width=4.3cm]{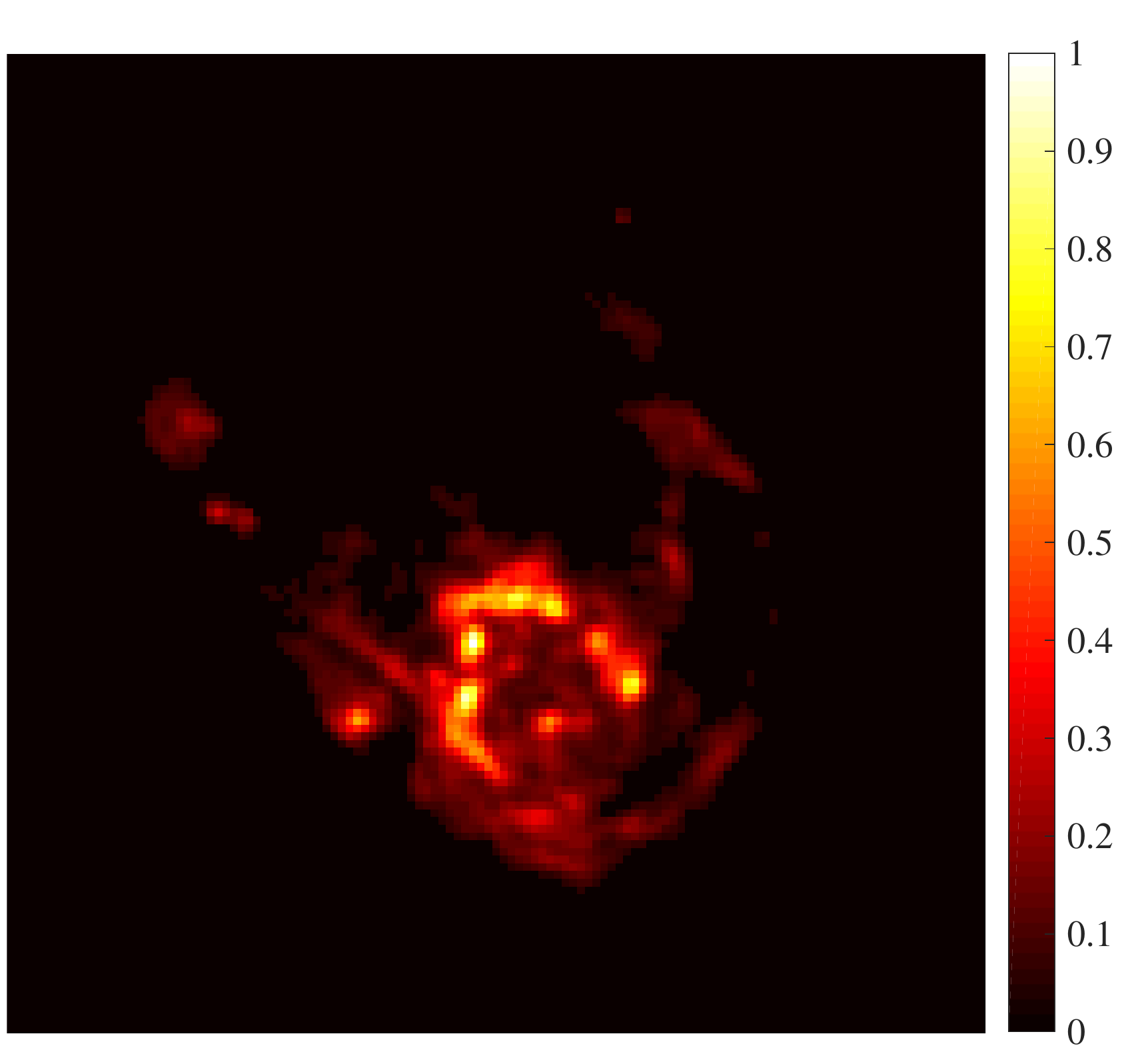}
&	\includegraphics[width=4.3cm]{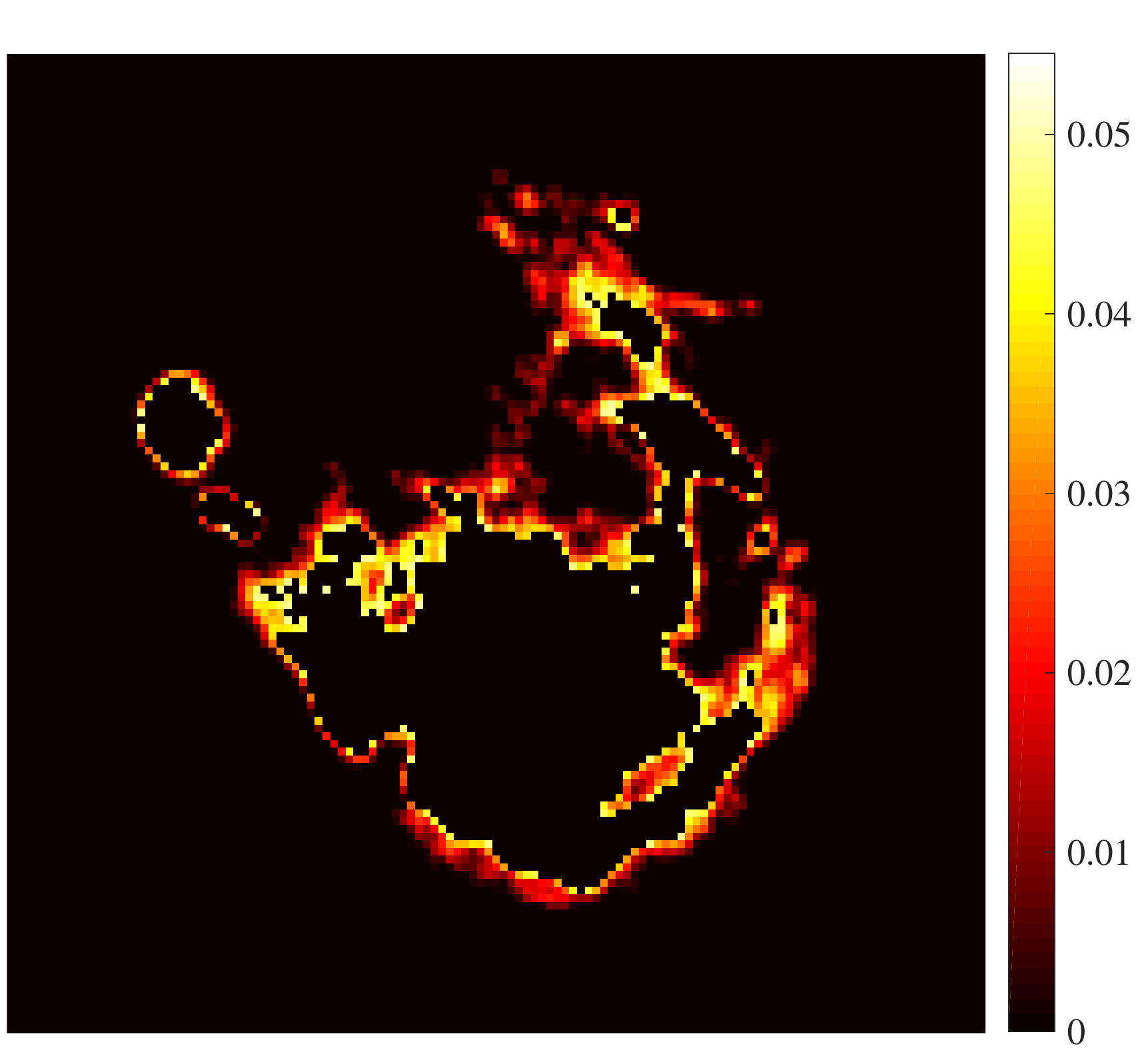}		\\
\hspace*{-0.3cm}
	\includegraphics[width=4.3cm]{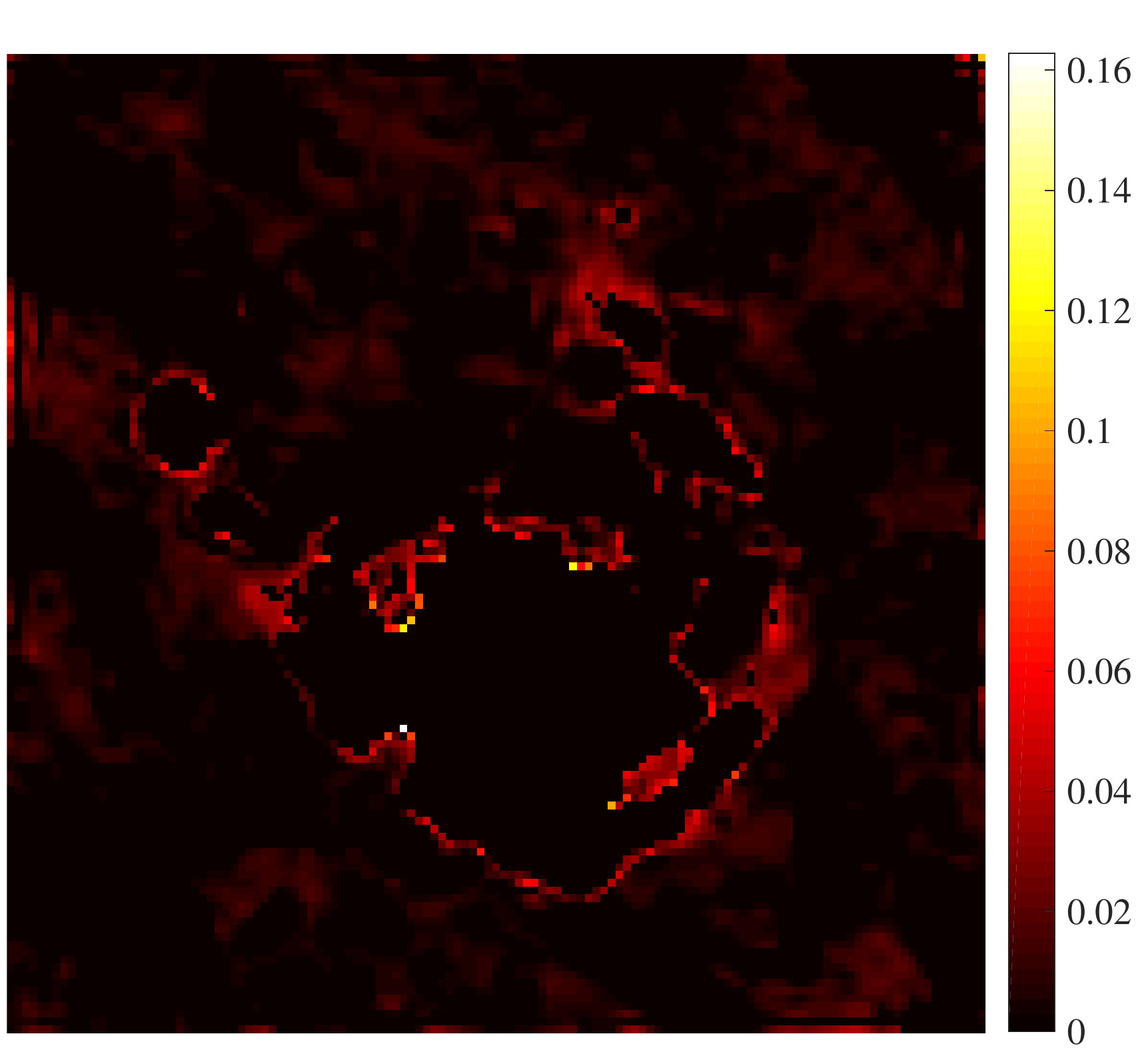}
&	\includegraphics[width=4.3cm]{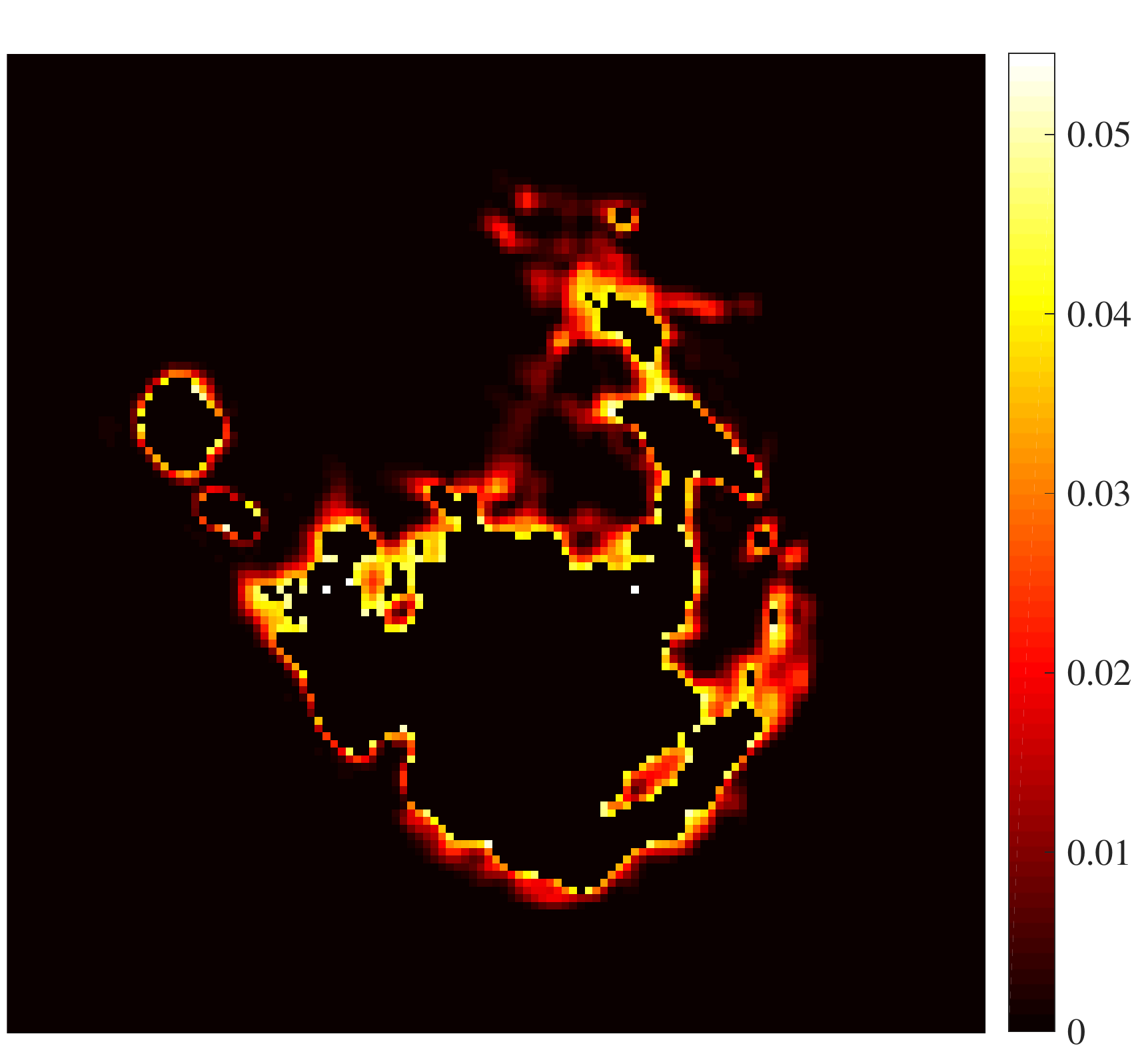}	\\
\hspace*{-0.3cm}
	\includegraphics[width=4.3cm]{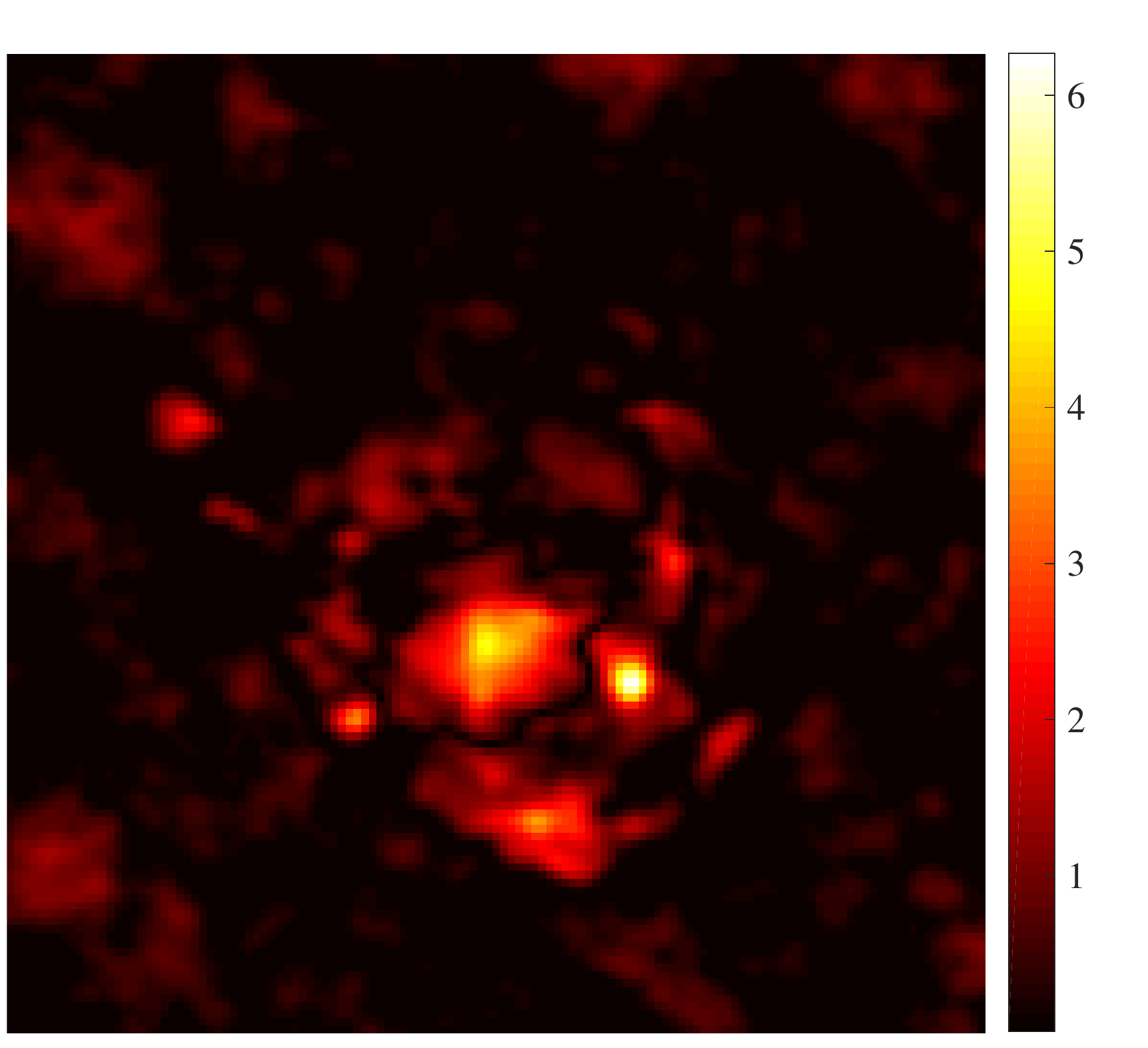}	
&	\includegraphics[width=4.3cm]{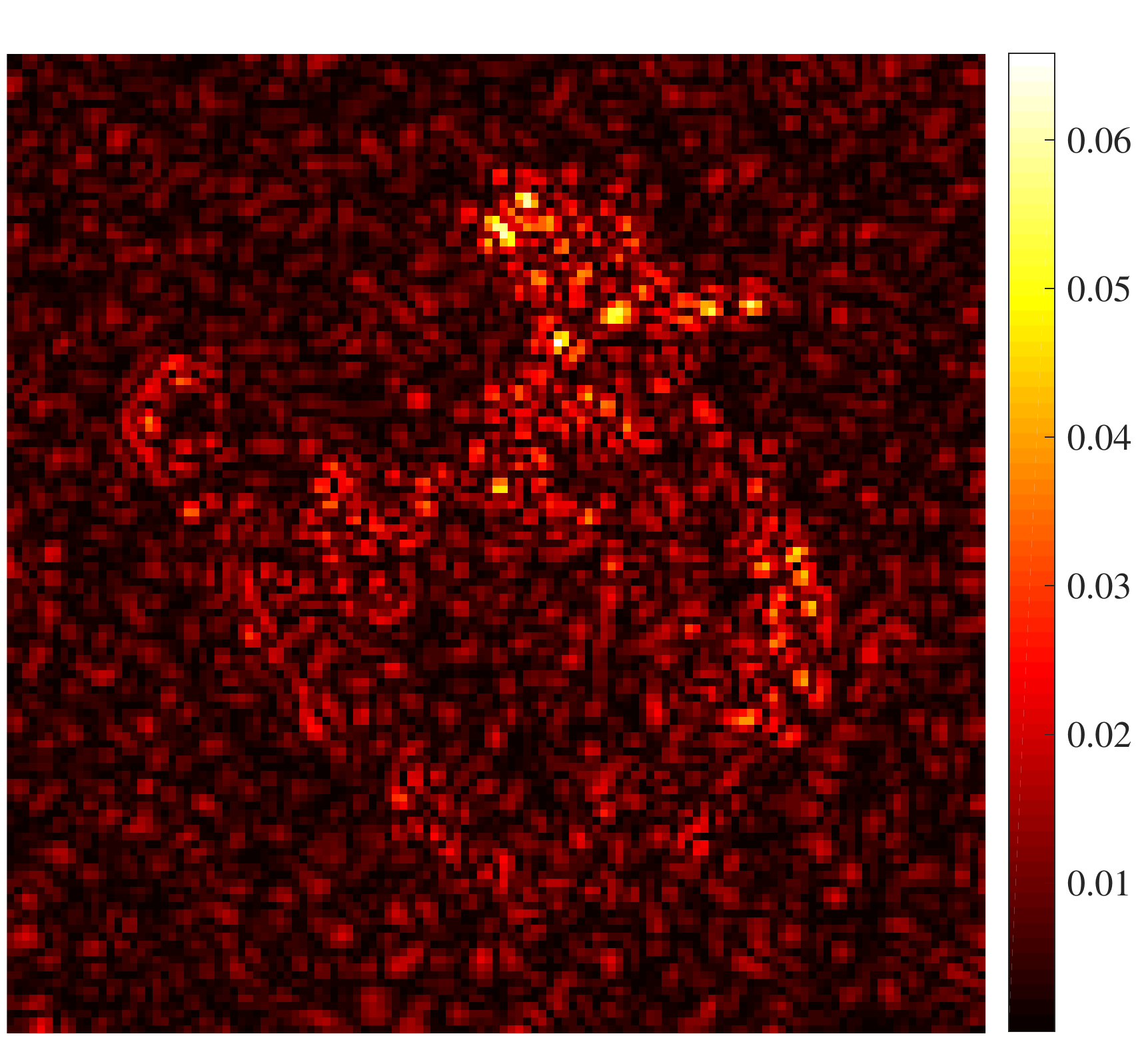}	\\
\hspace*{-0.3cm}
	\includegraphics[width=4.3cm]{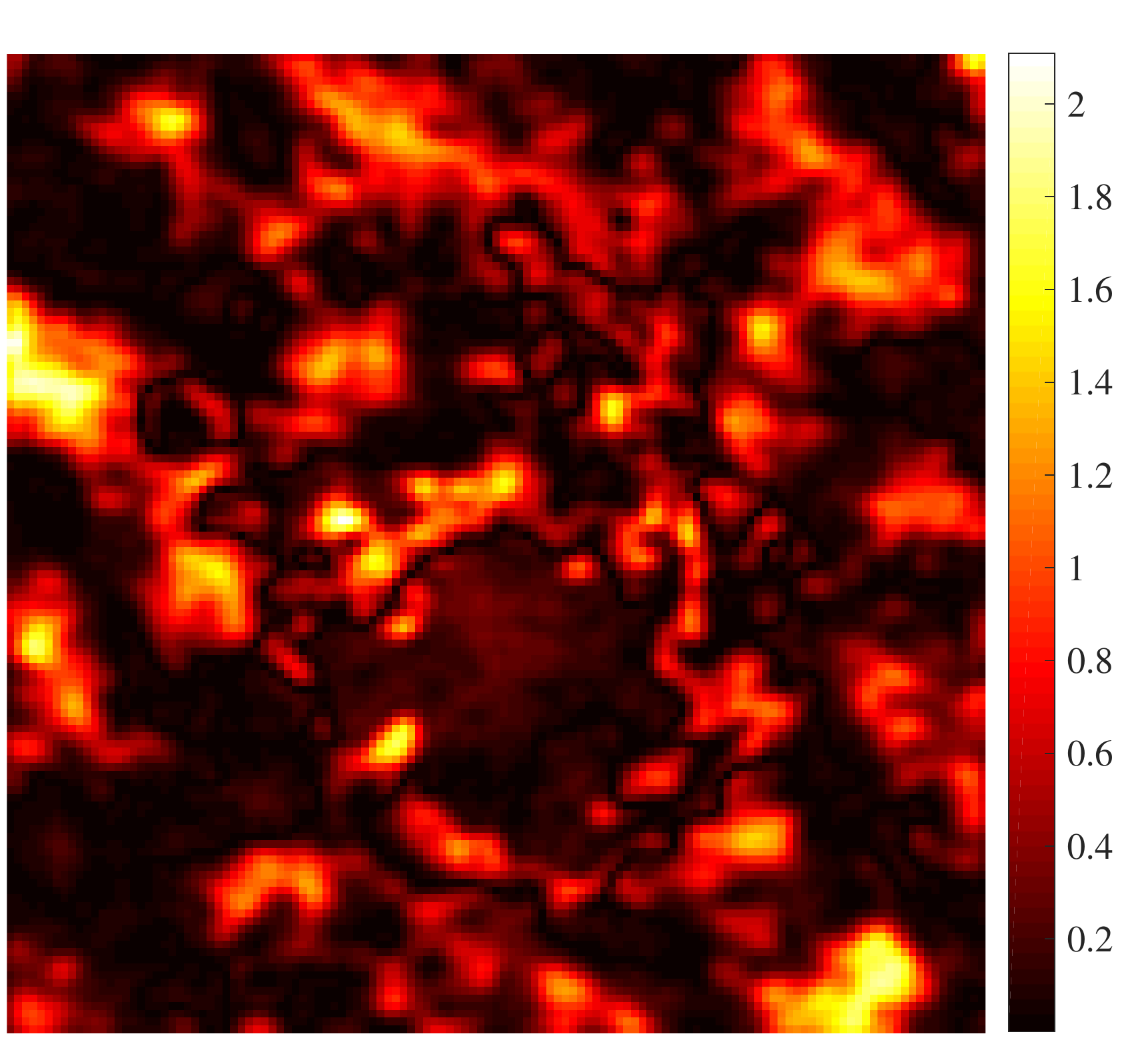}	
&	\includegraphics[width=4.3cm]{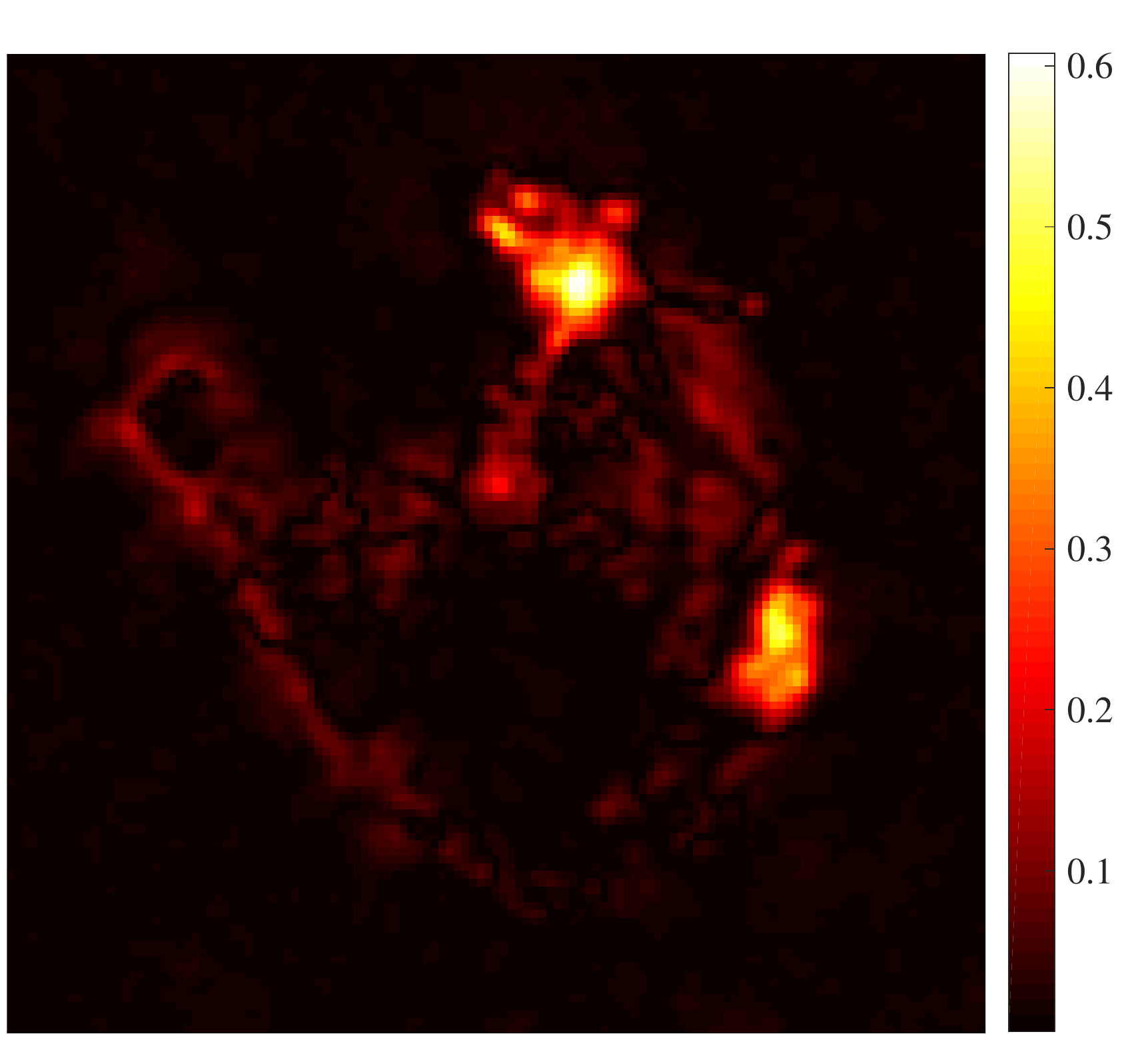}	
\end{tabular}
\vspace*{-0.2cm}
\caption{\label{Fig:testM31:images_delta01}
Images corresponding to simulations performed in Section~\ref{Ssec:sim:tests2} with $\kappa = 0.1$. 
The first row corresponds to (left) the known bright sources $\xb_o$ of the original image $\overline{\xb}$ and (right) the unknown image $\overline{\epsilonb}$. 
The second row corresponds to the estimates $\epsilonb^\star$ of $\overline{\epsilonb}$ obtained using (left) StEFCal-FB solving for DIEs, and (right) the proposed algorithm estimating the full DDEs. 
The third row shows the residual images considering the estimated $\Gbs^\star$ and $\epsilonb^\star$ obtained using (left) StEFCal-FB solving for DIEs, and (right) the proposed algorithm estimating the full DDEs. 
The fourth row corresponds to the residual images considering the true DDEs and the estimated image $\epsilonb^\star$ obtained using (left) StEFCal-FB solving for DIEs, and (right) the proposed algorithm estimating the full DDEs. 
}
\end{figure}

\begin{figure}
\begin{tabular}{c@{}c@{}}
\hspace*{-0.3cm}
	\includegraphics[width=4.3cm]{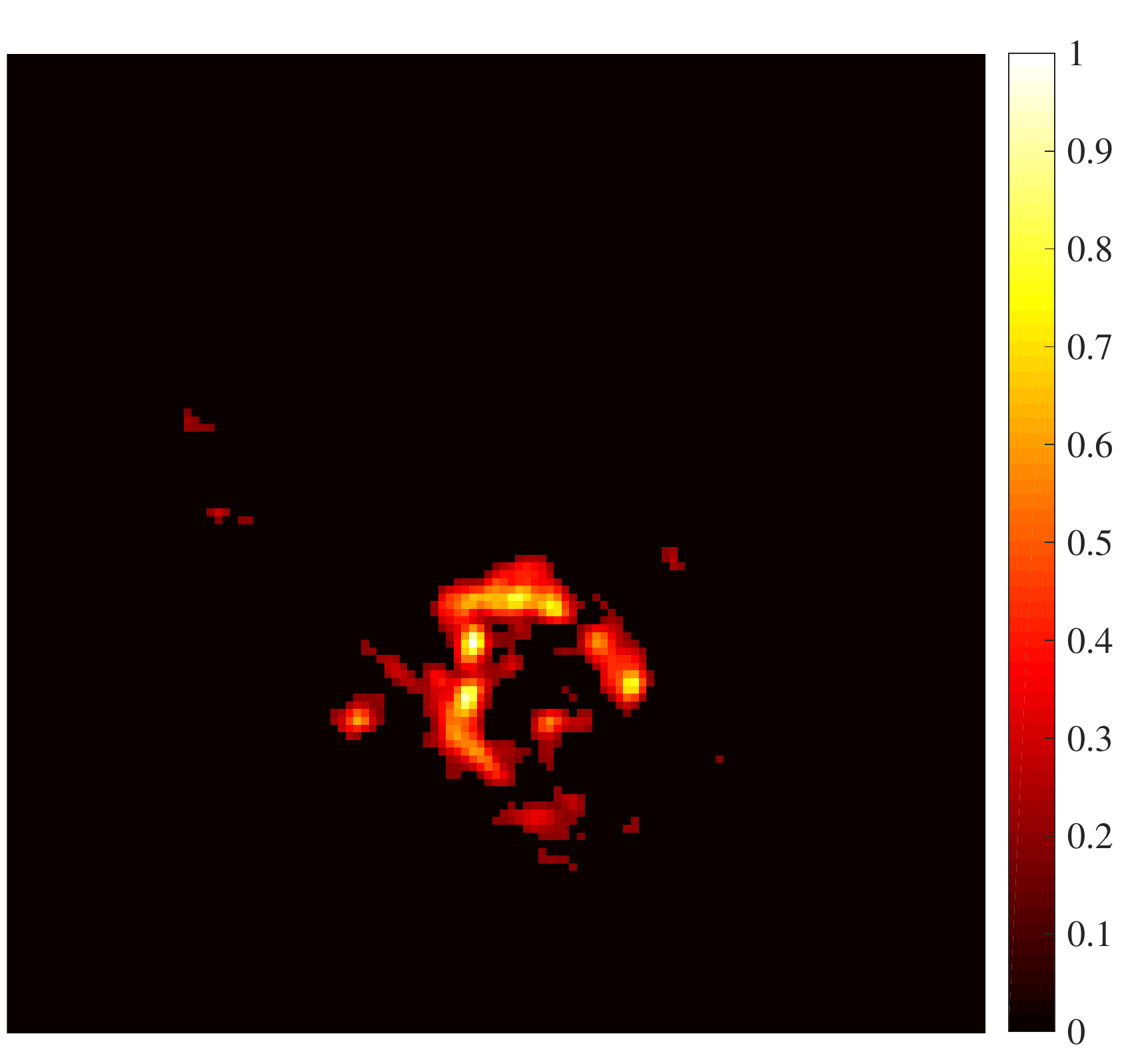}
&	\includegraphics[width=4.3cm]{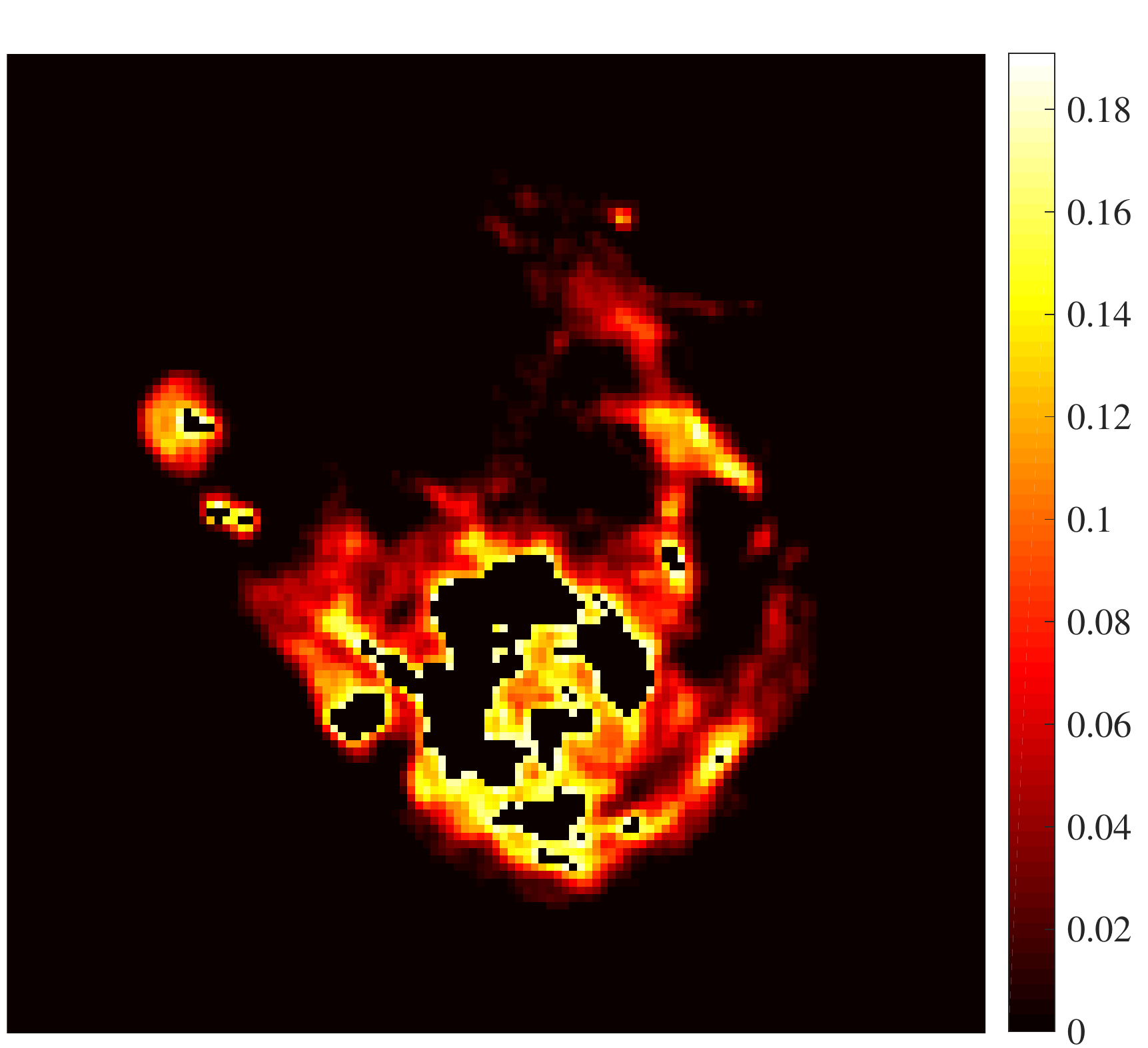}		\\
\hspace*{-0.3cm}
	\includegraphics[width=4.3cm]{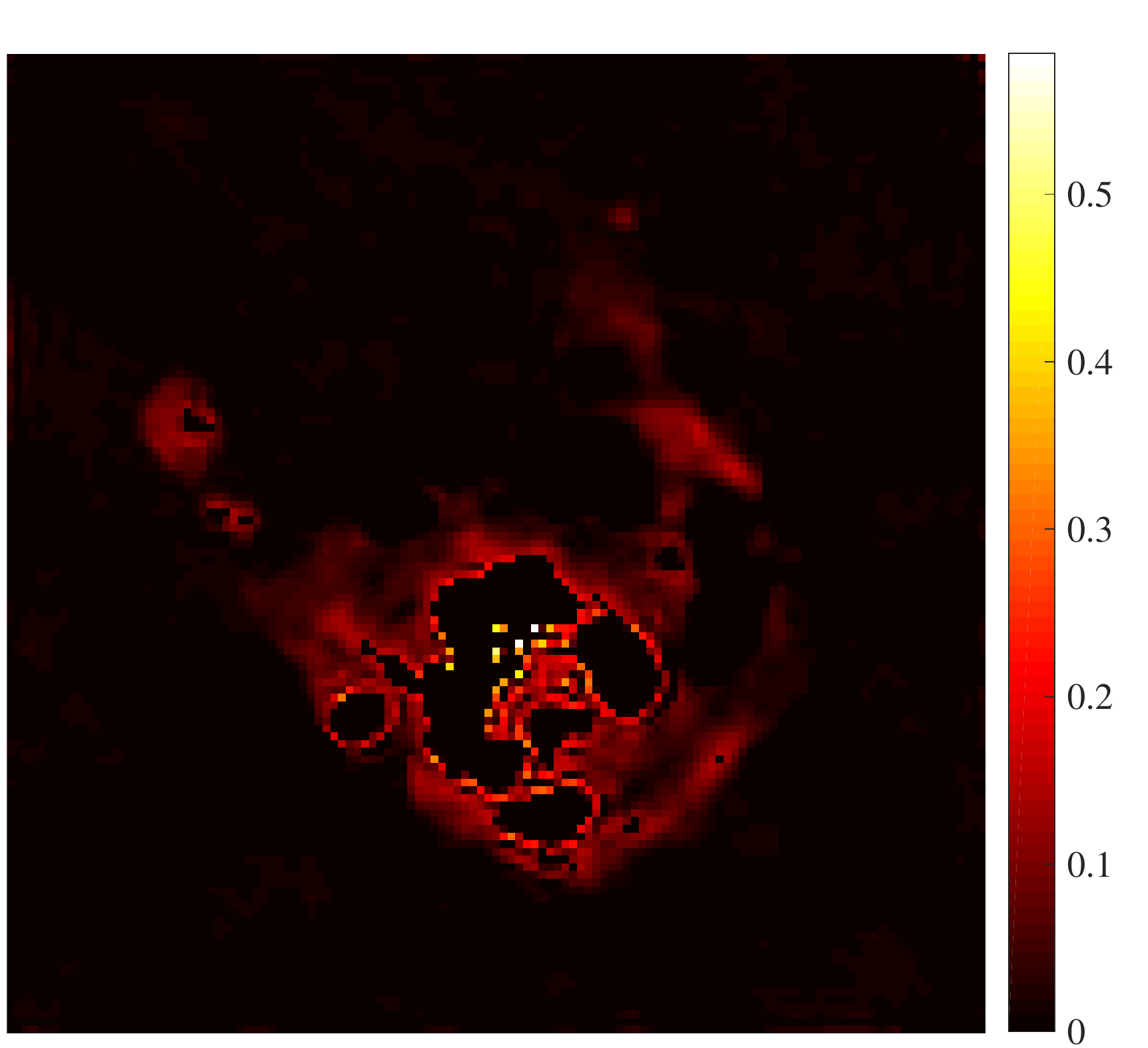}
&	\includegraphics[width=4.3cm]{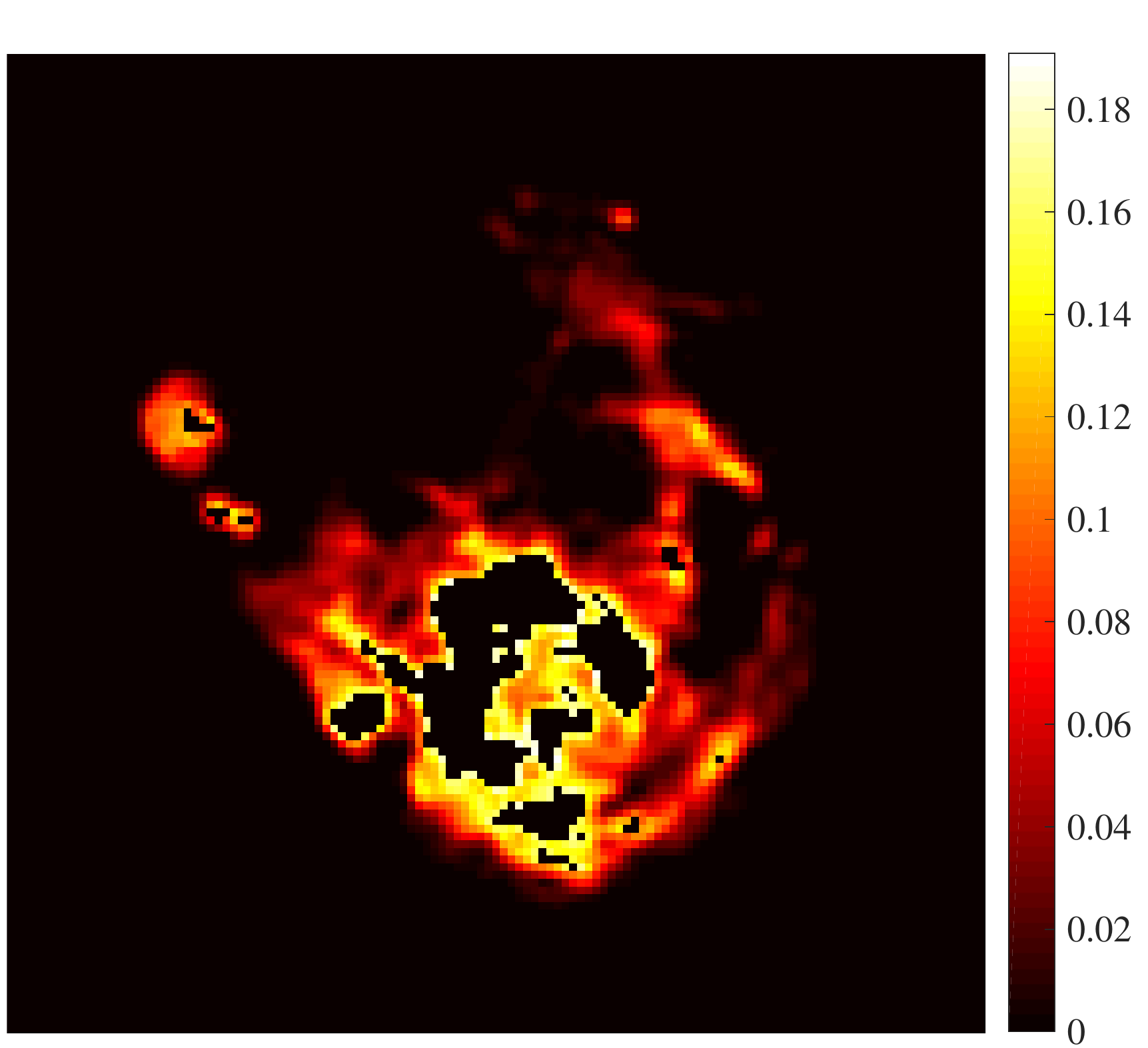}	\\
\hspace*{-0.3cm}
	\includegraphics[width=4.3cm]{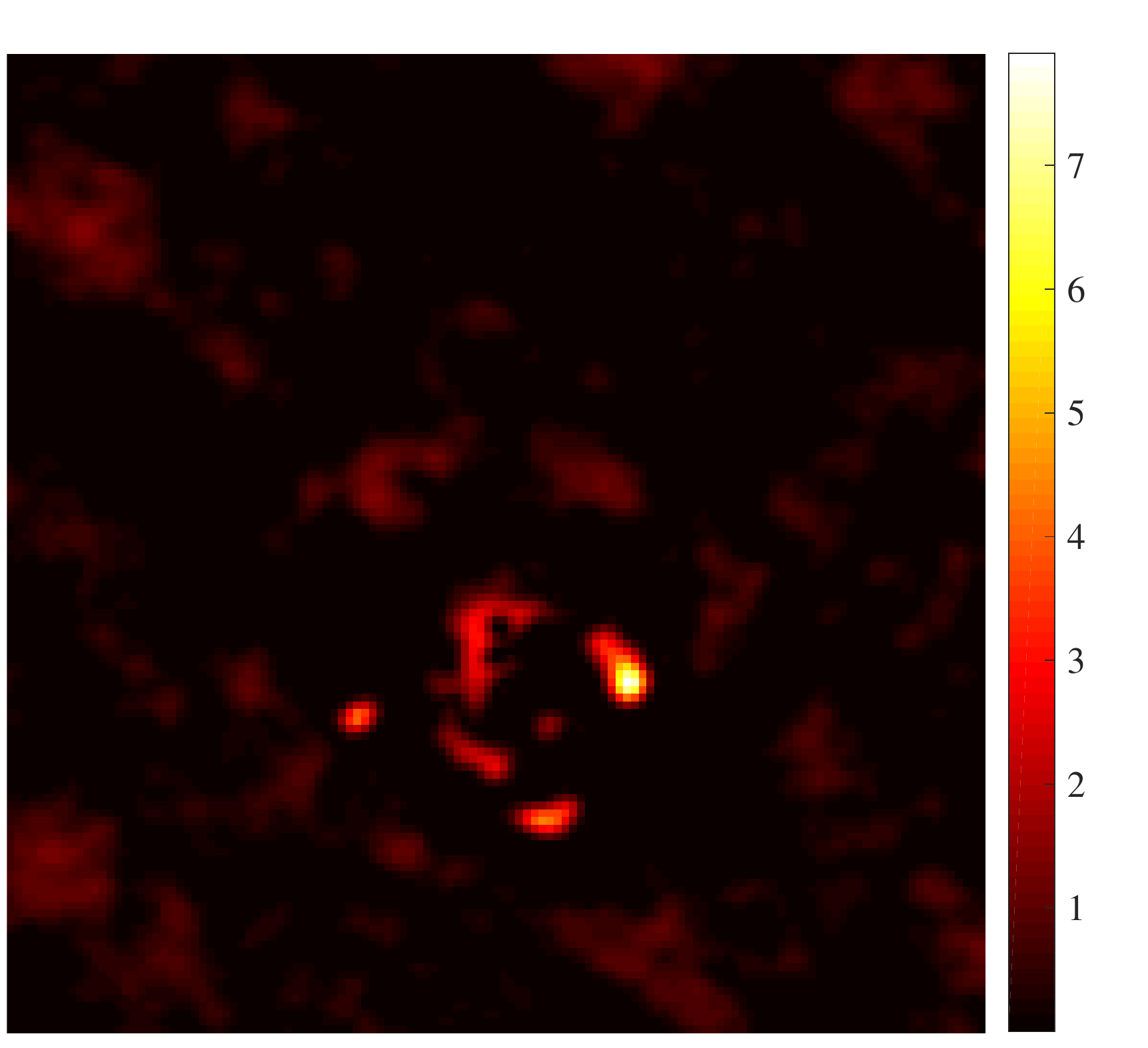}	
&	\includegraphics[width=4.3cm]{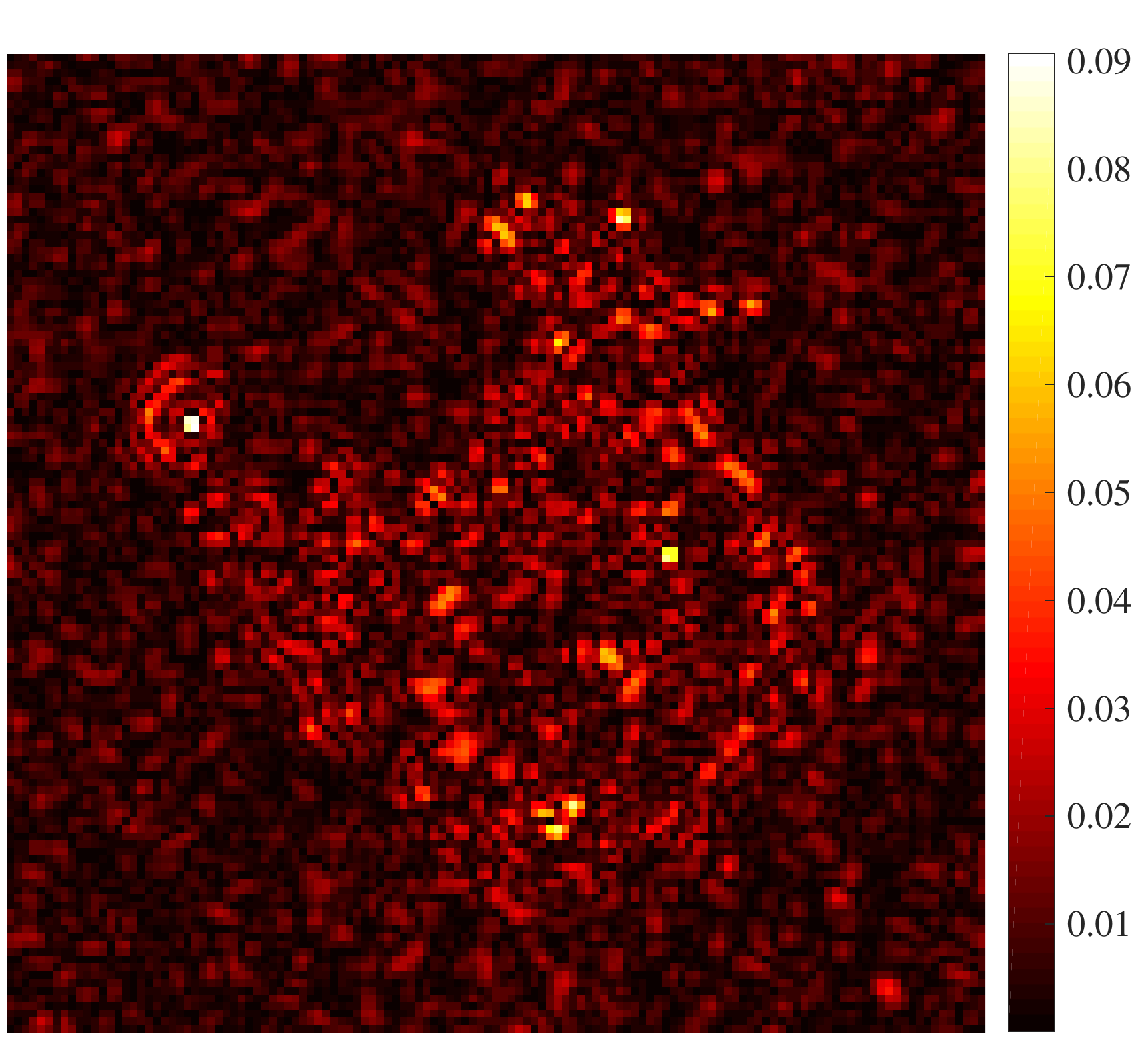}	\\
\hspace*{-0.3cm}
	\includegraphics[width=4.3cm]{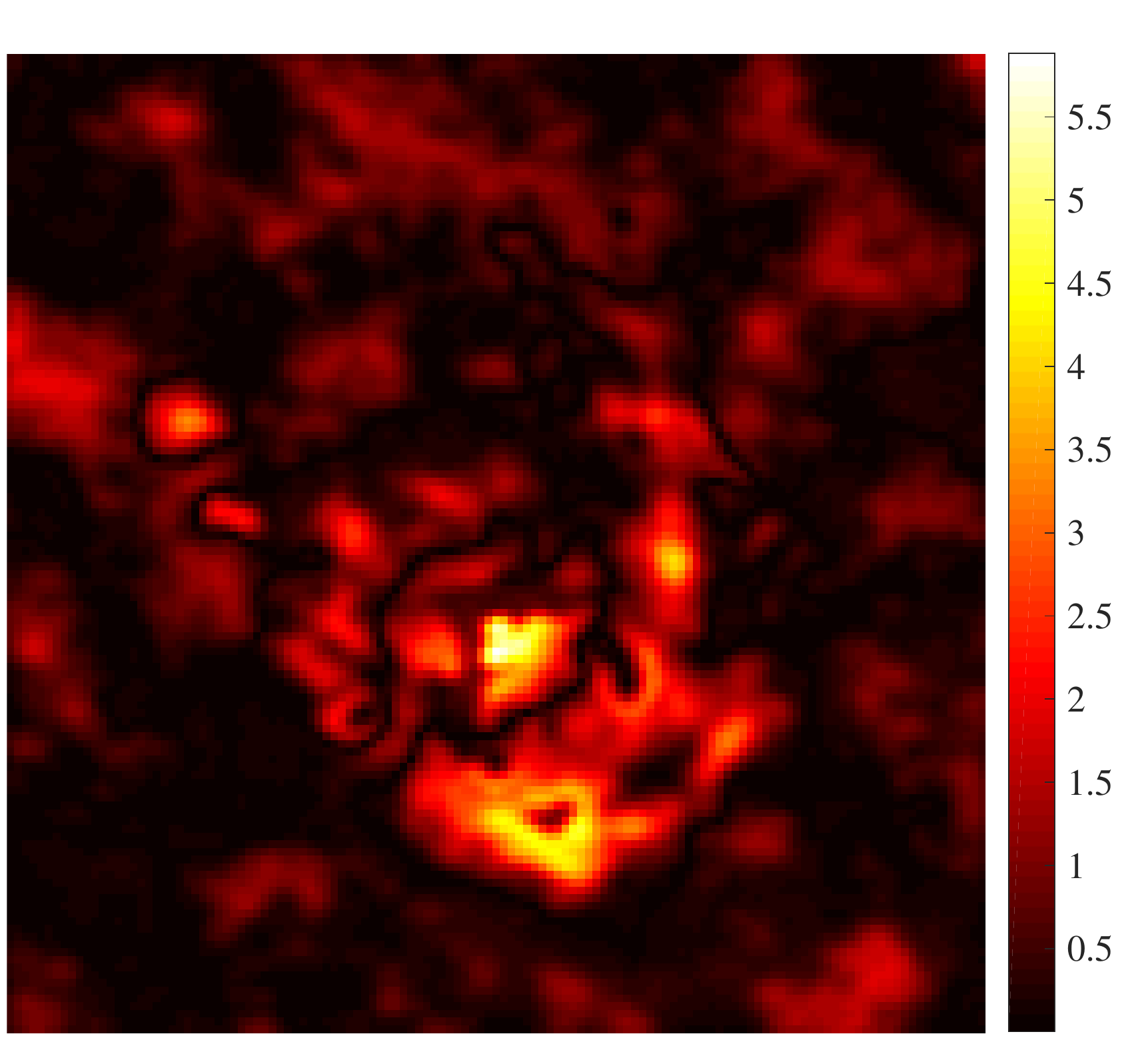}	
&	\includegraphics[width=4.3cm]{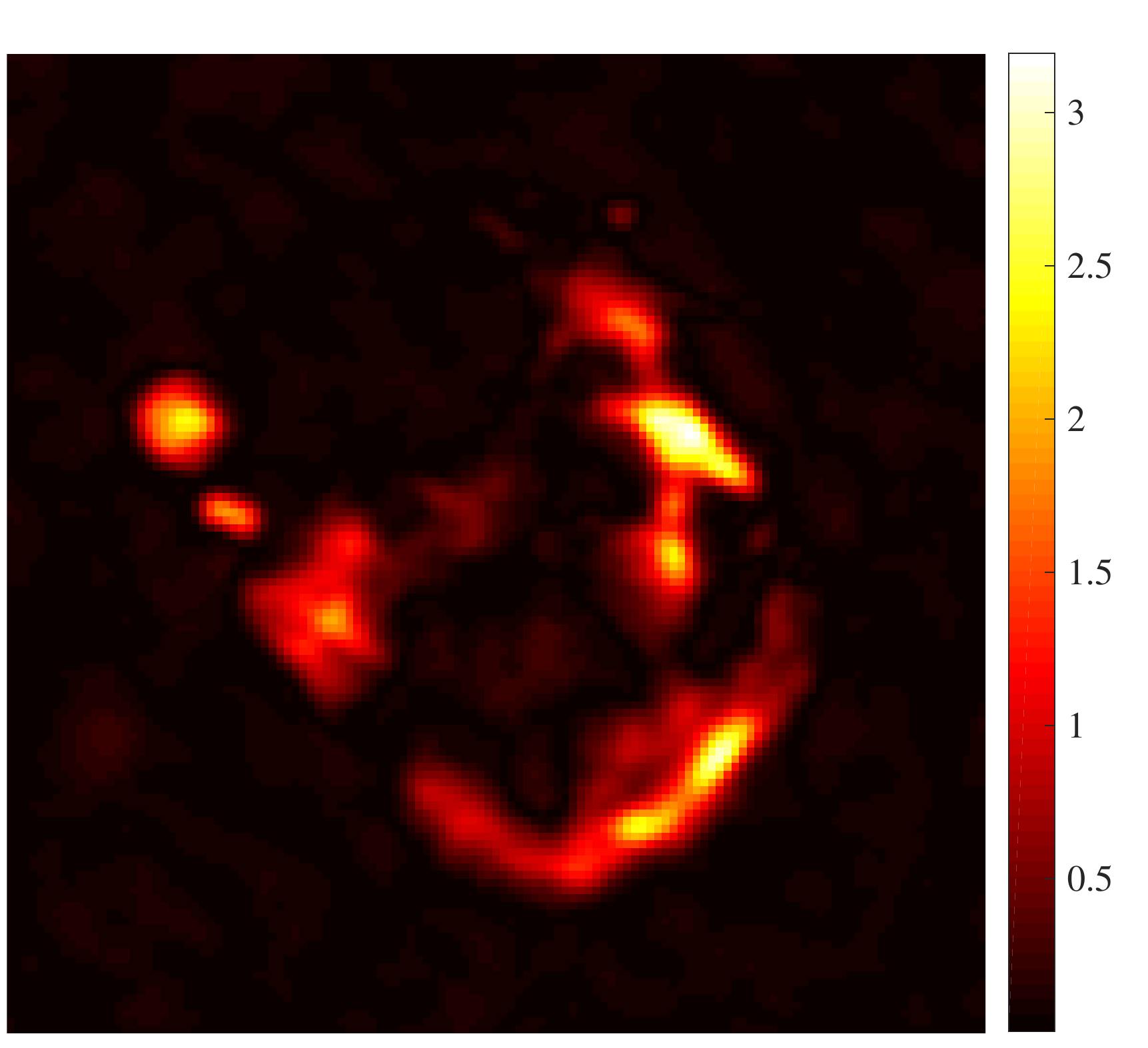}	
\end{tabular}
\vspace*{-0.2cm}
\caption{\label{Fig:testM31:images_delta05}
Images corresponding to simulations performed in Section~\ref{Ssec:sim:tests2} with $\kappa = 0.5$. 
The first row corresponds to (left) the known bright sources $\xb_o$ of the original image $\overline{\xb}$ and (right) the unknown image $\overline{\epsilonb}$. 
The second row corresponds to the estimates $\epsilonb^\star$ of $\overline{\epsilonb}$ obtained using (left) StEFCal-FB solving for DIEs, and (right) the proposed algorithm estimating the full DDEs. 
The third row shows the residual images considering the estimated $\Gbs^\star$ and $\epsilonb^\star$ obtained using (left) StEFCal-FB solving for DIEs, and (right) the proposed algorithm estimating the full DDEs. 
The fourth row corresponds to the residual images considering the true DDEs and the estimated image $\epsilonb^\star$ obtained using (left) StEFCal-FB solving for DIEs, and (right) the proposed algorithm estimating the full DDEs. 
}
\end{figure}

\begin{figure}
\begin{tabular}{c@{}c@{}}
\hspace*{-0.3cm}
	\includegraphics[width=4.3cm]{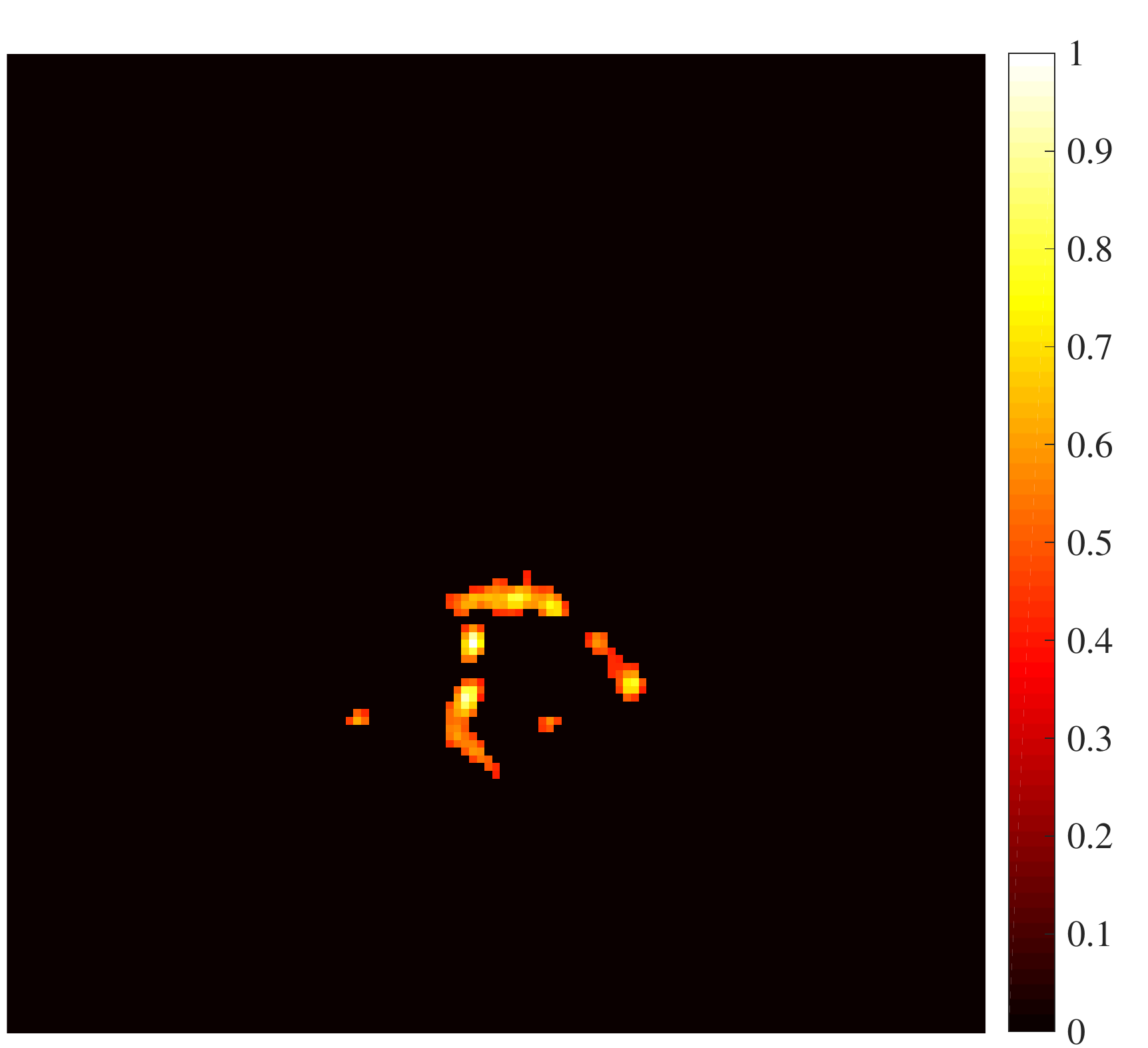}
&	\includegraphics[width=4.3cm]{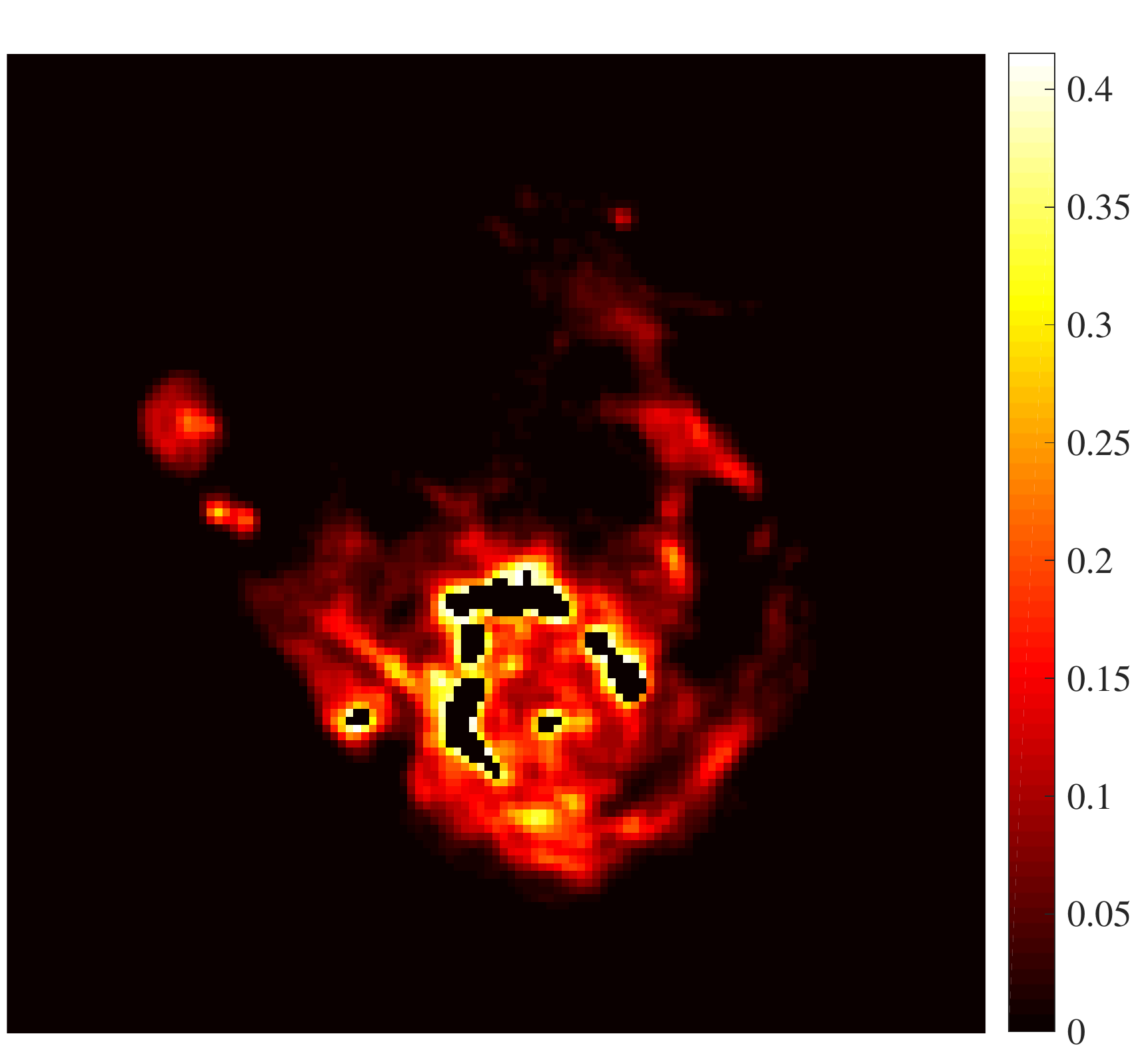}		\\
\hspace*{-0.3cm}
	\includegraphics[width=4.3cm]{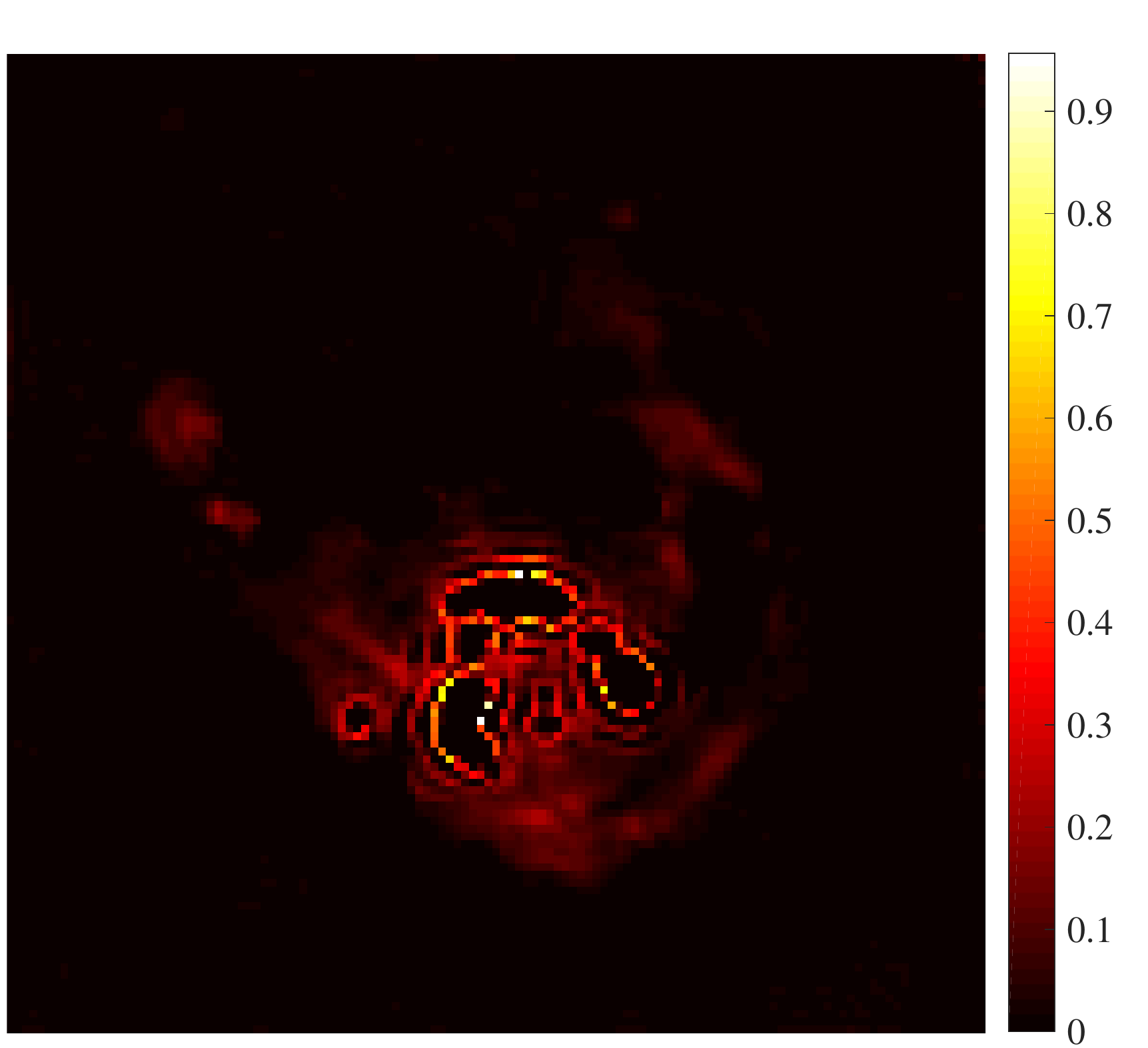}
&	\includegraphics[width=4.3cm]{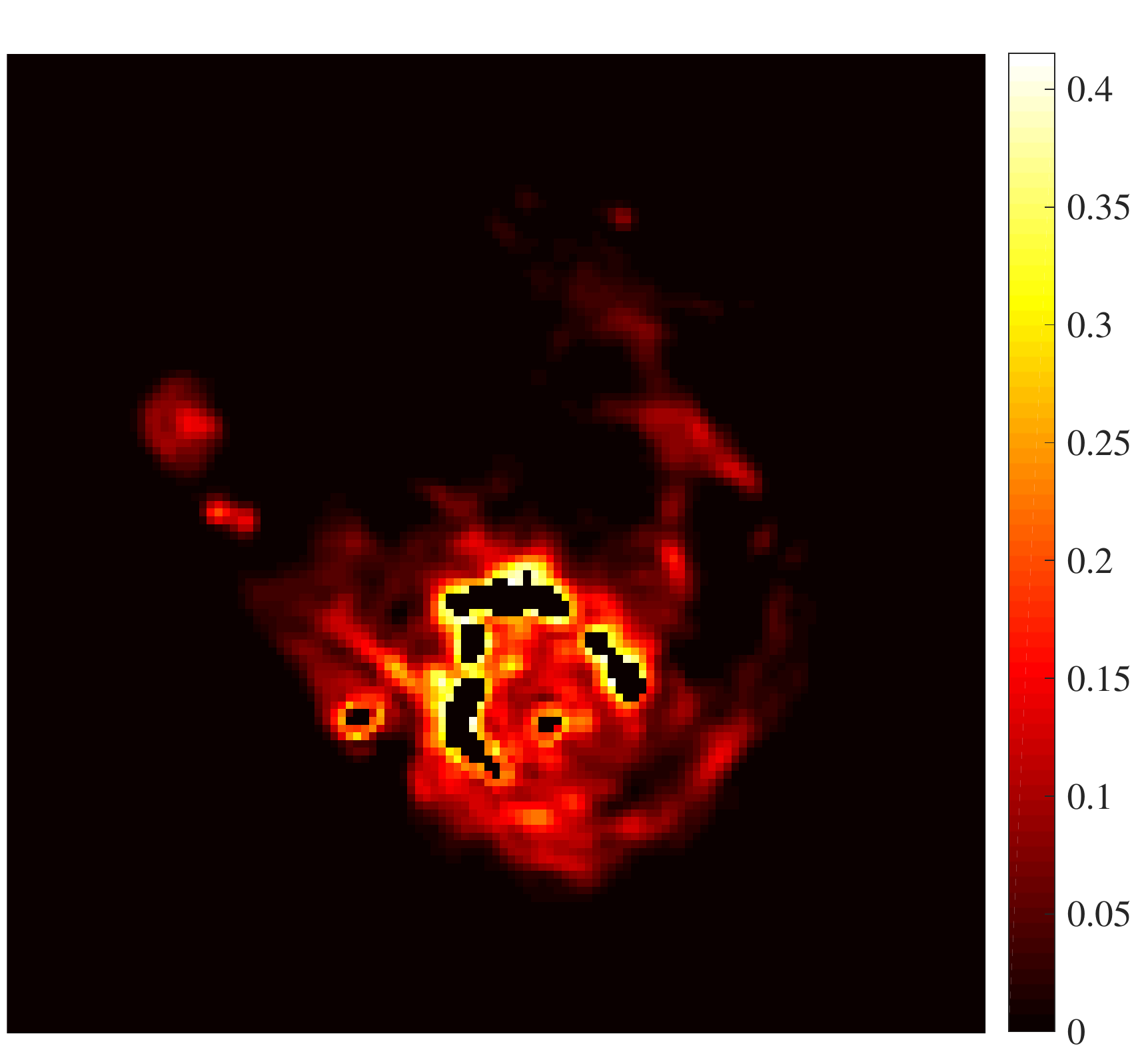}	\\
\hspace*{-0.3cm}
	\includegraphics[width=4.3cm]{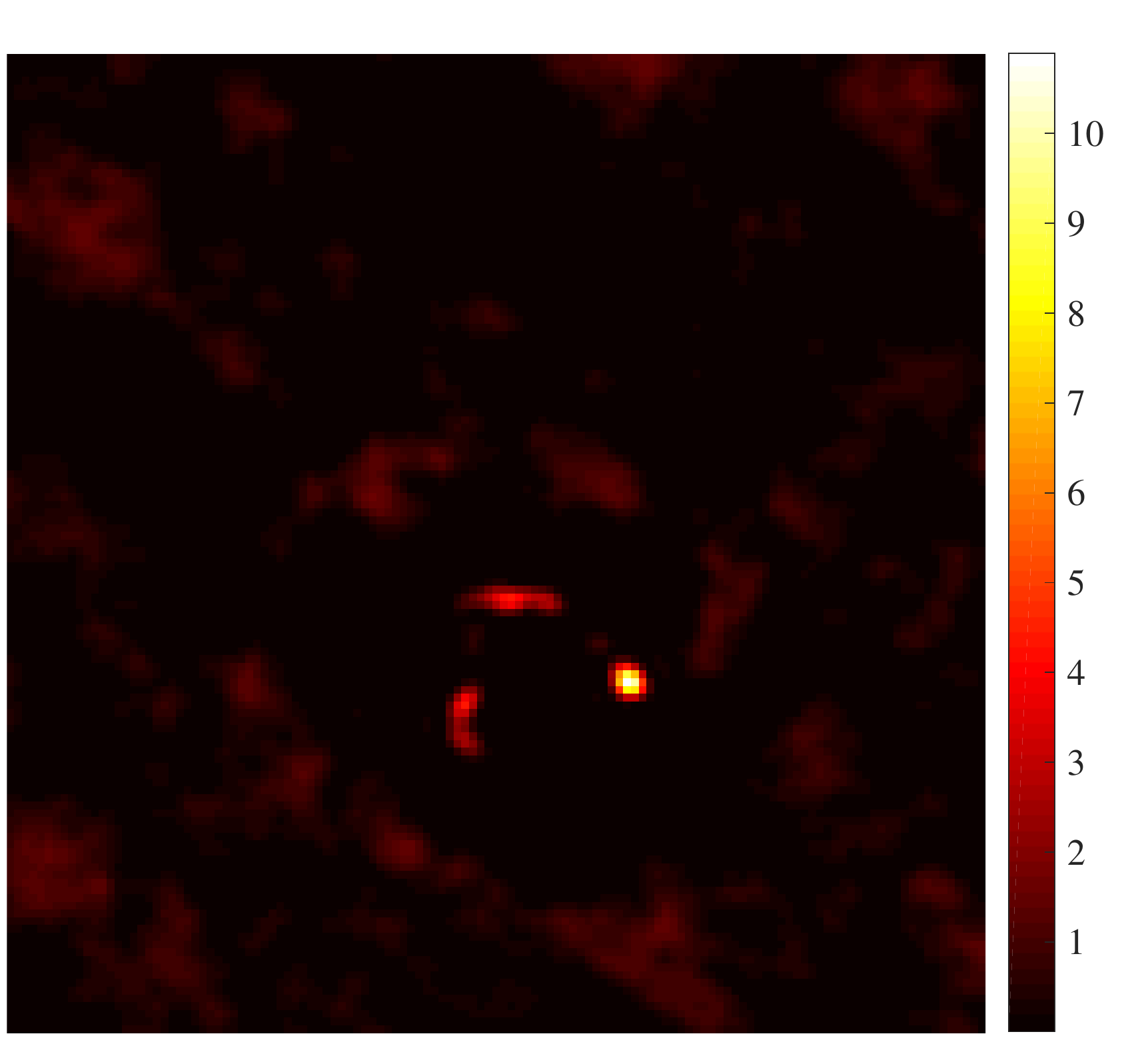}	
&	\includegraphics[width=4.3cm]{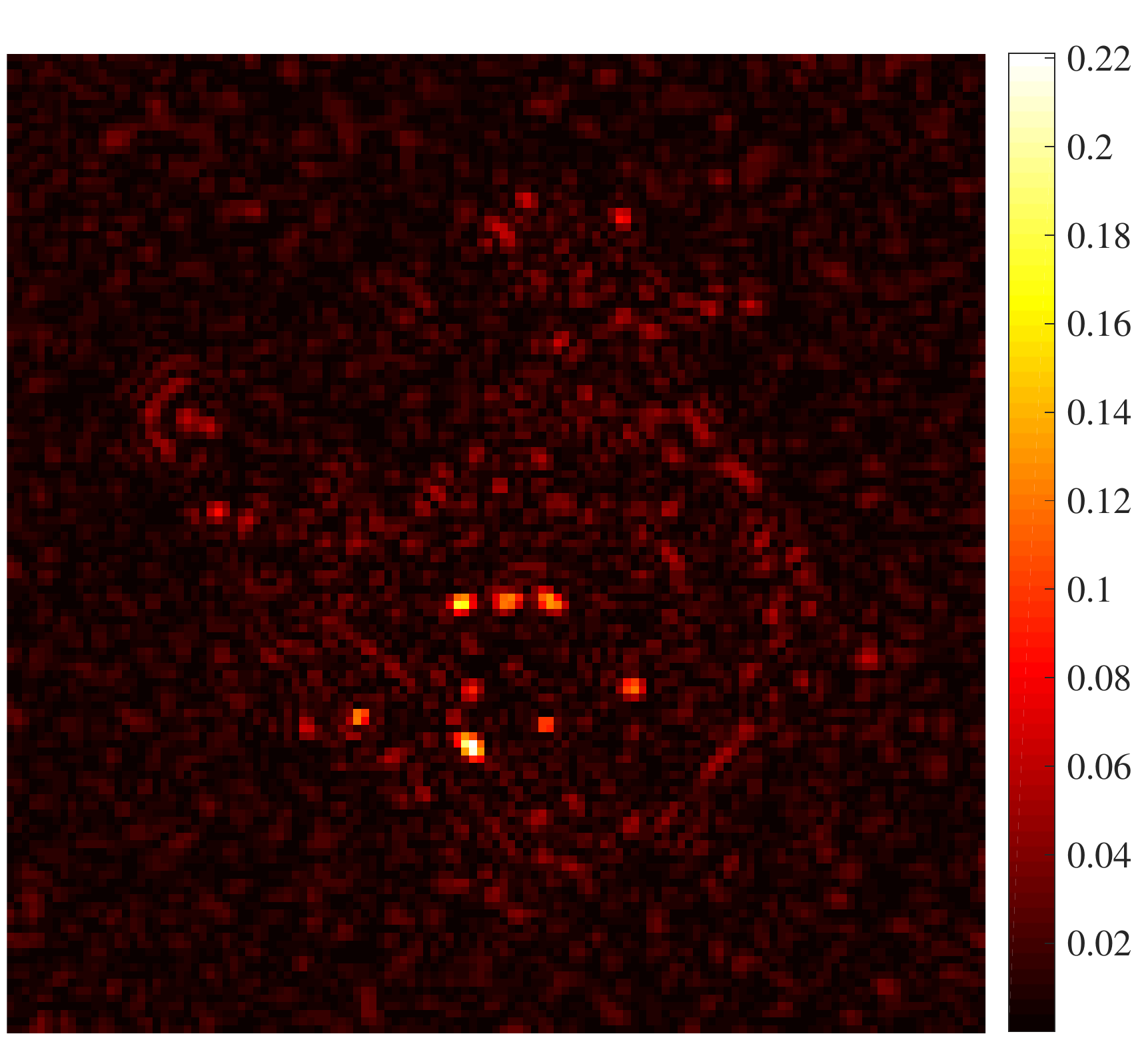}	\\
\hspace*{-0.3cm}
	\includegraphics[width=4.3cm]{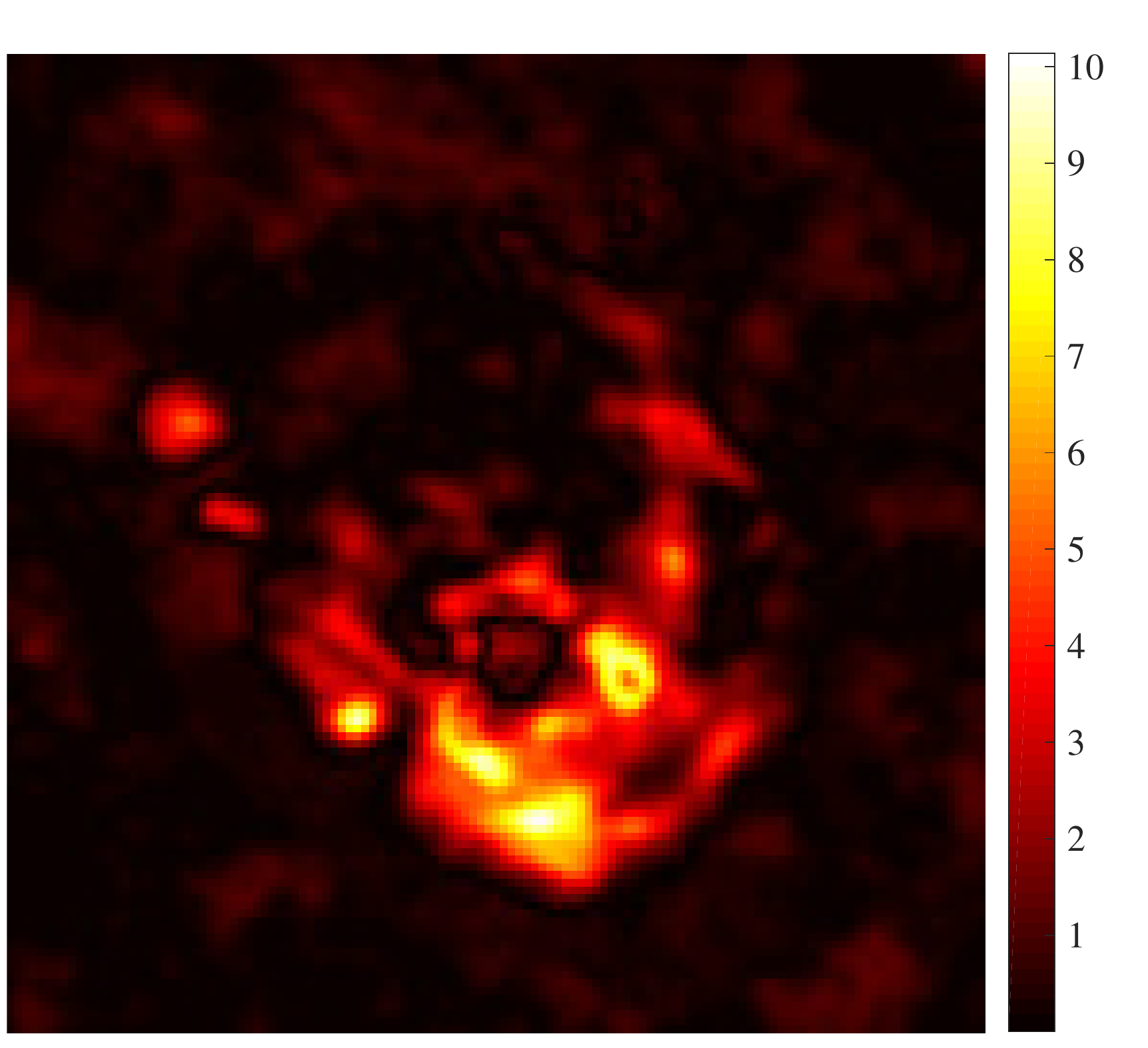}	
&	\includegraphics[width=4.3cm]{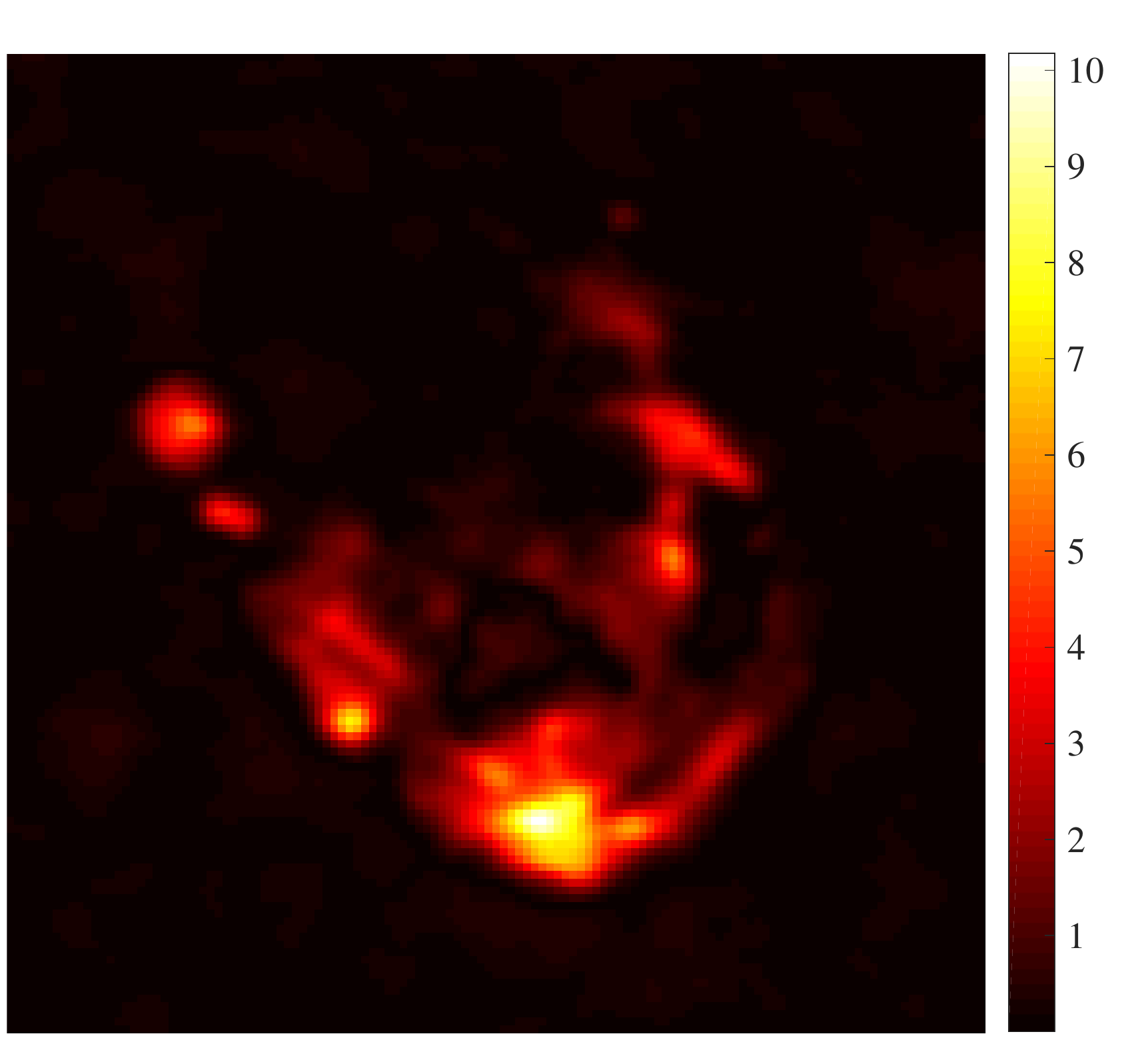}	
\end{tabular}
\vspace*{-0.2cm}
\caption{\label{Fig:testM31:images_delta1}
Images corresponding to simulations performed in Section~\ref{Ssec:sim:tests2} with $\kappa = 1$. 
The first row corresponds to (left) the known bright sources $\xb_o$ of the original image $\overline{\xb}$ and (right) the unknown image $\overline{\epsilonb}$. 
The second row corresponds to the estimates $\epsilonb^\star$ of $\overline{\epsilonb}$ obtained using (left) StEFCal-FB solving for DIEs, and (right) the proposed algorithm estimating the full DDEs. 
The third row shows the residual images considering the estimated $\Gbs^\star$ and $\epsilonb^\star$ obtained using (left) StEFCal-FB solving for DIEs, and (right) the proposed algorithm estimating the full DDEs. 
The fourth row corresponds to the residual images considering the true DDEs and the estimated image $\epsilonb^\star$ obtained using (left) StEFCal-FB solving for DIEs, and (right) the proposed algorithm estimating the full DDEs. 
}
\end{figure}

The results obtained considering $\kappa \in \{0.1, 0.5, 1\}$ are displayed in Figures~\ref{Fig:testM31:images_delta01}, \ref{Fig:testM31:images_delta05}, and \ref{Fig:testM31:images_delta1}, respectively. 
For each of these figures, the first row shows in the left image the known approximation $\xb_o$, and in the right image the unknown background $\overline{\epsilonb}$ to be estimated. 
The second row shows the estimated image $\epsilonb^\star$ of $\overline{\epsilonb}$, obtained (left) using the StEFCal-FB method solving only for DIEs, and (right) after the complete estimation of $\overline{\epsilonb}$ and the DDEs using the proposed method. 
Note that, as mentioned before, the StEFCal-FB methods gives similar reconstruction results as those obtained after the initialization of our method considering only the zero spatial frequency coefficients of the DDEs described in Section~\ref{Ssec:Algo:Init} (\textbf{Step 1}).
The SNR between the true $\overline{\epsilonb}$ and its estimate $\epsilonb^\star$ obtained using StEFCal-FB is equal to $0.65$ dB (resp. $5.52$ dB and $3.64$ dB) for $\kappa = 0.1$ (resp. $\kappa = 0.5$ and $\kappa = 1$). 
Similarly, the estimate $\epsilonb^\star$ obtained after the complete estimation of the DDEs using our method has an SNR equal to $16.83$ dB (resp. $17.47$ dB and $13.54$ dB)  for $\kappa = 0.1$ (resp. $\kappa = 0.5$ and $\kappa = 1$).
These results suggest that our method is efficient not only for the reconstruction of point sources, but also to estimate the background of extended sources. 
Similarly to the simulations presented in Section~\ref{Ssec:sim:test1-1}, the accuracy of the reconstruction for extended sources depends on the total energy of the known approximation $\xb_o$ with respect to the total energy of the unknown image $\overline{\epsilonb}$. As observed previously, if $\Es( \overline{\epsilonb} ) \ll \Es( \xb_o )$, our method reconstructs accurately the unknown sources. However, in the case when $ \Es( \overline{\epsilonb} ) $ is of the same order as $ \Es( \xb_o)$, the reconstruction is more difficult. Nonetheless, in the worst case considered when $\Es( \overline{\epsilonb} ) = \Es( \xb_o)$ (i.e. $\kappa = 1$) presented in Figure~\ref{Fig:testM31:images_delta1}, visually it can be observed that our method gives a good estimate $\epsilon^\star$ of $\overline{\epsilonb}$.
The two last rows of  Figures~\ref{Fig:testM31:images_delta01}, \ref{Fig:testM31:images_delta05} and \ref{Fig:testM31:images_delta1} are dedicated to the residual images obtained considering $\kappa \in \{0.1, 0.5, 1\}$. 
More precisely, the third rows show the residual images $| \Fbs^\dagger \Gbs^{\star \, \dagger} \big( \Gbs^{\star} \Fbs \xb^\star - \yb \big) |$ considering the estimated $\Gbs^{\star}$ and $ \xb^\star$ obtained (left) using the StEFCal-FB method solving only for DIEs, and (right) after the complete estimation of the image and the DDEs using our method. 
Similarly, the last rows show the residual images $ | \Fbs^\dagger \overline{\Gbs}^{ \dagger} \big( \overline{\Gbs} \Fbs \xb^\star - \yb \big) |$ where $\overline{\Gbs}$ corresponds to the true unknown DDEs. 
As expected, it can be observed that estimating the full DDEs with respect to estimating only the DIEs leads to residual images with smaller amplitudes for all the cases presented. 
Moreover, for the final results with DDEs estimation, the residual images have smaller amplitude considering the estimated DDEs (third rows) than the original DDEs (fourth rows). This observation suggests that there may be ambiguity errors between the image and the direction-dependent Fourier kernels which need to be corrected.

\section{Conclusions}
\label{Sec:Conc}

We propose a non-convex optimization algorithm to jointly calibrate DDEs and estimate the sky intensity. 
The algorithm is based on a block-coordinate forward-backward approach,  making use of suitable priors on both the image and the DDEs. 
More precisely, we use an $\ell_1$ based regularization term to promote sparsity on the image and the bright sources are assumed to be known, while we model the DDEs as smooth functions of the sky, i.e. spatially band-limited. 
Our method presents several advantages. 
Firstly, it benefits from convergence guarantees for both the image and the DDEs. 
Secondly, in contrast with DDE calibration methods developed recently, our method does not require a selection of calibrator directions since it constructs a smooth DDE screen applied to all the sources across the image. 
Finally, the proposed method is very general and can be easily adapted to the nature of the considered image. Indeed, we present two cases: the first case corresponds to images with point sources where the regularization term is chosen to promote sparsity of the image itself, while the second case deals with sophisticated extended sources where sparsity is promoted in a given dictionary. 
It is important to emphasize as well that, in the proposed algorithm, the DDEs are updated in parallel for all the antennas and at all time instants. 

We study the performance of the proposed method considering a large class of simulations, varying the images, their dynamic range, the size and the variations of the direction-dependent Fourier kernels, the antennas distribution and taking into account the time dependency. 
In all the presented simulations, we show that our method leads to better reconstruction quality than obtained by only estimating the DIEs. 
Moreover, our simulations suggest that using our method to jointly estimate the DDEs and the image result in significant improvement of the the dynamic range, which is orders of magnitude higher when compared to accounting for DIEs only.
Remark that all the parameters of the proposed method are fixed for all the different simulations presented in this paper. Therefore, it is worth noting that our method leads to very good reconstructions using a unique general framework converging in a complete automatic manner. 
The MATLAB code of the proposed method is available on GitHub (\texttt{https://basp-group.github.io/joint-img-dde-calibration/}).

Even though the proposed method is very promising, we intend to improve some points in future works. 
Firstly, in the presented simulations the bright sources in the images are assumed to be known exactly. Thus, we intend to extend our method to be able to take into account possible errors in the amplitudes and/or the positions of the bright sources. 
Secondly, we plan to investigate further regularization terms in order to reduce the ambiguity error appearing between the DDEs and the image. 
Moreover we aim to perform simulations on realistic data simulations, incorporating the gridding steps in the algorithm to use the non-uniform FFT, considering unknown size support for the direction-dependent Fourier kernels, and modelling time dependence in order to investigate the robustness of the proposed method. 
Finally, we plan to compare our approach to other state of the art algorithms.

\section*{Acknowledgements}

This work was supported by the UK Engineering and Physical Sciences Research Council (EPSRC, grant EP/M008843/1). 
We would like to thank Oleg Smirnov for insightful discussions.


\appendix

\section{Tables of notations}
\label{App:notations}

In this paper, we use italic bold letters to denote vectors, italic letters to denote scalar variables, and capital bold letters for matrices. 
For a given variable, e.g. $\xb$, $x(n)$ denotes its $n$-th coefficient, $\widehat{\xb}$ denotes its Fourier transform, $\overline{\xb}$ denotes the original unknown associated variable, and $\xb^\star$ denotes the estimated associated variable.

\begin{table*}
\begin{center}
\begin{tabular}{| l | l | l |}
\hline
$N$	&	Image dimension	&	\hspace*{0.5cm}\\
$n_a$	&	Number of antennas	\\
$M = n_a(n_a-1)/2$	& Number of measurements at each time instant	\\
$T$	&	Number of time instants	\\
\hline
$\overline{\xb} \in \eR^N$	&	Original unknown sky intensity image	\\
$\xb_o \in \eR^N$	&	Image containing the known brightest sources of $\overline{\xb}$	\\
$\overline{\epsilonb} \in \eR^N$	&	Image containing the unknown sources of $\overline{\xb}$	\\
$\Fbs \in \eC^{N \times N}$	&	Fourier matrix	\\
$\overline{\db}_{t,\alpha}\in \eC^{N}$ 	&	Original unknown DDE related to antenna $\alpha$ at time instant $t$	\\
$\overline{\Gbs} \in \eC^{TM \times N}$	&	Convolution matrix containing the antenna-based gains	\\
$\yb \in \eC^{TM}$	&	Complex visibility vector	\\
$y_{t,\alpha,\beta} \in \eC$	&	Element of the vector $\yb$ associated with the antenna pair $(\alpha, \beta)$ at time instant $t$	\\
$\Ybs_t \in \eC^{n_a \times n_a}$	&	Complex visibility matrix at instant $t$	\\
$\Ys_{t, (\alpha,\beta)} \in \eC$	&	Element of the matrix $\Ybs_t$ associated with the antenna pair $(\alpha,\beta)$ at time instant $t$	\\
$\Ybs_{t, \alpha} \in \eC^{n_a-1}$	&	Column of $\Ybs_t$ without its $\alpha$-th element, i.e. $\Ys_{t, \alpha} = \big( \Ybs_{t, (\alpha,\beta)} \big)_{\underset{\beta \neq \alpha}{\beta \in \{1,\ldots,n_a\}}}$	\\
$\overline{\Dbs}_{t,1} \in \eC^{n_a \times N}$	&	Matrix containing on each row the reordered Fourier transforms of the DDEs at time instant $t$	\\
$\overline{\Dbs}_{t,2} \in \eC^{n_a \times N}$	&	Matrix containing on each row the complex conjugates of the Fourier transforms of the DDEs at time instant $t$	\\
$\overline{\Xbs} \in \eC^{N \times N}$	&	Convolution matrix containing the Fourier transform of the original image	\\
\hline
\end{tabular}
\end{center}
\caption{\label{Tab:not:problem}
Notations used to define the radio interferometric inverse problem.}
\end{table*}

Table~\ref{Tab:not:problem} gives the notations used for the main variables describing the radio interferometric inverse problem, 
Table~\ref{Tab:not:min} gives the notations used for the imaging and calibration minimization problem, 
and Table~\ref{Tab:not:algo} gives the notations used for the proposed algorithm.

\begin{table*}
\begin{center}
\begin{tabular}{| l | l | l |}
\hline
$\eS$ 	&	Support of the Fourier transform of the DDEs		\\
$S$		&	Cardinal of $\eS$, i.e. number of non-zero Fourier coefficients of each DDE	\\
$\eD$	&	Constraint set for the DDEs defined in eq.~\eqref{eq:set_D}	\\
$(\vartheta_1, \vartheta_2) \in ]0, +\infty[^2$	&	Bounds on the amplitude of the Fourier coefficients of the DDEs defined in eq~\eqref{ass:ddes:ii:1}-\eqref{ass:ddes:ii:2}	\\
$\eE$	&	Constraint set for the image defined in eq.~\eqref{eq:def_C}	\\
\hline
$\overline{\Ubs}_{t,1} \in \eC^{n_a \times S}$	&	Matrix containing on each row the non-zero reordered Fourier coefficients of the DDEs at time instant $t$	\\
$\overline{\Ubs}_{t,2} \in \eC^{n_a \times S}$	&	Matrix containing on each row the non-zero reordered Fourier coefficients of the DDEs at time instant $t$	\\
$\overline{\Ubs}_1 \in \eC^{Tn_a \times S}$	&	Concatenation of matrices $(\overline{\Ubs}_{t,1})_{1 \le t \le T}$	\\
$\overline{\Ubs}_2 \in \eC^{Tn_a \times S}$	&	Concatenation of matrices $(\overline{\Ubs}_{t,2})_{1 \le t \le T}$	\\
$\overline{\ub}_{t,\alpha,1} \in \eC^S$	&	Row $\alpha$ of matrix $\overline{\Ubs}_{t,1}$	\\
$\overline{\ub}_{t,\alpha,2} \in \eC^S$	&	Row $\alpha$ of matrix $\overline{\Ubs}_{t,2}$	\\
$\Hbs_{t,\alpha,1} \in \eC^{(n_a-1) \times S}$	&	Convolution matrix defined in eq.~\eqref{eq:data_fid_d_C1}	\\
$\Hbs_{t,\alpha,2} \in \eC^{(n_a-1) \times S}$	&	Convolution matrix defined in eq.~\eqref{eq:data_fid_d_C2}	\\
$\Psib \in \eR^{N \times D}$	&	Sparsity basis used for the image estimation	\\
\hline
$\Gc \colon \eC^{Tn_a \times S} \times \eC^{Tn_a \times S} \to \eC^{TM \times N}$	&	Function defined such that $\Gc(\overline{\Ubs}_1, \overline{\Ubs}_2) = \overline{\Gbs}$	\\
$\Dc_{t,1} \colon \eC^{n_a \times S} \to \eC^{n_a \times N}$	&	Function defined such that $\Dc_{t,1}( \overline{\Ubs}_{t,1} ) = \overline{\Dbs}_{t,1} $	\\
$\Dc_{t,2} \colon \eC^{n_a \times S} \to \eC^{n_a \times N}$	&	Function defined such that $\Dc_{t,2}( \overline{\Ubs}_{t,2} ) = \overline{\Dbs}_{t,2} $	\\
$\Dc_{t,\alpha,1} \colon \eC^{n_a \times S} \to \eC^{(n_a-1) \times N}$	&	Selection operator to build matrix $\Hbs_{t,\alpha,1}$ in eq.~\eqref{eq:data_fid_d_C1}	\\
$\Dc_{t,\alpha,2} \colon \eC^{n_a \times S} \to \eC^{(n_a-1) \times N}$	&	Selection operator to build matrix $\Hbs_{t,\alpha,2}$ in eq.~\eqref{eq:data_fid_d_C2}	\\
$\Lc_{t,\alpha,1} \colon \eC^{N \times N} \to \eC^{N \times S}$	&	Selection operator to build matrix $\Hbs_{t,\alpha,1}$ in eq.~\eqref{eq:data_fid_d_C1}	\\
$\Lc_{t,\alpha,2} \colon \eC^{N \times N} \to \eC^{N \times S}$	&	Selection operator to build matrix $\Hbs_{t,\alpha,2}$ in eq.~\eqref{eq:data_fid_d_C2}	\\
$\Xc_{t,\alpha,1} \colon \eC^{N} \to \eC^{N \times S}$	&	Function defined such that $\Xc_{t,\alpha,1}(\Fbs \overline{\xb}) = \Lc_{t,\alpha,1}(\overline{\Xbs})$	\\
$\Xc_{t,\alpha,2} \colon \eC^{N} \to \eC^{N \times S}$	&	Function defined such that $\Xc_{t,\alpha,2}(\Fbs \overline{\xb}) = \Lc_{t,\alpha,2}(\overline{\Xbs})$	\\
$\sigma \colon \eC^S \to \eC^S$	&	Bijection defined such that $\overline{\ub}_{t,\alpha,2} = \sigma(\overline{\ub}_{t,\alpha,1})$	\\
$h\colon \eR^N \times \eC^{Tn_a \times S} \times \eC^{Tn_a \times S} \to \eR$	&	Data-fidelity term defined in eq.~\eqref{eq:h_x_final}, \eqref{eq:h_u1_final} and \eqref{eq:h_u2_final}	\\
$r \colon \eR^N \to ]-\infty,+\infty] $	&	Regularisation term for the image defined in eq.~\eqref{eq:def_reg_x_sparse} or \eqref{eq:def_reg_x_dict}	\\
$p \colon \eC^{Tn_a \times S} \times \eC^{Tn_a \times S} \to ]-\infty, + \infty]$	&	Regularisation term for the DDEs defined in eq.~\eqref{eq:reg_DD}	\\
\hline
\end{tabular}
\end{center}
\caption{\label{Tab:not:min}
Notations used to define the proposed minimization problem associated with the imaging and calibration problem.}
\end{table*}

\begin{table*}
\begin{center}	
\begin{tabular}{| l | l | l |}
\hline
$\eta \in ]0, +\infty[$ \hspace*{1.5cm}	&	Regularization parameter for the image		&	\hspace*{5cm}\\
$\nu \in ]0, +\infty[$	&	Regularization parameter for the DDEs	\\
$ i \in \eN $	&	Iteration number	\\
$J_{\epsilonb}^{(i)} \in \eN$	&	Number of sub-iterations to update the image at iteration $i$	\\
$J_{\Ubs_1}^{(i)} \in \eN$	&	Number of sub-iterations to update $\Ubs_1$ at iteration $i$	\\
$J_{\Ubs_2}^{(i)} \in \eN$	&	Number of sub-iterations to update $\Ubs_2$ at iteration $i$	\\
$J_{\text{cyc}} \in \eN $ ($J_{\text{cyc}} \ge 2$)	&	Number of iterations to perform a cycle of Algorithm~\ref{algo:global} 	\\
$\gamma_{\epsilonb}^{(i)} \in [0, +\infty[$	&	Step-size for the update of the image at iteration $i$	\\
$\gamma_{\ub_{t,\alpha,1}}^{(i)} \in [0, +\infty[$	&	Step-size for the update of $\ub_{t,\alpha,1}$ at iteration $i$	\\
$\gamma_{\ub_{t,\alpha,2}}^{(i)} \in [0, +\infty[$	&	Step-size for the update of $\ub_{t,\alpha,2}$ at iteration $i$	\\
$\xi_{\Ubs_{\text{tot}}} \in ]0,+\infty[$	&	Stopping criterion for the DDE estimation defined in eq.~\eqref{crit_stop:Utot}	\\
$\xi_{\epsilonb} \in ]0,+\infty[$	&	Stopping criterion for the image estimation defined in eq.~\eqref{crit_stop:im}	\\
$\zeta \in ]0,+\infty[$	&	Stopping criterion for the global algorithm defined in eq.~\eqref{crit_stop:obj}	\\
$(\underline{\tau}, \overline{\tau}) \in ]0, +\infty[^2$		&	Parameters used to determine the initialisation of the image	\\
\hline
\end{tabular}
\end{center}
\caption{\label{Tab:not:algo}
Notations used for Algorithm~\ref{algo:global}.}
\end{table*}



\bibliographystyle{mnras}
\bibliography{refs} 


\bsp	
\label{lastpage}
\end{document}